\documentclass{amsart}  
\usepackage{geometry}  
\usepackage{graphicx}
\usepackage{amssymb}
\title{Is it possible to discover a dark matter particle with an accelerator?}
\author{Vadim A. Bednyakov \\ {\em DLNP, JINR, Russia}}
\date{27.06.2015}  
\renewcommand{\baselinestretch}{1.2}

\begin{document} %%%%%%%%%%%%%%%%%%%%%%%%%%%%%
\begin{abstract}  \normalsize     
	The paper contains description of the main properties of the galactic dark matter (DM) particles, 
	available approaches for detection of DM, 
	main features of direct DM detection, 
 	ways to estimate prospects for the DM detection,
	the first collider search for a DM candidate within an Effective Field Theory, 
	complete review of ATLAS results of the DM candidate search with LHC RUN I,
	and less complete review of "exotic" dark particle searches with other accelerators and not only.

	From these considerations it follows that one is unable to prove, especially model-independently, 
	a discovery of a DM particle with an accelerator, or collider.
	One can only obtain evidence on existence of a weakly interacting neutral particle, which could be, 
	or could not be the DM candidate.  

	The current LHC DM search program uses only the missing transverse energy signature. 
	Non-observation of any excess above Standard Model expectations 
	forces the LHC experiments to enter into the same fighting for the best exclusion curve, 
	in which (almost) all direct and indirect DM search experiments permanently take place. 
	But this fighting has very little (almost nothing) to do with a real possibility of discovering a DM particle.
	The true DM particles possess an exclusive galactic signature --- annual modulation of a signal, 
	which is accessible today only for direct DM detection experiments. 
	There is no way for it with a collider, or accelerator.

	Therefore to prove the DM nature of a collider-discovered candidate one must find the candidate 
	in a direct DM experiment and demonstrate the galactic signature for the candidate. 
	Furthermore, being observed, the DM particle must be implemented into a modern theoretical framework.
	The best candidate is the supersymmetry, 
	which looks today inevitable for coherent interpretation of all available DM data.		
\end{abstract}
\maketitle
\def\met{$E^{\rm miss}_{\rm T}$}
\def\pt{$p^{ }_{\rm T}$}
\def\mt{$m^{ }_{\rm T}$}
\large 

%%%%%%%%%%%%%%%%%%%%%%%%%%%%%%%%%%%%%%%
\section{Dark matter particles are stable, neutral, with a clear galaxy feature}\label{sec:GalacticDM}
%\section{Dark matter particles are stable, neutral, with a clear galaxy feature}
	Galactic Dark Matter (DM) particles do not emit or reflect any detectable electromagnetic radiation 
	and clearly manifest themselves today only gravitationally by affecting other astrophysical objects.
	Numerous observational indications at astronomical and cosmological scales 
\cite{Zwicky:1933gu,Livio:2014gda,Drees:2012ji,Saab:2012th,Bertone:2004pz,Famaey:2015bba,Iocco:2015xga,Durazo:2015zzzz,McGaugh:2015tha,Iocco:2015bja,Sofue:2015xpa},
 	as well as results from very sophisticated numerical many-body simulations
	of genesis of cosmic large- and small-scale structures (see, for example, 
\cite{Kuhlen:2012ft}), indicate the presence of this new form of matter in the Universe. 
 
	In particular, stars and gas clouds in galaxies and galaxies in clusters move 
	faster than can be explained by the pull of visible matter alone.
	Light from distant objects may be distorted by the gravity of intervening dark material.
	The pattern of the large-scale structures across the Universe is largely dictated by DM.
	In fact, about 85\% of the Universe's mass is dark, 
	accounting for about a quarter of the total cosmic energy budget
\cite{Livio:2014gda}.
	Some of recent reviews on the subject can be found also in 
\cite{Gelmini:2015zpa,Hoeneisen:2015rva}.	

	For further consideration the local density (nearby the solar system) and local distribution of 
	the relic DM-particle population are both very important
\cite{Frere:2015xba}. 
	To allow a measurable direct detection event rate in a modern DM-detector
	it must be not very low 
\cite{Famaey:2015bba,Iocco:2015xga,Iocco:2015bja}.
	According to estimates based on a detailed model of our Galaxy
\cite{Kamionkowski:1997xg}
 	the local density of DM amounts to about
$\rho_{\rm local}^{\rm DM} \simeq 0.3 \ {\rm GeV/cm}^3
\simeq 5 \cdot 10^{-25} {\rm g/cm}^3$.
	Recent studies argue that the current best-fit value for the local DM density, 
	which should be used as a benchmark for direct DM detection searches, 
	is 0.4--0.5 GeV/cm$^3$
\cite{Famaey:2015bba,Pato:2015dua}. 
	The local flux of DM particles, which can cross the Earth, is expected to be 
$\displaystyle \Phi_{\rm local}^{\rm DM} \simeq 10^5 \, \frac {100 \ {\rm GeV}} {m^{}_{\rm DM}}  
{\rm cm}^{-2} {\rm s}^{-1},$ where $m^{}_{\rm DM}$ is the DM particle mass.
	In other words, one can expect that  1\,cm$^2$ of the Earth's surface 
	meets about $10^5$ DM particles each second, 
	provided their mass equals to 100 GeV/$c^2$.
	This value is often considered as a promising basis for 
	laboratory direct DM search experiments.

	Furthermore, the Big Bang conception of the early Universe
\cite{Kolb:1990vq} strongly supports the idea of non-gravitational coupling 
	of the DM particles with the ordinary matter.
	This interaction could be very weak, but not completely vanishing. 
	 
	Despite many other possibilities
\cite{Feng:2010gw} the Weakly Interacting Massive Particle (WIMP) is among the
	most popular candidates for the relic DM.
	Being electrically neutral and interacting rather weakly, the WIMPs 
	naturally reproduce the correct relic DM abundance, 
	if their masses coincide with a typical new physics TeV-scale --- 
	$M_{\rm WIMP} \le \frac{g^2}{0.3}\ 1.8~{\rm TeV}$
\cite{Christensen:2014yya}.
	This coincidence adds extra interest to the search for the DM particles, especially 
	directly with the LHC, which is nowadays the best artificial TeV-scale-physics explorer.
	 
	These particles are non-baryonic and there is no room for them in the 
	Standard Model of particle physics (SM), 
	in particular due to the Big Bang nucleosynthesis, which successfully predicts 
	the abundances of light elements such as deuterium, helium and lithium 
	arising from interactions in the early Universe.
	Furthermore, to explain the way in which galaxies form and cluster, 
	these massive DM particles should be non-relativistic, or so-called "cold DM" particles. 
	If they were relativistic, they could easily travel beyond the typical scale of a protogalaxy,
	and galaxy-scale structures would not have chance to appear 
\cite{Livio:2014gda}.

	Therefore one needs a New Physics beyond the SM (BSM).
      The lightest supersymmetric (SUSY) particle (LSP), the neutralino, 
      in many R-parity conserving model realizations of SUSY (MSSM, NMSSM, mSUGRA, etc) ---   
       being massive, neutral, stable and possessing correct relic abundance ---
       very naturally plays the role of the WIMP DM particle.

	The primary goal of modern particle physics and astrophysics is to detect 
	the DM particles that constitute the massive invisible halo of the Milky Way. 

	One believes
\cite{Livio:2014gda} that in spite of decades of compelling efforts, 
	all attempts to detect DM particles have failed so far
	(with one important exception of DAMA/LIBRA results).
	This "DM Problem" is a real challenge for modern physics and 
	experimental technology.
	To solve the problem, i.e. {\em at least}\/ to detect these DM particles, 
	one simultaneously needs to apply the front-end knowledge 
	of modern particle physics, astrophysics, cosmology and nuclear physics.
	Furthermore one should develop and have in long-term usage an 
	extremely high-sensitive setup, to say nothing about 
	complex data analysis methods (see, for example, discussion in 
\cite{Cerulli:2012dw}).

	It is clear why so many papers concerning the DM problem are published. 
	The papers discuss new DM candidates, new DM detection methods and 
	new technologies for the DM search, new strategies and new models, etc.
	Nevertheless, on this way one point almost always stays in shadow,  
	and this is the galactic origin and galactic belonging of the searched-for 
	and to a much greater extent, ever registered DM candidates. 
	The nature must be clearly used for the search strategy and
	especially to prove the registration of the "true" DM particles.
%%%%%
   %GalacticDM.tex

%%%%%%%%%%%%%%%%%%%%%%%%%%%%%%%%%%%%%%%%%
\section{How does one want to detect the DM?}\label{sec:DM-detections}
% 150627 GAMMA-400 + ...
%\section{How one wants to detect the DM?}\
	There is a common belief that the DM problem can be solved 
	by means of {\em a complete and balanced research program}\/ based on 
	the following four categories. \\
\underline{Direct Detection experiments} look for a direct DM interaction in an underground terrestrial 
	low-background laboratory, where a DM particle scatters off a (nuclear) substance of a detector, 
	producing a detectable recoil and/or ionization signal.  \\
\underline{Indirect Detection experiments} are unable to detect DM directly,
	 but with (huge) terrestrial setups try to register products of DM particle annihilations 
	 in cosmic objects like the Earth, the Sun, our own and/or another galaxy.
	It is assumed that pairs of DM particles annihilate each other producing high-energy 
	ordinary particles (antimatter, neutrinos, photons, etc).  
	In some models, the DM particles can be metastable and eventually decay 
	with production of SM particles. \\
\underline{Particle Collider experiments}  can help one to understand the properties of the DM particles. 
	The LHC and future lepton and hadron colliders 
	can produce energetic DM particles that obviously will escape detection, 
	but could be registered by means of an excess of events with missing energy or momentum. \\
\underline{Astrophysical Probes} provide one with information about
	non-gravitational interactions of DM particle populations, 
	such as DM densities in the centers of galaxies and cooling regimes of stars.
	The particle properties of DM are constrained here through observation of 
	their joint impact on astrophysical observables.  
  	Examples include self-interaction of DM particles affecting central DM densities in galaxies 
	(inferred from the rotation velocity or velocity dispersion), mass of DM particle affecting the  
	DM substructure in galaxies (inferred from strong lensing) 
	and annihilation of DM in the early Universe affecting the Cosmic 
	Microwave Background (CMB) fluctuations
\cite{Bauer:2013ihz}.

\begin{figure}[!ht] 
\includegraphics[width=\columnwidth]{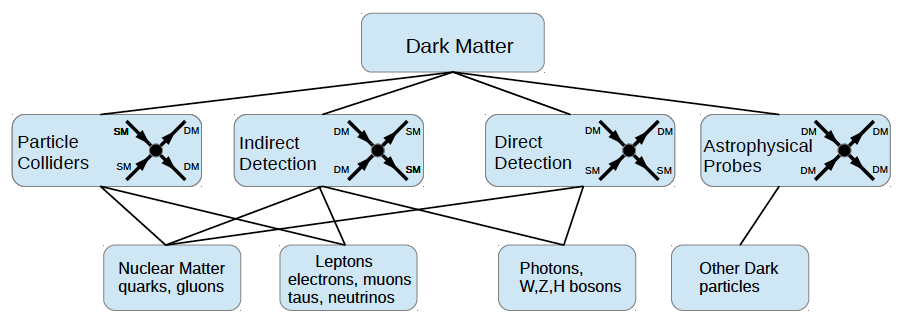}
\caption{The idea of DM particle registration with a terrestrial setup 
	relies on the common belief that DM can interact non-gravitationally with 
	nuclear matter, leptons, gauge and other bosons, and with other possible dark particles.  
	These interactions may be studied nowadays with
  	direct and indirect detection techniques, particle colliders, and via astrophysical probes 
 \cite{Bauer:2013ihz}. } \label{fig:2013ihz-DM-interplay}
\end{figure}

	These search strategies are schematically shown in 
Fig.~\ref{fig:2013ihz-DM-interplay}. 
	Each of these approaches has its own advantages and disadvantages.
	Therefore, one speaks about different types of complementarities of them
\cite{Gelmini:2015zpa,Bauer:2013ihz}.

	Below some words are said about the indirect detection experiments and the astroparticle probes. 
	The direct detection technique and searches with colliders are discussed in 
	sections \ref{sec:DirectDetectionDM} and \ref{sec:FirstColliderDM}, \ref{sec:DM-ATLAS-searches} 
	in more detail.
\smallskip
	
	The idea of the {\em indirect detection} relies on a set of reasonable assumptions. 
	One believes that a DM particle on its way through the space can be attracted by a rather 
	massive cosmic body, like the Sun, and one day can be trapped by the gravitational potential of the body.
	Afterwards this particle oscillates in the potential and passes many-many times through 
	the ordinary matter of the body. 
	In the course of this inevitable travel the WIMP nature of the DM plays a crucial role,
	the particle eventually loses its kinetic energy due to a non-vanishing weak interaction and 
	after some time sinks into the deepness of the body. 
	After accumulation of a critical amount of the DM particles within the body 
	they start a permanent process of mutual annihilation (again via weak-scale interaction).
	The annihilation products in a due course leave the body in the form of 
	fluxes of $\gamma$-rays, X-rays, neutrinos, electrons and 
	positrons, to a much less extent, antiprotons and antideuterons.
	Registration of these fluxes with a proper terrestrial setup
	constitutes the indirect detection of DM. 
	Therefore the aim of the indirect DM search is registration of the  
	processes in outer space that were responsible for modern relic DM  density. 
	These (mainly annihilation) processes are still under way,   
	especially in some space regions where the local DM density can strongly exceed 
	the average relic DM density
\cite{Askew:2014kqa}. 

	The indirect detection technique includes 
\cite{Zitzer:2015uta,Ibarra:2015tya,Ackermann:2015zua}
	space and ground-based gamma-ray telescopes like the Fermi Large Area Telescope (Fermi-LAT) 
\cite{Conrad:2015bsa}, cosmic ray detectors, large underground, under-ice and underwater Cherenkov 
	neutrino telescopes like Super-Kamiokande (see, for example, 
\cite{Choi:2015ara}), IceCube (see, for example,
\cite{Kasuya:2015uka}), ANTARES
\cite{Zornoza:2014dma} and BAIKAL (see, for example,
\cite{Avrorin:2014swy}).

	The main disadvantage of the indirect detection technique 
	is the "ordinary" astrophysics, which can easily produce irreducible backgrounds 
\cite{Askew:2014kqa}.
	Indeed, one must conclude that until now all attempts to detect DM indirectly have been inconclusive
\cite{Livio:2014gda}.

	In particular, an excess of positrons in the cosmic-ray spectrum up to 350 GeV, reported by 
	the Alpha Magnetic Spectrometer (AMS-02) on the International Space Station (ISS)
\cite{Hooper:2012gq}, 
	can be due to the DM particle annihilations.
	The excess strengthened previous results from the 
	Payload for Antimatter Matter Exploration and Light-nuclei Astrophysics (PAMELA) satellite
\cite{Adriani:2008zr}. 
	Nevertheless these "extra" positrons can be produced by other sources, 
	such as winds from rapidly rotating neutron stars
\cite{Livio:2014gda}. 
	Further observations of positrons with AMS-02 at higher energy  
	might distinguish between these hypotheses
\cite{Giesen:2015ufa}.

	Recently, the AMS-02 collaboration has reported results of the cosmic-ray antiproton measurement.
	Antiprotons can be a product of annihilation or decay of galactic DM.
	 It was claimed in 
\cite{Giesen:2015ufa,Evoli:2015vaa,Lin:2015taa}  
 	 that this $\bar{p}$ flux does not exceed uncertainties of the astrophysical secondary $\bar{p}$ flux.
	 Nevertheless the expected secondary antiproton flux has a tendency to be smaller than the observed one at 
	energy $\ge$ 100 GeV, allowing one to still think about a DM contribution in that energy range
\cite{Hamaguchi:2015wga,Chen:2015cqa}.

	Another excess of $\gamma$-rays, in the form of a narrow line at 130 GeV, from the Galactic center
	where DM particles can concentrate and annihilate,  
	was observed by the Fermi Gamma-ray Space Telescope 
\cite{Abdo:2010nc,Ackermann:2012qk}.
	But a similar line from the Earth's atmospheric limb implies that at least part of the 
	signal must be instrumental by origin
\cite{Livio:2014gda,Calore:2014nla}. 
	The High Energy Stereoscopic System (HESS) $\gamma$-ray telescope in Namibia
(see, for example, \cite{Abramowski:2013ax}), 
	which is observing the inner Galaxy in the 100 GeV to 1 TeV energy range, 
	can resolve the situation in coming years.
	Preferred in these cases, DM particle masses are expected to be in the range 10--40 GeV$/c^2$ and
	are rather sensitive to details of the electron interstellar propagation
\cite{Livio:2014gda}.
	The existence of DM with these masses can also be probed 
	by the temperature fluctuations of the CMB. %cosmic microwave background.
	The fluctuations would be damped 
	by any delay in recombination, which is expected 
	for such mass DM particles annihilation. 
	The lower the mass, the more ionizing photons are produced for any 
	cosmologically specified DM density. 
	It is anticipated that the Planck satellite 
	will soon set a more definitive constraint 
	on self-annihilating DM particle masses below 30--40 GeV$/c^2$
\cite{Livio:2014gda}.  

	There are also plans for a variety of sensitive gamma-ray telescopes 
	in the energy range 1--100 MeV  aimed at indirect study of low-mass DM annihilation or decay 
\cite{Boddy:2015efa,Ghorbani:2014gka,Ghorbani:2014qpa}. 

	Simultaneously, one looks for more massive (TeV-scale) particles, which  
	could be difficult to detect directly, because one should expect much smaller number density 
	of them to fit current DM density. 
	The gamma-ray astronomy has no such a limitation.
	The international Cherenkov Telescope Array (CTA) with more than 
	100 ground-based dedicated telescopes is planned to capture Cherenkov light flashes 
	from $\gamma$-rays scattered by the atmosphere.
	The CTA will be able to measure $\gamma$-rays with 100-TeV energy, 
	which could be generated by annihilations or decays of DM with 100-TeV scale masses. 
	This energy scale reaches the highest limit on the DM mass 
	expected from fundamental physics arguments 
\cite{Ibarra:2015tya}. 
%% 150627 %%%%%%%%%%%%%%%%%%%%%%%%%%%%%%%%%%%%%%
	With GAMMA-400 
\cite{Cumani:2015ava} $\gamma$-ray telescope one expects 
	new results in the energy range 0.1 -- 3000 GeV
\cite{Bhattacherjee:2014dya,Campbell:2013rua,Bergstrom:2012bd,Bringmann:2012ez}.
%%%%%%%%%%%%%%%%%%%%%%%%%%%%%%%%%%%%%%%%%%%%%

	An investigation of the angular cross-correlation of non-gravitational signals 
	with low-redshift gravitational probes is proposed in 
\cite{Regis:2015zka} as a most powerful technique to detect DM signal outside the Local Group.  
	This technique is more sensitive than other extragalactic $\gamma$-ray probes, 
	such as the energy spectrum and angular autocorrelation of the extragalactic background, 
	and emission from clusters of galaxies. 
	In particular, the measured cross-correlation can be explained by a DM particle, 
	with thermal annihilation cross section and mass between 10 and 100 GeV$/c^2$
\cite{Regis:2015zka}. 

	More information about indirect DM search experiments can be found, for example, in 
\cite{Gelmini:2015zpa,Khlopov:2014nva}. 
%\smallskip

	Due to unprecedented accuracy of astronomical observations 
	there are today many different variants for the 
	{\em astrophysical probes}\/ (assays) of the DM existence in the sky.
 
 	For example, one can look for a large number of starless DM halos surrounding the Milky Way, 
	which are not yet detected but well expected in modern cosmology and astrophysics as the cold DM.  
	These DM halos, in the form of clumps or streams, can move through or orbit the Milky Way and can
	increase substantially the direct DM detection rates together with   
	production of rather small, but detectable, velocity changes of the stars in the galaxy disk.
	These kinematic signatures  will be detected by  
	the new space telescope GAIA
\cite{Barstow:2014dda}  provided the starless DM halos take place in the sky
\cite{Feldmann:2013hqa}.

	Next, the amount of DM accumulated in the  
	neutron stars together with the energy deposition rate from the DM decays
	could set a limit on the neutron star survival rate against transitions to more compact objects 
	provided that DM is indeed unstable. 
	This limit sets constraints on the DM lifetime  
\cite{Perez-Garcia:2014dra}.

	Furthermore, one should look for some unusual signals in old neutron stars and white dwarfs. 
	In particular, the DM particles   
	accumulated in the core of a neutron star in the course of its long-term travel through the Galaxy  
	might form a tiny black hole that could eventually devour the home star, 
	causing a very unusual explosion
\cite{Fuller:2014rza}.  
	The influence of DM particles collected in the Sun 
	on the solar temperature variation could also be probed by helioseismology   
\cite{Livio:2014gda}.
	Merging galaxy clusters such as the Bullet Cluster can provide a powerful testing ground for 
	galactic DM observation \cite{Graham:2015yga}.   
	There are also proposals to consider a possibility of galactic DM interactions 
	with cosmic rays and different kinds of interstellar matter 
\cite{Carlson:2015daa}.
	In the case of multicomponent galactic DM sector with at least two DM species with different masses 
	one can use the DM-to-DM decays as a new complementary tool for investigation of the DM properties  
\cite{Dienes:2015qqa}.
     
	The potential of microwave background lensing to probe the DM distribution 
	in galaxy group and galaxy cluster halos was demonstrated in 
\cite{Madhavacheril:2014slf}.           
	Evidence was presented of the gravitational lensing of the 
	cosmic microwave background by $10^{13}$ solar mass DM halos. 
	The mean lensing signal is consistent with simulated DM halo profiles. 

	From the cosmological point of view one can put the following few constraints on DM.
	DM must have the correct cosmological energy density, 
	it must be massive so that it can act  as pressureless matter. 
	DM particles should not interact so strongly as to either disturb the well-understood CMB 
	or to fail to collapse sufficiently to explain the observed large-scale structure of the Universe
\cite{Askew:2014kqa}.

	In order to convincingly establish the nature of a DM candidate,
	one must reach consistency between all possible DM searches  
	for the common DM candidate parameters of mass, spin, and coupling strengths
\cite{Christensen:2014yya}.

%%%%%
  %DM-detections.tex

%%%%%%%%%%%%%%%%%%%%%%%%%%%%%%%%%%%%%%%
\section{Main features of the direct DM detection} \label{sec:DirectDetectionDM}  
%\section{Main features of the direct DM detection}
	 The direct DM detection has a bit exceptional 
	 status among the other DM search techniques discussed above. 
	 The reasons could be a rather old history of this approach, 
	 existence of the DAMA evidence (see below),
	  and a possibility of supporting us with the clearer and most decisive information on the DM problem
\cite{Cushman:2013zza}.	 	

	One should absolutely agree with the authors of % the "Nature" paper 
\cite{Livio:2014gda}, that 
"... the goal is to detect the particles that {\em constitute the massive halo} of dark matter that surrounds the Milky Way, {\em as they pass through our detectors} ..."  
	Another argument is 
\cite{Famaey:2015bba} that  "... {\em until dark matter particles are detected in the laboratory}, 
	it is also healthy to remember that there are hints that the dark sector might be more complicated ..." 
	Perhaps unconsciously, the decisive role of the laboratory 
	direct DM detection experiments was stressed here.

        Due to elastic WIMP-nucleus scattering the nuclear recoil energy is the main quantity to be measured 
        by a terrestrial detector in direct DM detection laboratory experiments
\cite{Goodman:1985dc}. 
        Detection of very rare events of such WIMP interactions 
        is a quite complicated task because of very weak WIMP coupling with ordinary matter. 
        The rates expected, for example, in SUSY models range from 10 to 10$^{-7}$ 
        events per kilogram detector material a day
\cite{Jungman:1996df,Lewin:1996rx,Bednyakov:2002mb,Bednyakov:1998is,Bednyakov:1997jr}. 
         Moreover, for WIMP masses between a few GeV$/c^2$ and 1 TeV$/c^2$ the energy
         deposited by the recoil nucleus is much less than 100 keV.

         Therefore, in order to be able to detect a WIMP, 
         an experiment  should have an "ideal" detector with 
1) a large target mass of an isotope (preferably with a high mass number $A$),
2) a low energy threshold (several keV),
3) an ultra-low radioactive background,
4) a possibility of distinguishing the signal (nuclear recoils) from the background (electronic recoils), and 
5) in order to reduce the cosmic-ray induced background, this detector has to be installed in a deep-underground laboratory, where the overburden almost completely eliminates the hadronic component 
of the cosmic rays and reduces the muon flux by 5-7 orders of magnitude
\cite{Schumann:2014uva}.
	Additionally one should have a possibility of running the setup for several years. 

	All these complications, especially the non-zero energy threshold, 
	the need for a large target mass and rather long exposure time, 
	are  sometimes considered as a kind of limitations in the sensitivity of the direct search experiments
\cite{Askew:2014kqa}.

\begin{figure}[h!]
\begin{center}
\includegraphics[width=0.7\textwidth]{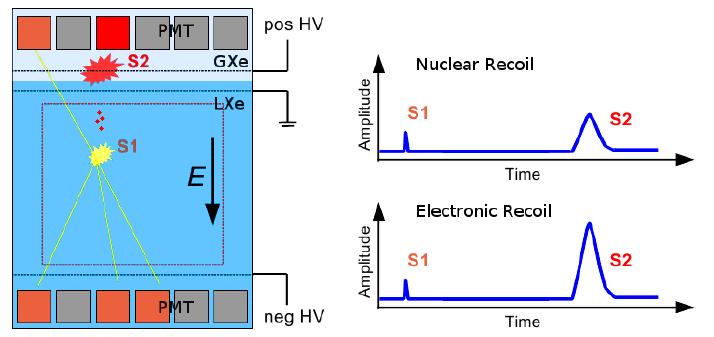}
\caption{
\label{fig:2014uva-DualTPCidea}
A dual phase time projection chamber measures scintillation light (S1) and the ionization charge
signal, which is converted to a proportional scintillation signal (S2) in the xenon gas phase (GXe). The time
distance between the two signals and the pattern on the top PMT array is used to reconstruct the event vertex.
The ratio S2/S1 is different for nuclear and electronic recoils and used for background discrimination
From \cite{Schumann:2014uva}.}
\end{center}
\end{figure}

	According to 
\cite{Schumann:2014uva,Schumann:2015wfa}
	two-phase time projection chambers (TPCs) filled with liquid (and gaseous) %noble gas  
	xenon (LXe) are considered today as the most sensitive technique for 
	direct WIMP search in an underground experiment.
	This statement relies on a (very) large homogeneous target of a very low background and 
	a possibility of localizing the interaction vertex. 
	The latter allows fiducial volume control and rejection of multiple-scatterring events. 
	The background events in the setup are further rejected by the simultaneous measurement 
	of the scintillation and ionization signals 
(Fig.~\ref{fig:2014uva-DualTPCidea}).
	
	Perhaps, one of the best examples of such a technique today is the XENON100 experiment, which 
	constitutes (with 161 kg of ultra pure LXe) the second phase of the XENON program 
\cite{Aprile:2012nq} for the direct DM detection.
	 After 224.6 live days of running no indication for an excess of events above the expected 
	 background has been observed with the XENON100, leading to strong upper limits for relevant 
	 WIMP-nucleon scattering cross sections
\cite{Aprile:2013doa,Orrigo:2015cha}.

	In the nearest future the LUX experiment is expected to reach its design goals
\cite{Akerib:2013tjd,Savage:2015xta}. 
	The other dual-phase LXe experiment is the Chinese experiment PandaX
\cite{Cao:2014jsa,Xiao:2014xyn} with a flat pancake-shaped target of 120 kg LXe (25 kg fiducial mass).
	At the famous LNGS underground laboratory, the XENON collaboration constructs
	a new XENON1T setup 
\cite{Aprile:2012zx,Aprile:2015lha} with a target mass of 2.2 t, 	
	which will be the first TPC for the WIMP search with a mass of about 1 t.
	The goal of XENON1T is to perform a background-free DM search run with an 
	exposure of 2 tonne$\,\times\,$years. 
	The expected sensitivity is at a level of $2\times 10^{-47}$ cm$^2$ at a WIMP mass of 50 GeV/$c^2$. 
	One believes that all sub-systems of XENON1T  
	can be re-used in a later upgrade phase of the experiment,  XENONnT, with about 7 t of LXe
\cite{Schumann:2014uva}.
	These projects will significantly improve the sensitivity to WIMP-nucleon interactions by 1--2 orders of magnitude 
	compared to the present status 
\cite{Schumann:2015wfa}.

	There are many other running direct DM search experiments 
 \cite{Angloher:2011uu,Angloher:2015eza, %CRESST
    Akerib:2013tjd, %LUX
    Agnese:2013rvf, %CDMS low mass
    Agnese:2014aze,Agnese:2015sga, %CDMS final
    Aalseth:2014jpa, %CoGeNT
    Bernabei:2008yi, %DAMA
    Aprile:2013doa, %XENON100 
    D'Angelo:2015lia, %DS-50, } and spin-dependent \cite{
  Archambault:2012pm, %PICASSO
    Desai:2004pq, %SuperK
    Abbasi:2009uz, %IceCube
    Behnke:2010xt, %COUPP
    Felizardo:2011uw}. %SIMPLE
    	The results of these experiments are given in the figures of section
\ref{sec:DM-ATLAS-searches} in connection with the collider DM search results of the ATLAS collaboration. 
	The experimental status of the direct DM experiments is comprehensively discussed in 
\cite{Saab:2012th,Bi:2014hpa,Cushman:2013zza,Schumann:2014uva,Schumann:2015wfa,Gelmini:2015zpa}.

%%%%%%%%%%%%%%%%%%%%%%%%%%%%%%%%%%
\subsection{Current situation --- fight for the best exclusion curve}
%%%%%%%%%%%%%%%%%%%%%%%%%%%%%%%%%%
	The nuclear recoil energy $E_{\rm R}$ produced by a DM WIMP with mass $m_\chi$
	is measured by a DM detector. 
	The differential event rate  (the spectrum) has the form
\begin{equation}
\label{Definitions.diff.rate}
S(t) \equiv	\frac{dR}{dE_{\rm R}} = N_A \frac{\rho_{\rm local}^{\rm DM}}{m_\chi} 
	\int^{v_{\max}}_{v_{\min}} dv f(v) v
	{\frac{d\sigma^A}{dq^2}} (v, q^2). 
\end{equation}
%Assuming that WIMPs are the dominant component of the DM halo of our Galaxy, one has $\rho_{\chi} = \rho_{\rm local}^{\rm DM}$. 
        The nuclear recoil energy $E_{\rm R} = q^2 /(2 M_A )$ is typically about $10^{-6} m_{\chi}$, 
	$N_A$ is the number density of target nuclei with mass $M_A$, 
	$v_{\max} = v_{\rm esc} \approx 600$~km/s, and
	$v_{\min}=\left(M_A E_{\rm R}/2 \mu_{A}^2\right)^{1/2}$ is
	the minimal WIMP velocity which still can produce the recoil energy $E_{\rm R}$.
	The WIMP-nucleus differential elastic scattering cross section 
	for spin-non-zero ($J\neq 0$) nuclei contains the coherent (spin-independent, or
	SI) and axial (spin-dependent, or SD) terms
\cite{Engel:1992bf,Ressell:1993qm}
\begin{eqnarray} 
\frac{d\sigma^A}{dq^2}(v,q^2) 
\label{Definitions.cross.section}
= \frac{\sigma^A_{\rm SD}}{4\mu_A^2 v^2}F^2_{\rm SD}(q^2)
           +\frac{\sigma^A_{\rm SI}}{4\mu_A^2 v^2}F^2_{\rm SI}(q^2).
\end{eqnarray}
 	The normalized  nuclear form-factors $ F^2_{\rm SD,SI}(q^2)$  
	are expressed in terms of the nuclear structure functions given in 
\cite{Engel:1992bf,Ressell:1993qm}. 
	For $q=0$ the nuclear SD and SI cross sections can be represented as
\begin{eqnarray} \nonumber  
\sigma^A_{\rm SI} &=& \frac{\mu_A^2}{\mu^2_p}A^2 \sigma^{p}_{{\rm SI}}, \\ 
\sigma^A_{\rm SD} &=&  \frac{4\mu_A^2}{\pi}\frac{(J+1)}{J}
             \left\{a_p\langle {\bf S}^A_p\rangle  + a_n\langle {\bf S}^A_n\rangle\right\}^2 = 
\frac{\mu_A^2}{\mu_p^2}\frac43 \frac{J+1}{J} \sigma^{}_{\rm SD}	
	\left\{ \langle {\bf S}^A_p\rangle \cos\theta
               +\langle {\bf S}^A_n\rangle \sin\theta \right\}^2.  \nonumber 
\end{eqnarray} 
	The effective spin cross section $\sigma^{}_{\rm SD}$ and the coupling mixing 
	angle $\theta$ were introduced \cite{Bernabei:2003za}:
$$\sigma^{}_{\rm SD}  =\frac{\mu_p^2}{\pi}\frac43 \Bigl[ a_p^2 +a_n^2 \Bigr], \quad
\tan\theta = \frac{{a}_{n}}{{a}_{p}}; \quad 
\sigma^p_{\rm SD}  = \sigma^{}_{\rm SD} \cdot \cos^2 \theta, \quad 
\sigma^n_{\rm SD}=\sigma^{}_{\rm SD} \cdot \sin^2 \theta.
$$
	Here, $\displaystyle \mu_A = \frac{m_\chi M_A}{m_\chi+ M_A}$
	is the reduced mass and $\mu^2_{n}=\mu^2_{p}$ is assumed.
	The dependence on the effective WIMP-quark   
	scalar ${\rm C}_{q}$ and axial-vector ${\rm A}_{q}$ couplings 
$\left( {\rm A}_{q}\cdot
      \bar\chi\gamma_\mu\gamma_5\chi\cdot
                \bar q\gamma^\mu\gamma_5 q + 
	{\rm C}_{q}\cdot\bar\chi\chi\cdot\bar q q\right)$	
	and on the spin ($\Delta^{(p,n)}_q$) and the scalar ($f^{(p)}_q \approx f^{(n)}_q$) 
	structure of the nucleons enter  via the zero-momentum-transfer 
	WIMP-proton and WIMP-neutron SI and SD cross sections 
\begin{eqnarray} \label{SI-cs-Deff.at-zero}
\sigma^{p}_{{\rm SI}}  = 4 \frac{\mu_p^2}{\pi}c_{0}^2, &\qquad&
	c^{}_0 = c^{p,n}_0 = \sum_q {\rm C}_{q} f^{(p,n)}_q; \\
\sigma^{p,n}_{{\rm SD}} =  12 \frac{\mu_{p,n}^2}{\pi}{a}^2_{p,n}  &\qquad&
	a_p =\sum_q {\rm A}_{q} \Delta^{(p)}_q, \quad 
	a_n =\sum_q {\rm A}_{q} \Delta^{(n)}_q.
\label{SD-cs-Deff.at-zero}
\end{eqnarray}
	The factors $\Delta_{q}^{(p,n)}$, which parameterize the quark 
	spin content of the nucleon, are defined as
	$ \displaystyle 2 \Delta_q^{(n,p)} s^\mu  \equiv 
          \langle p,s| \bar{\psi}_q\gamma^\mu \gamma_5 \psi_q    
          |p,s \rangle_{(p,n)}$.
The quantity $\langle {\bf S}^A_{p(n)} \rangle 
     = \langle A \vert  \sum_i^A {\bf s}^i_{p(n)} \vert A \rangle $ 
	is  the total spin of protons 
	(neutrons) averaged over all $A$ nucleons of the nucleus $(A,Z)$.
       For the direct DM detection 
	isotopes either $\langle{\bf S}^A_{p}\rangle$ or $\langle{\bf S}^A_{n}\rangle$ dominates:
	$\langle{\bf S}^A_{n(p)}\rangle \ll \langle{\bf S}^A_{p(n)}\rangle$
\cite{Ressell:1997kx,Bednyakov:2004xq,Bednyakov:2006ux}.
           The differential event rate 
(\ref{Definitions.diff.rate}) can be rewritten  in the form 
\cite{Bernabei:2003za,Bednyakov:2004be}
\begin{eqnarray}
\label{Definitions.diff.rate1}
\frac{dR({E_{\rm R}})}{d{E_{\rm R}}} 
	&=& \kappa^{}_{\rm SI}({E_{\rm R}},m_\chi)\,\sigma_{\rm SI}
         +\kappa^{}_{\rm SD}({E_{\rm R}},m_\chi)\,\sigma_{\rm SD}. \\
\nonumber
\kappa^{}_{\rm SI}({E_{\rm R}},m_\chi)
&=&       N_T \frac{\rho_\chi M_A}{2 m_\chi \mu_p^2 } 
          B_{\rm SI}({E_{\rm R}}) \left[ M_A^2 \right],
          \qquad 
B_{{\rm SI},{\rm SD}}({E_{\rm R}}) =
        \frac{\langle v \rangle}{\langle v^2 \rangle}
        F^2_{{\rm SI},{\rm SD}}({E_{\rm R}})I({E_{\rm R}}),
          \\
\kappa^{}_{\rm SD}({E_{\rm R}},m_\chi)
&=& \label{structure} \nonumber
     N_T \frac{\rho_\chi M_A}{2 m_\chi \mu_p^2 } B_{\rm SD}({E_{\rm R}}) 
        \left[\frac43 \frac{J+1}{J}
        \left(\langle {\bf S}_p \rangle \cos\theta
            + \langle {\bf S}_n \rangle \sin\theta   
       \right)^2\right]. 
\end{eqnarray}
        The dimensionless integral $I({E_{\rm R}})$   
        accumulates properties of the DM velocity distribution
$$ I({E_{\rm R}})= \frac{ \langle v^2 \rangle}{ \langle v \rangle }
 \int_{x_{\min}} \frac{f(x)}{v} dx 
    = \frac{\sqrt{\pi}}{2}
\frac{3 + 2 \eta^2}{{\sqrt{\pi}}(1+2\eta^2)\mbox{erf}(\eta) + 2\eta e^{-\eta^2}}
                [\mbox{erf}(x_{\min}+\eta) - \mbox{erf}(x_{\min}-\eta)].
$$ 
        Here one assumes the Maxwell-Boltzmann DM velocity distribution in the rest frame of the Galaxy,
	$\eta$ denotes the dimensionless Earth's speed with respect to the halo, and 
        $\displaystyle x_{\min}^2 =  \frac{3}{4}\frac{M_A{E_{\rm R}}}{\mu^2_A{\bar{v}}^2}$
\cite{Freese:1987wu,Lewin:1996rx}. 
        The error function is $\displaystyle \mbox{erf}(x) = \frac{2}{\sqrt{\pi}}\int_0^x dt e^{-t^2}$.
        The velocity variable is the dispersion $\bar{v}\simeq 270\,$km$/$c.
        The mean velocity ${\langle v \rangle} = \sqrt{\frac{5}{3}} \bar{v}$.
	Integrating the differential rate 
(\ref{Definitions.diff.rate}) from the recoil energy threshold $\epsilon$ 
	to some maximal energy $\varepsilon$,
        one obtains the total detection rate $R(\epsilon, \varepsilon)$
	as a sum of the SD and SI terms
\begin{eqnarray}\label{Definitions.total.rate}\nonumber
R(\epsilon, \varepsilon)=
     R_{{\rm SI}}(\epsilon, \varepsilon) + R_{{\rm SD}}(\epsilon, \varepsilon) = 
      \int^{\varepsilon}_{\epsilon} d{E_{\rm R}} \kappa^{}_{\rm SI}({E_{\rm R}},m_\chi)\,\sigma_{\rm SI}
    + \int^{\varepsilon}_{\epsilon} d{E_{\rm R}} \kappa^{}_{\rm SD}({E_{\rm R}},m_\chi)\,\sigma_{\rm SD}.
\end{eqnarray}
	To accurately estimate the event rate $R(\epsilon, \varepsilon)$,
	one needs to know a number of quite uncertain 
	astrophysical and nuclear structure parameters
	as well as very specific characteristics of the experimental setup
\cite{Cerulli:2012dw} (see also 
\cite{Schneck:2015eqa}).

        Finally, to estimate the expected direct DM detection rates --- 
        via formulas (\ref{Definitions.diff.rate}) or (\ref{Definitions.diff.rate1}) ---
        one should calculate the WIMP-proton and WIMP-neutron spin $\sigma^{p,n}_{{\rm SD}}$   
       and scalar $\sigma^{p,n}_{{\rm SI}}$ cross sections at $q=0$. 
       To this end a SUSY-like model or some measured data 
       (for example, from the DAMA/LIBRA experiment) can be used.  
       It is these calculations that are usually compared with experimental results,  
       which are presented in the form of exclusion curves --- upper limits 
       of the cross sections as functions of the WIMP mass.
 
\begin{figure}[h!]
\begin{center}
\includegraphics[width=\textwidth]{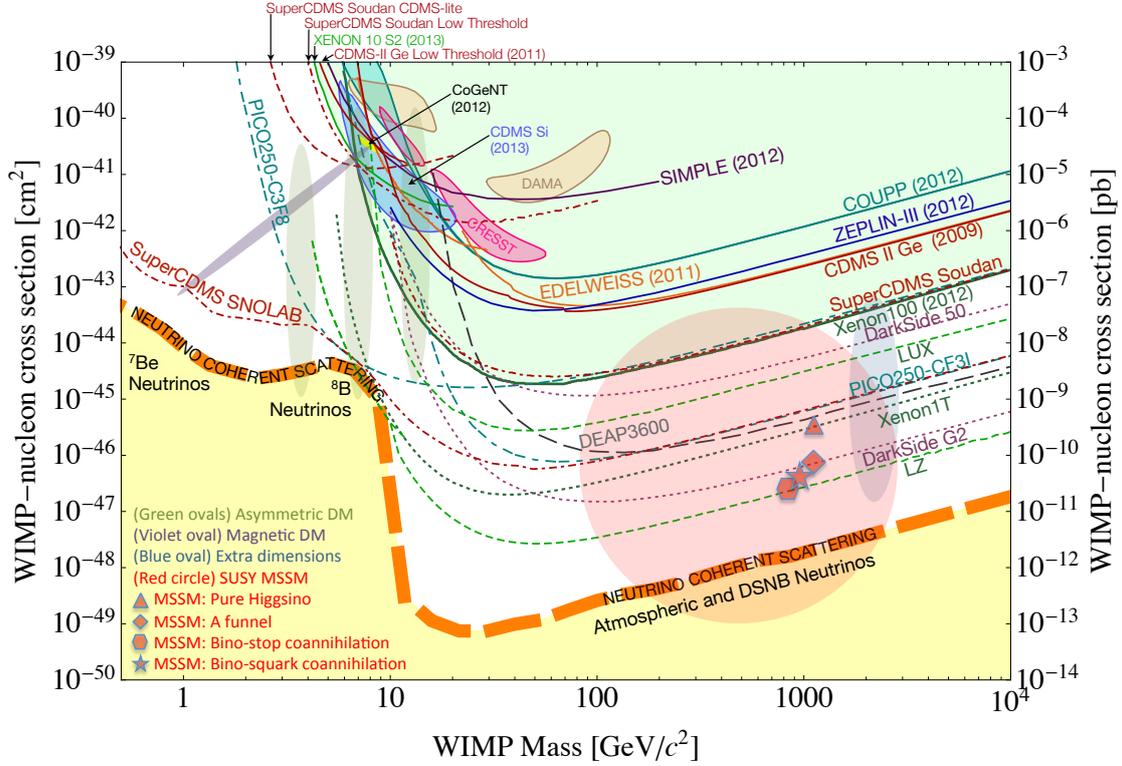}
\caption{\label{fig:SI-overview}
A compilation of WIMP-nucleon spin-independent cross section limits (solid curves), 
hints for WIMP signals (shaded closed contours) and projections (dot and dot-dashed curves) for  
direct detection experiments that are expected to operate over the next decade.  
Also shown is an approximate band where coherent scattering of $^8$B solar neutrinos, 
atmospheric neutrinos and diffuse supernova neutrinos with nuclei will begin to limit the sensitivity of direct detection experiments to WIMPs. 
Finally, a suite of theoretical model predictions is indicated by the shaded regions, with model references included.
From \cite{Cushman:2013zza}.}
\end{center}
\end{figure}

\begin{figure}[h!]
\begin{center}
\includegraphics[width=0.49\textwidth]{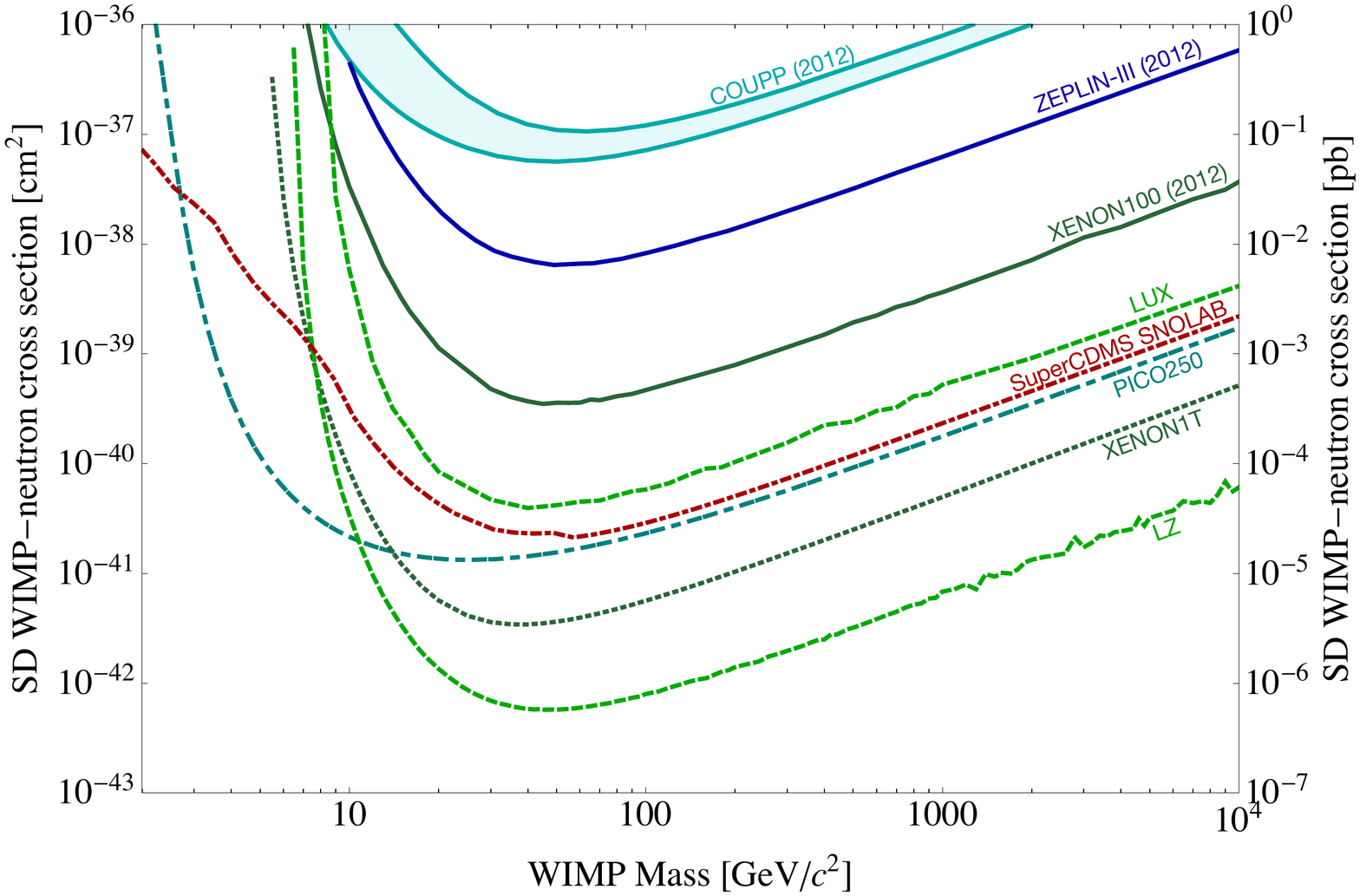}
\includegraphics[width=0.49\textwidth]{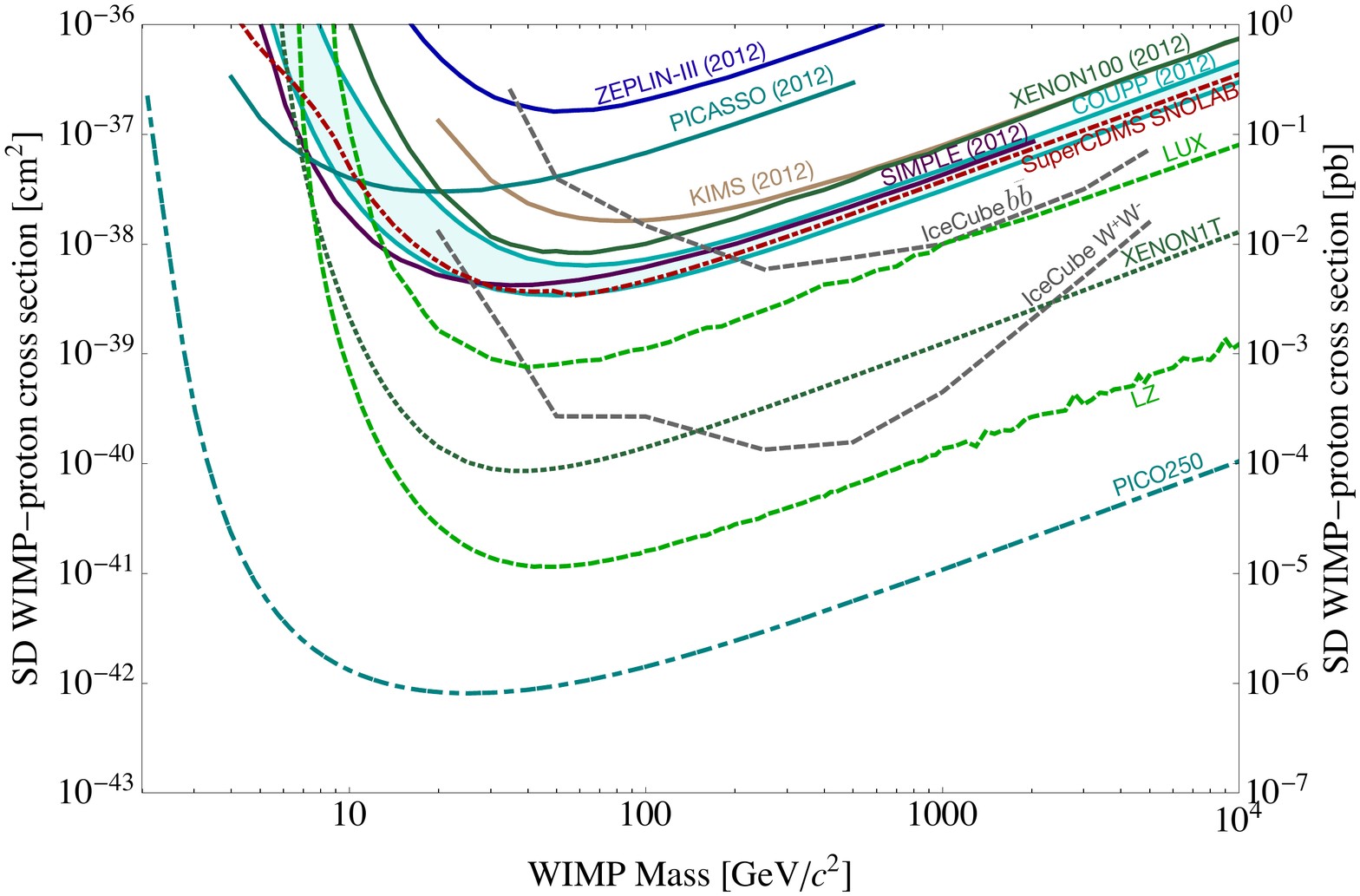}
\caption{\label{fig:SD-overview}
A compilation of WIMP-nucleon spin-dependent cross section limits (solid curves) 
and projections (dot and dot-dashed curves) for 
direct detection experiments that are expected to operate over the next decade.
From \cite{Cushman:2013zza}.}
\end{center}
\end{figure}

	Figures~\ref{fig:SI-overview} and \ref{fig:SD-overview} from 
\cite{Cushman:2013zza}  illustrate the typical situation in the field of direct DM search experiments.
	They contain a lot of exclusion curves that demonstrate the recent achievements 
	and expected improvements in sensitivity for both  $\sigma^{}_{\rm SI}$ and $\sigma^{}_{\rm SD}$ 
	WIMP-nucleon cross sections together with a range of theoretical benchmarks 
\cite{Cushman:2013zza}.

       In the case {\em of non-observation}\/ of a DM signal 
       the exclusion curve simply reflects the sensitivity 
       of a given direct DM search experiment and potentially 
       allows one to constrain some version of the SUSY-like model, 
      provided the curve is sensitive enough.
       Therefore the best exclusion curve is currently the only aim of almost all  
       direct DM search experiments (DAMA/LIBRA is a clear exception).
       The main competition between these experiments
       takes place right in the field of the exclusion curves.

	It is important to note that without proper knowledge of the
	nuclear and nucleon structure it is not possible to extract reliable and useful information  
	from direct DM search experiments
	(at least in the form of the $\sigma^{n,p}_{{\rm SD}}$ and
        $\sigma^{}_{{\rm SI}}$ cross sections).
	However,  astrophysical uncertainties, in particular the DM distribution in the vicinity of the Earth,
	make it far more difficult to interpret the results of the DM search experiments.
	To have a chance to compare sensitivities of different experiments, people adopted a
	common truncated Maxwellian DM particle distribution but nobody can prove its correctness.
	In the case of undoubted direct DM detection one can make some conclusions about the real
	DM particle distribution in the vicinity of the Earth.

       Furthermore, almost by definition (from the very beginning), 
       a modern experiment aiming at the best exclusion curve 
       is doomed to non-observation of the DM signal. 
       This is due to the fact that a typical expected DM signal 
       spectrum exponentially drops with recoil energy  and it is practically  
       impossible to single it out from the background non-WIMP
       spectrum of a typical  detector. 
       One needs a clear signature
       of interactions between WIMP particles and target nuclei.
       Ideally, this signature should be a unique feature of such an interaction
\cite{Spooner:2007zh}.
       Without these signatures one can hardly convince anybody that 
       the final measured spectrum is saturated by the DM particles.
       Furthermore, on the basis of measured recoil spectra, with the help of these signatures, 
       one can estimate the WIMP mass. 

	Therefore, exclusion curves hardly help to prove an observation of a DM signal.
        Nevertheless {\em an exclusion curve is at least something from nothing observed.}\/
        It allows a sensitivity comparison of different experiments and therefore one can  
        decide who is the best "excluder".
       Unfortunately, the SUSY model paradigm is very flexible, it has a lot of parameters, 
	and one hardly believes that an exclusion curve can ever impose any decisive constraint on it. 
	Over many years the experimental groups 
	compare obtained exclusion curves with SUSY (MSSM, NMSSM, etc) predictions. 
	These comparisons permanently show that some extra domains of relevant parameter
	space are excluded.  
	However, this long-term "exclusion business" has no practical benefit either for SUSY
	model construction, or planning of a new DM search.  	
        The situation is much worse due to the  
        famous nuclear and astrophysical uncertainties involved
        in the evaluation of the exclusion curves. 
        This is why it does not look very decisive to use refined data and methods 
        and spend big resources fighting only for the best exclusion curve  
        (a new information on the subject can be found in
\cite{Catena:2015uua}).
	This fighting could only be accepted when one tries to strongly improve the sensitivity of
	a small DM detector with a view to use many copies of it in a huge detector array with a total tonne-scale mass
\cite{Bednyakov:2008gv}. 

	In fact, all modern very sensitive and very low-background direct DM detectors 
	are aimed at the best exclusion curve from the very beginning. 
	They are looking for some event excess above a very low background, provided it is well understood.   
	If one is the best in seeing nothing, a new best exclusion curve appears. 
 
	Furthermore, if one manages to find some events pretending to be the wanted excess,  
	the powerful statistical likelihood technique allows one to quantify existence 
	of this excess (in terms of 1-3$\sigma$) 
	but does not allow one to prove the real detection of the DM interactions.
	Indeed, if a DM particle with mass $m^{}_{\rm DM}$ 
	exists and can interact with ordinary matter 
	with some not-very-small cross section $\sigma^{}_{\rm DM}$, 
	then an excess induced by these DM particles should be seen 
	(better in a measured recoil spectrum). 
	But the opposite is not correct. 
	Any (or very specifically selected) excess (in the recoil spectrum) is not yet 
	a proof of the existence of the DM-particle interaction with the detector material. 
	It is not enough.
	One needs an unambiguous signature of the galactic nature of the observed interactions. 
	The annual modulation component in the excess 
	events could prove the DM nature of the excess.

\subsection{Positive signatures and recoil spectra}
%%%%%%%%%%%%%%%%%%%%%%%%%
	The problem of DM particle detection is also very complicated due to a lack of 
	any reliable indication where one should look for a signal of the DM interactions --- 
	the DM masses and DM-SM couplings are (in general) unknown. 
	Therefore, to certainly detect a DM particle one has to unambiguously 
	register some "positive signature" of the DM interaction with a target material.
	This positive signature must be very specific (ideally unique) only for the true DM interaction,
	and reliable experimental observation of the signature will serve as 
	unambiguous proof of the DM detection. 
	
	There are some typical characteristics of WIMP DM particle interactions with a nuclear target which 
       	can potentially play the role of these positive DM signatures
\cite{Bednyakov:2008gv}. 
   	First of all, WIMPs produce nuclear recoils, 
   	whereas most radioactive backgrounds produce electron recoils. 
   	Nevertheless, for example, neutrons 
   	(and any other heavy neutral particle) can also produce nuclear recoils. 

   	Due to the extremely rare event rate of the WIMP-nucleus interactions
   	(the mean free path of a WIMP in matter is of the order of a light year),
   	one can expect two features.
   	One is the negligible probability of two consecutive interactions in a single detector or two 
   	closely located detectors. 
   	Multiple interactions of photons, $\gamma$-rays or neutrons under the
   	same conditions are much more common. 
   	Therefore only non-multiple interaction events can be from the DM WIMPs.
   	The other feature is a uniform distribution of the WIMP-induced events throughout a detector.
   	This feature can be used to identify background 
   	events (from photons, neutrons, beta and alpha particles) 
   	in rather large-volume position-sensitive detectors.

    	The shape of the DM WIMP-induced recoil energy spectrum
    	can be predicted rather accurately (for given WIMP mass, 
    	fixed nuclear structure functions, and astrophysical parameters). 
    	The observed energy-recoil spectrum, claiming to be from DM particles, 
    	must be consistent with the expectation. 
    	However, this shape is exponential, right as it is for many background sources.
   	Unfortunately, the nuclear-recoil feature, the non-multiple interaction,
   	the uniform event distribution throughout a detector, 
	and the shape of the recoil energy spectrum 
	could not be the clear ``positive signature'' of the DM interactions.  

\begin{figure}[h!] 
\includegraphics[width=0.8\textwidth]{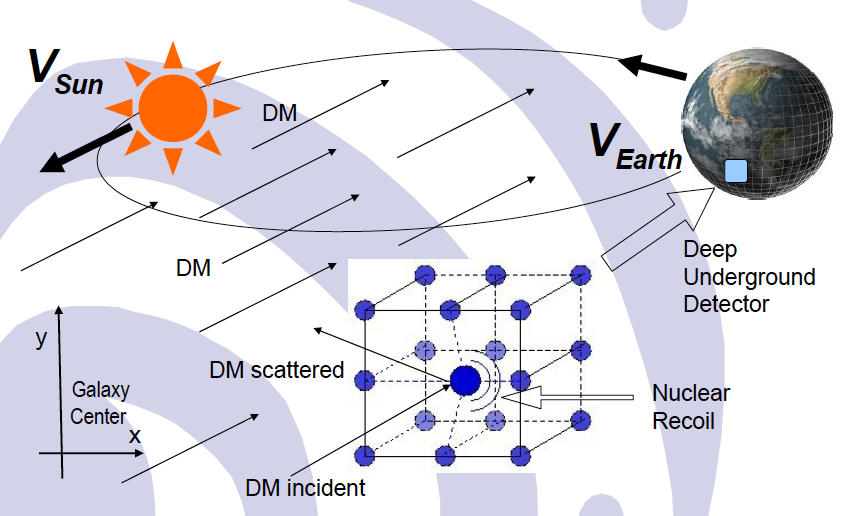}
\caption{Detection of DM particles via elastic scattering from target nuclei in 
a detector located deep underground. Due to the expected annual modulation signature 
of the recoil measured event rate the Sun-Earth system is a particularly proper setup for 
successful direct DM detection \cite{Bednyakov:2008gv}.}
\label{fig:2008gv-DDMD}
\end{figure}

   The currently most promising, technically feasible 
   and already used (by the DAMA collaboration) ``positive signature'' is
   the annual modulation signature
 \cite{Freese:1987wu,Lewin:1996rx,Freese:2012xd}. 
   The DM flux and its average kinetic energy vary annually 
   due to the combined motion of the Earth and the Sun relative to the Galactic center.
   The impact DM energy increases (decreases) when 
   the Earth velocity is added to (subtracted from) the velocity of the Sun
 (see Fig.~\ref{fig:2008gv-DDMD}).
	The recoil spectrum produced from the DM-nucleus scattering in a
	target detector is therefore expected to show this annual modulation effect
\cite{Freese:1987wu,Freese:2012xd}.

      The velocity of the Earth relative to the
      Galaxy is $v_E(t)=v_S+v_O \cos \gamma \cos \omega(t-t_0)$, where
      $v_S$ is the Sun's velocity relative to the Galaxy,  
      $v_O$ is the Earth's orbital velocity around the Sun  
      and $\gamma$ is the angle of inclination of the
      plane of the Earth's orbit relative to the galactic plane.  
      One has $\omega=2\pi/T$ ($T=1$ year) and
      the maximum velocity occurs at day $t_0=155.2$ (June 2).
      The change in the Earth's velocity relative to the incident DM particles 
      leads to a yearly modulation of the scattering event rate 
(\ref{Definitions.diff.rate}) in the form   
$$ S(t) =S_0+S_m \cos{\omega (t-t_0)},
$$
     where $S_0$ is the constant part and $S_m$ is the amplitude of the modulated signal.
     Both of them depend on the target nucleus $(A,Z)$, DM particle mass and    
     density $\rho_{\rm local}^{\rm DM}$, 
    velocity distribution of the DM particles in the solar vicinity $f(v)$, 
     and cross sections of the DM-nucleus scattering
(see, for example, 
\cite{Jungman:1996df,Lewin:1996rx,Bednyakov:1999yr,Bednyakov:1998is} and a new recent paper 
\cite{DelNobile:2015tza}).
 
      In general, the expected modulation amplitude is rather small (about 7\%) 
\cite{Bernabei:2003za,Belli:2011kw,Cerulli:2012dw} 
      and to observe it, one needs huge (at best tonne-scale) detectors 
      which can continuously operate during 5--7 years (to have a chance to observe the annual modulation effect).
      Of course, to reliably use this signature one should prove 
      the absence of annually modulated backgrounds.
      This question is crucially investigated by the DAMA/LIBRA collaboration
\cite{Belli:2011kw,Cerulli:2012dw}.

	Unfortunately, mainly due to not-yet-enough target mass and short running time, the 
	above-mentioned promising LXe based experiments seems to be unable to see a positive 
       annual modulation signature of the DM interactions. 
	Furthermore, to the best of our knowledge, they not ever aimed at fulfilling the goal. 

      Another potentially promising positive DM signature is connected with a possibility 
      of registering the direction of the recoil nuclei induced by a DM particle.
      One plans to measure the correlation of the event rate with the Sun's motion. 
      Unfortunately, the task is extremely complicated (see for example,  
\cite{Vergados:2003pk,Morgan:2004ys,Mohlabeng:2015efa}). 

        The third potentially useful positive WIMP DM signature 
	is related with the coherence of the WIMP-nucleus spin-independent interaction. 
	Due to a rather low momentum transfer, a WIMP coherently scatters by the whole target nucleus and 
	the elastic cross section of this interaction should be proportional to $A^{2}$, where $A$ is 
	the atomic number of the target nucleus.
	Contrary to the $A^2$ behavior, the cross section of neutron scattering by nuclei 
	(due to the strong nature of this interaction)
	is proportional to the geometrical cross-section of the target nucleus 
	($A^{2/3}$ dependence). 
	To reliably use this $A^2$ signature, one has to satisfy at least two conditions.
	First, one should be sure that the spin-independent WIMP-nucleus interaction really 
	dominates over the relevant spin-dependent interaction. 
	This is far from being obvious
\cite{Bednyakov:2008zz,Bednyakov:2004be,Vergados:2004hw}.
         Second, one should rather accurately measure the recoil spectra (in the worst case integrated event rates) 
	 under the same background conditions, at least for two targets with a different atomic number $A$.
	 Currently, this goal looks far from being achievable. 

	At the level of our knowledge the DM problem could not be solved independently 
     	from the set of other related problems, such as proof of SUSY, astrophysical dark matter properties, etc.
     	Furthermore, due to the huge complexity of the DM search (technical, physical, astrophysical, 
     	necessity for positive signatures, etc),   
     	one should, perhaps,  deal with the DM problem boldly using a reliable 
	{\em model-dependent}\/ framework --- 
     	for example the framework of SUSY, where the same LSP neutralino 
     	should be seen coherently (or lead to coherent effects) in all available experimental observables 
     	(direct and indirect DM searches, rare decays, high-energy searches at LHC, astrophysics, etc).
     	A success of the SUSY framework  
	will be a proof of the SUSY existence and simultaneous solution of the DM problem.
     	In some sense, this SUSY framework can serve as a specific and 
	{\em very decisive and positive} DM signature. 

       Ideally, in order to be convincing, an eventual DM signal should combine
        more than one of these positive DM signatures
\cite{Spooner:2007zh}. 

        A physical reason (if one forgets about the above-mentioned competition of experiments) 
	to {\em improve}\/ the exclusion curve is usually an attempt to constrain a SUSY-like model. 
	Unfortunately, this is almost hopeless due to the huge flexibility of 
	these models and uncertainties from the nuclear structure and astrophysics. 
        At the present level of experimental accuracy, 
	simple fighting for the best exclusion curve is almost useless 
	either for real DM detection or for substantial restriction of the models.
       One should go beyond the exclusion curve and
       try to obtain a reliable {\em recoil energy spectrum}.
       Very accurate off-line investigation of the measured spectrum 
       allows one to single out different non-WIMP background
       sources and to perform controllable background subtractions.  
       The spectrum allows one to look for the annual modulation effect, 
       the only currently available positive signature 
       of DM particle interactions with terrestrial nuclei.
       The effect is not simply a possibility of background rejection 
       (among many others as claimed, for example, in 
\cite{Saab:2012th}), it is a {\bf unique signature} 
        which demonstrates the galactic nature of the DM interaction with matter. 
	This is inevitable for the laboratory proof of the DM population around the Earth 
\cite{Bednyakov:2012cu}.

	Finally, the future requirements for 100-t scale direct DM detector, 
	will meet a severe restriction ---
	further sensitivity is strongly limited by an irreducible neutrino background
(Fig.~\ref{fig:SI-overview}), 
	mainly from supernovae, the Sun and neutrinos from cosmic rays 
\cite{Livio:2014gda,Gutlein:2014gma}. 
	In this situation it is clear that to have progress in the DM detection 
	a positive DM signature --- the annual modulation --- 
	it is inevitable to fight against the irreducible $\nu$-background. 
  
 %%%%%%%%%%%%%%%%%%%% 
\subsection{The DAMA/LIBRA evidence}
%%%%%%%%%%%%%%%%%%%%%
        Till now only the DAMA (DArk MAtter) collaboration 
       	has certainly given the evidence (at 9.3$\sigma$ CL) 
	for the presence of DM particles in the halo of our Galaxy.
	The DAMA/LIBRA setups have in use a highly radio-pure NaI(Tl) target during 14 annual cycles
	(corresponding to cumulative exposure of 1.33 tonne$\times$years)
	at the deep underground Gran Sasso National Laboratory 
\cite{Bernabei:2003za,Belli:2011kw,Cerulli:2012dw,Bernabei:2014maa}. 
	The evidence is based on model-independent registration of 
	changes in the flux of DM particles hitting the DAMA setup  ---
	the predicted annual
        modulation of specific shape and amplitude due to the combined motions 
	of the Earth and the Sun through the Galaxy 
\cite{Freese:1987wu}.
	
\begin{figure}[!h] 
\begin{center}\includegraphics[width=0.50\textwidth]{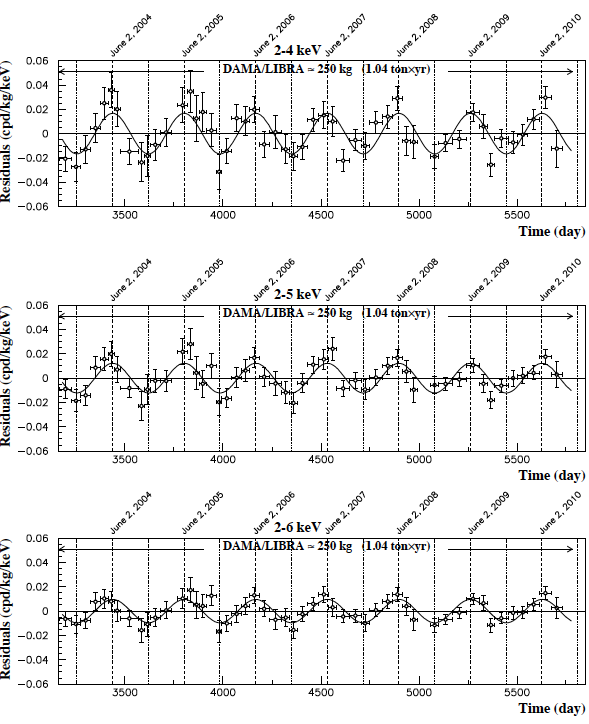}
\end{center}
\caption{Experimental residual rate of the single-hit scintillation events measured
by DAMA/LIBRA--phase1 in the (2--4), (2--5) and (2--6) keV energy intervals 
as a function of the time.  
The data points present the experimental errors as vertical bars and the 
associated time bin width as horizontal bars. 
The superimposed curves are 
$A\cos\omega(t- t_0)$ with a period $T = 2\pi/\omega=1\,$yr, 
a phase $t_0 = 152.5\,$day (June 2nd) and modulation amplitudes, $A$, equal to the central
values obtained by best fit on the data points of the entire DAMA/LIBRA--phase1.
The dashed vertical lines correspond to the maximum expected for the DM signal
(June 2nd), while the dotted vertical lines correspond to the minimum.
From \cite{Bernabei:2014maa}. } 
\label{fig:DAMA1}
\end{figure}

	The DAMA/LIBRA measured modulation amplitude of the single-hit events  
	is (0.0112 $\pm$ 0.0012) counts/day/kg/keV. 
	The measured phase is (144 $\pm$ 7) days and 
	the measured period is  (0.998 $\pm$ 0.002) yr 
(Fig.~\ref{fig:DAMA1}). 
	These values are well in agreement with those expected for true DM particles.
	This intriguing evidence of the DM detection is still very debated,
	simply because of its severe conflict with numerous null-results of 
	almost all other direct DM search experiments
\cite{Schumann:2014uva}.
	People widely discussed many other possible 
	annual phenomena, such as, for example, 
	neutrons leached from the rocks surrounding the 	
	underground experiment in response to seasonal temperature variations.
	Nevertheless, after many years of dedicated investigations of the 
	DAMA collaboration no systematics or side reactions able to mimic 
	the signature (that is, able to account for the measured modulation amplitude and 
	simultaneously satisfy all the requirements of the signature) 
	has been found or suggested
\cite{Bernabei:2013xsa,Bernabei:2014maa}.
 
	For completeness one should point out that indications on the 
	similar annual cycles, consistent in phase with DAMA observations, 
	were also seen (with much low statistical significance) 
	by the CoGeNT (Coherent Germanium Neutrino Technology) 
\cite{Aalseth:2011wp,Aalseth:2012if}, 
	CRESST-II \cite{Angloher:2011uu}, and CDMS-Si \cite{Agnese:2013rvf} 
	DM-search experiments
\cite{Schumann:2015wfa}.

	Despite the strong and reliable belief of the DAMA collaboration 
	in the observation of the annual modulation signature,  
	it is obvious that such a serious claim
	should be verified by at least another one completely independent experiment.  
	If one wants to confirm (more important, if one wants to reject) the DAMA result, 
	one should perform a new experiment which would have 
	the same or better sensitivity to the annual modulation signature.
	Furthermore, it would be reasonable  to locate a new 
	setup in another low-background underground laboratory, and
	perhaps, in the Southern Hemisphere, where this new 
	DM-modulation experiment 
	would gauge the extent of the Earth's seasonal effects, 
	which would be out of phase relative to ones in the Northern Hemisphere 
\cite{Livio:2014gda}.

{\em Finally}, 
	one can believe or cannot believe in the DAMA/LIBRA DM evidence, 
	but to have a final scientific decision on the subject, 
	one should prove experimentally the non-existence of the annual modulation effect 
	observed by the DAMA/LIBRA collaboration at the 9$\sigma$ level.
	To this end one should first {\em be able to ever see the effect in principle}, 
	to prove the proper sensitivity of a used or proposed setup. 	
%%%%%%

%%%%%%%%%%%%%%%%%%%%%%%%%%%%%%%%%%%%%%%%%%%%%%%  
\section{How can one estimate the prospects for the DM detection?}\label{sec:EstimateDM}  
%\section{How one can estimate the prospects for the DM detection?}
%%%%%%%%%%%%%%%%%%%%%%%%%%%%%%%%%%
	Before designing a future DM detector one reasonably wants to have a feeling concerning  
	how many events one could register during meaningful time of measurements with this detector
	(the expected event rate, section \ref{sec:DirectDetectionDM}). 
	A source of this information can be one of many recently appeared DM-models 
(section \ref{sec:ExoticsDMsearches}), or available experimental and astrophysical data.
	In the latter case one again should have a model that could simultaneously describe 
	all the data and could give predictions for the DM rate 
	via a unique set of parameters (as a SUSY-model can do). 
	
	Rather thorough investigations of prospects for direct DM detection have been carried out on the border of 
	millennia on the basis of different versions of SUSY extension of the SM (MSSM, mSUGRA, NMSSM, etc).
	The main reason was the LSP --- stable, neutral, massive, weakly-interacting particle --- 
	an excellent (by-product) DM candidate in SUSY models with R-parity conservation. 
	There are several main possibilities for the LSP in the MSSM --- neutralino, gravitino, sneutrino, etc. 
	The neutralino LSP was a particularly well-studied case (see, for example, 
\cite{Berlin:2015aba}).

	The modern SUSY framework is very appropriate for the goal.
	 It allows one to describe simultaneously (with one set of SUSY parameters) 
	 all available experimental observables from very low up to very high energies.
	 Therefore expected direct DM detection rates can be rather easily connected 
	with probabilities of rare processes, results of high-energy measurements  (LEP, FNAL),  
	cosmological observations and restrictions on relic DM abundance (density).  

	For example, one can investigate the MSSM parameter space
	taking into account available accelerator, non-accelerator and cosmological constraints
	on the SUSY particle spectrum (upper bounds on their masses).
	On this basis one can obtain a reliable prediction for the expected SUSY DM detection rate
	in any high-accuracy DM detector (such as one with LXe). 
	Non-observation of the relevant DM candidates allows one
	to exclude some domains of the MSSM parameter space and make, for instance,  
	predictions for charged Higgs boson search with a collider 
\cite{Bednyakov:1999vh}. 
	Furthermore, in this approach the MSSM Higgs bosons 
 	produced in the LSP-neutralino self-annihilation in the Earth and the Sun 
	could severely contribute to the total indirect DM detection rate.
	This contribution results in some lower bound for the muon flux from the Sun  
	and one can expect an energetic $\tau$-neutrino flux from the Sun at
	a level of $10^2$ m$^{-2}$ yr$^{-1}$ due to these charged Higgs bosons 
\cite{Bednyakov:2000uw}.

	Nevertheless, at that time the high-energy colliders and accelerators were not yet considered 
	(perhaps, due to relatively low energy, etc)
	as useful tools for "direct" collider DM search. 
	 
	The unique LHC energy and the SUSY paradigm 
	have opened new possibilities for connection of 
	LHC experiments with the other DM searches. 
	One can use available results from one kind of these experiments 
	for goal-pointing for another one by means of 
	some important hints on where one should search for a signal. 
	
	An example can be found in 
\cite{Bednyakov:2009zz}
	where prospects for observation of a SUSY-like signal from two gluinos ,
	were investigated within a certain region of the mSUGRA parameter space.
	In the region the lightest neutralinos $\chi_{1}^{0}$ are DM particles with $m_{\chi_1}\simeq 60$ GeV$/c^2$ 
	and naturally explain the excess of diffuse galactic $\gamma$-rays	
	observed by the EGRET space apparatus. 
	Additionally the cross-section of the two gluino production in $pp$-collisions
	at the LHC (14~TeV) was estimated at a level of 5--10~pb. 
	Rather high transverse missing energy carried away by the two $\chi_{1}^{0}$
	is the essential signature of these events that allows significant background reduction. 
	Furthermore, distributions of the invariant masses of two opposite
	charged lepton pairs produced by the $\chi_{2}^{0}\rightarrow\chi_{1}^{0}l^{+}l^{-}$
	three-body decays have kinematic endpoints which measure the difference
	between the masses of $\chi_{2}^{0}$ and $\chi_{1}^{0}$.
	These signatures still demonstrate good prospects for discovery of the gluinos at the LHC. 

	The same SUSY-based interplay between the direct DM search and the LHC search for DM 
	was considered earlier, for example in 
\cite{Eifert:2008zza} and in 
\cite{Polesello:2004qy}. 	
 	In the latter paper, under an assumption that R-Parity conserving SUSY
	 is already discovered at the LHC, a question of the LSP to be a real DM particle was addressed. 
	 In particular, the consistency of the observed SUSY realization (mSUGRA as an example) 
	 with astrophysical  and non-accelerator constraints was studied  
	 and requirements for statistical and systematic accuracy of LHC measurements 
	 were formulated quantitatively.  
	A discussion was given in 
\cite{Barberio:2009xca} how one can constrain the underlying SUSY model 
	and hence extract information about the nature of DM particles on the 
	basis of techniques used to reconstruct decays of SUSY particles
	assumed to be observed at the LHC. 
  
	Recently a question of DM complementarity was studied in the 19-parameter 
	phenomenological Minimal Supersymmetric Standard Model
	(pMSSM) including all available experimental constraints
\cite{Ismail:2014fca}. 
	 The ability to investigate the neutralino DM properties
	 by the direct DM experiments XENON1T and LUX-ZEPLIN,  
	 by indirect DM searches with Fermi-LAT, CTA and IceCube, as well as by the LHC studies  were examined. 
	 In particular, it was shown that  expected LHC constraints on the pMSSM models are 
	directly sensitive to many other sparticles beyond the LSP. 
	Other sensitive tests of the pMSSM models are spin-independent 
	direct detection with LUX-ZEPLIN and indirect detection through CTA. 
	It was found that all discussed DM search techniques 
 	work well together within the MSSM and  
	on this basis form a comprehensive program to determine DM properties
\cite{Roszkowski:2014iqa}.

	Another investigation of the prospects of indirect and direct DM searches 
	within the MSSM approach with 9  parameters (MSSM-9) was reported in 
\cite{Catalan:2015cna}. 
	A Monte Carlo scan of the parameter space was performed with allowance for  
	all available particle physics constraints including the Higgs mass of 126 GeV$/c^2$.  	
	Two regions for DM were found with a TeV mass  LSP neutralino. 
	Prospects for future indirect (with the CTA) and direct
	(with XENON-1T) searches of these LSP were discussed and some search strategies were proposed
\cite{Catalan:2015cna}. 
	Phenomenological study of the CMSSM (mSUGRA)  with non-thermal neutralino DM candidate was carried out in
\cite{Aparicio:2015sda}. 

	Contrary to the MSSM in a GUT-defined SUSY model their parameters are no longer independent.  
	The LHC prospects to probe a broad class of GUT-inspired SUSY models were studied in 
\cite{Kowalska:2015zja} with all available experimental constraints including 
	the bounds from the muon $g$--2 anomaly, the DM relic density and the Higgs mass measurement.
	
	Some version of the pMSSM was also used in 
\cite{Caron:2015wda} 
	for explanation of an excess in $\gamma$-rays from the center of
	our Galaxy due to DM annihilation. 
	The LSP neutralinos with mass about 90 GeV$/c^2$ 
	 provide a reasonable description of the excess, 
	 have correct relic density  and are consistent with the other astroparticle and collider experiments.
	If the pMSSM explanation of the excess seen by Fermi-LAT is correct, 
	a DM signal might be discovered soon --- claimed the authors of 
\cite{Caron:2015wda}. 	
	
	Prospects of the detection of a GeV-scale neutralino as DM 
	in the Next-to-Minimal Supersymmetric Standard Model (NMSSM) at the 14 TeV LHC
	were studied in 
\cite{Han:2015zba} by means of dedicated scans 
	of the relevant parameter space including all available constraints. 
	It was demonstrated in  
\cite{Gherghetta:2015ysa} that with a neutralino as DM candidate, 
	the gamma-ray excess observed in the Fermi-LAT data
	cannot be accommodated in the MSSM. 
	To reach the goal one needs the NMSSM with an extra singlet superfield 
	and specific collider phenomenology. 

	There are some other investigations of DM detection prospects within SUSY-like models. 
	For example, one believes 
\cite{Arina:2015uea} that in some SUSY models 
	a right-handed sneutrino can be LSP and can play the role of a good DM candidate
\cite{An:2011uq,BhupalDev:2012ru,Banerjee:2013fga}.
	Being a scalar particle this LSP will have the LHC signatures 
	quite distinct from those one expects for the LSP-neutralino.
	Constraints on the sneutrino-LSP scenario were studied on the basis of general results of 
	SUSY searches at RUN I of the LHC and most promising sneutrino signatures were proposed for 
	further searches at RUN II.
	Phenomenological constraints on a light GeV-scale sneutrino as a DM particle were also investigated in  
\cite{Kakizaki:2015nua}.
	How to find a "natural" supersymmetry via the interplay
       between the LHC and direct DM detection was discussed in
\cite{Barducci:2015ffa}.

	Therefore, the "indirect" constraints on DM properties from collider physics were studied and used 
	during many years. 
	One of recent examples can be found also in 
\cite{Akula:2011dd},  
	where constraints on DM from the first CMS and ATLAS supersymmetry 
	searches are investigated. 
	It was shown that within the minimal supergravity model (mSUGRA), 
	the early search for SUSY (superpartners) at the LHC excluded a  
	remarkable portion of the parameter space available for DM direct detection experiments. 
	In particular, the prospects for detecting the SUSY DM (being the neutralino LSP) 
	in the XENON and CDMS 
	experiments are significantly affected in the low LSP mass region. 
	 In the case of non-universal soft breaking the regions excluded by minimal SUGRA, 
	 are not excluded at all. 
	 The authors allow an optimistic 
	 conclusion that an observation of DM (by direct detection) 
	 in the LHC excluded regions of mSUGRA might indicate non-universalities 
	 in soft breaking SUSY. 	 
	 Another important example of modern LHC influence on prospects of DM direct and
	 indirect detection within SUSY can be found in a recent paper of Pran Nath
\cite{Nath:2015dza}. 
	This is a canonical way of collider data usage for direct/indirect DM search studies.

	These few examples show that a properly developed model, 
	especially the one like supersymmetric MSSM, mSUGRA, or NMSSM, 
	can accurately unify information from all available energy scales and experiments and 
	produce constraints or predictions for each of   
	the DM search approaches listed in section \ref{sec:DM-detections}.
	Therefore it is very reasonable to have in mind the 
	SUSY-like background under the {\em complementarily (or interplay)}\/   
	of all above-mentioned DM search opportunities.  	
%%%%%%
   % EstimateDM.tex

%t%%%%%%%%%%%%%%%%%%%%%%%%%%%%%%%%%%%%
\section{First "direct" DM search with Tevatron}\label{sec:FirstColliderDM}  
%\section{First "direct" DM search with Tevatron}
	Due to exciting results of the Tevatron and big expectations with the LHC, about 5 yeas ago
	a "direct" approach to the DM problem at colliders was proposed 
\cite{Goodman:2010yf} and rapidly accepted  
	by {\em the collider} community
\cite{Bai:2010hh,Goodman:2010ku,Fox:2011pm,Busoni:2014uca,Busoni:2013lha}. 
	The main goal was to relate the pair production rate of DM candidates at colliders 
	to the annihilation and scattering rates at direct and indirect DM experiments. 
	It was believed that on this way (see, for example,
\cite{Goodman:2010yf,Goodman:2010ku})
	 as few assumptions as possible should be made about unknown underlying new physics. 

	A true DM particle (section \ref{sec:GalacticDM}) is electrically
	neutral, massive enough, weakly interacting (WIMP), and stable. 
	Therefore, the WIMP feature of the DM particle defines the very strategy 
	of "direct" DM search with a collider.
	WIMPs, once produced, like neutrinos
	do not interact either electromagnetically or strongly 
	with ordinary matter and pass numerous detectors layers without a trace.
	Their signature is invisibility. 
	However the experiments ATLAS and CMS were designed to see this invisibility 
	by means of an accurate measurement of energy deposited in the detectors by all
	other "visible" products of a hard LHC $pp$-collision.
	Hence, exploiting this hermeticity, one can judge the WIMP presence 
	from the disbalance of the measured energy/momentum. 
	Since the longitudinal momenta of the colliding partons are unknown, 
	only the transverse missing energy, \met\/, can be reliably used to search for 
	the WIMP traces.
	In other words, hunting for WIMPs
	one should look for events with remarkable momentum imbalance, 
	which is transverse to the initial proton beam line
\cite{Lowette:2014yta}.

	The collider searches for DM have their own advantages and disadvantages
\cite{Mitsou:2014wta,Askew:2014kqa}.
	Obviously, a collider search for WIMP does not suffer from astrophysical uncertainties.
	{\em The search does not care about the existence of the Milky Way at all.}\/ 
	Due to the famous low-recoil problem (small energy deposition) sensitivity of 
	direct and indirect detection techniques almost 
	vanishes with reduction of the WIMP mass.
	In contrast, colliders are able to copiously produce the light WIMPs 
	(if production cross section is not very small),
	and one has no problem with sensitivity up to very low-mass WIMPs
	(if one finds a way to suppress neutrino background in this case).
  	However, collider searches suffer from parton distribution function suppression
	for high WIMP masses (above hundreds of GeV), 
	where the other two search techniques are more robust. 

	Already at this stage a very important precaution  
\cite{Mitsou:2014wta,Askew:2014kqa} says that a great disadvantage of any collider DM search
	 is absence of any possibility of proving whether the observed \met\ signal is actually 
	 caused by the true DM particle or produced by a particle that is stable only on collider timescales, 
	but not cosmological ones. 
	The other techniques deal with the true DM particles produced long ago.
	Unfortunately this precaution almost always is forgotten or ignored 
	(see discussion in section \ref{sec:Discussion}).

\smallskip
	It looks like that the first quantitative discussions of a "direct" collider search for DM 
	were based on the first mono-jet (+ large \met) results of the CDF at the Tevatron 
	and were given in \cite{Bai:2010hh}  and in \cite{Goodman:2010yf,Goodman:2010ku}.
	These papers are rather remarkable. 
	First, they have started to use the Effective Field Theory (EFT) approach in the subject, 
	and next, from the very beginning many very serious assumptions have been 
	used without any (or reasonable) proof.
	
	Everybody will agree that true DM particles might be created at a collider.	 	   
	The very idea that direct DM detection requires an interaction of DM 
	particles with ordinary matter (nucleons, nuclei, etc) is correct. 
	But the point following intuitively from 
Fig.~\ref{fig:2013ihz-DM-interplay}, that  {\em the same interaction} can lead to the 
	DM particle pair production at a hadron collider 
\cite{Bai:2010hh} far to be obvious. 
	It needs to be well justified.   
	Furthermore, in general it is wrong within such a complete theoretical model as SUSY.
	For example, in a R-conserved SUSY model the DM WIMPs are LSPs 
	and usually constitute final products 
	of cascade decays of heavier unstable SUSY-particles 
	accompanied by SM particles, in particular, 
	with high transverse momentum $p^{}_{\rm T}$.
	Nevertheless as soon as one assumes that 
	{\em the same couplings lead to direct DM particle production at hadronic colliders} 
	such as the Tevatron, one {\em can}\/ 
	investigate the interplay between the two DM searches (within the EFT).
	One should also bear in mind that although the basic processes 
	that work in direct detection and in collider production of DM  
	occur through $s$- and $t$-channel exchange of the same mediator, 
	the regimes probed in the two types of the experiments are very different
\cite{Bai:2010hh}. 

	Due to the assumed DM nature (electric neutrality and stability), 
	these WIMPs will leave any detector tracelessly, otherwise they could be detected 
	(by ionization, interaction, or decay) and do not have a chance to serve as DM. 
	Therefore signature of such escaping particles is (large) missing energy and momentum.
	More strictly, the latter is missing vector transverse momentum, 
	$p^{\rm miss}_{\rm T}$, the magnitude of which is called \met\/
\cite{ATLAS:2012ky}. %%%%%%%%%%%%%%%%%%%%%%%%%%%%%%%%%%%%	
	Unfortunately, only large transverse missing energy \met\/ could be measured.
	To catch the signature one should search for visible particles recoiling against the WIMPs 
	and triggering a relevant DAQ system.

	This signature can be used to set constraints on the WIMP couplings to the 
	constituents of nuclei, which in turn can be translated to constraints 
	on direct detection cross sections
\cite{Goodman:2010ku}.

	The CDF collaboration has performed a search for one-jet events with large missing transverse
	energy ($\bar{p}p\to j+ \chi\chi +X$) using 1\,fb$^{-1}$ of data
\cite{Aaltonen:2008hh}. .
Here $\chi$ denotes the WIMP.
	These mono-jet searches at the Tevatron within the above-mentioned assumptions 
	can be connected (via the same effective operators) to the DM direct detection searches
	and can place limits on the expected rates of the latter. 
		
	 The analysis performed in 
\cite{Bai:2010hh} showed for the first time 
	that in many cases the Tevatron provided the best limits, 
	particularly for light WIMPs with mass below 5 GeV$/c^2$, 
	and for SD WIMP-nucleon  interactions. 
	The bounds on the strength of the various effective operators were translated
	into bounds on direct detection rates. 
	This enables one to plot the Tevatron limits --- exclusion curves --- 
	in the $\sigma^{\rm SI, SD}_{\rm DM}$ -- $m^{}_{\rm DM}$ planes
(as in Figs.~\ref{fig:SI-overview} and \ref{fig:SD-overview} from section 
\ref{sec:DirectDetectionDM}). 
	The relevant figures can be found in  
\cite{Bai:2010hh}.

	In general these results confirmed the main expectations that collider bounds 
	could be more promising when DM direct detection scattering is suppressed,
	for example, by kinematics due to (very) light DM mass, 
	or due to non-coherent WIMP-nucleus SD  interactions.
	Nevertheless, with this good result, the authors of  
\cite{Bai:2010hh} have mentioned that an introduction of a light mediator 
	significantly weakens the collider bounds. 
	Furthermore, any direct detection discovery that is in apparent conflict with
	obtained mono-jet limits will thus point to a new light state coupling 
	the SM to the dark sector
\cite{Bai:2010hh}. 
	Some suspicions appear after these words concerning usefulness of 
	the Tevatron constraints for direct detection experiments.
	For example, if one plans a new direct DM search experiment with 
	ambitious goals, he should take into account the already achieved results, 
	especially those, which teach one where is nothing to see. 
	The Tevatron limits excluded some regions of DM properties, 
	but, nevertheless, the authors of the results openly allowed a direct DM
	experiment to look for signal in the excluded region. 	
	
\smallskip
	Most complete theoretically justified investigations of the collider limits for the DM properties 
	were given in 
\cite{Goodman:2010yf,Goodman:2010ku}.	
	Models, where a WIMP DM candidate $\chi$  is a  
	fermion or a scalar interacting with quarks and/or gluons, 
	 were analysed within the EFT. 
	The authors assumed that {\em the WIMP is the only new particle} in the energy 
	ranges relevant for current experiments.
	(It looks like a very strong assumption, because in almost all reliable BSM theories it is not true).
	Next WIMPs appear in pair only. 
	Under this assumption, the WIMP will couple to the SM particles through 
	higher dimensional operators in the EFT,  
	presumably mediated by particles of the dark sector which are 
	somewhat heavier than the WIMP itself. 
	The WIMP is assumed to be a singlet under the SM gauge groups, 
	and thus possesses no tree-level couplings to the electroweak gauge bosons. 

	Contrary to only 4 effective operators discussed in 
\cite{Bai:2010hh} a complete list of 24 operators was considered in 
  \cite{Goodman:2010ku}. 
	The operators for Dirac fermions and scalars fall into six categories 
(Table~\ref{tab:2010ku}) with characteristic \met\ spectral shapes
\cite{Aad:2015zva}. 
%%%
\def\g{\gamma}
\begin{table}[h!]
\caption{Operators coupling WIMPs, denoted as $\chi$, to SM particles. 
The operator names beginning with D, C, R  apply to WIMPs that are Dirac fermions, 
complex scalars or real scalars respectively.
From \cite{Goodman:2010ku}. }
\begin{center}
\begin{tabular}{|l|c|l|c|l|c|} \hline
Name&  Operator &Name&  Operator &Name&  Operator\\  \hline
D1 & $m_q\, \bar{\chi}\chi\bar{q} q /M_*^3$  & 
D9 & $\bar{\chi}\sigma^{\mu\nu}\chi\bar{q}\sigma_{\mu\nu} q /M_*^2$  &
C3 &  $\chi^\dagger\partial_\mu\chi\bar{q}\g^\mu q /M_*^2$   \\
D2 & $im_q\, \bar{\chi}\g^5\chi\bar{q} q /M_*^3$ &   
D10 & $i\, \bar{\chi}\sigma_{\mu\nu}\g^5\chi\bar{q}\sigma_{\alpha\beta}q /M_*^2$  &
C4 &  $\chi^\dagger\partial_\mu\chi\bar{q}\g^\mu\g^5q /M_*^2$    \\
D3 & $im_q\, \bar{\chi}\chi\bar{q}\g^5 q /M_*^3$  &
D11 & $\alpha_s\, \bar{\chi}\chi G_{\mu\nu}G^{\mu\nu} /4M_*^3$  &
C5 & $\alpha_s\, \chi^\dagger\chi G_{\mu\nu}G^{\mu\nu} /4M_*^2$   \\
D4 & $m_q\, \bar{\chi}\g^5\chi\bar{q}\g^5 q /M_*^3$ &
D12 & $i\alpha_s\, \bar{\chi}\g^5\chi G_{\mu\nu}G^{\mu\nu} /4M_*^3$ & 
C6 & $i\alpha_s\, \chi^\dagger\chi G_{\mu\nu}\tilde{G}^{\mu\nu} /4M_*^2$   \\ 
D5 & $\bar{\chi}\g^{\mu}\chi\bar{q}\g_{\mu} q /M_*^2$  &
D13 & $i\alpha_s\, \bar{\chi}\chi G_{\mu\nu}\tilde{G}^{\mu\nu} /4M_*^3$  &
R1 & $m_q\, \chi^2\bar{q}q /2M_*^2$    \\
D6 & $\bar{\chi}\g^{\mu}\g^5\chi\bar{q}\g_{\mu} q /M_*^2$ &   
D14 & $\alpha_s\, \bar{\chi}\g^5\chi G_{\mu\nu}\tilde{G}^{\mu\nu} /4M_*^3$  &
R2 & $\chi^2\bar{q}\g^5 q$  $im_q/2M_*^2$   \\
D7 & $\bar{\chi}\g^{\mu}\chi\bar{q}\g_{\mu}\g^5 q /M_*^2$  &
C1 & $m_q\, \chi^\dagger\chi\bar{q}q /M_*^2$ &  
R3 & $\alpha_s\, \chi^2 G_{\mu\nu}G^{\mu\nu} /8M_*^2$   \\
D8 & $\bar{\chi}\g^{\mu}\g^5\chi\bar{q}\g_{\mu}\g^5 q /M_*^2$  &
C2 & $im_q\, \chi^\dagger\chi\bar{q}\g^5 q /M_*^2$  &  
R4 & $i\alpha_s\, \chi^2 G_{\mu\nu}\tilde{G}^{\mu\nu} /8M_*^2$   \\
 \hline
\end{tabular}
\end{center}
\label{tab:2010ku}
\end{table}%
	For each operator the 
	parameter $M_*$ can be determined (or constrained) from comparison of (initiated by this operator) 
	WIMP-pair hadro-production ($pp, p\bar{p}\to \chi\chi+X$) with relevant measurements. 

	This set of high dimensional contact operators defines the 
	Effective Field Theory (EFT) description of the WIMP-hadron interactions.
	It is a nonrenormalizable field theory and it breaks down at some energy scale 
	represented by the masses of those particles which have been integrated out. 
	In the operator definitions, $M_*$ is the suppression scale of the 
	heavy mediator particles that are integrated out. 
	The quantities $M_*$ which characterize the interaction strength of the interactions are 
	functions of the masses and the coupling strengths of the mediating particles to WIMPs 
	and SM fields and can be computed in terms of the fundamental parameters 
	for any full-scale theory
\cite{Goodman:2010ku}.

\begin{figure}[h!] 
\includegraphics[width=0.6\textwidth]{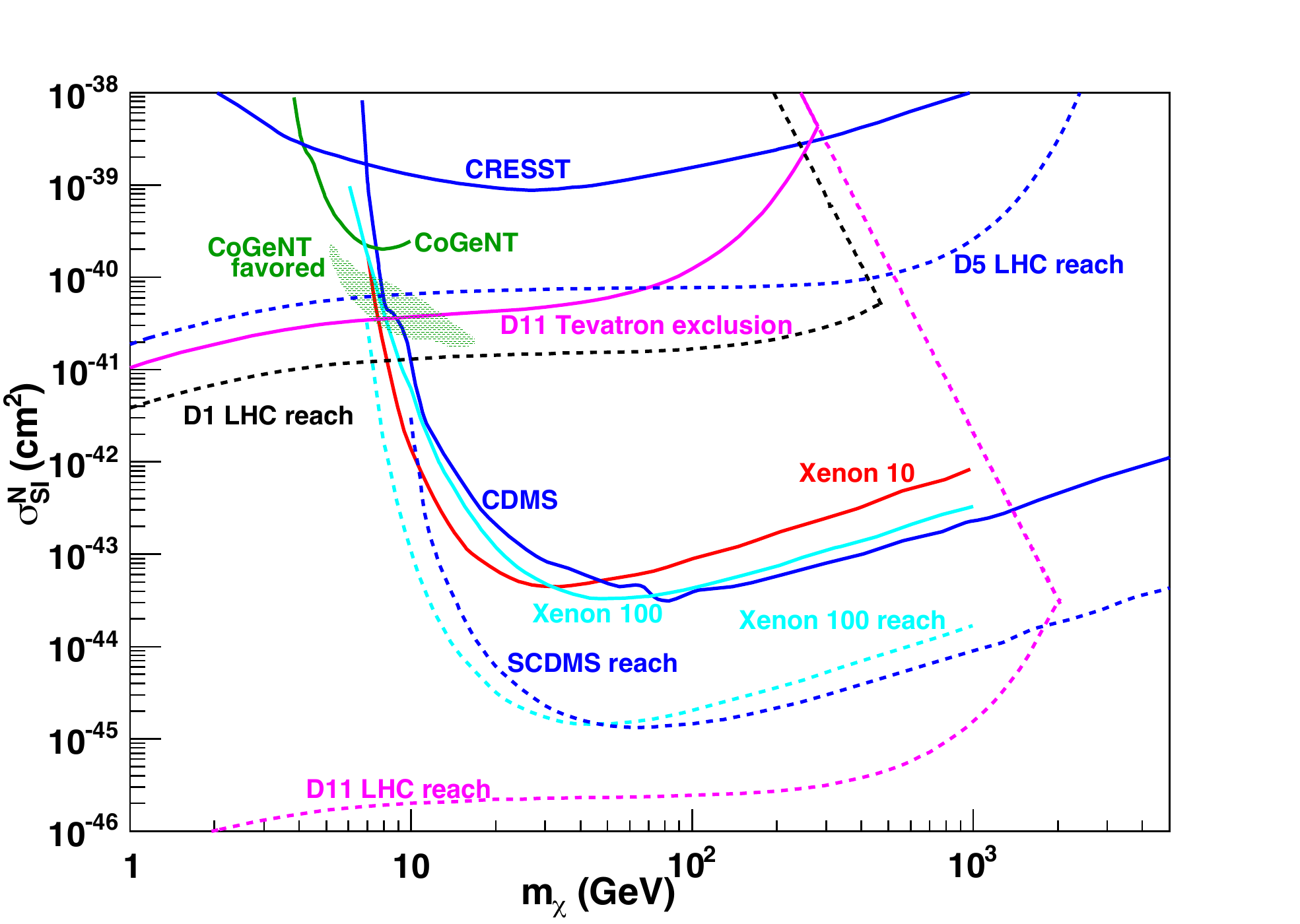}
\caption{\label{fig:DSIplot}
Experimental limits on spin-independent WIMP direct detection 
from several direct DM search experiments 
\cite{Angloher:2002in,Ahmed:2009zw,Angle:2007uj,Aalseth:2010vx,Aprile:2010um,Akerib:2006rr,Aprile:2009yh} 
in comparison with the Tevatron exclusion curves (for the operator D11 --- solid magenta line) 
and LHC discovery reaches (dashed lines) for relevant operators.  
Figure from \cite{Goodman:2010ku}. }
\end{figure}

\begin{figure}[h!]
\includegraphics[width=0.6\textwidth]{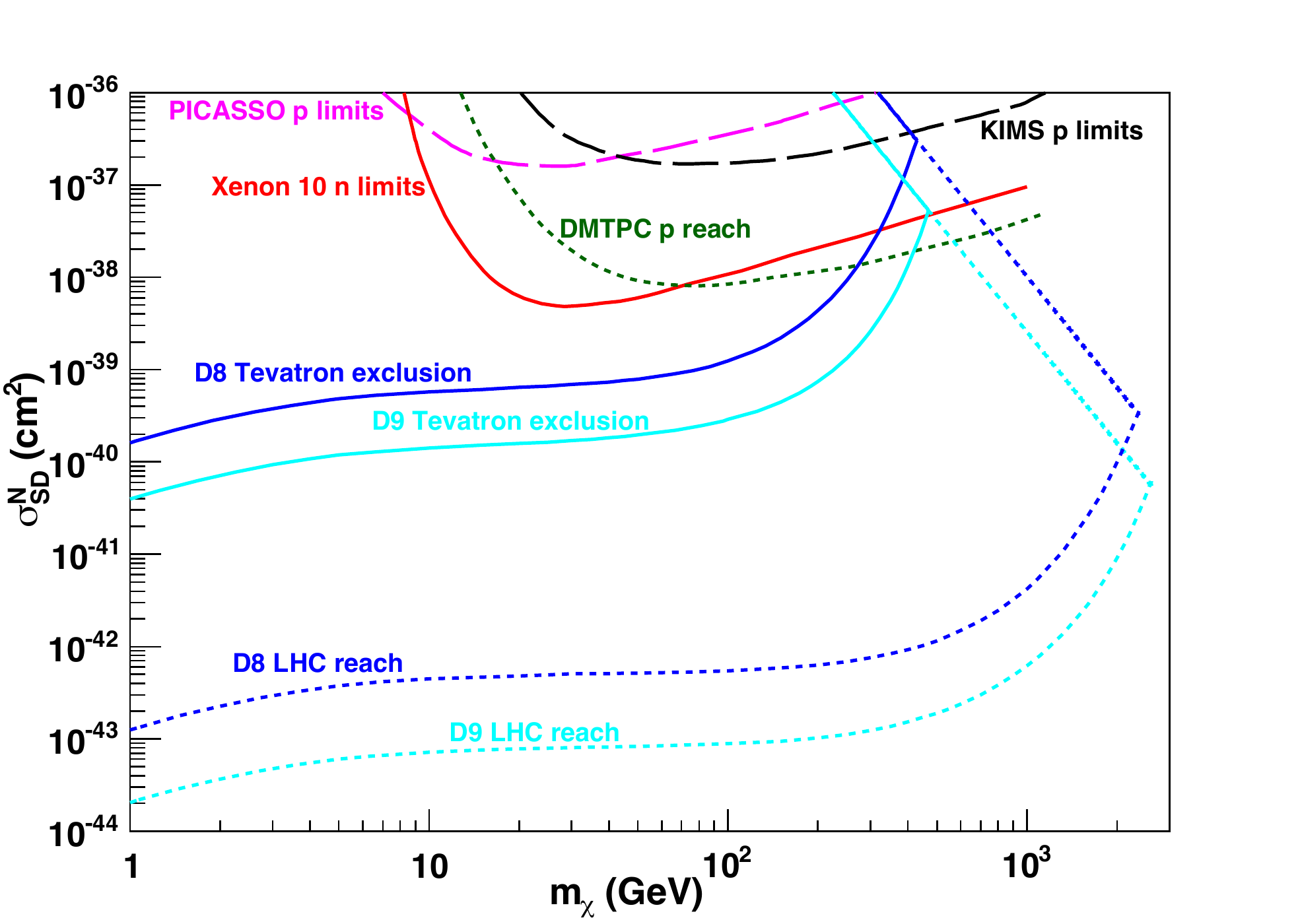}
\caption{\label{fig:DSDplot}  
Experimental limits on spin-dependent WIMP direct detection from several direct DM search experiments  
 \cite{Archambault:2009sm,Lee.:2007qn,Angle:2007uj,Sciolla:2009fb}   
in comparison with the Tevatron exclusions curves (solid lines) and 
LHC discovery reaches (dashed lines) for relevant operators.
Figure from \cite{Goodman:2010ku}.}
\end{figure}

	The authors focused on the mono-jet event search at the Tevatron, 
	where the WIMPs recoil against a single jet ($\bar{p}p\to \chi\chi+j+X$), 
	with restrictions on any additional SM radiation.
	These mono-jet searches were used to determine the constraints 
	on the coefficients of the effective operators as a function of the WIMP mass.
	Many relevant figures are given in
\cite{Goodman:2010ku}.	 
	The bounds on the strength of effective-operator interactions of 
	WIMPs with hadrons were translated into constraints on the possible 
	contributions to direct detection cross sections 
	for each of those operators in the form of numerous exclusion curves 
	(for each operator from 
Table~\ref{tab:2010ku}).
	In the $\sigma^{}_{\rm DM}$--$m^{}_{\rm DM}$ planes
(Figs. \ref{fig:DSIplot}--\ref{fig:DSDplot})
	these curves are superimposed with the relevant direct detection exclusion curves
	for comparison (and competition).  

	As authors 
\cite{Goodman:2010ku} noted, in all cases colliders can probe regions of very light WIMP masses more effectively than direct detection experiments. 
 	Furthermore, it was shown that for many considered operators 
	the direct detection rates are expected to be very small because of the 
	velocity suppression, and colliders become the only way to effectively probe 
	such kind of possible WIMP-hadron interactions.
	The question concerning applicability of these operators for true DM 
	interaction still remained unanswered.     
	In the case of a WIMP whose dominant recoil is through a SD interaction, 
	collider constraints are already much stronger even than the expected 
	reaches of near-future direct detection experiments. 

\smallskip
	The important question on validity of the EFT approach was discussed already in
 \cite{Goodman:2010ku}.
 	Obviously, the validity of EFT itself and what happens above the regime of its validity 	
	depends on the underlying complete model, and therefore is inevitably model dependent. 
	The specifics of a complete model makes collider bounds stronger or weaker.
	In some sense the fact reincarnates the model-dependence in the usage of EFT. 
	To allow the EFT approach, 
	some lower bound for any $M_*$ should exist depending on a given WIMP mass.
	These constraints are shown in 
Figs. \ref{fig:DSIplot}--\ref{fig:DSDplot} as straight lines for large $m_\chi$. 

	People believe 
\cite{Gramling:2013wra} the EFT approach allows a valuable comparison of complementary results on DM detection. 
	If any of the experiments sees a signal, the interpretation in this approach can lead to further insights into the 
	nature of DM as well as the underlying physics by a comparison of different techniques and observables.

	In general, the EFT approach is considered conservative, 
	but in regimes where the validity might be questioned the cross section is mostly underestimated, 
	compared to the full theory, which leads to more conservative limits
\cite{Goodman:2010ku}.
	Finally, the authors wrote {\em that while effective theories may not always capture our 
	favorite parameters of our favorite complete models, 
	they do provide a language to describe WIMP-SM interactions which captures a wide 
	class of theories in a fairly model-independent fashion}
\cite{Goodman:2010ku}. 

	A reader could ask --- for what? 
	Being sooner or later detected, the DM particle must be implemented into one of our 
	(perhaps, new) "favorite complete models" like, for example, SUSY. 
	It is not clear how this "fairly model-independent fashion" could help. 

\smallskip
	{\em Concluding this section} one can first say that the competition (comparison) 
	between direct DM and collider DM search experiments {\em has started to run} 
	only in the field of exclusion curves. 
	Unfortunately this competition helps almost nothing on the way to real DM detection
(section \ref{sec:DirectDetectionDM}). 
	Next, one can point out that a huge bulk of publications on the subject appeared 
	after the famous papers
\cite{Goodman:2010yf,Bai:2010hh,Goodman:2010ku}.
	A start of the "direct" DM search at the LHC was given, 
	followed by many proposals for very exotic DM searches with other accelerators, 
	and not only (section \ref{sec:ExoticsDMsearches}).

	And finally, speaking about DM search or {\em ever detection with a collider}\/ one 
	must not forget that a WIMP {\em is not yet} a DM particle.
	As nicely mentioned recently in 
\cite{Bauer:2013ihz,Cushman:2013zza,Gelmini:2015zpa},  
{\em the discovery of a DM signal at particle colliders only establishes the production 
of a particle with lifetime greater than about 100 ns. 
The assumption that this particle contributes to DM requires an extrapolation in 
lifetime of 24 orders of magnitude! 
It is only by corroborating a particle collider discovery through another method that 
one can claim that the collider discovery is relevant for cosmology.}
Colliders cannot say anything about the stability of WIMPs 
which is the essential property of the true DM particle
\cite{Diehl:2014dda}. 
Therefore only the direct DM detection can play the  
decisive role in the DM problem (section \ref{sec:Discussion}).

%%%%%% 

%%%%%%%%%%%%%%%%%%%%%%%%%%%%%%%%%%%%%%
\section{ATLAS results on the DM search at LHC}\label{sec:DM-ATLAS-searches}
%\section{ATLAS results on the dark matter search at LHC}
%%%%%%%%%%%%%%%%%%%%%%%%%%%%%%%
	As already discussed in section 
\ref{sec:FirstColliderDM}, the missing large transverse energy, \met, 
	(caused by the escaping GeV-scale-mass WIMPs) is the key  
	signature on which the idea of the LHC search for DM particles strongly relies. 
	The events possessing the \met\/ can be produced in association with 
	ordinary matter (which tag the event) --- photons, jets (from quarks or gluons), 
	$W$-, $Z$-, Higgs-bosons and heavy quarks ($b$- and top-quark). 
	One assumes these SM particles back-to-back 
	recoil against the undetected WIMPs, making the latter "visible" due to 
	a large value of measured \met. 
	More generally, these events (with large \met) 
	constitute a low-background sample 
	that provides powerful sensitivity to new physics phenomena
\cite{Aad:2012fw,Chatrchyan:2012tea}.	

	Despite the point that the LHC experiments cannot establish whether {\em a WIMP candidate is stable on
	cosmological time scales and hence is a true DM candidate} 
	the terms WIMP and DM-particle are used as {\em synonymous} 
\cite{ATLAS:2012ky} in all experimental papers in the section.
	
\begin{figure}[!htpb] \begin{center}
    \includegraphics[width=0.6\textwidth]{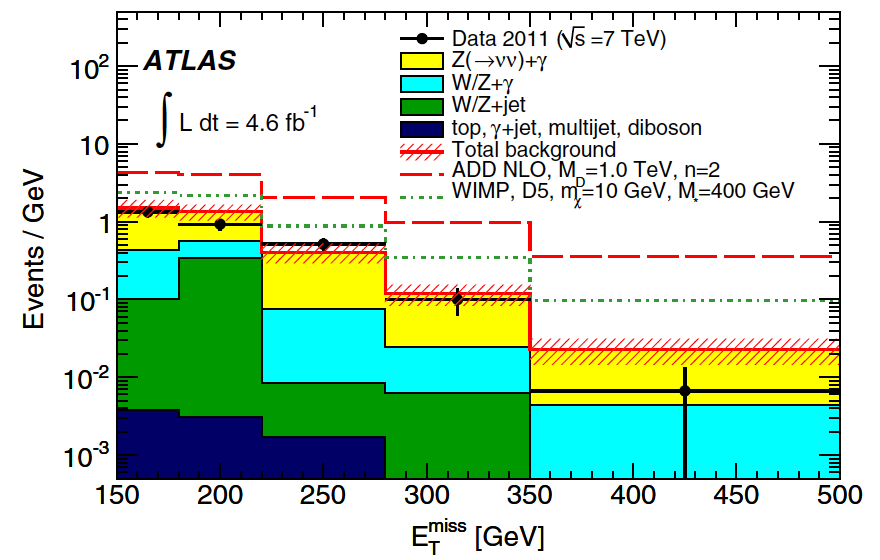}
  \end{center}\caption{Measured \met\/ distribution in comparison with 
  the relevant SM background predictions and some expectations. 
  From \cite{Aad:2012fw}. }\label{fig:2012fw}
\end{figure}

\begin{figure}[!ht]
\begin{center} \includegraphics[width=0.4\textwidth]{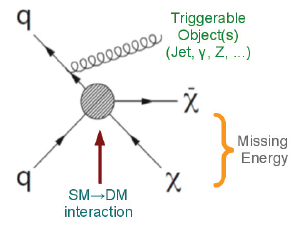} \end{center} 
\caption{Production of WIMP candidates tagged by a measurable object from initial-state radiation. 
 The final state monojet and/or monophoton together with (large) \met\/  provides a new physics collider signature.
From \cite{Diehl:2014dda}.} \label{fig:Diehl}
\end{figure}

\smallskip %\hrule\bigskip %\noindent %---------------------------------------------------
	The first ATLAS paper on "direct"
{\em Search for dark matter candidates and large extra dimensions in events with 
a photon and missing transverse momentum in $pp$ collision data at $\sqrt{s}=7$ TeV with the ATLAS detector} 
	was published at the very beginning of 2013
\cite{Aad:2012fw}. 
	New phenomena were looked for in events with an 
	energetic photon and large missing transverse momentum
	($pp\to \gamma+E^{\rm miss}_{\rm T} + X$)   
	at 7 TeV and integrated luminosity of 4.6 fb$^{-1}$.
	The measurements were found to be 
	in good agreement with the SM predictions for the background 
(Fig.~\ref{fig:2012fw}). 
%%%%%%%%%%%%%%%%%
 	The obtained results were converted into 
	model-independent 90\% CL upper limits on the	
	visible (new physics) cross section $\sigma\times A\times \epsilon$ 
	(cross section$\times$acceptance$\times$efficiency) of 5.6 fb. 	
	The results were further translated into exclusion curves  
	on the pair-production cross section of WIMP candidates in $pp\to \chi\chi+\gamma+X$
	under the assumption that the pair can be traced (optimistically, detected) 
	due to energy imbalance with an energetic photon (from initial-state radiation).  
	The generic graph in 
Fig.~\ref{fig:Diehl} shows typical production of two WIMPs in  
	a collision of quarks from initial protons 
\cite{Diehl:2014dda}.

	To obtain the limits on WIMP-SM interaction 
	one needs a tool to describe interaction of WIMPs with the SM particles. 
	To this end the EFT approach with several operators from
\cite{Goodman:2010ku} was used. 
\begin{figure}[!h] \begin{center}
\includegraphics[width=0.8\textwidth]{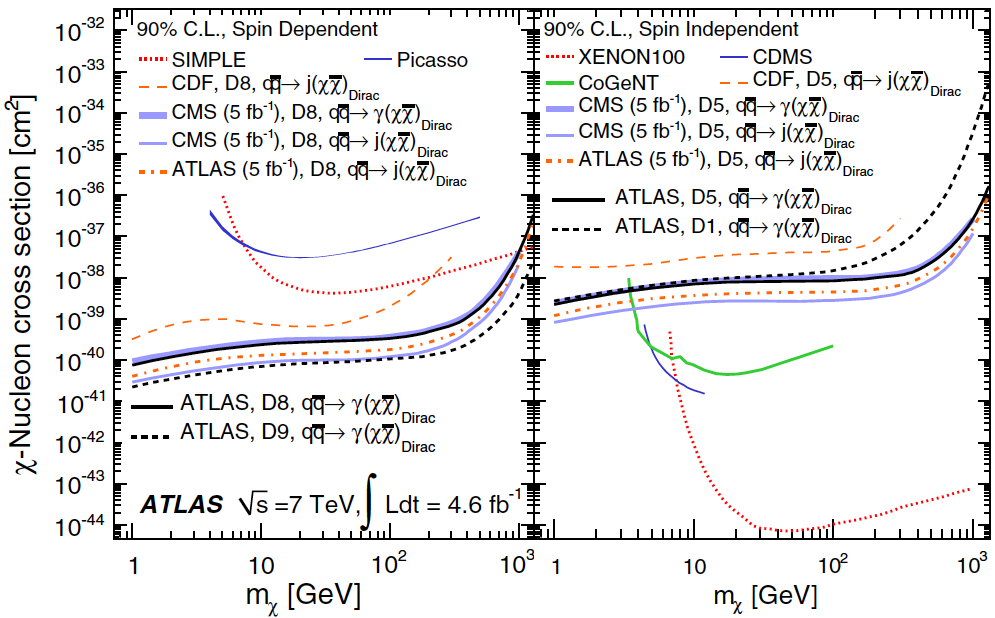}\end{center}\caption{
Upper limits (90\% CL) on the nucleon-WIMP cross section as a function of WIMP mass 
for spin-dependent (left) and spin-independent (right) interactions. 
The results are compared with previous mono-jet and mono-photon results at colliders  
\cite{ATLAS:2012ky,Chatrchyan:2012tea,Chatrchyan:2011nd}
and results from direct detection experiments. From \cite{Aad:2012fw}. 
}\label{fig:2012fw-SDSI}
\end{figure}
	The WIMPs were assumed to be Dirac fermions with $m_\chi$ between 
	1 GeV$/c^2$ and 1.3 TeV$/c^2$ and the effective operators D1 (scalar), D5 (vector), 
	D8 (axial-vector), and D9 (tensor) from 
Table~\ref{tab:2010ku} were used. 
	These operators correspond to spin-independent (D1 and D5) and 
	spin-dependent (D8 and D9) interactions. 
	In the case of the D1 (D5) spin-independent operator, 
	values of $M_*$ below 31 and 5 GeV (585 and 156 GeV) were 
	excluded at 90\% CL for $m_\chi$ equal to 1 GeV$/c^2$ and 1.3 TeV$/c^2$, respectively. 
	Values of $M_*$ below 585 and 100 GeV (794 and 188 GeV) were 
	excluded for the D8 (D9) spin-dependent operator
	for $m_\chi$ equal to 1 GeV$/c^2$ and 1.3 TeV$/c^2$, respectively. 
%%%%%%%%%%%%%%
	On the basis of the prescription of 
\cite{Goodman:2010ku} these results were translated into exclusion 
	curves --- upper limits   for the nucleon-WIMP interaction cross section as a function of $m_\chi$
(Fig.~\ref{fig:2012fw-SDSI}).
 	From the figure one can conclude 
\cite{Aad:2012fw} that obtained results (under validity of the EFT) 
	 gave the best exclusion limits for spin-independent nucleon-WIMP interactions 
	 in the small-mass region ($1<m_\chi < 3.5$ GeV$/c^2$) and for 
	 spin-dependent interactions for all masses (1 GeV$/c^2$ $<m_\chi <$1 TeV$/c^2$). 
 	These results confirmed the general expectations 
 	(discussed in section \ref{sec:FirstColliderDM}), and  
 	improved previous CDF-based achievements. 

\smallskip %\hrule\bigskip %\noindent %--------------------------------------------------- 
The second ATLAS paper on  
 {\em Search for dark matter candidates and large extra dimensions in events with 
 a jet and missing transverse momentum with the ATLAS detector} was published in the middle of 2013 
 \cite{ATLAS:2012ky}.
	Again, new phenomena were searched in events with a high-energy jet and 
	large missing transverse momentum 
	($pp\to {\rm jet} +E^{\rm miss}_{\rm T} + X$) 
	using data at 7 TeV and an integrated luminosity of 4.7 fb$^{-1}$. 
 %%%%%%%%%%%%%%%%%
\begin{figure}[!htpb] \begin{center}
    \includegraphics[width=0.9\textwidth]{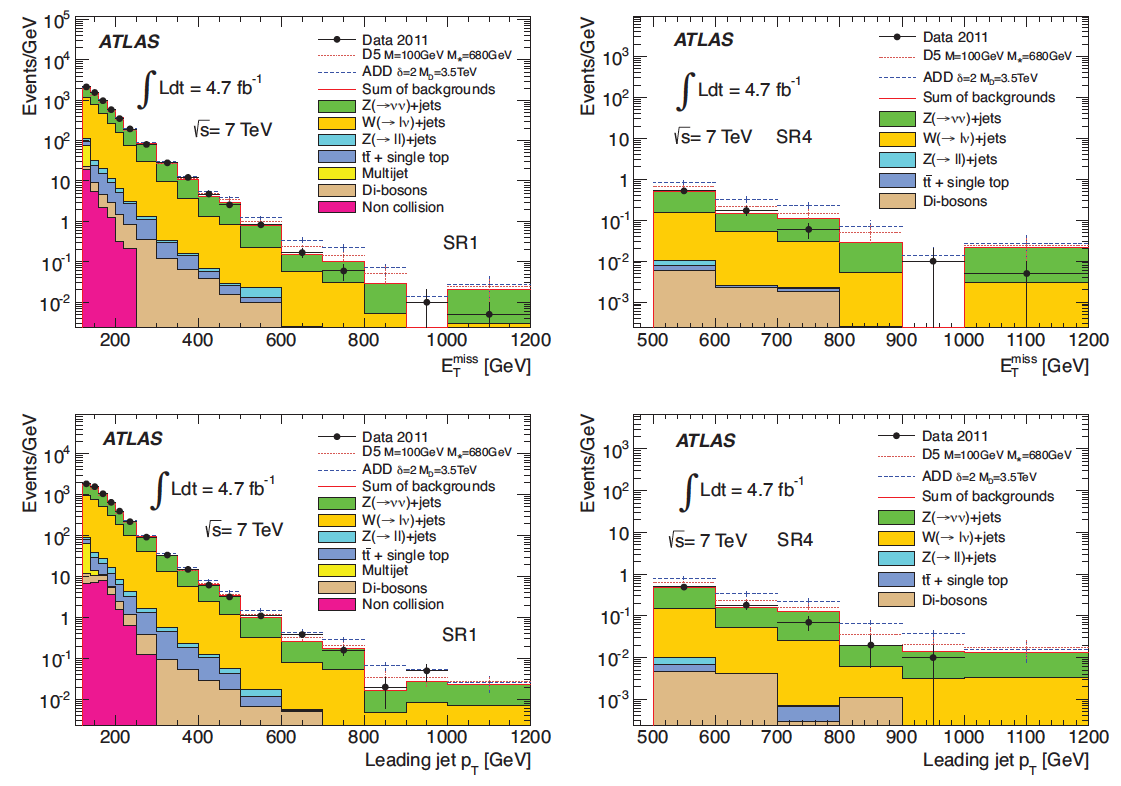}
  \end{center}\caption{Measured \met\/ and leading jet $p^{}_{\rm T}$ 
  distributions in comparison with 
  the relevant SM (background) predictions. From \cite{ATLAS:2012ky}. 
	}\label{fig:2012ky}
\end{figure}

	The main physics results of the search are given in 
Fig.~\ref{fig:2012ky} as the measured \met\/ and leading  
	jet $p^{}_{\rm T}$ distributions 
	in comparisons with the SM predictions. 
	No excess of events beyond expectations from SM processes was observed
	and some limits on the visible cross sections of new physics processes were obtained. 

	The EFT approach (section \ref{sec:FirstColliderDM})  
	was used for comparison of the LHC DM-search 
	results with the results of direct and indirect DM searches. 
	According to 
Table~\ref{tab:2010ku}, interactions of SM particles and WIMPs are described
	by only two parameters, the scale $M_*$ and the WIMP particle mass $m_\chi$.
	From measured distributions  
(Fig.~\ref{fig:2012ky}) lower limits on the mass parameters $M_*$ of effective operators 
	associated with the above-mentioned new processes can be first derived as functions of the 
	WIMP mass $m^{}_\chi$ 
\cite{Gramling:2013wra}.
	To this end, under assumptions of WIMP pair-production 
	and absence of any other possible new particles, 
	5 effective WIMP-SM contact operators from 
Table~\ref{tab:2010ku} were considered, and for each of them 
	constraints on $M_*$ as functions of $m_\chi$ were derived and presented in 
Fig.~4 of paper \cite{ATLAS:2012ky}. 

\begin{figure}[!ht]\center
\includegraphics[width=0.48\linewidth]{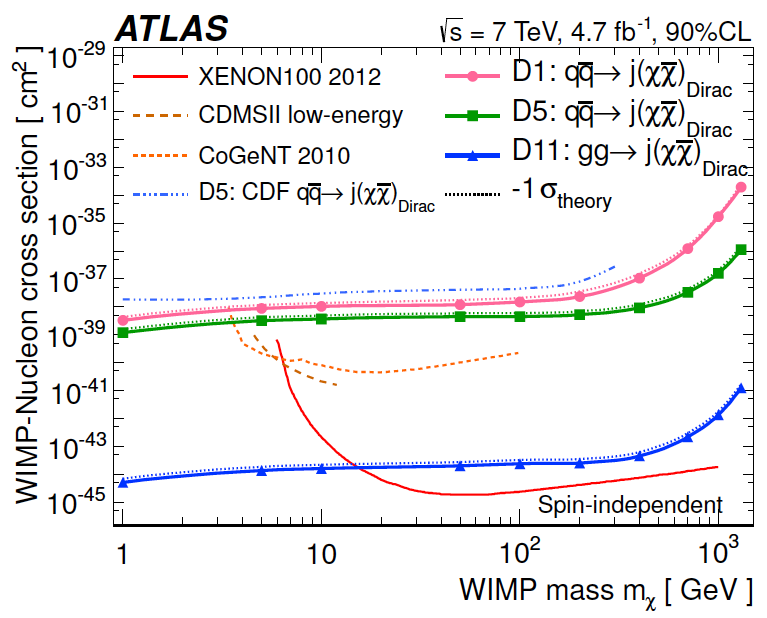}
\includegraphics[width=0.48\linewidth]{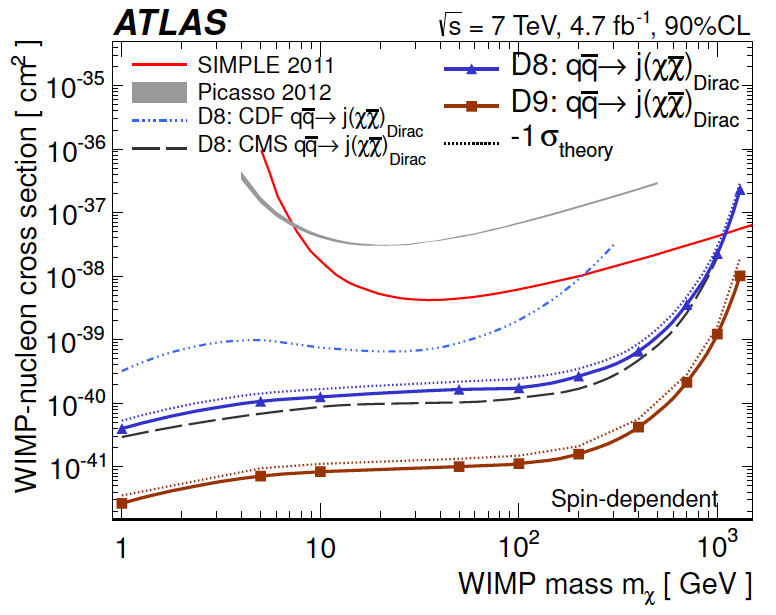}
\caption{ATLAS limits (90\% CL) on spin-independent (left) 
and spin-dependent (right) WIMP-nucleon scattering cross sections. 
More details are in \cite{ATLAS:2012ky}.}
\label{fig:2012ky-SISD}
\end{figure}

	Next, these bounds on $M_*$ (for a given $m_\chi$) can be converted within EFT 
	into the bounds on WIMP-nucleon cross sections, which are probed by
	direct DM detection experiments (section \ref{sec:DirectDetectionDM}). 
	Depending on the type of interaction, contributions to SD- or SI-  
	WIMP-nucleon interactions are expected. 
	This translation  
	into bounds on WIMP-nucleon cross sections is shown in 
Fig.~\ref{fig:2012ky-SISD}  
\cite{ATLAS:2012ky}. 
	Again, the spin-independent ATLAS-based exclusion curves are particularly relevant in
	the low $m_\chi$ region ($< $10 GeV$/c^2$) where the direct DM  
	limits could suffer from a technical problem with a very low recoil energy deposition.   
	The bound looks "especially powerful"  in the case of the gluon D11-operator, 
it is competitive to direct detection bounds up to $m_\chi$ of 20 GeV$/c^2$, 
provided this gluon operator is relevant for the direct DM search technique.
Some of the limits are substantially better than the limits set by direct and indirect DM 
detection experiments, in particular at small WIMP masses $m_\chi<10$ GeV$/c^2$
\cite{ATLAS:2012ky}.

	The EFT approach  
\cite{Goodman:2010ku} also allows one to interpret the obtained  
	collider limits in terms of the relic abundance of WIMPs, 
	or WIMP self-annihilation rate, which is defined as the product of 
	the annihilation cross section times the relative velocity, averaged over the DM velocity distribution.
	This interpretation was shown in Fig. 7 of 
\cite{ATLAS:2012ky}, where the limits on vector and axial-vector effective operators 
	were translated into upper limits on the WIMP annihilation rate to the four
 	light quark flavours.   
	The complementarity between collider and indirect 
	searches for DM was demonstrated by the figure
\cite{ATLAS:2012ky}. 
	Nevertheless, it is important to remember 
\cite{Diehl:2014dda} this complementarity  
	remains valid or makes sense only under a number of important assumptions ---
	the EFT must be valid, WIMPs must interact with SM particles exclusively via only one of 
	the EFT operators 
	(since a mix of operators with potential interference effects is not considered), 
	and the interactions must be flavour-universal for the four light quarks. 
	
	In general, the ATLAS mono-jet results are somewhat better than the ATLAS mono-photon one
	overall due to the higher statistics in the data samples
\cite{Diehl:2014dda}. 

\begin{figure}[!h] 
\center
\includegraphics[width=0.2\linewidth]{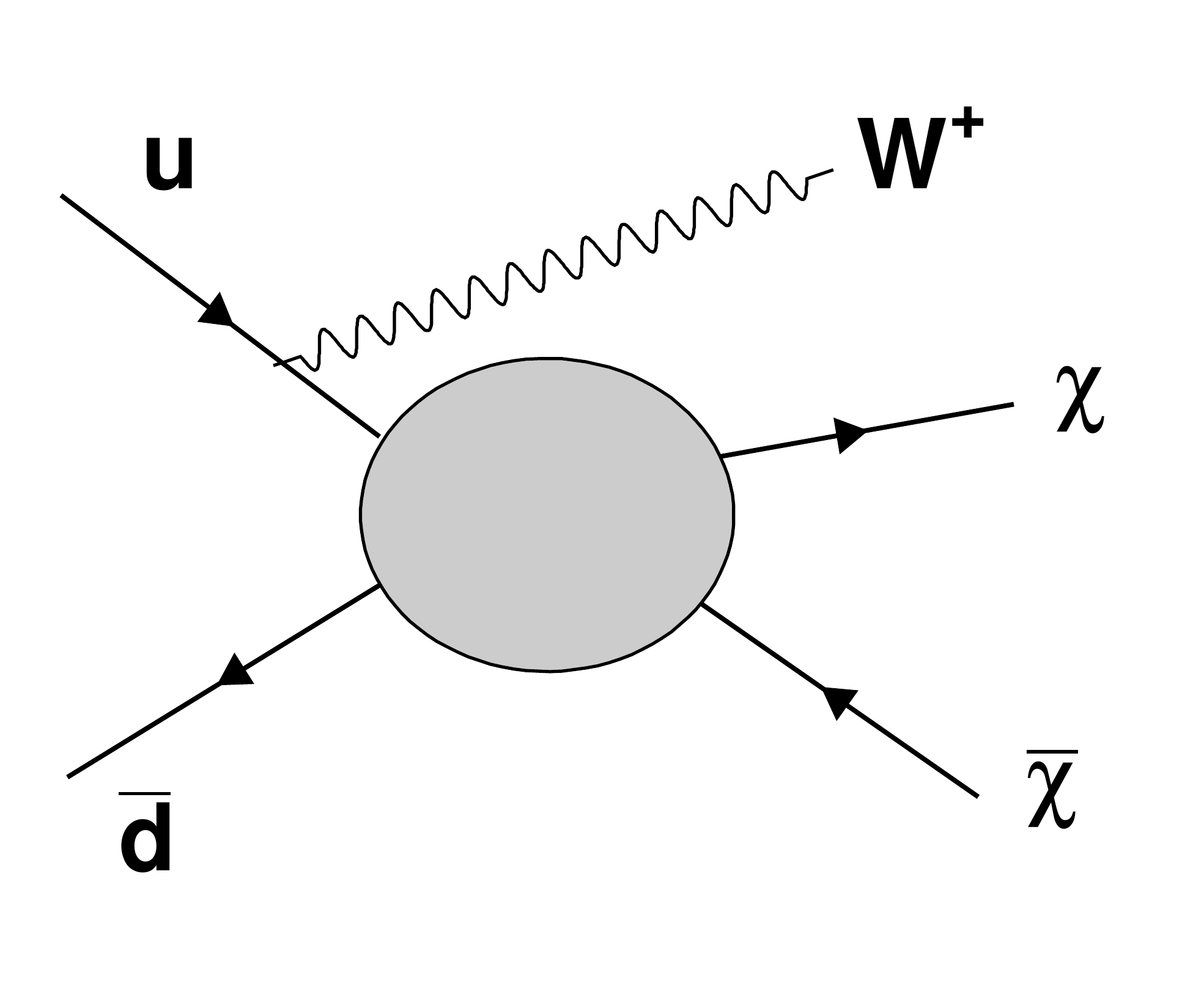} \ \ \
\includegraphics[width=0.2\linewidth]{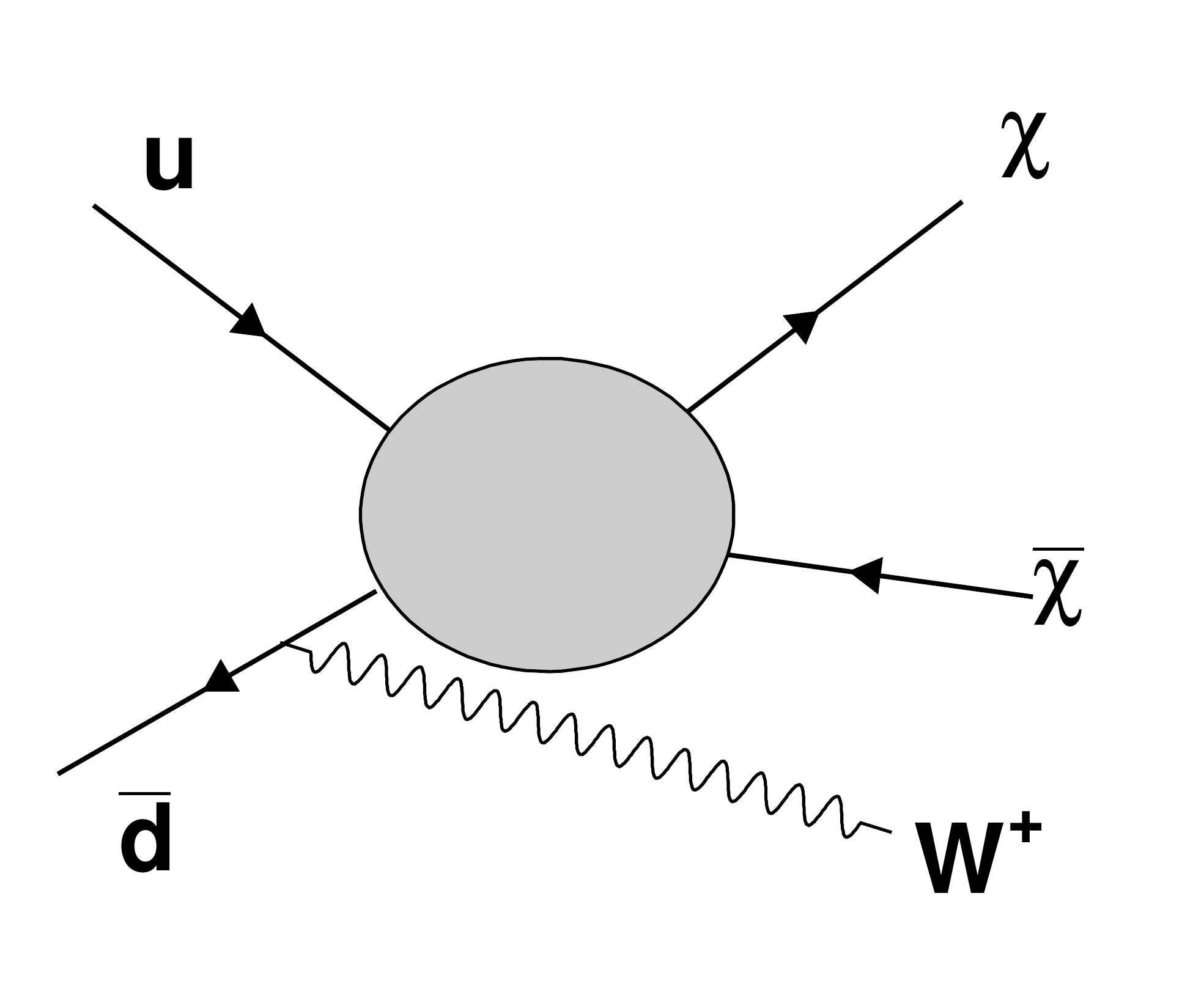}
\caption{General graph for pair production of WIMPs ($\chi\bar{\chi}$) in $pp$ collisions 
with initial-state radiation of a $W$ boson.  
The same graphs can be given for initial-state radiation of a $Z$ boson. 
From \cite{Aad:2013oja}.}
\label{fig:2013oja-W}
\end{figure}

\smallskip %%\hrule\bigskip %\noindent %--------------------------------------------------- 
	The 3rd ATLAS paper about {\em Search for dark matter in events with a 
hadronically decaying W or Z boson and missing transverse momentum ...} 
	was published on 29 January 2014
\cite{Aad:2013oja} and already used data at 8 TeV and 20.3 fb$^{-1}$.   
	Events with a hadronic jet with the jet-mass consistent with a $W$- or $Z$-boson, 
	and with large missing transverse momentum were analyzed. 
	For the first time, the goal was to look for the presence for a WIMP pair as DM candidates ($\chi\bar{\chi}$), 
	produced in $pp$ collisions in association with a $W$ or $Z$ boson 
	($pp\to W/Z +E^{\rm miss}_{\rm T} + X$, see Fig.~\ref{fig:2013oja-W}).	
	As before, one assumed that the undetected $\chi\bar{\chi}$-pair 
	produces large missing transverse momentum 
(\met$\, >150$~GeV in \cite{Aad:2013oja}), which can be "detected" (balanced) 
	 by means of {\em a single massive jet} reconstructed from the 
	 hadronic decays $W\to q\bar{q}'$ or $Z\to q\bar{q}$.

	The proposal to use the signature with 
	hadronical decays of $W$- or $Z$-boson is based on the results of 
\cite{Bai:2012xg}, where a source of enhancement was found.
	Usually, due to a large rate of gluon/quark initial-state radiation relative 
	to $\gamma$-, $W$- or $Z$-boson radiation, 
	the strongest limits mainly come from mono-jet analyses. 
	But the EFT operators used in the mono-jet search (Table~\ref{tab:2010ku}) 
	have equal couplings of the $\chi$-particles to $u$-type and $d$-type quarks.  
	From 
Fig.~\ref{fig:2013oja-W} one can see two ways of the $W$-boson radiation --- 
	from the initial $u$-quark or $\bar{d}$-quark. In the case of equal coupling, 
	this interference is destructive and gives a small $W$-boson emission rate. 
	Otherwise, due to constructive interference,   
	the mono-$W$-boson production can be strongly enhanced 
\cite{Bai:2012xg}
	in comparison with other possibilities (moon-jet, mon-photon, etc).
 
\begin{figure}[!h] \begin{center}
    \includegraphics[width=0.5\textwidth]{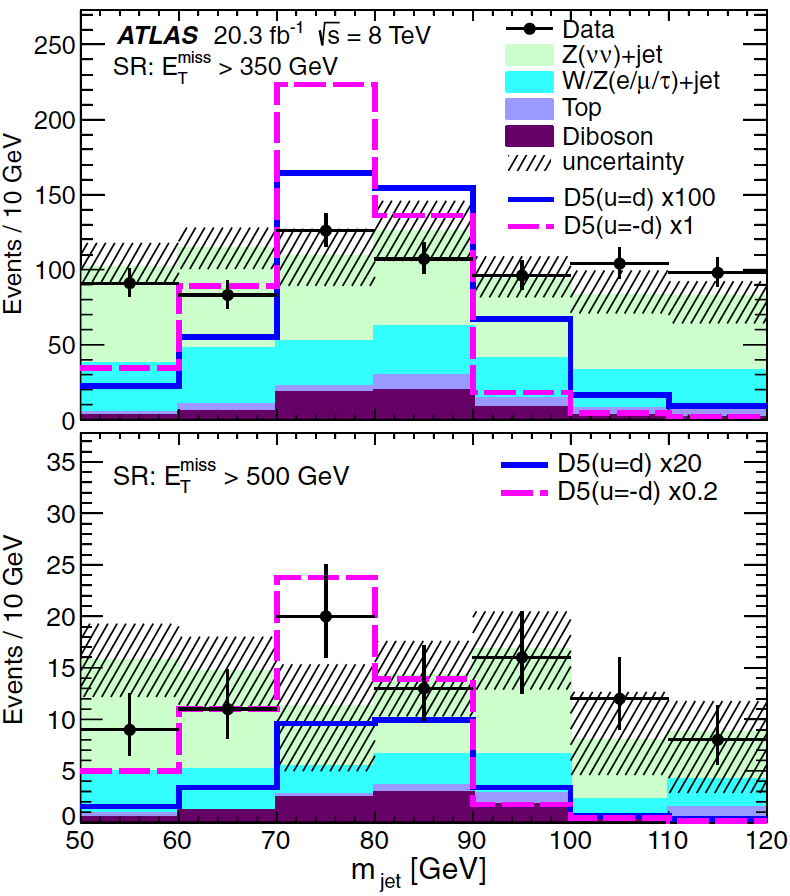}
  \end{center}\caption{Distribution of $m_{\textrm{jet}}$ in the data and for the 
  background in the signal regions (SR) with \met$>350$ GeV (top) and 
\met$> 500$ GeV (bottom). 
 Also shown are the combined mono-$W$-boson and mono-$Z$-boson
signal distributions with $m_\chi =1$ GeV$/c^2$ and $M_* = 1$ TeV for the
D5 destructive and D5 constructive cases. 
From \cite{Aad:2013oja}. }\label{fig:2013oja-Main}
\end{figure}

	After sophisticated analysis, the finally obtained results 
	and predicted backgrounds in the two signal regions 
	are shown in
Fig.~\ref{fig:2013oja-Main} for the $m^{}_{\rm jet}$ distribution.  
	The data agreed well with the background SM estimate for each \met\/ threshold.
	
\begin{figure}[!h] \begin{center} 
\includegraphics[width=0.4\textwidth]{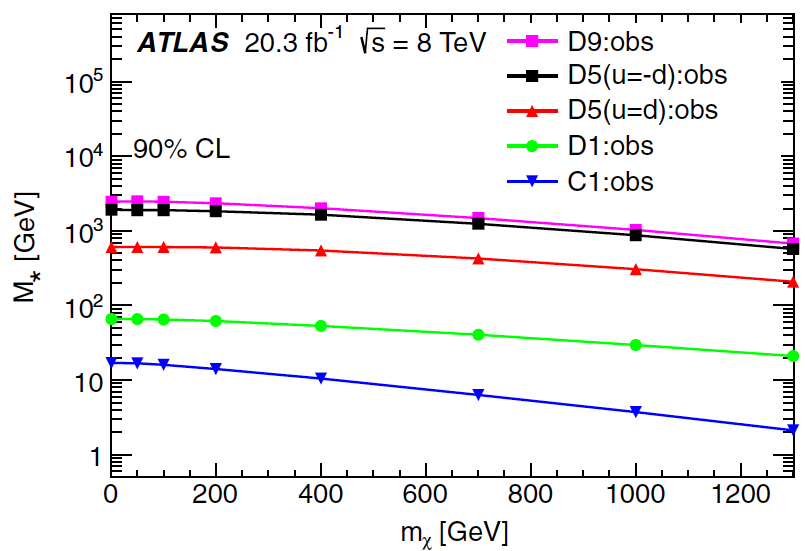}
 \end{center}\caption{Observed limits on the effective theory
mass scale $M_*$ as a function of $m_\chi$ at 90\% CL from combined
mono-$W$-boson and mono-$Z$-boson signals for various operators.
For each operator, the values below the corresponding line are excluded. 
 From \cite{Aad:2013oja}. }\label{fig:2013oja-DDDC}
\end{figure}

	On the basis of this result new limits (as a function of $m_\chi$) 
	were set on the mass scales $M_*$ of the C1 (scalar), D1 (scalar), D5 (vector),  
	and D9 (tensor) effective operators from Table~\ref{tab:2010ku}.
	Figure~\ref{fig:2013oja-DDDC} shows the excluded regions 
	in the $M_*$--$m_\chi$ plane for these operators.	
	These limits were set on the WIMP DM signals 
	using the expected shape of the "signal"
$m_{\textrm{jet}}$ distribution given in Fig.~\ref{fig:2013oja-Main}.  
	It is remarkable, how much these $M_*$-limits differ from each other for different operators.

\begin{figure}[!h] \begin{center} 
\includegraphics[width=0.7\textwidth]{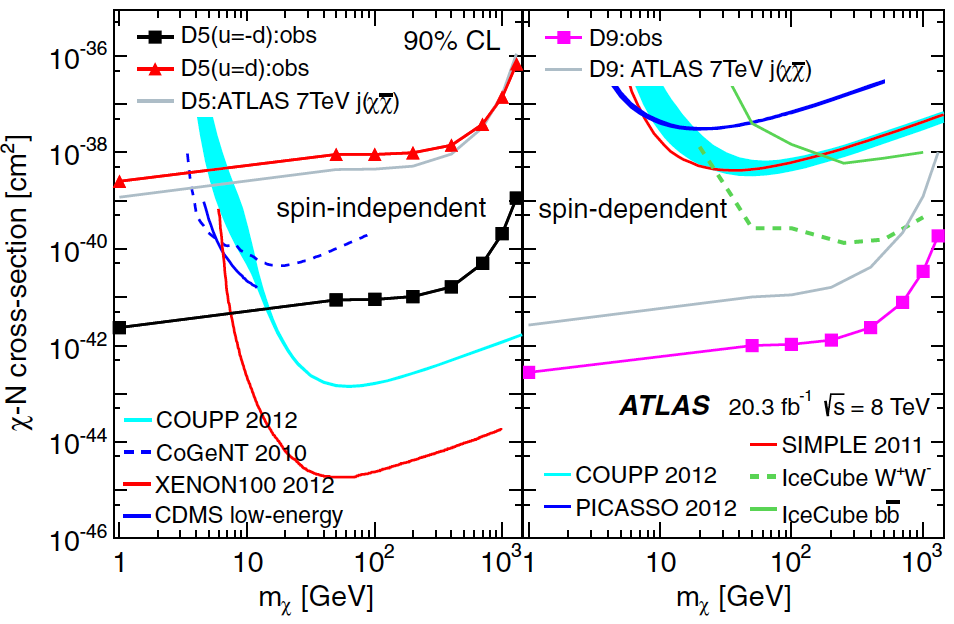}
 \end{center}
 \caption{Limits on $\chi$--nucleon cross sections as a function of WIMP mass $m_\chi$
at 90\% CL for spin-independent (left) and spin-dependent
(right) effective operators, compared to some previous limits
\cite{Aprile:2012nq,Ahmed:2010wy,Aalseth:2010vx,Archambault:2012pm,Felizardo:2011uw,Aartsen:2012kia,Behnke:2012ys}.
 From \cite{Aad:2013oja}. }
 \label{fig:2013oja-SISD}
\end{figure}

	In Figure~\ref{fig:2013oja-SISD} for both the
	spin-independent (C1, D1, D5) and the spin-dependent interaction (D9)
	exclusion curves for the WIMP--nucleon cross sections are 
	given following method of \cite{Goodman:2010ku}.  
	The results are traditionally compared with measurements from direct detection experiments.
	One can conclude from this figure
\cite{Aad:2013oja}  that (as generally expected for a collider DM search) 
	the search for WIMP pair production in association with $W/Z$-boson 
	extends the limits on the WIMP--nucleon cross section 
	in the low mass region $m_{\chi}<10$~GeV$/c^2$.  
	Next, the comparison of these new limits 
	with the limits set by ATLAS in the 7~TeV mono-jet analysis
	demonstrates improvement by about 3 orders of magnitude.
	The reason is the constructive interference 
	(due to the opposite sign of $u\chi$-type and $d\chi$-type couplings),
	which has led to a very large increase 
	in the $W$-boson-associated production cross section.
	For other cases, the limits are similar.
	
	Assuming a simple model of WIMP production via the Higgs boson, 
	a re-analysis of the data  
\cite{Aad:2013oja} was carried out. 
	The upper limit on the  Higgs boson production through
	$WH$ and $ZH$ modes and decay to invisible particles is 
	obtained to be 1.3~pb at 95\% CL for $m_H=125$ GeV$/c^2$. 
\begin{figure}[!h] \begin{center} 
\includegraphics[width=0.45\textwidth]{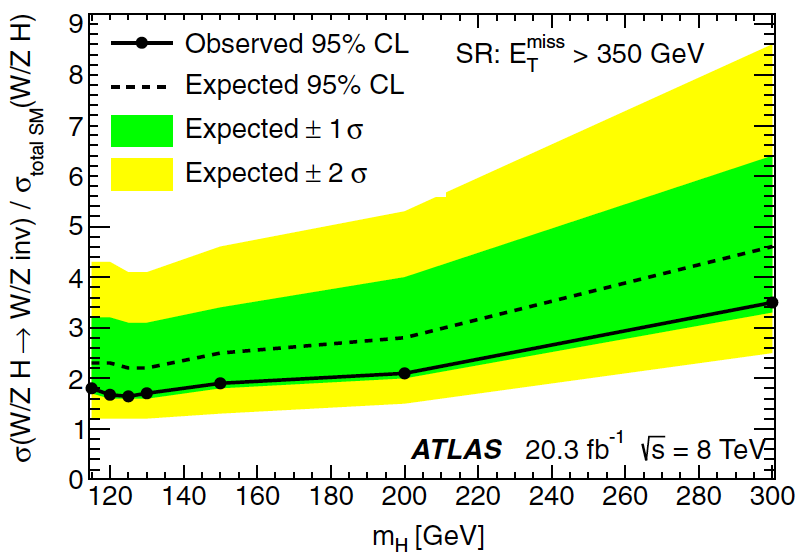}\end{center}
 \caption{Limit on the Higgs boson cross section for decay to invisible particles divided 
 by the cross section for decays to SM particles as a function of $m_H$ at
95\% CL, derived from the signal region (SR) with \met$\, > 350$ GeV.
From \cite{Aad:2013oja}. }
\label{fig:2013oja-H-inv}
\end{figure}
Figure~\ref{fig:2013oja-H-inv} 
shows the upper limit
of the total cross section of $WH$ and $ZH$ processes with $H\rightarrow\chi\bar{\chi}$, normalized to the SM next-to-leading order prediction for the $WH$ and $ZH$ production cross section (0.8~pb is predicted for $m_H=125$ GeV$/c^2$ in SM), which is 1.6 at $95\%$ CL for $m_H=125$ GeV$/c^2$.
This was the first new experimental result on the subject. 

\smallskip %%%\hrule\bigskip %\noindent %--------------------------------------------------- 
	Continuation of the ATLAS collaboration search for DM with the help of the 
	$Z$ and Higgs bosons was carried out in the dedicated paper 
\cite{Aad:2014iia}. 
	Here the \met\/ was used as a signal ---  
	the momentum of the reconstructed $Z$ boson is expected to be balanced 
	by the momentum of the invisibly decaying Higgs boson
	(with its mass allowed in the range $110<m_H<400$~GeV$/c^2$).
	The measured \met-distribution in events with an electron or a muon pair 
	consistent with $Z$-boson decay can be used to constrain the $ZH$ production cross 
	section times the branching ratio of the Higgs boson decaying to invisible particles.  
	
	The total cross section for the associated production of the SM Higgs boson, 
	with $m_H=125.5$~GeV$/c^2$,  and a $Z$ boson, according to 
\cite{Heinemeyer:2013tqa} is 
	331~fb at $\sqrt{s} = 7$~TeV and 410~fb at $\sqrt{s} = 8$~TeV. 
	The branching ratio of the Higgs boson decay to invisible SM particles
	($H\to ZZ^*\to 4\nu$) is $1.2\times 10^{-3}$.
	The performed search could not be sensitive to this value,
	and the main idea was to look for some enhancement 
	in the invisible decay mode due to a contribution of not-SM particles. 
	
\begin{figure}[!ht]\begin{center}
\includegraphics[width=0.6\textwidth]{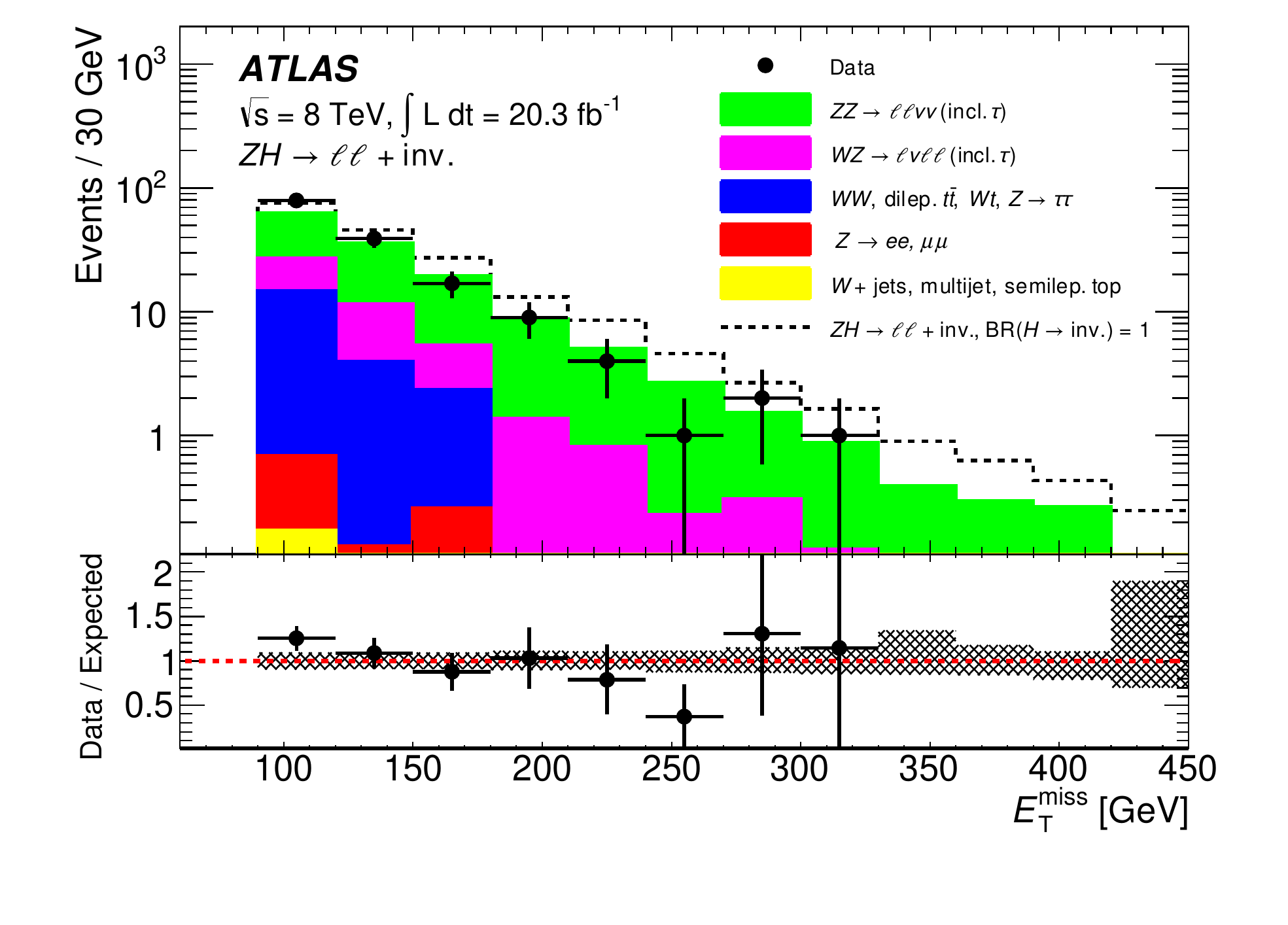}
\end{center}\caption{
\label{fig:2014iia-Main} 
Distribution of \met\ after the full selection in the 8 TeV data (dots). 
The filled stacked histograms represent the background expectations. 
The signal expectation for Higgs boson with $m_H=125.5$~GeV$/c^2$,
a SM $ZH$ production rate and  $\text{BR}(H\to \text{invisible})=1$ is
stacked on top of the background expectations. 
From \cite{Aad:2014iia}. }
\end{figure}

	The numbers of observed and expected events  
	are shown in Table~\ref{tab:eventstable_llinv}. 
\begin{table*}[!ht] 
 \caption{\label{tab:eventstable_llinv} 
 	Number of events observed in data and expected from the signal 
 ($m_H=125.5$ GeV$/c^2$, $\sigma_{ZH}^\mathrm{SM}$, $\text{BR}(H\to\text{invisible})\!=\!1$)
 and from each background source for the 7 and 8 TeV data-taking periods. 
 From \cite{Aad:2014iia}.  }
\centering
\begin{tabular}{|l|c|c|} \hline 
Data Period    & 2011 (7 TeV+4.5/fb) & 2012 (8 TeV+20.3/fb) \\ \hline
$ZZ \rightarrow \ell\ell\nu\nu$  &  $20.0 \pm 0.7 \pm 1.6$    & $91 \pm 1 \pm 7$  \\ 
$WZ \rightarrow \ell\nu\ell\ell$  &  $4.8 \pm 0.3 \pm 0.5$     & $26 \pm 1 \pm 3$  \\ 
Dileptonic $t\bar{t}$, $Wt$, $WW$, $Z\to\tau\tau$ &  $0.5 \pm 0.4 \pm 0.1$ & $20 \pm 3 \pm 5$  \\ 
$Z\to ee$, $Z\to\mu\mu$               &  $0.13 \pm 0.12 \pm 0.07$  & $0.9 \pm 0.3 \pm 0.5$  \\ 
$W$ + jets, multijet, semileptonic top  &  $0.020 \pm 0.005 \pm 0.008 $     & $0.29 \pm 0.02 \pm 0.06$  \\ 
Total background &  $25.4 \pm 0.8 \pm 1.7$     & $138 \pm 4 \pm 9$  \\  \hline
Signal & $8.9 \pm 0.1 \pm 0.5$&$44 \pm 1 \pm 3$\\ \hline
      Observed        & 28 & 152  \\  \hline 
  \end{tabular}
\end{table*}
Figure~\ref{fig:2014iia-Main} shows the \met\ distribution after the full 
event selection for the 8 TeV data and the expected backgrounds.
	No significant deviation from the SM expectation was observed in both sets of 
	data collected at 7 TeV and 8 TeV by the ATLAS experiment. 

	Assuming the SM rate for $ZH$ production, an upper limit of 75\% (at 95\% CL) 
	was set on the branching ratio of the SM Higgs boson decay to 
	invisible particles.  
	The expected limit in the absence of beyond-SM decays to invisible 
	particles is 62\% (at 95\% CL).
	Limits were also set on the cross section times branching ratio for additional neutral Higgs boson, 
	with $110<m_H<400$ GeV$/c^2$, 
	produced in association with a $Z$ boson and decaying to invisible particles. 

	The obtained limit on BR$(H\to\text{invisible})$ for the 125-GeV Higgs boson 
	can be interpreted in terms of an upper limit on the WIMP-nucleon cross section, 
	when the Higgs boson serves as a mediator between WIMP and SM particles
	and therefore can decay to the WIMP pair
\cite{Fox:2011pm}.
	The formalism used to interpret the $\text{BR}(H\to \text{invisible})$ limit in terms 
	of the spin-independent WIMP-nucleon cross sections is described in
\cite{Kanemura:2010sh,Djouadi:2011aa}.

\begin{figure}[h] 
\begin{center} \includegraphics[width=0.6\textwidth]{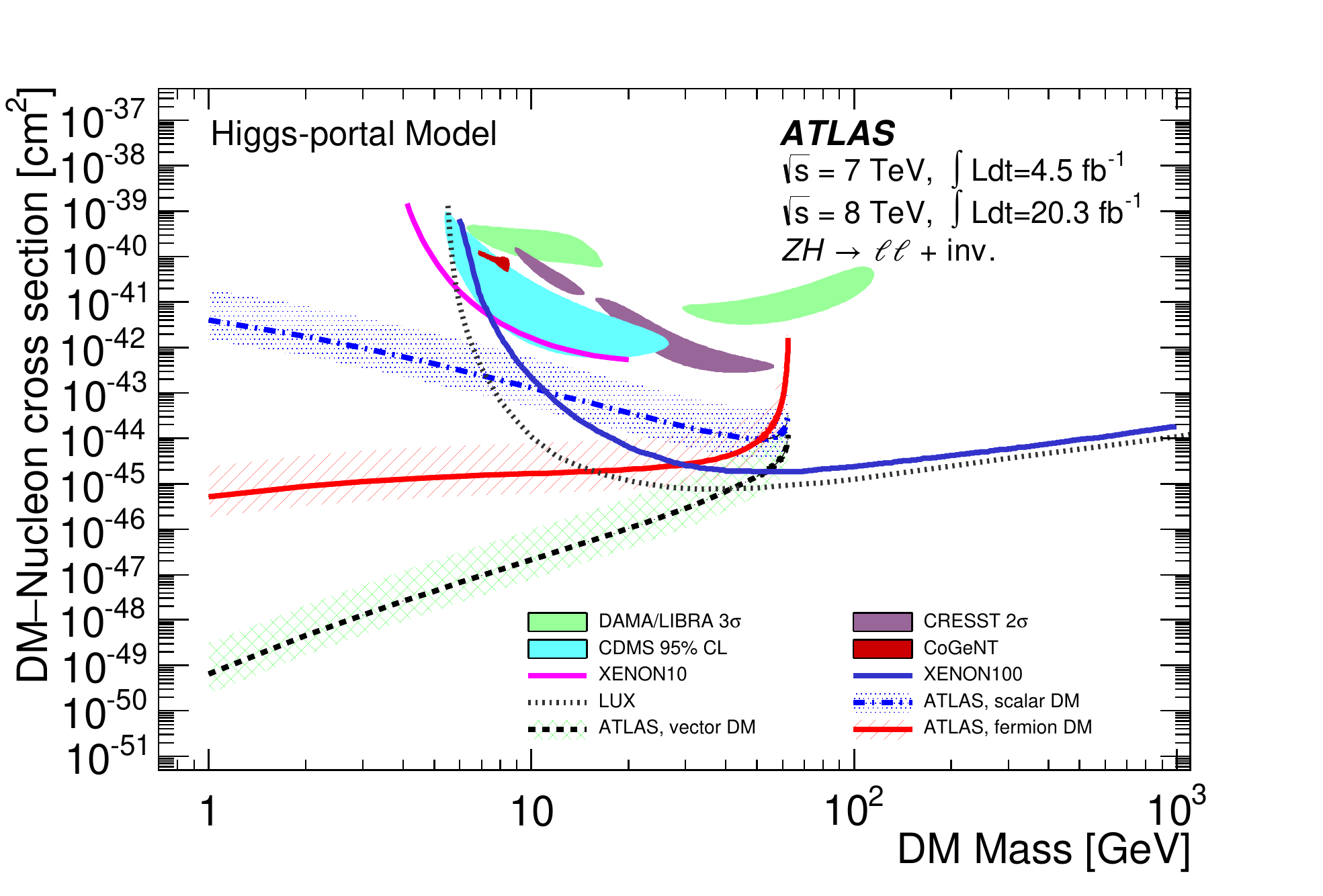} \end{center}
\caption{Limits on the WIMP-nucleon scattering cross section (90\% CL), extracted from the 
BR$(H\to \text{invisible})$ limit, compared to results from direct-search experiments.
Spin-independent results from direct-search experiments are shown from 
\cite{Bernabei:2008yi,Angle:2011th,Aprile:2012nq,Angloher:2011uu,Aalseth:2011wp,Fox:2011px,Agnese:2013jaa,Akerib:2013tjd}.
From \cite{Aad:2014iia}.} \label{fig:2014iia-fig4} 
\end{figure}

Figure~\ref{fig:2014iia-fig4} 
	shows upper limits on the WIMP-nucleon cross 
	section for three cases in which a single WIMP candidate 
	was considered, being a scalar, a vector or a Majorana fermion. 
	There is no sensitivity to these models once the mass of the WIMP 
	candidate exceeds $m_H/2$. 
	The Higgs--nucleon coupling was taken as $0.33^{+ 0.30}_{- 0.07}$
	\cite{Djouadi:2011aa}, 
	the uncertainty of which is expressed by the bands in the figure. 
	It is seen  that one has 
\cite{Aad:2014iia} the strongest limits for low-mass WIMP DM candidates.

\begin{figure}[!ht]
\begin{center}
\includegraphics[width=0.5\textwidth]{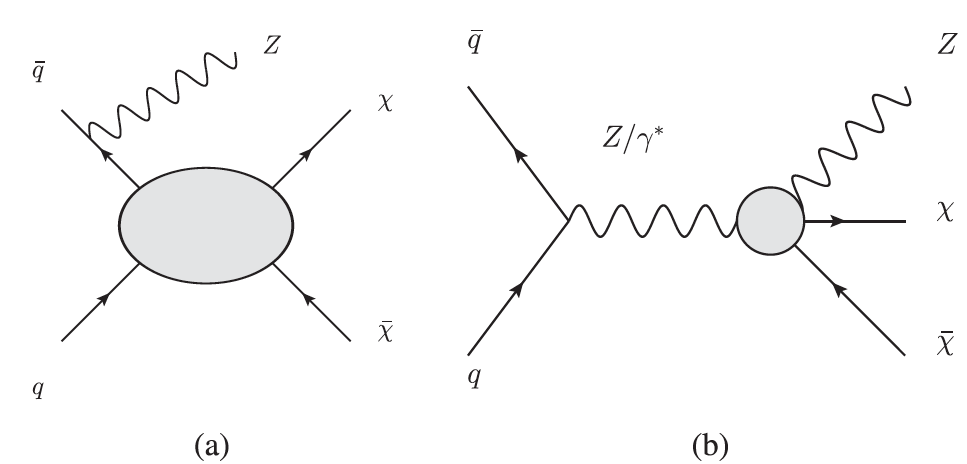}
\caption{Graph (a) shows 2 WIMPs and $Z$ boson production $pp\rightarrow\chi\bar{\chi}+Z$ with the ISR operator, 
and graph (b) shows the same process via the $ZZ\chi\chi$ vertex. From \cite{Aad:2014vka}.
}\label{fig:2014vka-Z}
\end{center}
\end{figure}

\begin{figure}[!ht]
\begin{center}
\includegraphics[width=0.5\textwidth]{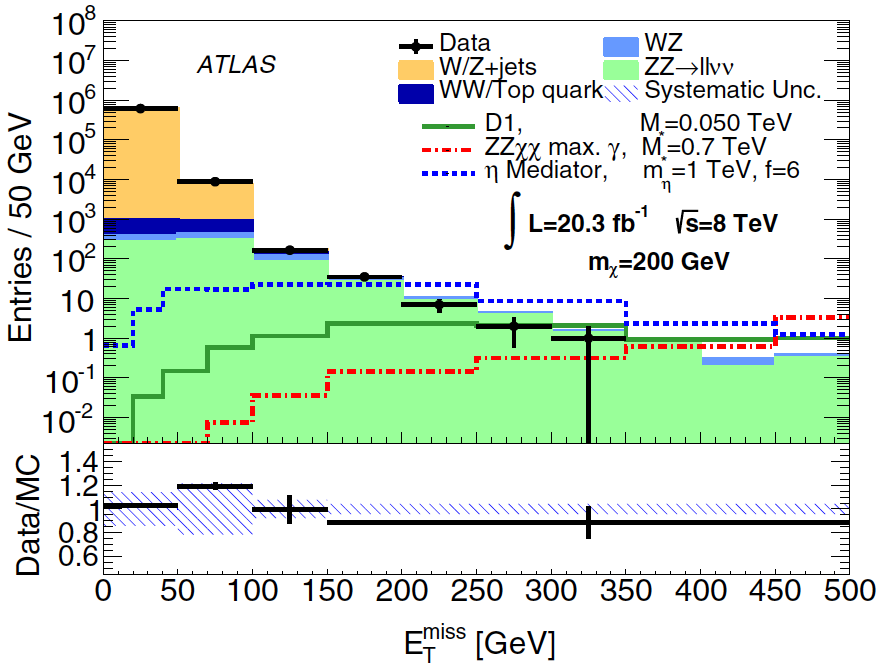}
\caption{\label{fig:2014vka-Main} 
\met\ distributions after all event selections. 
The hypothetical $pp\rightarrow Z\chi\bar{\chi}$ signals are given 
for various values of the mass scale, $M_{*}$. 
The WIMP  mass is $m_{\chi}=200$ GeV$/c^2$. 
From \cite{Aad:2014vka}}
\end{center}
\end{figure}

%%%%%%%%%%%%%%%%%%%%%%%%%%%%%%%%%%
\smallskip 
	The next step of the ATLAS DM program was 
{\em Search for dark matter in events with a Z boson and missing transverse momentum at $\sqrt{s}$=8 TeV ...} 
\cite{Aad:2014vka}.
	Events with large \met\/ 
	and $e^+e^-$- or $\mu^+\mu^-$-pairs consistent with 
	the decay of a $Z$ boson (Fig.~\ref{fig:2014vka-Z}) were analyzed
	at 20.3 fb$^{-1}$ statistics of 8-TeV data. 
	It was assumed that the large \met\ stemmed from the escaping $\chi\bar{\chi}$ particles.
	Several signal regions with different requirements on the 
\met\ were defined. 
	From Fig.~\ref{fig:2014vka-Main} one can conclude that 
	no excess above the SM prediction (the background) was observed. 
	This is the main model-independent result of 
\cite{Aad:2014vka}.          

	For interpretation of the obtained results the EFT framework
\cite{Goodman:2010ku} recently extended to describe interactions with electroweak bosons
\cite{Fox:2011pm} was used.
	Again, the $\chi$-particle is considered as {\em the only new particle}; 
	the mediator mass is assumed to be {\em larger than typical momentum transfer}, 
	and the WIMPs are produced {\em in pairs only}.
	For graph (a) in Fig.~\ref{fig:2014vka-Z} one used 
	contact operators D1, D5, D9 from 
Table~\ref{tab:2010ku}.  
	Following 
\cite{Fox:2011pm}, dimension-5 and dimension-7 operators were used to 
	describe the WIMP interaction with the electroweak bosons.
	The dimension-7 operator couples $\chi$ to $Z\gamma^{*}$ and $\chi$ to $ZZ$. 
	Since a $Z$ boson is in the final state for each operator,  
	intermediate states with a $Z$ or $\gamma^{*}$ each contribute to the matrix element. 
	The relative contribution of the $Z$ and $\gamma^{*}$ diagrams is a parameter of the theory.
	According to 
Fig.~\ref{fig:2014vka-Z}, 
	two models of the WIMP+$Z$ production were considered, 
	via initial state $Z$-boson radiation, and when $Z$ interacts directly with WIMPs. 
	The latter case was not previously investigated.

\begin{figure}[!ht]
\begin{center}
\includegraphics[width=0.6\textwidth]{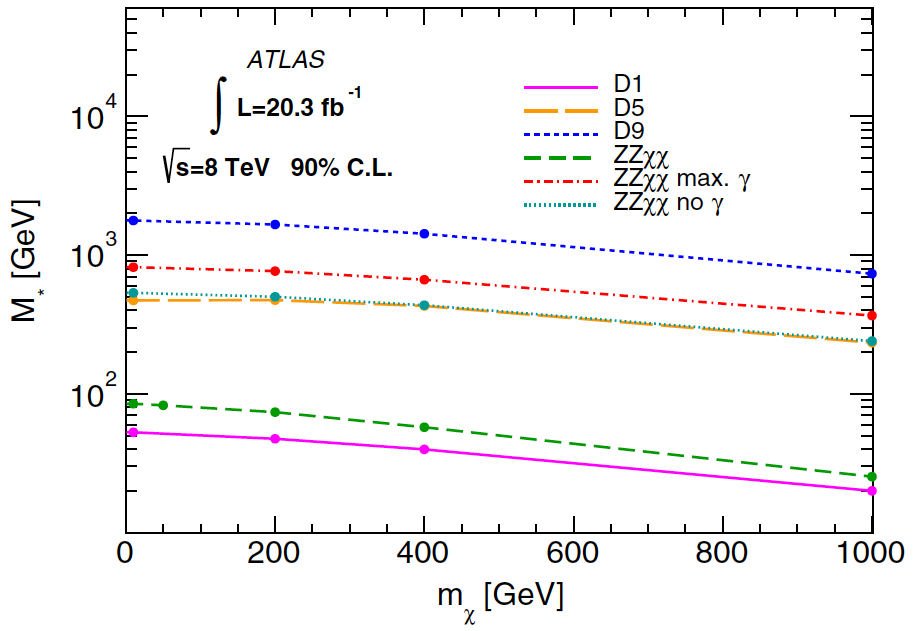}
\end{center}
\caption{Lower limits (90\% CL) on the mass scale, $M_{*}$, 
of considered {\em six} contact operators as a function of $m_{\chi}$. 
For each operator, the values below the corresponding line are excluded. 
From \cite{Aad:2014vka}.}
\label{fig:2014vka-EFTlimits}
\end{figure}

The \met\ region (Fig.~\ref{fig:2014vka-Main}) 
	with the best expected limit on the cross section for 
$pp\rightarrow Z\chi\bar{\chi}+X$ production
	was used to calculate the limits (given in Fig.~\ref{fig:2014vka-EFTlimits}) 
	on the mass scale $M_{*}$ as a function of $m_{\chi}$ 
	for each effective operator in both above-mentioned models.
	Again, one sees substantial variation of the $M_{*}$ as a function of the form of the effective operators.  

\begin{figure}[!ht]
\begin{center}
\includegraphics[width=0.7\textwidth]{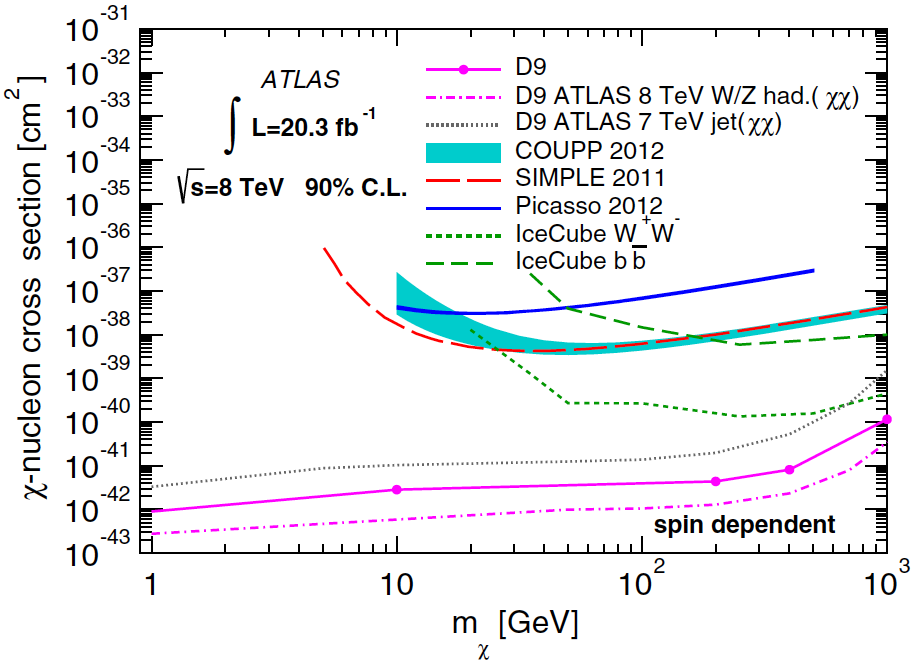}
\end{center}
\caption{\label{fig:2014vka-SD} 
Observed upper limits  (90\% CL) on the $\chi$-nucleon cross section as a function 
of $m_{\chi}$ for the spin-dependent D9 operator. 
The limits are compared with ATLAS results from hadronically decaying $W/Z$~
\cite{Aad:2013oja} and $j$ + $\chi\chi$~\cite{ATLAS:2012ky} searches, 
COUPP~\cite{Behnke:2012ys}, SIMPLE~\cite{Felizardo:2011uw}, 
PICASSO~\cite{Archambault:2012pm}, and IceCube~\cite{Aartsen:2012kia}.
From \cite{Aad:2014vka}.
}
\end{figure}

\begin{figure}[!h] 
\begin{center} 
\includegraphics[width=0.7\textwidth]{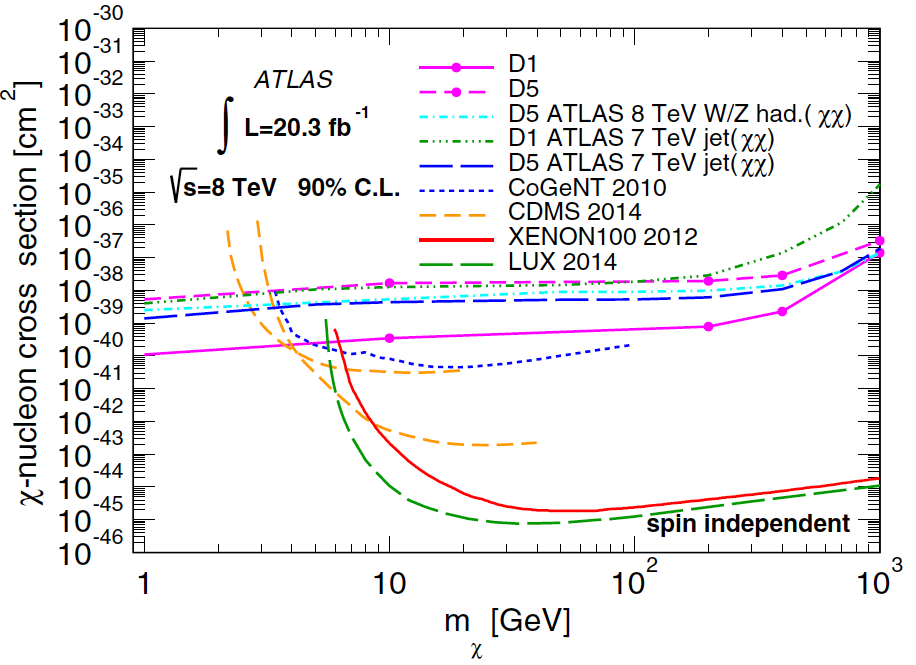} 
\caption{\label{fig:2014vka-SI}
Observed upper limits (90\% CL) on the $\chi$-nucleon cross section as a function 
of $m_{\chi}$ for spin-independent effective operators. 
The limits are compared with ATLAS results from  hadronically decaying 
$W/Z$~\cite{Aad:2013oja} and $j$ + $\chi\chi$~\cite{ATLAS:2012ky} searches,
CoGeNT~\cite{Aalseth:2010vx}, 
XENON100~\cite{Aprile:2012nq}, CDMS~\cite{Agnese:2013jaa,Agnese:2014aze}, 
and LUX~\cite{Akerib:2013tjd}.
From \cite{Aad:2014vka}.}
\end{center}
\end{figure}

	To complete the traditional "collider-DM-search" analysis one transformed EFT limits from 
Fig.~\ref{fig:2014vka-EFTlimits} into exclusion curves for 
	the $\chi$-nucleon cross sections.
	The comparison of these limits with the direct and indirect detection exclusion curves
	is shown in 
Figs.~\ref{fig:2014vka-SD} and~\ref{fig:2014vka-SI}.
	One can conclude from these figures that the ATLAS spin-dependent limits under discussion 
	are less stringent than the ATLAS limits for WIMP candidates 
	recoiling against a $W$ or $Z$ boson decaying to hadrons 
\cite{Aad:2013oja}.	
	The limits degrade by 13--23\%, 
	depending on the \met\ signal region under consideration.
	On the contrary, there is some improvement between the ATLAS results 
	in the case of the spin-independent $\chi$-nucleon interaction, 
	but, in general, all collider limits are still far to be competitive with direct DM limits.  
	
	In
\cite{Aad:2014vka} 	
	limits are also set on the coupling and mediator mass of a 
	model in which the interaction is mediated by a scalar particle
\cite{Bell:2012rg}. 
	In this model a scalar-mediator $\eta$, with mass $m_{\eta}$, and a 
	$\eta$-WIMP coupling strength $f$ is responsible for the production of the DM particles. 
	Limits on the cross section times branching ratio in the scalar-mediator model are shown in 
	Fig.~7 of \cite{Aad:2014vka}, and limits on $f$ as a function of mediator mass $m_{\eta}$ and $m_{\chi}$, as well as the exclusion region, are shown in Fig.~8 of \cite{Aad:2014vka}.
	These limits do not look very exciting, and neither does their further applicability. 

\begin{figure}[!ht] 
\includegraphics[width=0.2\textwidth]{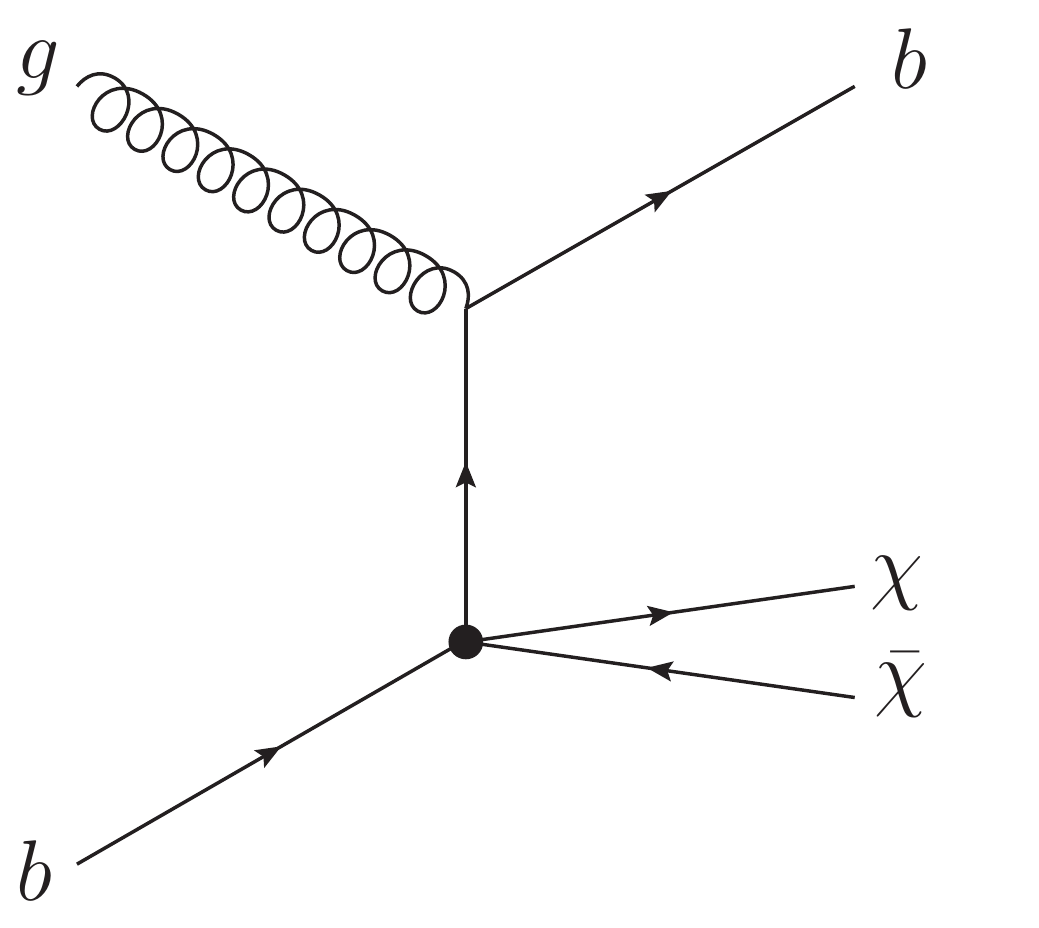} \hspace*{25pt}   
\includegraphics[width=0.2\textwidth]{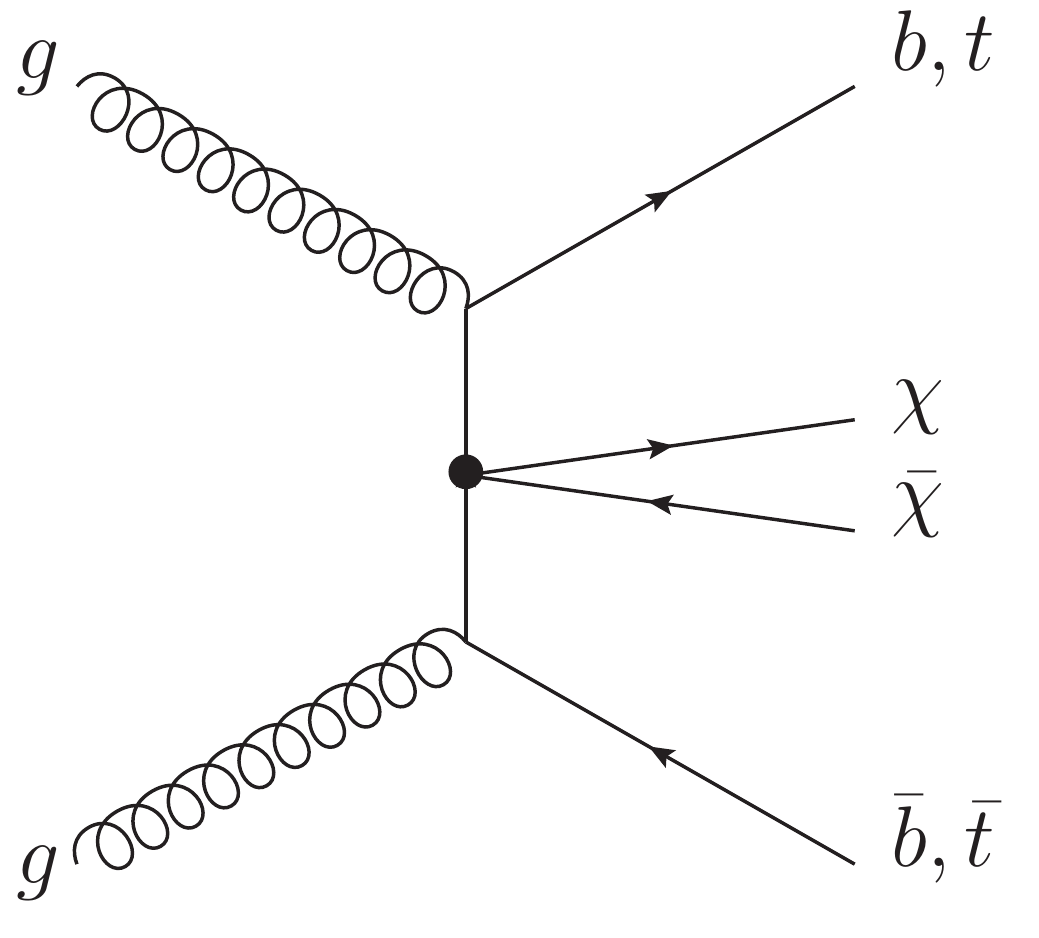} 
  \caption{Dominant graphs for WIMP $\chi$ production in conjunction with (left) 
  a single $b$-quark and (right) a heavy quark (bottom or top) pair. 
From \cite{Aad:2014vea}.} \label{fig:2014vea-DMb_tt}
\end{figure}

\smallskip
	Investigation of the DM problem with the ATLAS detector was continued by dedicated   
{\em Search for dark matter in events with heavy quarks and missing transverse momentum ...} 
\cite{Aad:2014vea} with 20.3 fb$^{-1}$ of $pp$ collisions collected at 8 TeV. 	
	This search for WIMP pair production in association with bottom or top quarks
($pp\to \chi\chi+ b(\bar{b}), t(\bar{t})+X$) was carried out
	in events with large \met\/ produced together with high-momentum jets 
	of which at least one was a b-quark-jet 
(Fig.~\ref{fig:2014vea-DMb_tt}). 
	Final states with top quarks were selected by requiring a high jet multiplicity and 
	in some cases a single lepton
\cite{Aad:2014vea}. 

\begin{figure*}[!ht]
\includegraphics[width=0.8\textwidth]{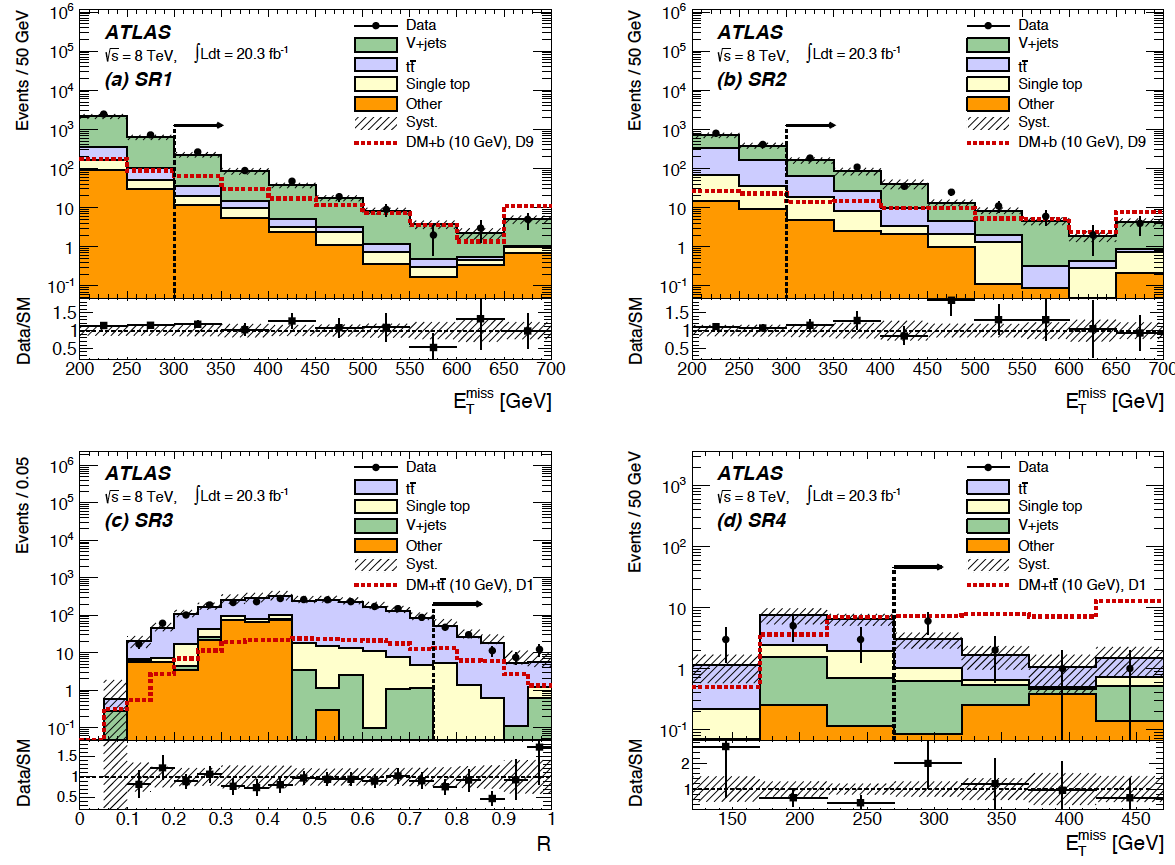}
\caption{Comparison between data and expected SM background.  
    \met\ distributions in SR1 (a) and SR2 (b) with expected signal (red line) 
    for $\chi \bar{\chi} + b(\bar{b})$ due to D9-operator. 
   $R$ distribution in SR3 (c) excluding the selection on $R$.  
   \met\ distribution in SR4 (d) excluding the selection on \met.  
   In (c), (d) the expected signal (in red) is given for
    $\chi \bar{\chi} + t\bar{t}$  due to D1-operator.  
    The final selection requirements are indicated by an arrow.
  Other backgrounds are composed of diboson and multijet production.  
 From \cite{Aad:2014vea}.}
 \label{2014vea-DataMC}
\end{figure*}

Figure~\ref{2014vea-DataMC} shows the measured \met\ distributions for 
	three signal regions and the so-called Razor variable
	$R$-distribution for the 3rd signal region. 
	Variable $R$ allowed one to utilize 
	both transverse and longitudinal information about the event
\cite{Rogan:2010kb}. 
	On this basis maximal rejection of the abundant $t\bar{t}$-background
	was achieved for $R >0.75$.	
	In the figure all expected signals of the WIMP production were calculated for $m_\chi=10\,$GeV/$c^2$. 
	The data distributions were found to be consistent with the relevant SM expectations.

\begin{figure*}[!ht]
\includegraphics[width=0.8\textwidth]{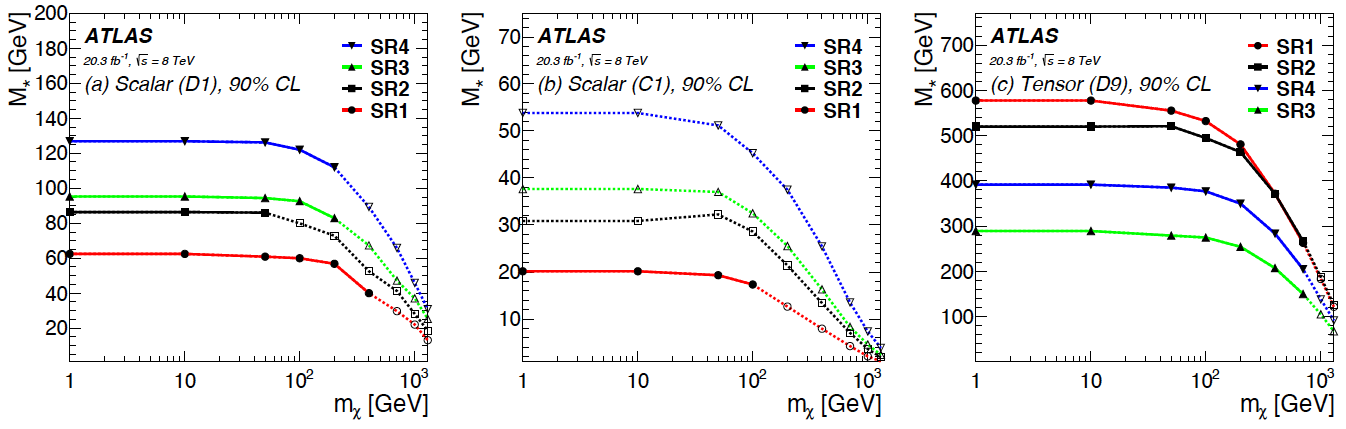}
\caption{Lower limits on $M_*$  (90\% CL) for the  SR1 (red), SR2 (black), 
SR3 (green), and SR4 (blue) as a function of $m_{\chi}$ for the operators 
(a) D1, (b) C1, and (c) D9. 
Solid lines and markers indicate the validity range of the EFT.  
the dashed lines and hollow markers represent the full collider constraints.
From \cite{Aad:2014vea}.}
\label{fig:2014vea-DCD}
\end{figure*}

	From the measured \met-distributions limits were set on the mass scale $M_*$ 
	of the scalar D1, C1 and tensor D9  EFT operators 
(Table~\ref{tab:2010ku}), that were used to describe WIMP-SM interactions
\cite{Goodman:2010ku}.
	For these operators
Fig.~\ref{fig:2014vea-DCD} 
	shows $M_*$--$m_{\chi}$ lower limits obtained from 4 different signal regions
	SR1--SR4, defined in Table~1 of  
\cite{Aad:2014vea}. 
	Typical rather substantial $M_*$-dependence on a type of EFT operator and signal region 
	is clearly seen. 

	From
Fig.~\ref{fig:2014vea-DCD}
	the traditional exclusion curves for the WIMP-nucleon cross-section 
	for SI  and SD interactions were obtained. 
\begin{figure}[!ht]
\includegraphics[width=0.7\textwidth]{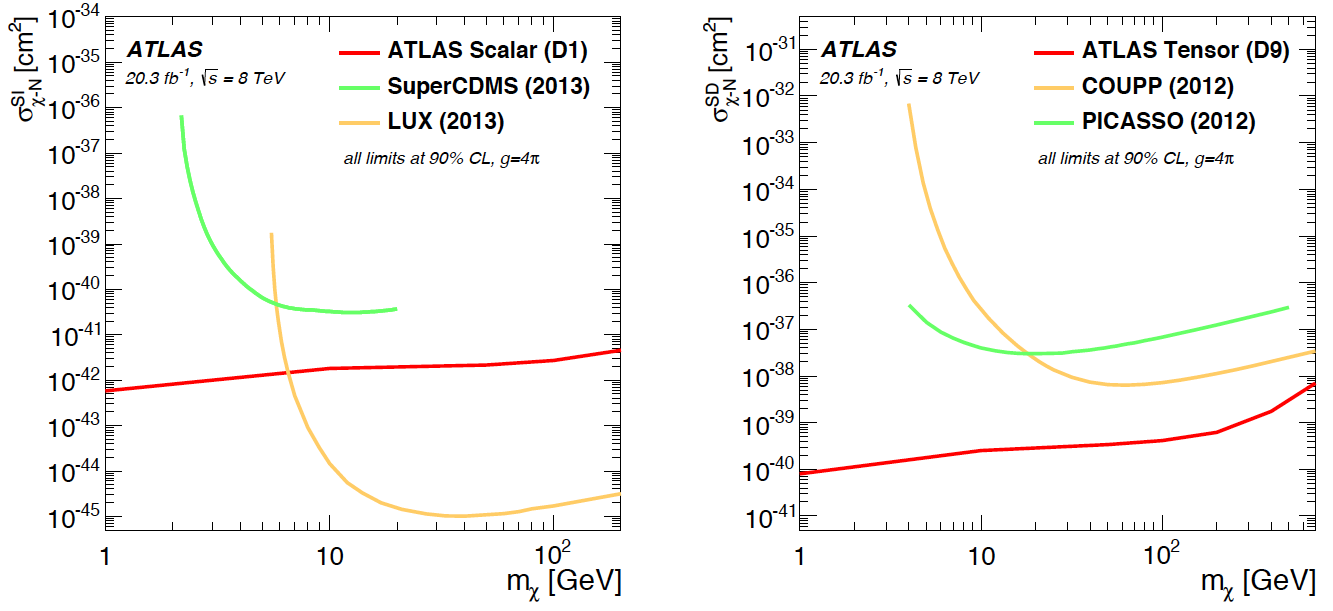}
\caption{90\% CL upper limits (red) on the 
$\chi$-nucleon cross-section as a function of $m_{\chi}$
 (left) for the spin-independent coupling (scalar operator D1), and  
(right) for the spin-dependent coupling  (tensor operator D9). 
The other  
curves show the exclusion curves from some of direct DM experiments
\cite{Akerib:2013tjd,Agnese:2013jaa,Behnke:2012ys,Archambault:2012pm}.  
From \cite{Aad:2014vea}.}
  \label{fig:2014vea-SISD}
\end{figure}
Figure~\ref{fig:2014vea-SISD} shows that, as expected, 
	the limits on the $\chi$-nucleon cross-section 
	improve rather significantly previous constraints in the regions of low-mass WIMPs.

\begin{figure}[!ht]\centering 
\includegraphics[width=0.3\textwidth]{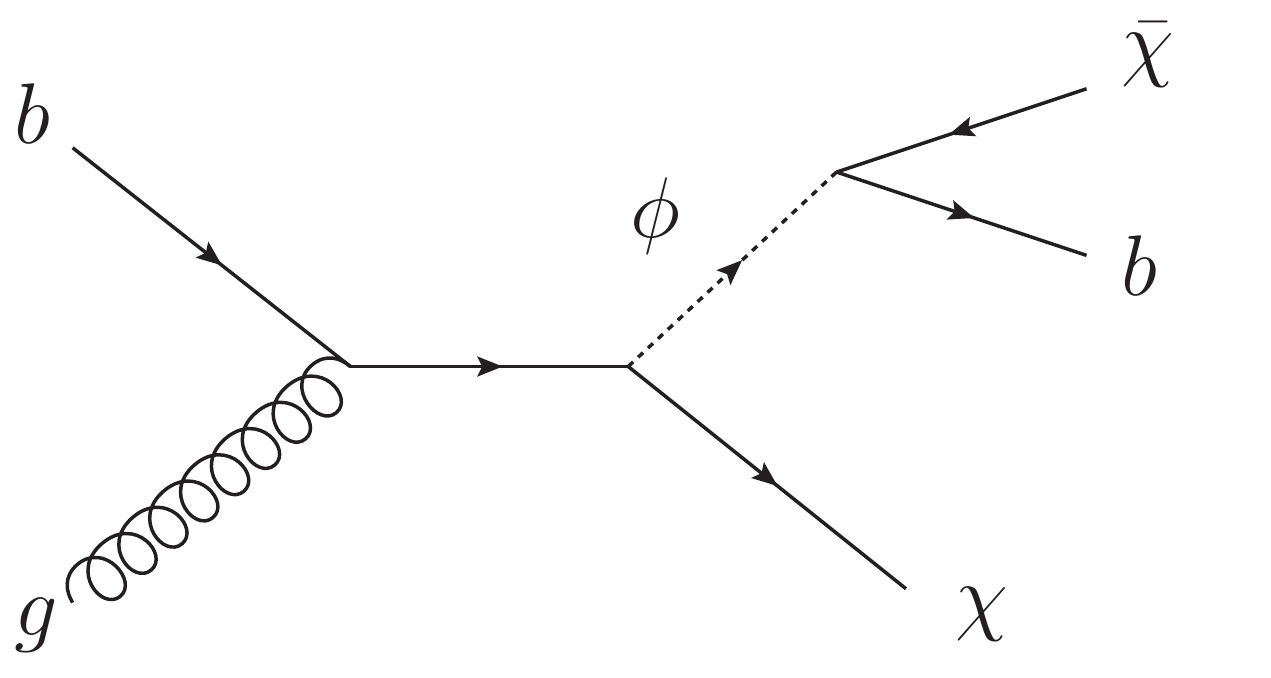}
\caption{Example of DM, $\chi$, production in the $b$-FDM model.
From \cite{Aad:2014vea}.  }
\label{fig:2014vea-bFDM}
\end{figure}

	The results of 
Fig.~\ref{2014vea-DataMC} were also interpreted in a bottom-Flavoured DM model 
($b$-FDM)~\cite{Agrawal:2014una}. 
	The model was proposed to explain the Fermi-LAT excess of $\gamma$-rays from the Galactic center 		
	in terms of specific DM annihilation, when  
	the DM particles with mass of about 35~GeV$/c^2$ 
	annihilated into $b$-quarks via a colored  
	new scalar field, $\phi$ 
(Fig.~\ref{fig:2014vea-bFDM}). 
	The DM $\chi$ particle is assumed to be a Dirac fermion that couples to right-handed 
	down-type quarks, mainly to the $b$-quarks.
	Therefore the collider signature is $b$-quarks produced in association with \met. 

	From the observed exclusion contour in the ($m_\chi$, $m_\phi$) plane given in Fig.~7 of 
\cite{Aad:2014vea}, one concluded that in the $b$-FDM model
	mediators with $300 < m_\phi< 500$ GeV$/c^2$ are excluded at 95\% CL
	for DM particles with $m_\chi\simeq 35$ GeV$/c^2$.  
	Unfortunately, this information is not yet enough to reject completely the $b$-FDM model. 
	The general conclusion of \cite{Aad:2014vea} is typical --- 
	the data are consistent with the SM, limits are the strongest at low DM masses. 

\smallskip  
	The ATLAS collaboration has produced new results of mono-photon, 
	mono-lepton and mono-jet study with 8 TeV data, and has 
	improved previously published 7-TeV-based DM constraints.
	Remarkably, the new "8-TeV" papers used 
	"new phenomena" or "new particles" instead of the words "dark matter" in their titles.	  

\begin{figure}[!ht]
\centering
\includegraphics[width=0.5\linewidth]{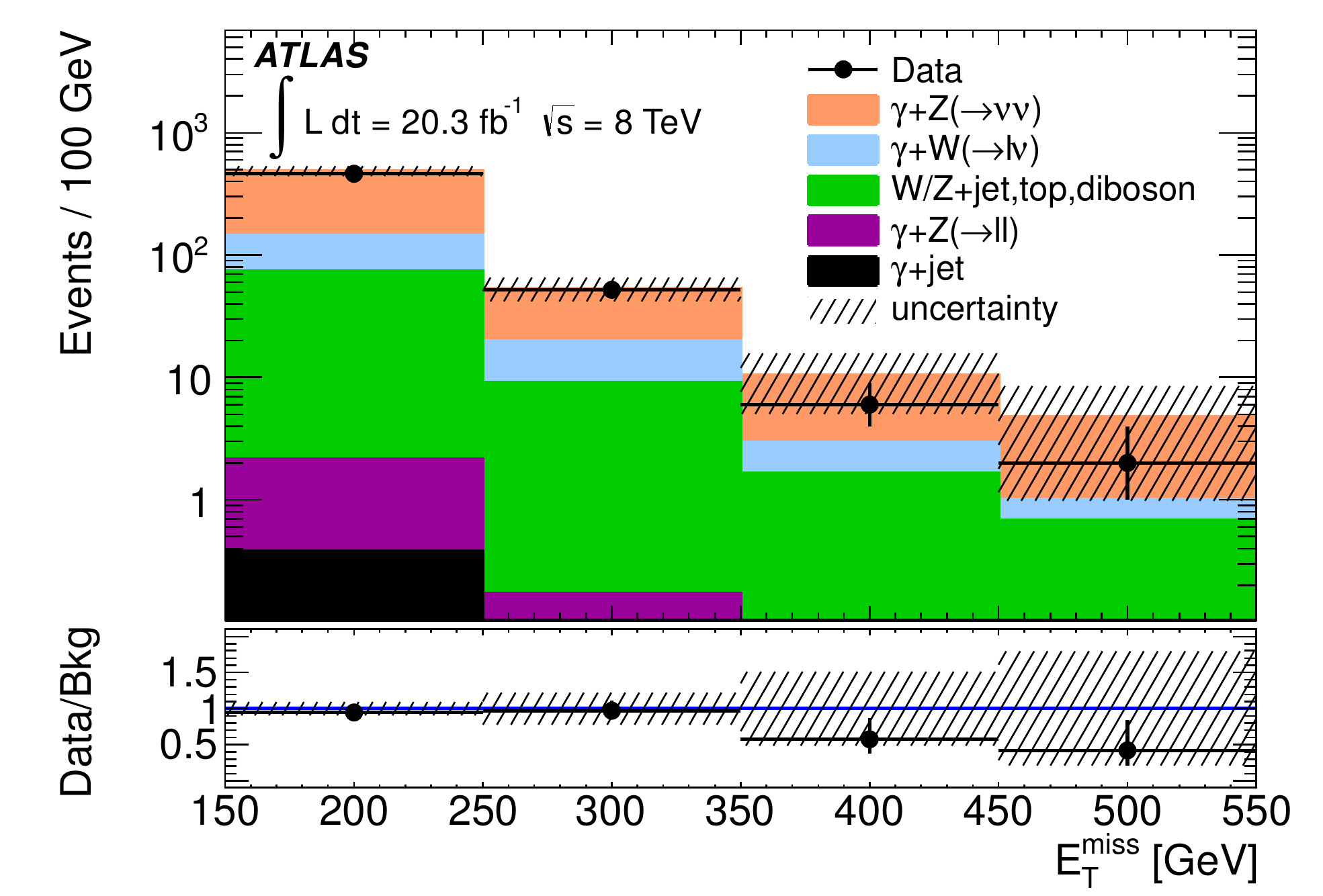} 
\caption{Distribution of \met\ in the signal region for the data and for the background.  
The lower part of the figure shows the ratios of the data to the expected-background event yields.
From \cite{Aad:2014tda}.}
\label{fig:2014tda-Main}
\end{figure}

	Results of a search for {\em new phenomena} in events with an energetic photon 
	and large missing transverse momentum  
	in $pp$ collisions at 8 TeV and integrated luminosity of 20.3 fb$^{-1}$
	were reported in 
\cite{Aad:2014tda}. 
	The obtained \met\ distribution  
	of events with an energetic photon is shown in Fig.~\ref{fig:2014tda-Main}.
	The observed (in signal region) 521 events were well described by the SM
	background prediction of $557 \pm 36 \pm 27$, extrapolated from control regions. 
	These results were interpreted in terms of exclusions on models 
	that would produce an excess of  the $\gamma+$\met\ events.

	If $\sigma$ denotes a cross section of a new physics process, 
	producing the $\gamma+$\met\ signature, 
	then the most model-independent limit can be set on the fiducial cross section $\sigma\times A$.
	The fiducial acceptance $A$ was defined in
\cite{Aad:2014tda}. 
	The limit on $\sigma\times A$ was derived from a limit on the visible 
	cross section $\sigma\times A \times \epsilon$, 
	where $\epsilon$ is the fiducial reconstruction efficiency. 
	A conservative estimate $\epsilon = 69$\% was computed using 
	WIMP samples with no quark/gluon produced from the main interaction vertex. 
	The finally observed upper limit on the fiducial cross section was 
	5.3 fb (95\% CL).   
	This limit is applicable to any model that produces $\gamma+$\met\ 
	events in the fiducial region and has similar reconstruction efficiency $\epsilon$
\cite{Aad:2014tda}.

\begin{figure}[!h]\centering
\includegraphics[width=0.3\linewidth]{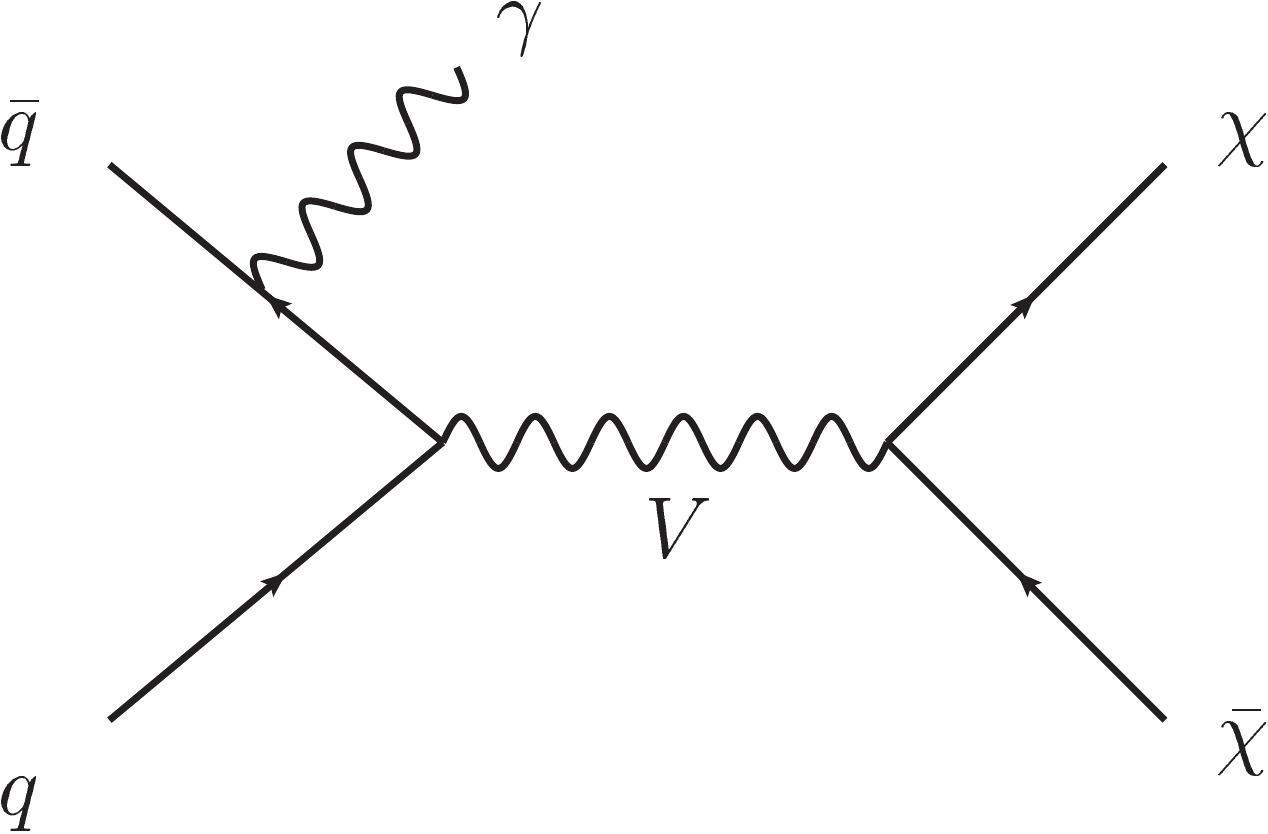} 
\caption{Production of $\chi\bar{\chi}$-pairs via an explicit $s$-channel  $V$-mediator. 
The mass suppression scale of contact interaction appears via 
$M_*=m_V/\sqrt{g_fg_\chi}$, where $m_V$ is the mediator mass, 
$g_f$ and $g_\chi$ represent the relevant coupling factors. 
 From \cite{Aad:2014tda}.}
  \label{fig:2014tda-Vgraph}
\end{figure}

	In 
\cite{Aad:2014tda} both the EFT approach 
\cite{Goodman:2010ku} with two parameters $m_\chi$ and $M_*$ 
	and a model with a $Z'$-like mediator
\cite{Fox:2011pm} were considered.
	In the latter case the mediator state $V$ can be explicitly produced
	when the typical momentum transfer in LHC collisions could reach 
	the scale of the microscopic interaction $Q \ge m_V$    
(Fig.~\ref{fig:2014tda-Vgraph}).
	The interaction is described in the model by four parameters ---
	$m_\chi$, $m_V$, the width of the mediator $\Gamma$, 
	and the overall coupling $\sqrt{g_fg_\chi}$. 
 
	From the main results given in 
Fig.~\ref{fig:2014tda-Main} traditional EFT constraints  
	were derived in the form of
	limits on the $M_*$ as a function of $m_\chi$.
	But at this time one has cared about validity of the EFT.   
	When the momentum transfer $Q$ 
	becomes comparable to the mass of the intermediate state 
	$m_V=M_*\sqrt{g_fg_\chi}$ the EFT approach fails 
\cite{Goodman:2010ku,Busoni:2013lha}.
	In order to have the situation under some control, limits obtained in
\cite{Aad:2014tda} were presented only when (in simulated events) $Q<m_V$, 
	for $\sqrt{g_fg_\chi}=1$, or $4\pi$ (i.e. when the perturbative approach was still valid). 
	This procedure was referred to as truncation.	
	For the effective operators D5 (vector), D8 (axial vector), and D9 (tensor) from Table~\ref{tab:2010ku} 
	these truncated limits are collected in Fig.~7, 8, and 9 
of \cite{Aad:2014tda}. 
	
	Finally these $M_*$-limits were translated, according to \cite{Goodman:2010ku},
	into exclusion curves for the WIMP-nucleon interaction cross section 
	as a function of $m_\chi$.  
%%%%%%%%%%%%%%%%%%%%%%%%
\begin{figure*}[!ht] \centering
\includegraphics[width=0.7\linewidth]{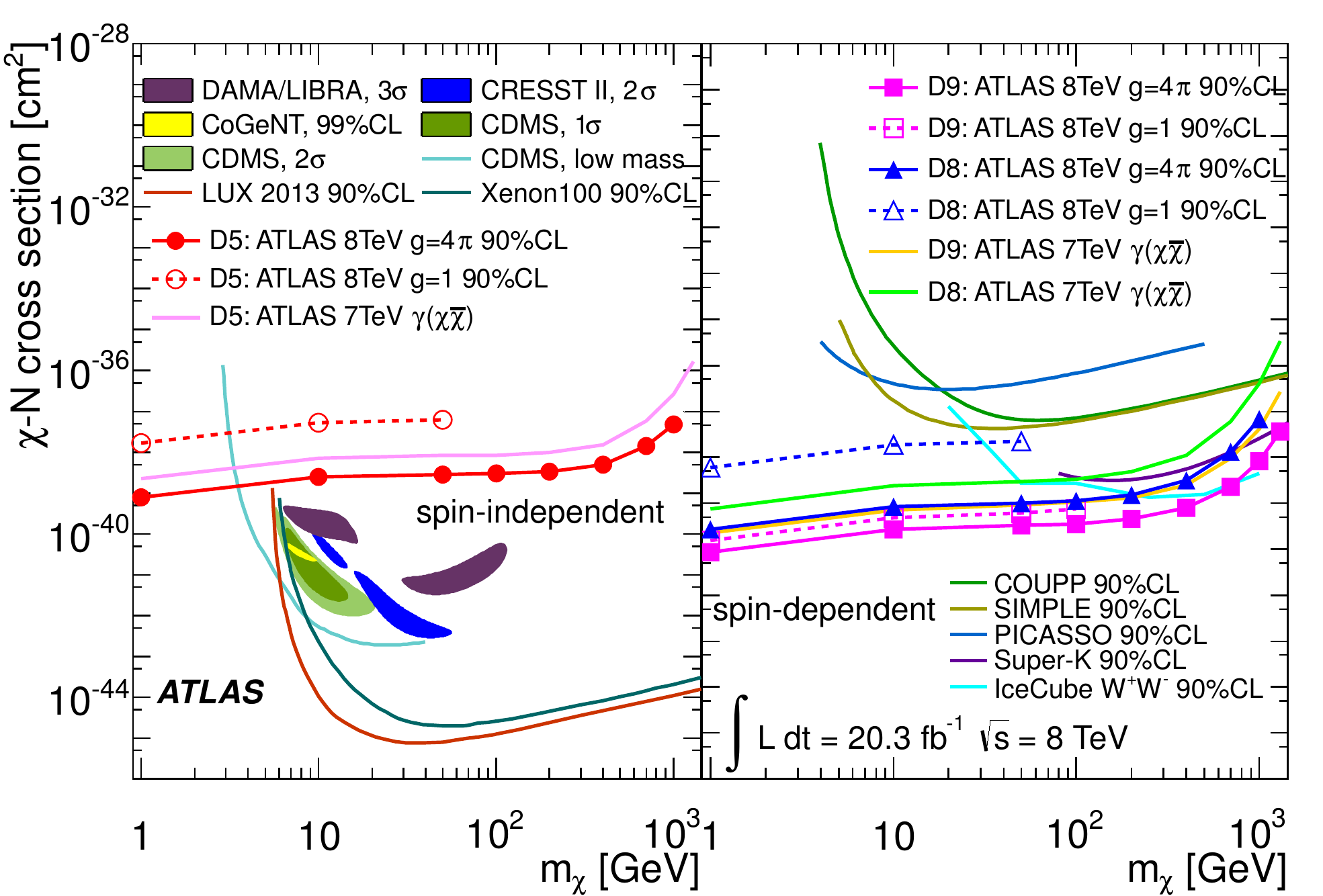} 
\caption{Upper limits (90\% CL) on the $\chi$-nucleon cross section as a function of $m_{\chi}$
for spin-independent (left) and spin-dependent (right) interactions.
 The truncation procedure is applied and coupling strength $g\!\equiv\!\sqrt{g_fg_\chi}\!=\!1$, or  $4\pi$.
 The previous ATLAS results obtained with 7 TeV data 
 and results from DM search experiments~
 \cite{Aprile:2012nq,Agnese:2013jaa,Aalseth:2014jpa,Archambault:2012pm,Felizardo:2011uw,Akerib:2013tjd,Desai:2004pq,Aartsen:2012kia,Behnke:2012ys,Angloher:2011uu,Agnese:2013rvf,Bernabei:2008yi,Abramowski:2013ax}
are also shown.
 From \cite{Aad:2014tda}.}
\label{fig:2014tda-SISD}
\end{figure*}
	The results are shown in 
Fig.~\ref{fig:2014tda-SISD} for spin-independent (D5) and spin-dependent (D8, D9) $\chi$--nucleon 
	interactions and are compared to other measurements from various DM search experiments.

	One can conclude from the figure that 
	the LHC search for WIMP pair production accompanied by an energetic $\gamma$-quanta 
	traditionally extends the limits on the $\chi$-nucleon scattering cross section 
	to the low-mass region of $m_{\chi}<10$ GeV$/c^2$. 
	In fact, it brings nothing new.

\smallskip
	The $\gamma+$\met\ data were used in 
 \cite{Aad:2014tda} to constrain another DM model, which coupled directly WIMPs and SM gauge bosons
\cite{Nelson:2013pqa}.
	The effective WIMP couplings to different bosons were 
	parametrized by the coupling strengths $k_1$ and $k_2$, 
	which correspond to the U(1) and SU(2) gauge sectors of the SM. 
\begin{figure}[!h]\centering
  \includegraphics[width=0.3\linewidth]{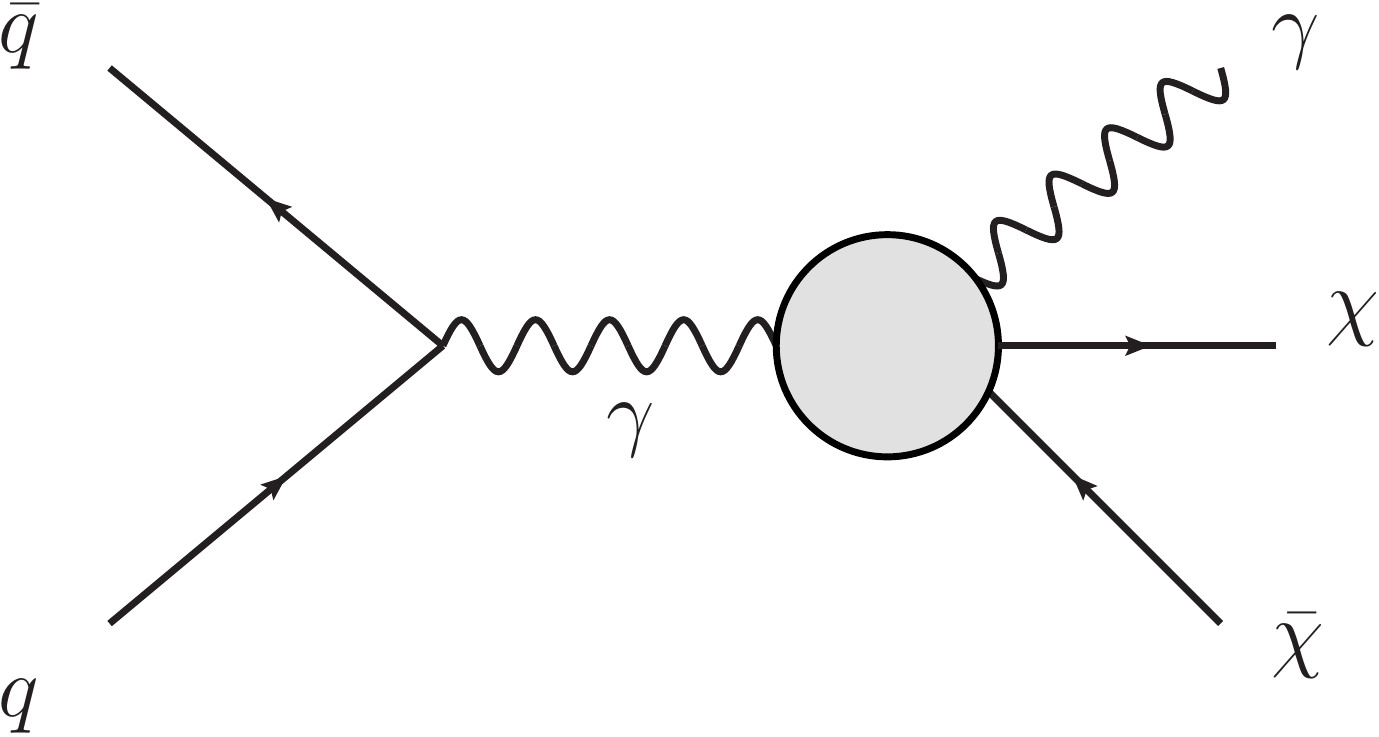} 
  \caption{Production ($s$-channel) of pairs of WIMP particles ($\chi\bar{\chi}$) via an 
  effective  $\gamma\gamma\chi\bar{\chi}$ vertex.
From \cite{Aad:2014tda}.}
  \label{fig:2014tda-Fermi}
\end{figure}
	WIMP production in the model via 
	$pp\rightarrow\gamma + X \rightarrow\gamma\chi\bar{\chi} + X'$, 
	does not require any initial-state radiation 
(Fig.~\ref{fig:2014tda-Fermi}). 
 	This model can also describe 
	the line near 130 GeV in the Fermi-LAT $\gamma$-ray spectrum
\cite{Weniger:2012tx}, 
	and allows a direct comparison of the Fermi-LAT and the ATLAS data in the same parameter space.	

\begin{figure}\centering
\includegraphics[width=0.5\linewidth]{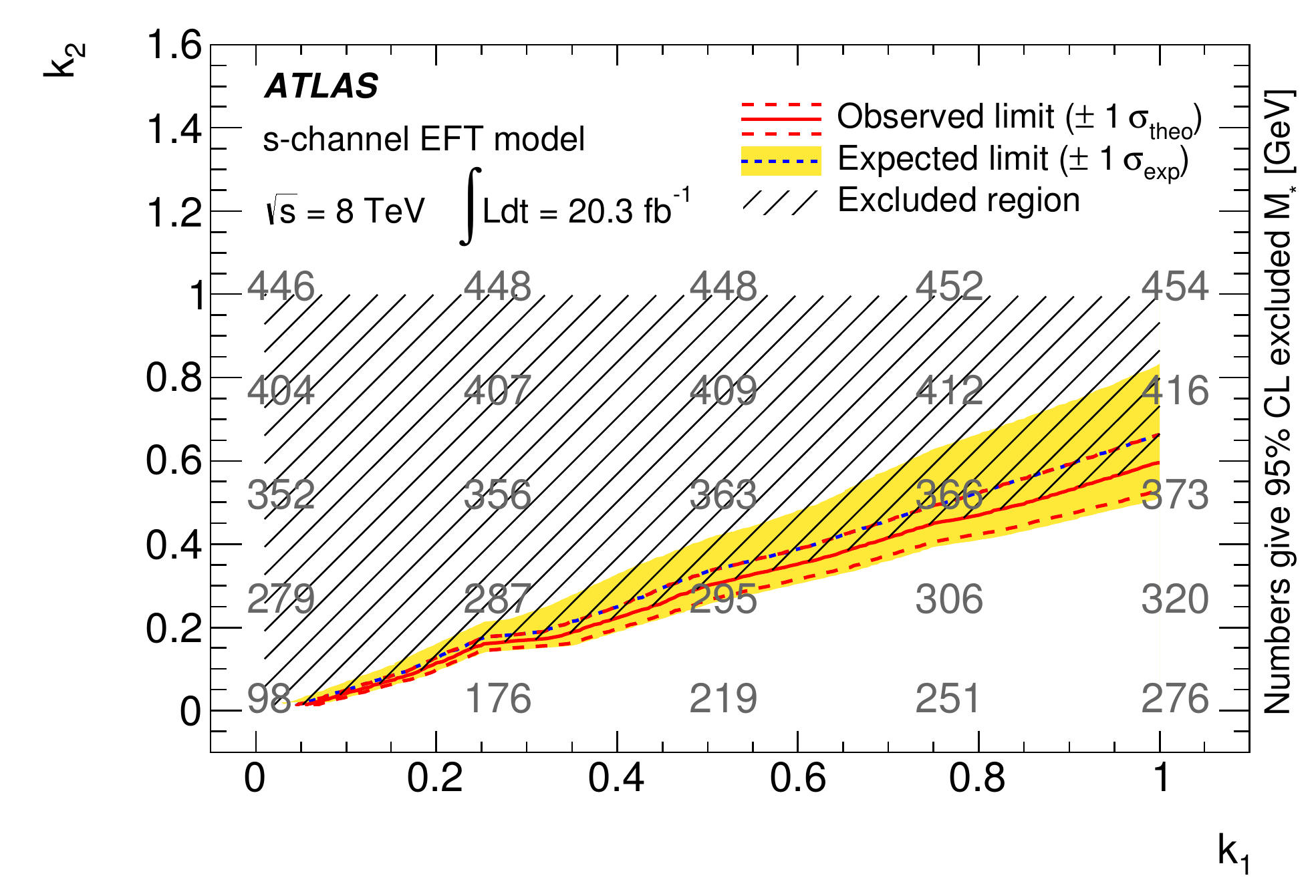} 
\caption{Limits at 95\% CL on the effective mass scale $M_*$  in the ($k_2$,$k_1$) 
parameter plane for the $s$-channel EFT model inspired by Fermi-LAT $\gamma$-ray line, for $m_{\chi} = 130$ GeV$/c^2$.  
 The upper part of the plane is excluded.
 From \cite{Aad:2014tda}.}  \label{fig:2014tda-s-channel} 
\end{figure}

	For this model limits were placed on the effective mass scale $M_*$
	 in the ($k_2$,$k_1$) parameter plane, 
	as shown in  Fig.~\ref{fig:2014tda-s-channel}.
	The exclusion line is drawn by considering the value of $M_*$ 
	needed to generate the $\chi\bar{\chi}\rightarrow\gamma\gamma$ 
	annihilation rate consistent with the observed 
	Fermi-LAT $\gamma$-ray line.

\smallskip
	Contrary to the EFT problem with verification of validity (by means of truncation), 
	any simplified model with explicit mediator is ultraviolet complete 
	and therefore robust for all values of $Q$. 
	For such a simplified $Z'$-like model with vector interactions 
	and mediator width $\Gamma=m_V/3$, Fig.~14
from \cite{Aad:2014tda} shows the 
	limits on the coupling parameter 
	$\sqrt{g_fg_\chi}$ calculated for various values of the WIMP 
	and mediator particle masses, and compared to the lower 
	limit resulting from the relic DM abundance 
\cite{Komatsu:2010fb}.
	This competition between the relic DM abundance limits and LHC-found limits 
	looks very complicated and not very impressive. 
	
\begin{figure}[htbp]\centering
\includegraphics[width=0.6\linewidth]{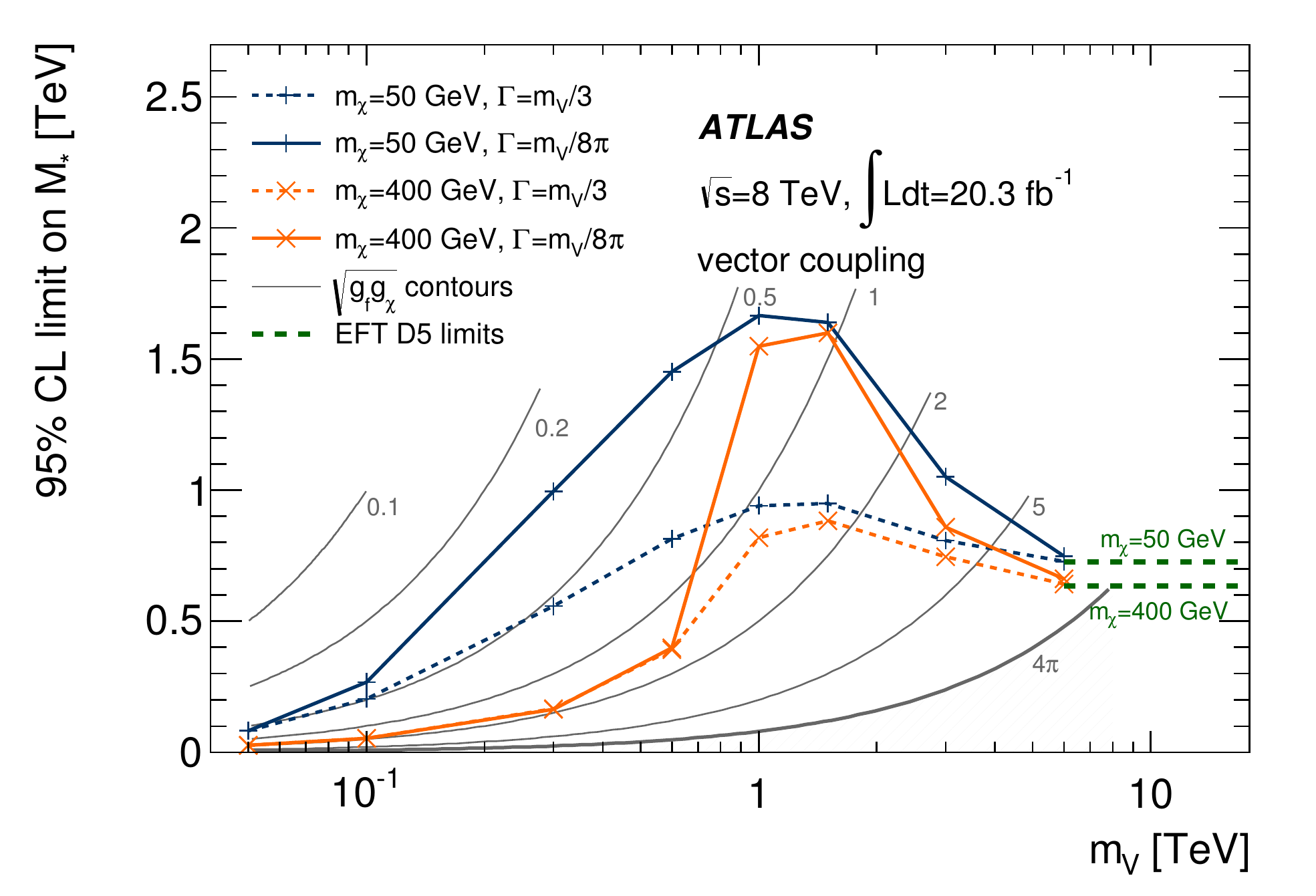}
\caption{Observed lower limits at 95\% CL on the EFT suppression scale $M_*$ as 
a function of the mediator mass $m_V$, for a $Z'$-like mediator with vector interactions.  
Results are shown for different values of the
mediator total decay width $\Gamma$ and compared to the EFT observed limit results
for a D5 (vector) interaction (dashed line).
From \cite{Aad:2014tda}.}
\label{fig:2014tda-VectorMediator} 
\end{figure}

\begin{figure}[htbp]\centering
\includegraphics[width=0.6\linewidth]{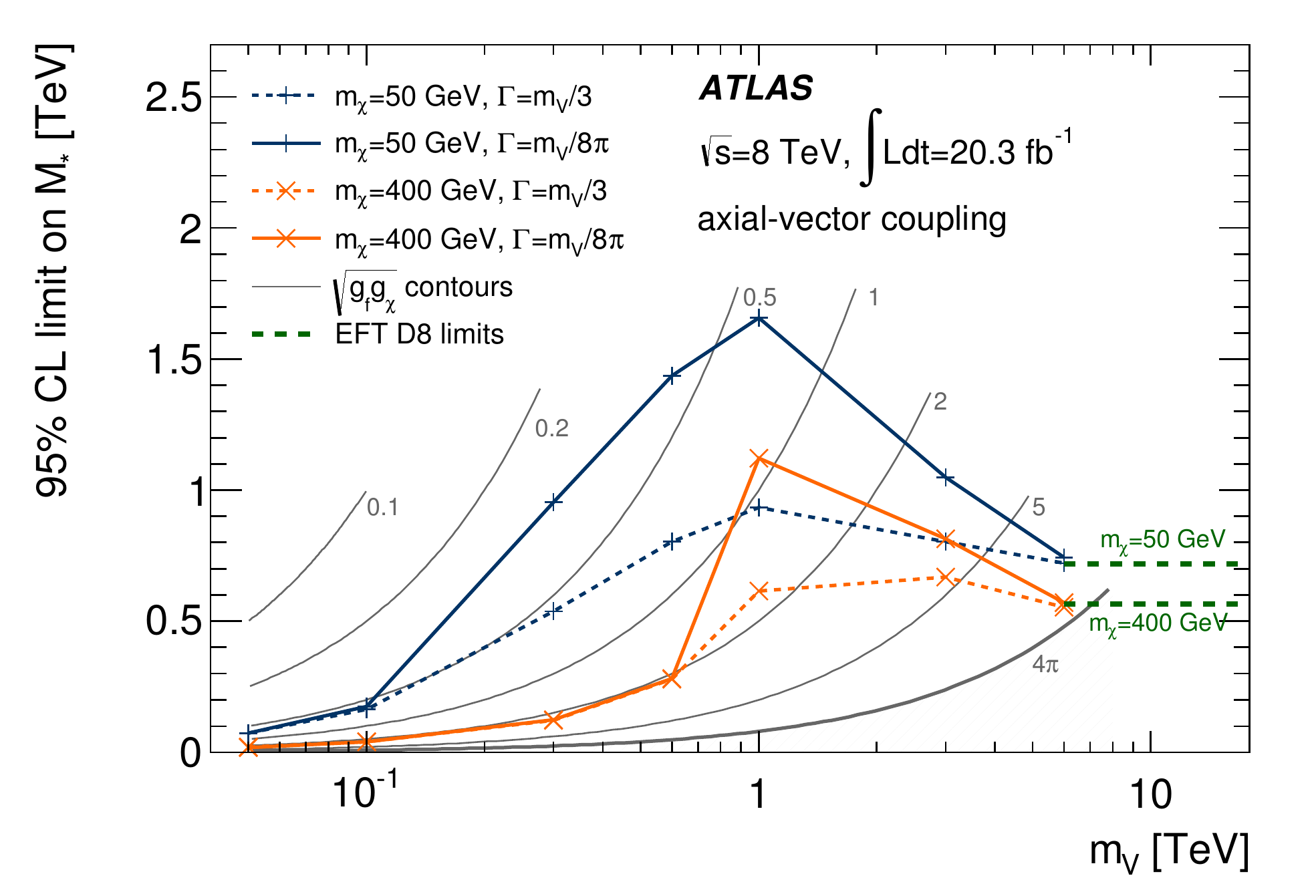} 
\caption{Observed limits at 95\% CL on the EFT suppression scale $M_*$ as a function of the 
mediator mass $m_V$, for a $Z'$-like mediator with axial-vector interactions. 
Results are shown for different values of the mediator total decay width 
$\Gamma$ and compared to the EFT observed limit results for a D8 (axial-vector)
 interaction (dashed line).
From \cite{Aad:2014tda}.}
   \label{fig:2014tda-AxialMediator} 
\end{figure}

	Furthermore, for this model with $Z'$-like mediator
\cite{Fox:2011pm} 
	limits on $M_*$ as a function of $m_V$ are shown
	for vector (Fig.~\ref{fig:2014tda-VectorMediator}) and 
	for axial-vector (Fig.~\ref{fig:2014tda-AxialMediator})  interactions.
	The limits are given for two representative WIMP masses $m_\chi$ of 50 and 400 GeV$/c^2$. 
	The $M_*$--$m_V$ contours (thin lines) for an overall coupling
	$\sqrt{g_fg_\chi}=0.1,0.2,0.5,1,2,5,4\pi$ are also shown in both figures.

	One can see that when $m_V$  
	is greater than the LHC reach, the EFT approach provides a good approximation of 
	the simplified model. 
	The EFT limits look always more conservative than those from the 
	simplified model as long as $m_V$ is greater than or equal to the value used for EFT truncation. 
	This can be seen by comparing the $M_*$ limits derived from the EFT approach 
	using truncation to those of the simplified model, recalling that $m_V=M_*\sqrt{g_f g_{\chi}}$.

\smallskip
	Finally observed $\gamma+$\met\ distributions  were used  in
\cite{Aad:2014tda} for supersymmetry constraints.
	Collisions of protons could result in pair production of squarks, $\tilde{q}$, 
	which could decay to a SM quark and a stable neutralino $\tilde{\chi}_1^0$.   
\begin{figure}[!ht]\centering
  \includegraphics[width=0.3\linewidth]{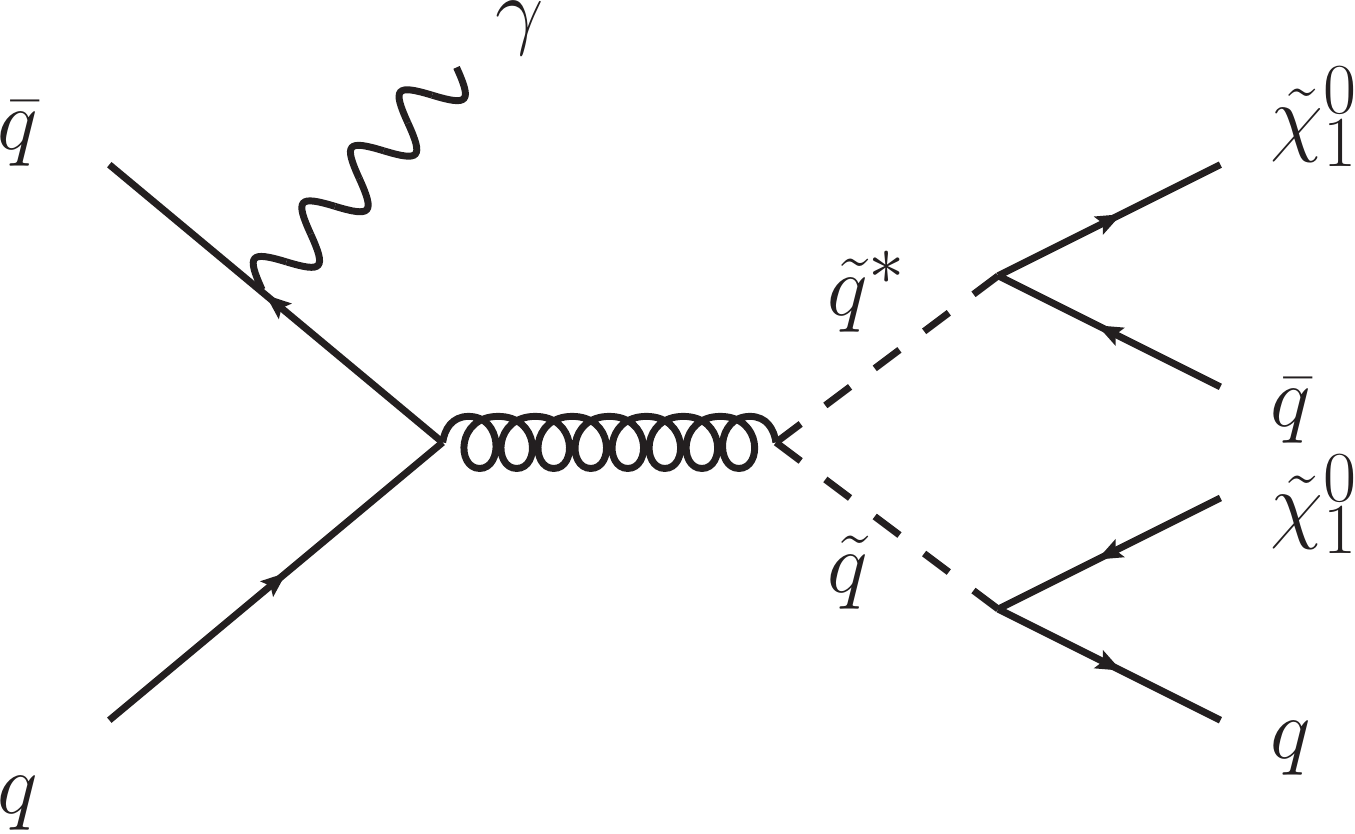} 
  \caption{Pair production of squarks ($\tilde{q}$), followed by decay into quarks and neutralinos ($\tilde{\chi}_1^0$). 
 From \cite{Aad:2014tda}.}
  \label{fig:2014tda-SUSY}
\end{figure}
If the mass difference $m_{\tilde{q}}-m_{\tilde{\chi}_1^0}$ is small, the SM quarks would have very low momentum and would therefore not be reconstructed as jets.   Again, the radiation of a photon either from an initial-state quark or an intermediate squark would result in $\gamma+$\met\ events, 
as shown in Fig.~\ref{fig:2014tda-SUSY}.

\smallskip
	One can conclude that the data of
\cite{Aad:2014tda} are well described by the expected SM backgrounds. 
	The observed upper limit on the fiducial cross section for the production of 
	$\gamma+$\met\ events is 5.3 fb (95\% CL).
	More sophisticated analysis was given, validity of the EFT was discussed.
	Exclusion limits were presented on models of new phenomena with large extra spatial dimensions, 
	supersymmetric quarks, and direct pair production of WIMP DM candidates
\cite{Aad:2014tda}.  
	Nevertheless, the paper leaves a feeling that one sinks into
	a set of quasi-model-independent receipts with a number of parameters
	of low physical meaning.   

\begin{figure*}[!ht] \centering
  \includegraphics[width=0.6\textwidth]{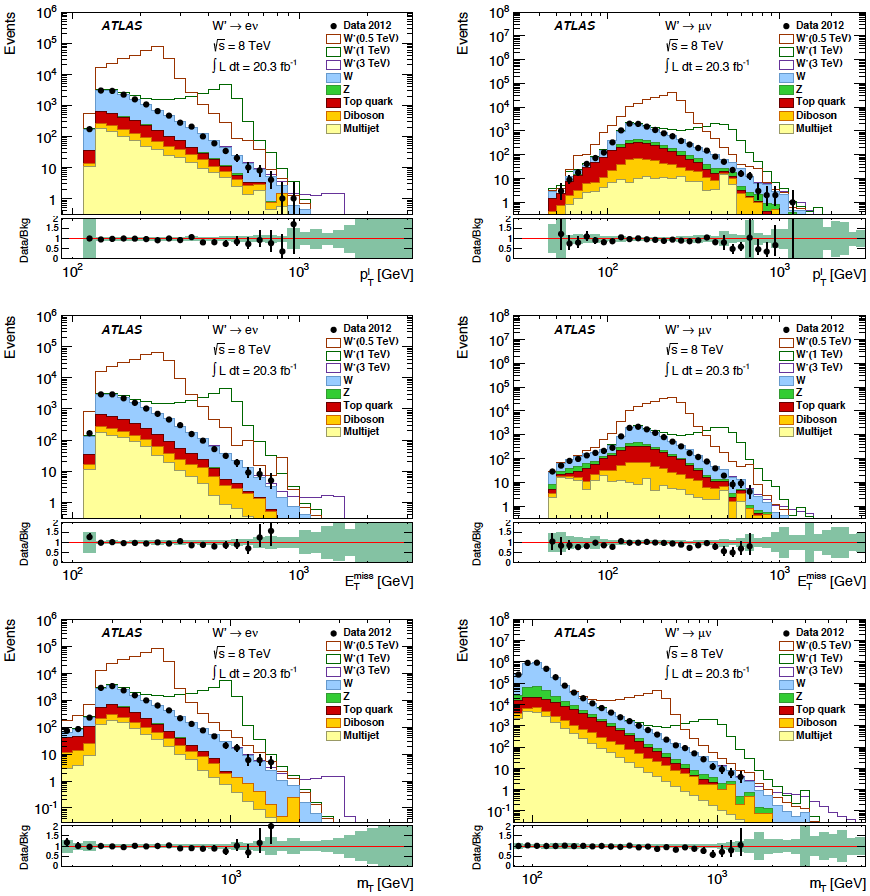}
  \caption{Spectra of lepton \pt\ (top), \met\ (centre) and \mt\ (bottom)
  for the electron (left) and muon (right) channels.  
 The spectra of \pt\  and \met\ are shown with the requirement \mt\ $>252$ GeV.
 The ratio of the data to the total background prediction is shown below each of the distributions.
From \cite{ATLAS:2014wra}.} \label{fig:2014wra-MainResult}
\end{figure*}

\smallskip 
	 Continuation of the 
	{\em Search for new particles in events with one lepton and missing transverse momentum in $pp$
	collisions at $\sqrt{s}$ = 8 TeV with the ATLAS detector} 
	using 20.3 fb$^{-1}$ of collected data was published in 
\cite{ATLAS:2014wra}.
	For measured electrons and muons   
Fig.~\ref{fig:2014wra-MainResult} shows the \pt, \met, and \mt\ spectra 
	after final event selection.  
	The expected SM background and 
	three examples of $W'$-boson signals at different masses are also given in the Figure.	

	The transverse mass variable
	$m_{\rm T} =\sqrt{ 2 p_{\rm T}  E^{\rm miss}_{\rm T} (1 - \cos \varphi_{\ell\nu})}$
	was used to identify the signal.
	Here \pt\ is the lepton transverse momentum, 
	\met\ is the magnitude of the missing transverse momentum vector and 
	$\varphi_{\ell\nu}$ is the angle between the \pt\ and \met\ vectors.
	The \met\ in each event is evaluated by summing over energy-calibrated physics objects
	(jets, photons and leptons) and adding corrections for calorimeter deposits not associated with these objects
\cite{ATLAS:2014wra}.

	One can see agreement between the data and the predicted SM background 
	for events with \mt\ $<\ $252~GeV, the lowest \mt\ threshold used to search for a new physics.
	The value of the \mt\ threshold was a result of an optimization procedure. 
	No significant excess beyond SM expectations is observed. 

	The mono-lepton data spectra 
(Fig.~\ref{fig:2014wra-MainResult}) were used to constrain direct production of WIMP DM candidates, 
	which expected to be pair-produced, $pp\rightarrow\chi\bar{\chi}+X$, via some non-SM intermediate state. 
	To obtain these constraints the EFT contact operators  
	D1 (scalar), D5c (vector, with constructive interference) and D5d (vector, with destructive interference) 
	and D9 (tensor) from 
Table~\ref{tab:2010ku} were used.

\begin{figure*}[!ht]
  \centering
   \includegraphics[width=0.4\textwidth]{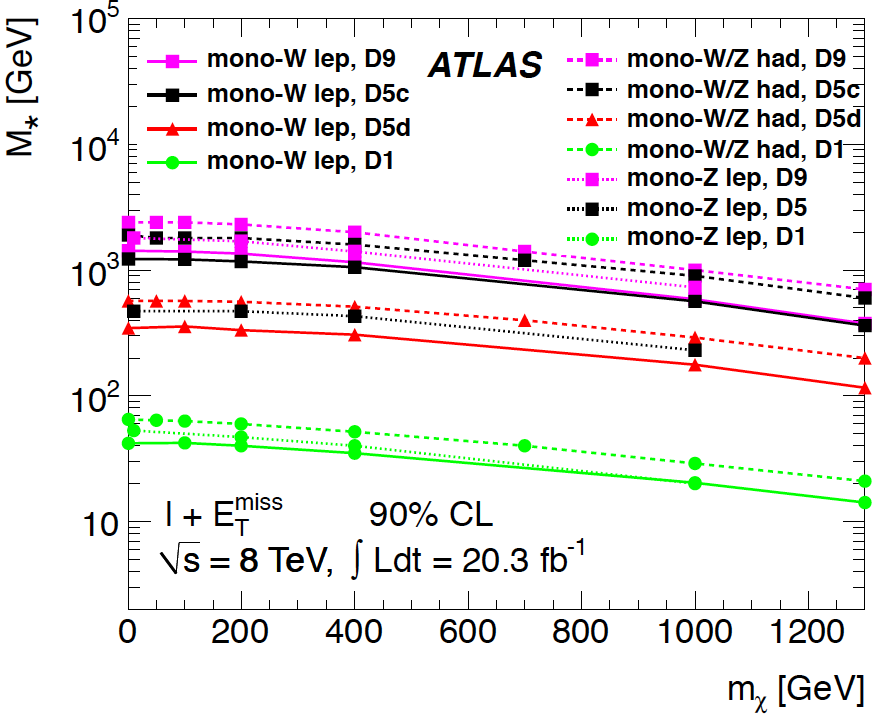}
  \caption{Observed limits for various EFT operators on $M_{*}$ as a function of the $m_{\chi}$ at 90\% CL 
  for the combination of the $e$- and $\mu$-channel.  
  The values below the corresponding line are excluded. 
 Results of the previous ATLAS searches  
  are also shown.
  \label{fig:2014wra-EFTlimits}
 From \cite{ATLAS:2014wra}. }
\end{figure*}

\begin{figure*}[!ht]
\centering\includegraphics[width=0.7\textwidth]{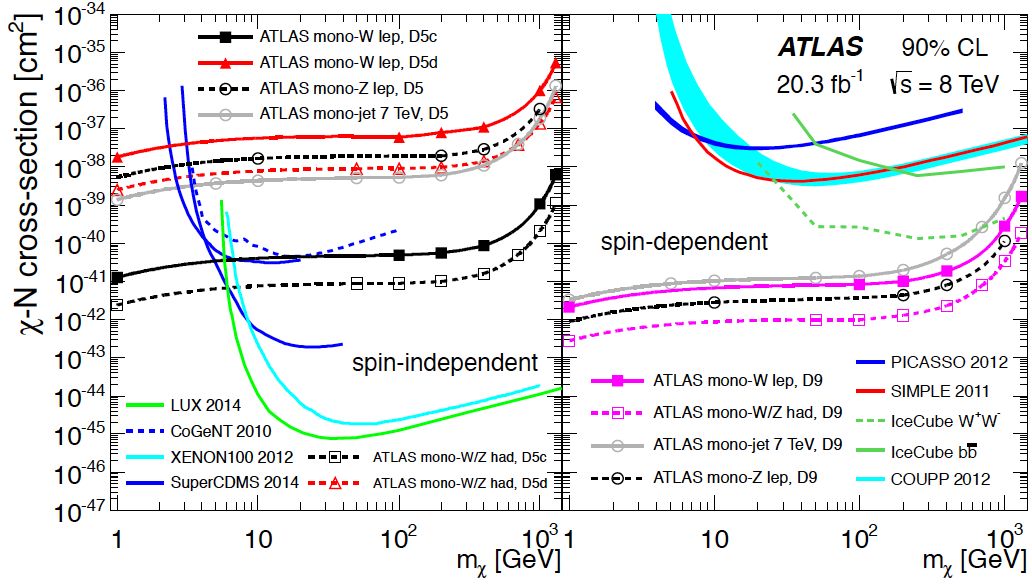}
\caption{Observed limits on the WIMP-nucleon scattering cross-section as a function of $m_{\chi}$ 
	at 90\% CL for spin-independent (left) and spin-dependent (right) EFT operators. 
	Results are compared with the previous ATLAS searches and with direct detection searches by
CoGeNT~\cite{Aalseth:2010vx}, 
XENON100~\cite{Aprile:2013doa},  
CDMS~\cite{Agnese:2013jaa,Agnese:2014aze}, 
LUX~\cite{Akerib:2013tjd},
COUPP~\cite{Behnke:2012ys}, 
SIMPLE~\cite{Felizardo:2011uw},
PICASSO~\cite{Archambault:2012pm} and 
IceCube~\cite{Aartsen:2012kia}.
\label{fig:2014wra-SISD}
From \cite{ATLAS:2014wra}.}
\end{figure*}

	On this basis the new results of the ATLAS search for pair production of WIMP particles 
	in association with a leptonically decaying $W$-boson at 8 TeV 
(Fig.~\ref{fig:2014wra-MainResult})
	were transformed into limits on $M_{*}$
	(Fig.~\ref{fig:2014wra-EFTlimits}) 
	and into exclusion curves for the WIMP-nucleon scattering cross-section
	(Fig.~\ref{fig:2014wra-SISD}).
	Both are shown as a function of the WIMP mass $m_{\chi}$.
	Results of the previous ATLAS searches for $W/Z$-boson decaying hadronically 
\cite{Aad:2013oja}, $Z$-boson decaying leptonically  
\cite{Aad:2014vka}, and $j+\chi\chi$
\cite{ATLAS:2012ky} are also given in the figures. 

	One can see that the WIMP production signature with hadronical $W$ decays gave 
	 a factor of 1.5 stronger limits on $M_{*}$ with respect to the signature with leptonical $W$ decays.
	The limits in Fig.~\ref{fig:2014wra-EFTlimits} were expected to be stable down to arbitrarily small values.
 	One should note that the comparison between direct detection and ATLAS results 
(given in Fig.~\ref{fig:2014wra-SISD}) is only possible within the  validity of the EFT formalism 
(section \ref{sec:Discussion}) which was not discussed in 
\cite{ATLAS:2014wra}.

	Besides looking for the WIMP pair production some other Beyond-SM results were obtained
in \cite{ATLAS:2014wra}. 
	In particular, a $W'$-boson with Sequential SM couplings was excluded  
	for masses up to 3.24 TeV$/c^2$, and excited chiral $W^*$-bosons 
\cite{Chizhov:2009fc,Chizhov:2008tp} 
	with equivalent coupling strengths were excluded for masses up to 3.21 TeV$/c^2$. 

\smallskip 
	Much more sophisticated analysis of the mono-jet signature was carried out 
	with 20.3 fb$^{-1}$ of data in the paper 
\cite{Aad:2015zva} titled as
{\em Search for new phenomena in final states with an energetic jet and large 
missing transverse momentum in $pp$ collisions at $\sqrt{s}=8$ TeV with the ATLAS detector}.

	Remember that  
	the signature with an energetic jet and large \met\/  
	is considered as a very distinctive tool for a new physics search at colliders.  
	These "monojet-like" (mono-$\gamma$, mono-$W/Z$, mono-$H$, etc) final states were  
	already studied
\cite{Abazov:2008kp,Aaltonen:2012jb,Chatrchyan:2011nd,Chatrchyan:2012me,Chatrchyan:2012tea,Aad:2011xw,ATLAS:2012ky,Aad:2012fw,Khachatryan:2014rra,Aad:2014vea,Aad:2014vka,Aad:2013oja,Khachatryan:2014tva}
	in the searches for SUSY, large extra dimensions, and WIMPs as candidates for DM. 

	The mono-jet events for the study were required
in \cite{Aad:2015zva}
	to have no leptons at all and at least one jet with \pt $>$ 120 GeV$/c$. 
	The missing transverse momentum of these events were 
	varied between \met $>$ 150 GeV and \met $>$ 700 GeV. 
	The full data selected and the expected SM background are presented 
	in Tables~4 and 5 of original paper  
\cite{Aad:2015zva}. 

\begin{figure}[!ht]\begin{center}
\includegraphics[width=0.7\textwidth]{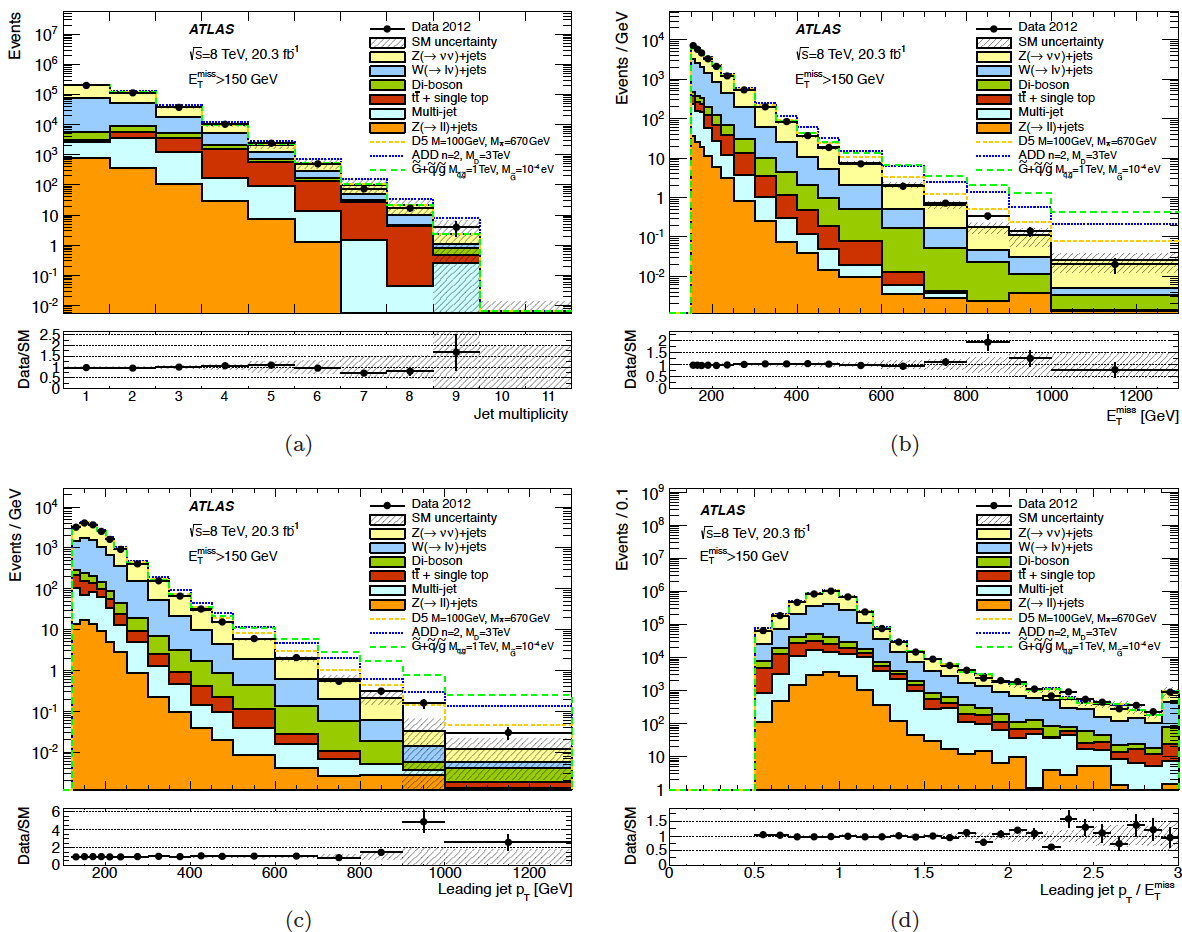}
\end{center}\caption{
Measured distributions of (a) the jet multiplicity,  (b) \met, (c) leading jet \pt, and (d) the leading jet 
\pt to \met\ ratio for the SR1  selection compared to the SM expectations.  
From \cite{Aad:2015zva}.}
\label{fig:2015zva-MainRes150GeV}
\end{figure}

\begin{figure}[!ht]
\begin{center}
\includegraphics[width=0.75\textwidth]{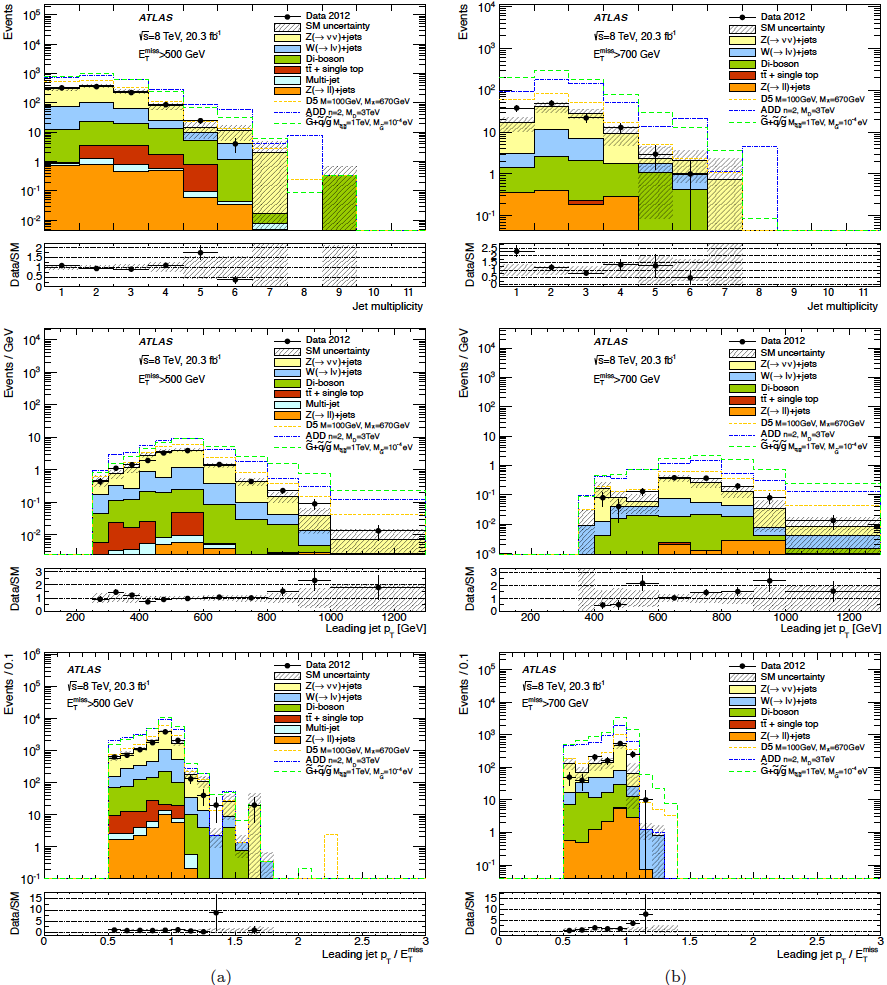}
\end{center}\caption{
Measured distributions of the jet multiplicity,  leading jet \pt, and the leading jet \pt to \met\ ratio
for (a) SR7 and (b) SR9 selections compared to the SM expectations. 
From \cite{Aad:2015zva}.}
\label{fig:2015zva-MainRes500GeV}
\end{figure}

	Several measured distributions for three signal regions  
	SR1 (\met $>$150 GeV), SR7 (\met $>$500 GeV) and SR9 (\met $>$700 GeV)
	together with the SM expectations 
	are shown in 
Fig.~\ref{fig:2015zva-MainRes150GeV}  and
Fig.~\ref{fig:2015zva-MainRes500GeV}. 
	For illustration purposes, the figures  
	include the impact of different Beyond-SM (ADD, WIMP, and GMSB SUSY) scenarios.  

	In general, good agreement was observed between the data and the SM expectations.  
	The largest difference (1.7$\sigma$ deviation) between the number of events in the data and the 
	expectations was observed for \met $>$700 GeV (signal region SR9).

	The agreement between the data and the SM expectations  
	was used to put model-independent  upper limits 
(Table~\ref{tab:2015zva})	
	on the visible BSM cross section $\sigma \times A \times \epsilon$
	(cross section $\times$ acceptance $\times$ efficiency), using approach of
 \cite{Read:2002hq} and the systematic uncertainties on the backgrounds 
	and the uncertainty on the integrated luminosity. 
\begin{table}[ht!]
\caption{Observed (expected) 90$\%$ and 95$\%$ CL upper limits on the
	visible cross section $\sigma \times A \times \epsilon$ in fb 
	for the SR1--SR9 selections. From \cite{Aad:2015zva}.}
\begin{center}
\begin{tabular}{|l|c|c|} \hline 
Signal Region & 90$\%$ CL  & 95$\%$ CL  
\\ \hline
SR1, \met $>$ 150 GeV & 599 (788)   & 726 (935)   \\
SR2, \met $>$ 200 GeV & 158 (229)   & 194 (271)   \\
SR3, \met $>$ 250 GeV& 74 (89)     & 90  (106)    \\
SR4, \met $>$ 300 GeV& 38 (43)     & 45  (51)    \\
SR5, \met $>$ 350 GeV&  17 (24)    &  21 (29)    \\
SR6, \met $>$ 400 GeV&  10 (14)     &  12  (17)      \\
SR7, \met $>$ 500 GeV &  6.0 (6.0)      &  7.2  (7.2)      \\
SR8, \met $>$ 600 GeV&   3.2 (3.0)     &   3.8 (3.6)     \\
SR9, \met $>$ 700 GeV &   2.9 (1.5)     &   3.4 (1.8)     \\ \hline
\end{tabular}
\label{tab:2015zva}
\end{center}
\end{table}
	The Table shows that visible cross sections  
	$\sigma \times A \times \epsilon$ above 599~fb--2.9~fb  
	are excluded at 90$\%$ CL 
	for SR1--SR9 selections, respectively.
	Simulation of background processes $Z(\to\nu\bar{\nu})$+jets
	allowed one to find that typical event selection efficiencies $\epsilon$ vary from $88\%$ 
	for SR1 and $83\%$ for SR3 to $82\%$ for SR7 and $81\%$ for SR9. 

	 The main model-independent results of 
\cite{Aad:2015zva} given in Table~\ref{tab:2015zva} and in 
Fig.~\ref{fig:2015zva-MainRes150GeV}  and
Fig.~\ref{fig:2015zva-MainRes500GeV} can be transformed into exclusion limits on
pair production of WIMP DM candidates, and further, for example, 
on models with large extra spatial dimensions
and production of very light gravitinos in a gauge-mediated SUSY model. 

	As already discussed above, any search for WIMPs 
	at a collider, in particular at the LHC
\cite{Steigman:1984ac}, is an important possibility of spreading some light on the DM problem.
	To fit the correct relic density for non-relativistic cold DM in the early universe
\cite{Kolb:1990vq} the WIMP masses are allowed to be between a few GeV$/c^2$ and a TeV$/c^2$, 
	and they are expected to interact, despite gravity, only (very) weakly.  
	Like ordinary neutrinos, WIMPs will escape detection 
	because they ca not deposit any measurable amount of energy in a calorimeter. 
	Therefore, one inevitably concludes that WIMP production at a collider can be 
	noticed only by means of large transverse momentum imbalance  ($p^{\rm miss}_{\rm T}$)
	of ordinary particles, the magnitude of which is called \met. 
 
\begin{figure}[!ht] %%%%%%%%%%%%%%%%%%%%%%%%%%%%%% 
\centering\includegraphics[width=0.5\textwidth]{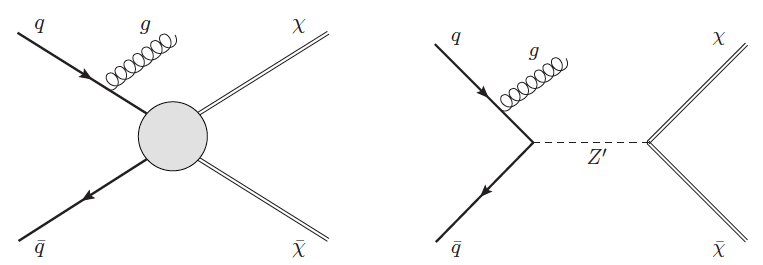}
\caption{Feynman graphs for production of WIMP pairs $\chi\bar{\chi}$ 
associated with a jet from initial-state radiation of a gluon, $g$,
via a contact effective operator (left), and in
a simplified model with a $Z^\prime$ intermediate boson (right). 
From \cite{Aad:2015zva}.}
\label{fig:2015zva-Graphs} 
\end{figure}

	To convert the main model-independent results of
\cite{Aad:2015zva} into numerical constraints on the DM problem, 
	one traditionally used the EFT approach, where the effective contact operators 
(Table~\ref{tab:2010ku}) describe WIMP-SM interaction,
	which in fact could be mediated by a single new heavy particle or 
	particles with mass too large to be produced directly at the LHC 
(Fig.~\ref{fig:2015zva-Graphs}(left)).

	The  representative set of seven operators (Table~\ref{tab:2010ku}) 
	included C1 (scalar), D1 (scalar), D5 (vector), D8 (axial-vector), 
	D9 (tensor), C5 (scalar), D11(scalar) operators. 
	The first five describe bilinear quark couplings to WIMPs, $q\bar{q}\to \chi\bar{\chi}$,
	the latter two describe $gg\to \chi\bar{\chi}$ couplings. 
 	Despite the serious questions about the validity of the EFT approach 
(section~\ref{sec:Discussion}), one is forced to use it due to a lack of a reasonable alternative approach 
	for comparing LHC results with results of direct and indirect DM searches. 

\begin{figure}[!h] \centering
\includegraphics[width=0.7\textwidth]{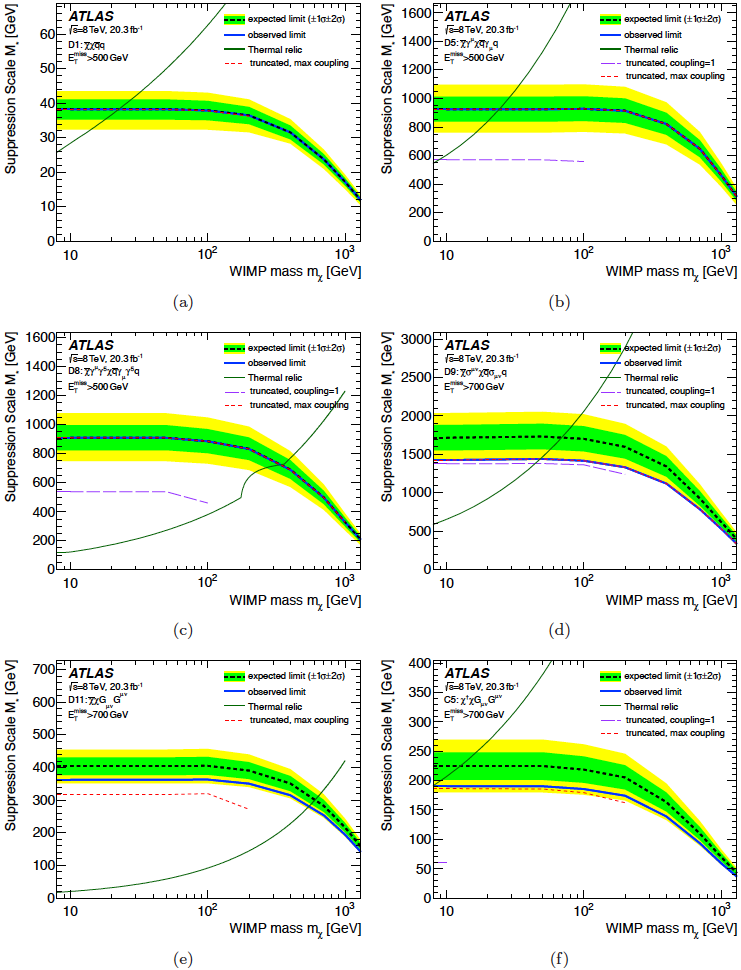}
\caption{Lower limits at 95\% CL on the scale $M_*$ are shown as a function of the WIMP mass $m_\chi$ for 
  (a) D1, (b) D5, (c) D8, (d) D9, (e) D11 and (f) C5 operators.
  The expected and observed limits are shown as dashed black and solid blue lines, respectively. 
  The rising green lines are the $M_*$ values at which WIMPs of the given mass 
  result in the relic density as measured by  WMAP
  \cite{Hinshaw:2012aka},  
  assuming annihilation in the early universe proceeded exclusively via the given operator. 
  The purple long-dashed line is the 95$\%$ CL observed limit on $M_*$ imposing a validity
  criterion with a coupling strength of 1, the red dashed thin lines
  are those for the maximum physical coupling strength. 
  From \cite{Aad:2015zva}.}
\label{fig:2015zva-EFTlimits}
\end{figure}

	From signal regions exhibiting the best expected sensitivity,
	for each EFT operator under consideration one first extracted the limits  
	on $M_*$ as a function of  $m_\chi$
(Fig.~\ref{fig:2015zva-EFTlimits}).
	The 1$\sigma$ and 2$\sigma$ error bands around the expected limit 
	are due to the acceptance uncertainties. 
	In particular, the main sources of experimental uncertainties were 
	the parton-shower matching scale (5\%), 
	jet and \met\ energy scale (up to 10\%), and PDF (5--29\%). 
	The main sources of theoretical uncertainties were variation of the
	renormalization and factorization scales of the EFT operators (up to 46\%),
	and uncertainty due to the PDF for the operators (20--70\%).
	The effect of the beam-energy uncertainty (2--9\%) 
	on the observed limit was negligible 
\cite{Aad:2015zva}.

	Looking at
Fig.~\ref{fig:2015zva-EFTlimits}   
 	one agrees with 
\cite{Aad:2015zva} that a demonstration of the EFT validity 
	could be done by relating the scale $M_*$ to the mass of a mediator  
	$m_V$ and the coupling constants $g_i$ by $m_V = f(g_i, M_*)$.
	The explicit form of the function $f$  
	depends on the concrete operator. 
	For a given operator, the validity criterion is $Q < m_V$, where $Q$ is a momentum transferred in 
	the hard interaction.
	Following this criterion, events were excluded from the analysis and omitted in 
Fig.~\ref{fig:2015zva-EFTlimits}.   
 	The criterion also depends on $g_i$, for which one considers two possibilities: 
	$g_i=1$, and the maximum possible coupling to stay in the perturbative regime ($\sqrt{g_i g_j} = 4\pi$). 

	After reducing the signal cross section according to the criterion,   
	the scale $M_{*}$ was recalculated in two expected truncated limit lines in 
Fig.~\ref{fig:2015zva-EFTlimits}. 
	The truncated limits fulfill the validity criteria wherever the lines are drawn in the figure.

 	Finally, using D9 as an example, one can see that  
	the maximum coupling criterion is fulfilled for all WIMP masses, 
	the $g_i=1$ criterion is fulfilled for $m_\chi < 200$~GeV$/c^2$. 
	For C5, the validity criterion for $g_i=1$ 
	is violated over almost the whole WIMP mass range, 
	and a truncated limit line is only drawn up to a WIMP mass of 10 GeV$/c^2$.
   
	For completeness, 
Fig.~\ref{fig:2015zva-EFTlimits} also includes WIMP thermal relic curves, 
	calculated for only one-in-time effective operator in question
\cite{Goodman:2010ku},
	with the WIMP-SM coupling, that for fixed $M_*$ and $m_\chi$    
	gives the correct relic abundance. 	 
	Remarkably, the D8-thermal-relic line has a bump at the top-quark mass where the 
	annihilation channel to top quarks opens
\cite{Aad:2015zva}. 
	Under the assumption that true DM is entirely composed of these thermal relics, 
	the limits on $M_*$ which are above the value required for the thermal relic density 
	exclude the case where WIMPs annihilate to SM particles via the
	corresponding operator. 
	
	This discussion, taken almost completely from  
\cite{Aad:2015zva}, demonstrates to an unbiased reader how complicated, poor controlled,
and, therefore, rather useless are all these attempts to stay within the EFT approach.
	Furthermore, one can notice that the variation of upper limits on the scale
	$M_*$ as a function of an effective operator looks very substantial.
	Several auxiliary curves (with truncated limits and for relic abundance) 
	in the figures convince the reader in the high accuracy level of the analysis, 
	but make the impression of the figures very complicated 
	and do not clarify the usefulness of these curves.

\begin{figure}[!h]
\centering
\includegraphics[width=\textwidth]{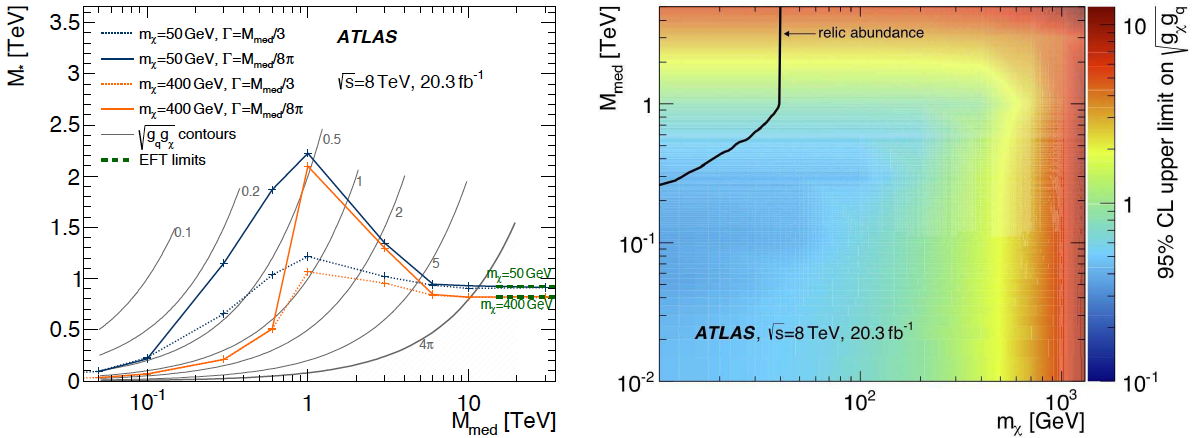}
\caption{(a) Observed 95$\%$ CL limits on the suppression scale 
$M_*$ as a function of the mediator 
mass $M_{\rm med} \equiv m_V $,
for $m_\chi = 50$ and 400 GeV$/c^2$. 
The width of the mediator $\Gamma=m_{V}/3$, or $m_{V}/8\pi$. 
The  corresponding limits from EFT models are shown as dashed lines;  
 contour lines indicating a range of values of 
 $\sqrt{{g}_q\, {g}_\chi}$ are also shown.
  (b) Observed 95\% CL upper limits on  
  $\sqrt{{g}_q\, {g}_\chi}$ in the plane of $M_{\rm med}$ versus $m_\chi$. 
  Values leading to the correct relic abundance~\cite{Hinshaw:2012aka} 
  are shown by the black solid line. 
From \cite{Aad:2015zva}.}
\label{fig:2015zva-Simple} 
\end{figure}

	To obtain quantitative limits on WIMP pair production
	a so-called simplified model was used {\em in addition to the EFT approach} in  
\cite{Aad:2015zva} 
	for alternative description of the WIMP-pair production.
	In the model the WIMPs couple to a $q$-pair explicitly via a new vector $Z^\prime$ boson  
(Fig.~\ref{fig:2015zva-Graphs}(right)) with mass $m_{V}$ and width $\Gamma$.   
	In this case only coupling of $q\bar{q}\to \chi\bar{\chi}$  
	can be probed and the product of the coupling constants  
	$\sqrt{{g}_q\, {g}_\chi}$ can be constrained. 
	Since in the model one explicitly has $M_{*} = {m_V}/{\sqrt{{g}_q\,{g}_\chi}}$,
	this equality can be given as a point in the $M_{*}$--$m_{V}$ plane in 
Fig.~\ref{fig:2015zva-Simple}(a).
	Therefore for a given $m_V$ and two representative values of $\Gamma$,
	one can compare the "true" above-mentioned value of the $M_{*}$ 
	with the $M_{*}$ value (shown as dashed line)
	derived assuming a contact interaction. 

	From the figure one confirms the rather obvious point that  
	the contact interaction can work only for $m_{V}\ge 5$~TeV$/c^2$ 
	because in the intermediate range
(700~GeV$/c^2$~$< m_{V} < 5$~TeV$/c^2$) the mediator can be produced resonantly 
	and the true $M_{*}$ is higher than the $M_{*}$ obtained in the contact interaction regime. 
	In this case the contact interaction limits will be pessimistic. 
 	Next, smaller mediator masses $m_V < 700$~GeV$/c^2$ give smaller true $M_{*}$ limits 
	because once $m_\chi > m_V$  
	the WIMP pair production via this mediator will be kinematically suppressed. 
	In this region, the contact interaction limits would be
	optimistic and overestimate the true $M_{*}$ values
\cite{Aad:2015zva}.
	In this situation a question survives:
{\em  How could one  know which mediator mass "works today"?}

	In Fig.~\ref{fig:2015zva-Simple}(b)  the observed 
	upper limits on the product of couplings of the simplified model vertex $\sqrt{{g}_q\, {g}_\chi}$ 
	are shown in the $m_V$--$m_\chi$ plane.  
	Within this model, the regions to the left of the correct-relic-density line
	lead to the values of the relic density larger than measured and are excluded.
	 
	Therefore the use of a simplified model allows one to avoid the poorly controlled 
	problems of the EFT validity, 
	to obtain some feeling about the EFT validity,   
	but as a price an extra parameter dependence appears 
	and the obtained constraints become more and more complicated. 

	Unfortunately, at the moment the EFT approach {\em looks inevitable}\/ if one wants to obtain from   
Fig.~\ref{fig:2015zva-EFTlimits}
	some valuable constraints on possible WIMP-nucleon SD and SI couplings. 
	Therefore, as before, for example in 
\cite{ATLAS:2012ky}, 
	these $M_*$--$m_\chi$ limits with relevant effective operators 
 	were converted into exclusion curves for the SD and SI 
	WIMP-nucleon scattering cross sections.
\begin{figure}[!h]
\centering
\includegraphics[width=\textwidth]{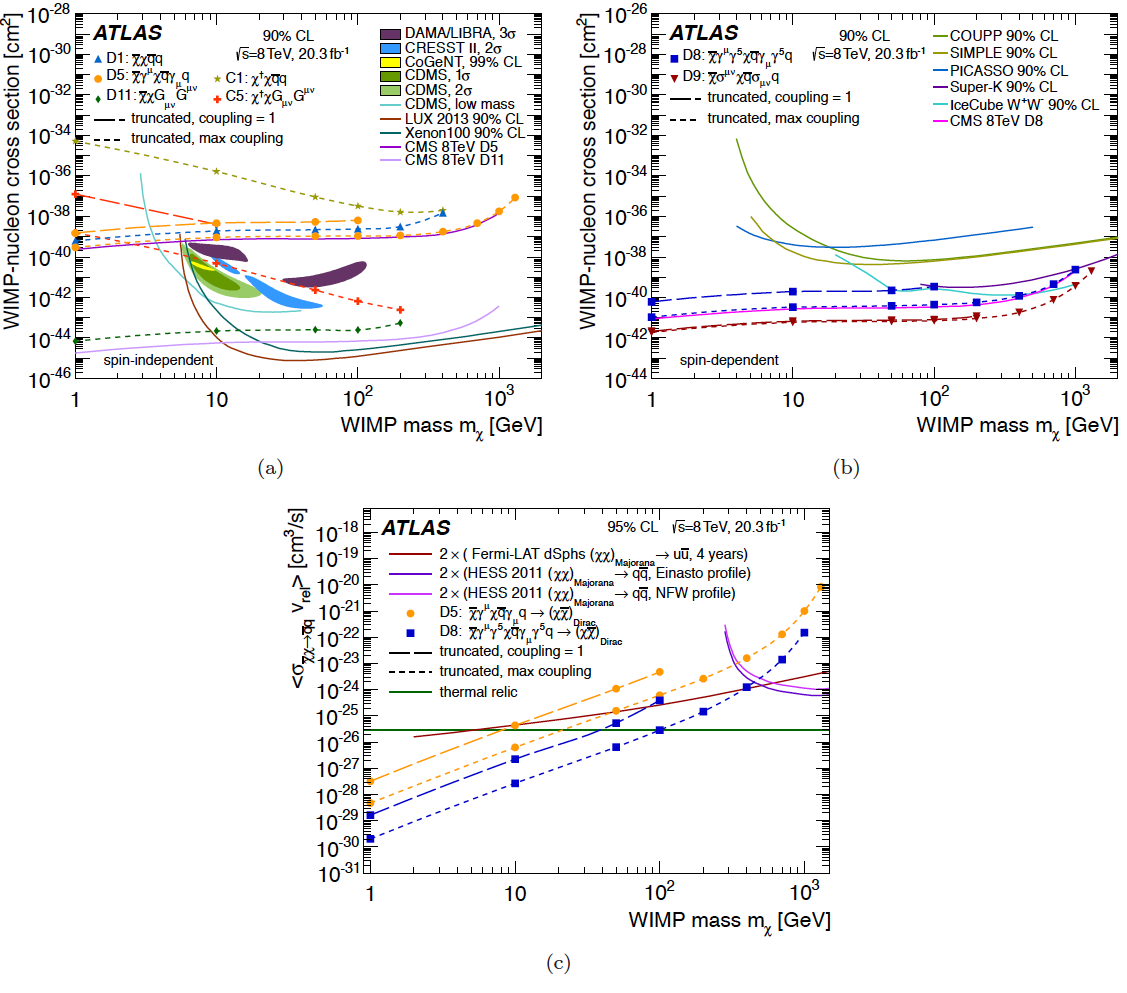}
\caption{The exclusion curves for 
(a) the spin-independent and (b) spin-dependent WIMP-nucleon cross section 
as a function of the WIMP mass $m_\chi$ for different effective operators. 
Results from direct-detection experiments for the spin-independent
  \cite{Angloher:2011uu, %CRESST
    Akerib:2013tjd, %LUX
    Agnese:2013rvf, %CDMS low mass
    Agnese:2014aze, %CDMS final
    Aalseth:2014jpa, %CoGeNT
    Bernabei:2008yi, %DAMA
    Aprile:2013doa} %XENON100
  and spin-dependent \cite{Archambault:2012pm, %PICASSO
    Desai:2004pq, %SuperK
    Abbasi:2009uz, %IceCube
    Behnke:2010xt, %COUPP
    Felizardo:2011uw} %SIMPLE
  cross section, and the CMS (untruncated) results~\cite{Khachatryan:2014rra} 
  are shown. % for comparison.  
  (c) The 95\% CL limits on the WIMP annihilation rate as a function of the WIMP mass.  
  Results from $\gamma$-ray telescopes \cite{Ackermann:2013yva,Abramowski:2011hc} 
  are also shown, 
  along with the thermal relic density annihilation rate~\cite{Hinshaw:2012aka,Ade:2013zuv}.
  From \cite{Aad:2015zva}.}
\label{fig:2015zva-DMlimits}
\end{figure}

	The ATLAS-obtained exclusion curves 
(Fig.~\ref{fig:2015zva-DMlimits}) look particularly relevant 
	for low WIMP masses and remain important for all $m_\chi$. 
	The spin-dependent ATLAS exclusion curves in 
Fig.~\ref{fig:2015zva-DMlimits}(b) 
	are based on limits from the D8 (axial-vector) and D9 (tensor) operators. 
	Both of them  are significantly stronger than those from direct-detection experiments.
	The spin-independent ATLAS exclusion curves
Fig.~\ref{fig:2015zva-DMlimits}(a) traditionally look less restrictive. 

Figure~\ref{fig:2015zva-DMlimits}(c) illustrates conversion of the $M_*$--$m_\chi$ constraints 
	into upper limits on the WIMP annihilation rate, calculated as the product of annihilation cross section 
	$\sigma$ and the relative WIMP velocity $v$ 
  	averaged over the DM velocity distribution $\langle\sigma v\rangle$. 
	Results of vector and axial-vector operators 
	describing the annihilations of WIMPs to the four light-quark flavors
	are only shown.
	One can compare these limits with those obtained earlier in  
\cite{ATLAS:2012ky}.
	Limits on the WIMP annihilation to $u\bar{u}$- and $q \bar{q}$-pairs 
	from galactic high-energy $\gamma$-ray
	observations by the Fermi-LAT~\cite{Ackermann:2013yva} and 
H.E.S.S.~\cite{Abramowski:2011hc} telescopes 
	and 
	the annihilation rate that follows from the thermal relic density measured by
WMAP~\cite{Hinshaw:2012aka} and PLANCK~\cite{Ade:2013zuv} satellites 
	are also shown for comparison. 

Figure~\ref{fig:2015zva-DMlimits} again concerns the EFT validity problem 
	and shows the effects of the truncation procedure  
	on the upper limits for the considered WIMP observables. 
	In general, the EFT limits remain valid for WIMP masses up to about 200~GeV$/c^2$. 
	The effect depends strongly on the operator and the values for the couplings. 
	The allowed variation of the coupling strengths 
	leads to changes in the limits of up to one order of magnitude.
{\em Strictly speaking, it is rather difficult to believe in usefulness of such limits.} 

\smallskip
	The mono-jet+\met\/ experimental data  
(Figs.~\ref{fig:2015zva-MainRes150GeV}  and \ref{fig:2015zva-MainRes500GeV})
	can be used (as before in  
\cite{Aad:2014iia}) for a study of the Higgs boson invisible decays
\cite{Fox:2011pm}.
	A sizable value of BR($H\to\,$invisible)  
	is not yet experimentally excluded.
	There are models connecting a "hidden" DM sector with SM particles via  
	direct DM-Higgs-SM-couplings
(for example, 
\cite{Bishara:2015cha,Dutra:2015vca,Fedderke:2014wda,Djouadi:2011aa}).  
%%%%%
	In this case the Higgs boson can decay into invisible WIMP DM candidates 
	producing a deviation of a measured SM Higgs branching ratio from the expected one  
\cite{Askew:2014kqa}.

\begin{figure}[!ht]
\begin{center}
  \includegraphics[width=0.6\textwidth]{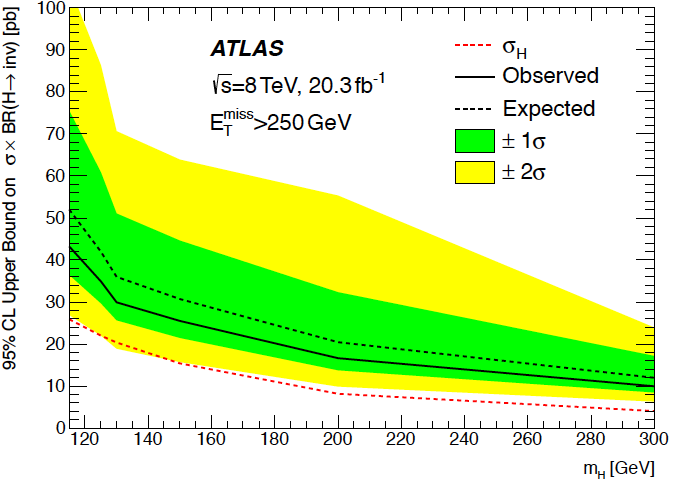}
\end{center}
\caption{The observed (solid line) and expected (dashed line) 95$\%$ CL upper limit on 
$\sigma \times {\rm{BR}}(H \to {\rm{invisible}})$ as a function of the boson mass $m_H$.
The shaded areas around the expected limit indicate the expected 
$\pm 1\sigma$  and $\pm 2\sigma$ ranges of limits in the absence of a signal.
The expectation for a Higgs boson with ${\rm{BR}}(H \to {\rm{invisible}}) = 1$, $\sigma_{H}$,   
is also shown. 
 From \cite{Aad:2015zva}.}
\label{fig:2015zva-Hinv} 
\end{figure}

	At the LHC, based on the associate Higgs-$Z$ boson production,
	the strong upper limits of 58--65\% (at 95\% CL) were already set    
	on the branching ratio for the Higgs invisible decay
\cite{Aad:2014iia,Chatrchyan:2014tja}.
	This investigation was continued in 
\cite{Aad:2015zva}, where the mono-jet+\met\/ final state was
	used to search for the production of an invisibly decaying  
	SM Higgs-like boson with an allowed mass range between 115~GeV$/c^2$ and 300~GeV$/c^2$.
	Figure~\ref{fig:2015zva-Hinv} shows the observed and expected limits 
	on the production cross section times branching ratio 
	$\sigma \times {\rm{BR}}(H \to {\rm{invisible}})$ as a 	function of the boson mass. 

	Values for $\sigma \times {\rm{BR}}(H \to {\rm{invisible}})$ above  44~pb for 
$m_H = 115$~GeV$/c^2$ and 10~pb for $m_H = 300$~GeV$/c^2$ are excluded.  
	Comparison with previous constraints from the analysis of
	$ZH (Z \to \ell^+ \ell^-)$ final states 
\cite{Aad:2014iia} shows that the obtained result is less sensitive and does not yet have the
sensitivity to probe the SM Higgs boson couplings to invisible particles, at least for Higgs boson 
 mass of 125~GeV$/c^2$
\cite{Aad:2015zva}.

	Concluding the very sophisticated study of 
\cite{Aad:2015zva} one can say that search for new phenomena in events with an energetic 
	jet and large \met\ at $\sqrt{s}=8$~TeV did not give evidence for disagreement with the SM expectations. 
	The results were translated into model-independent  
	upper limits on new physics contributions.
	Furthermore, a very sophisticated analysis of the EFT validity was carried out. 
	To this end a special Appendix was arranged in 
\cite{Aad:2015zva}. 
	The discussion of the subject runs to the higher level. 

%%%%%%%%%%%%%%%%%%%%%%%%%%%%%%%%%%%%%%%%%%%%%%%%%%%%%%%%%
	The latest, to our knowledge, results of the ATLAS search for invisible decays of the Higgs boson produced
        	in association with a hadronically decaying vector boson in $pp$ collisions at $\sqrt{s}$ = 8 TeV are given in 
\cite{Aad:2015uga}.

\smallskip
	Very similar analyses were carried out by CMS collaboration, and similar results were obtained.
	In particular, the review {\em Search for Dark Matter at CMS}
\cite{Lowette:2014yta} has presented the results from searches 
	for directly produced WIMP particles on the basis of the   
	full LHC RUN I dataset of 20 fb$^{-1}$ at 8 TeV. 
	Final states with a mono-jet, mono-photon, and mono-lepton signature were 
	considered, as well as processes with WIMP particles produced 
	in association with top quarks. 
	Most of these results were interpreted using the EFT
	approach, while results in simplified models were also reported.
	The latest results of the CMS collaboration DM search are given in 
\cite{Khachatryan:2015nua}.	
	Recent reviews of the subjects can also be found in 
\cite{Askew:2014kqa,Mitsou:2014wta}.

\smallskip
{\em In the conclusion}\/ 
	of this section (about ATLAS program for the WIMP search at the LHC)  one can 
{\em first}\/  point out that unfortunately, despite increasing complexity and increasing accuracy 
	of the analyses performed in 2010--2014, all searches gave no evidence for a signal of physics 
	beyond the SM based on the \met\/ signature. 
	This is the main physics result of the program.
	
{\em Second}, 
	the justification of connection of this program with the DM problem 
	relies on a key assumption that the WIMPs looked for at the LHC
	are equivalent to the true  DM particles.

{\em	Third}, 
	one believes,  
	with all known caveats on the EFT results in mind, 
	a few robust observations can be made on the complementarity between 
	the collider and direct searches for DM particles. 
	The first feature is the strength of the collider analyses searching for low-mass WIMPs
	with reasonable sensitivity up to zero mass.
	Nevertheless, a question survives about usefulness of these "zero-mass" results for solution of the DM problem.
	Nowadays it looks unrealistic that cosmological (cold, warm, etc) DM particle masses could be indeed so small. 
	Obviously, at a rather high WIMP mass, the collider WIMP search potential 
	vanishes due to the drop of the WIMP production cross section
\cite{Lowette:2014yta}.
	Furthermore, at least today, the direct DM detection technique is more constraining for spin-independent 
	scattering for WIMP masses above a few GeV$/c^2$
\cite{Askew:2014kqa}.

	Another complementarity follows from the point that collider people believe that 	
	the direct-detection experiments have typically reduced sensitivity to spin-dependent interactions, 
	and hence allows the collider searches to constraint severely this kind of interaction at
	intermediate WIMP masses as well.  
	Nevertheless, it is obvious that this statement has no power to reduce importance of modern 
	direct DM-search experiments especially sensitive to the spin-dependent DM-proton/neutron couplings. 
	The simple reason is that collider constraints concern WIMPs 
	escaping detection, but not true DM particles.
	This most crucial assumption forces one to be very careful when comparing 
	collider results with direct searches, 
	to say nothing about the specific assumption of EFT or/and many specific variants of a simplified model.

	As it pointed, in particular in 
\cite{Fox:2011pm}, if WIMP-SM interactions at LHC energies cannot be described by effective operators, 
	the obtained WIMP-SM constraints, depending on the mass and width of the intermediate particle, 
	can become either significantly stronger or considerably weaker. 
	This statement almost kills the meaning of the EFT-based analysis, 
	because today one is unable to know for sure (also a posteriori) which of the unknown intermediate 
	particles really contributed, to say nothing about its mass and width.
	 	 	
\begin{figure}[!h] 
\vspace*{-2pt}\begin{center}
\includegraphics[width=0.5\textwidth]{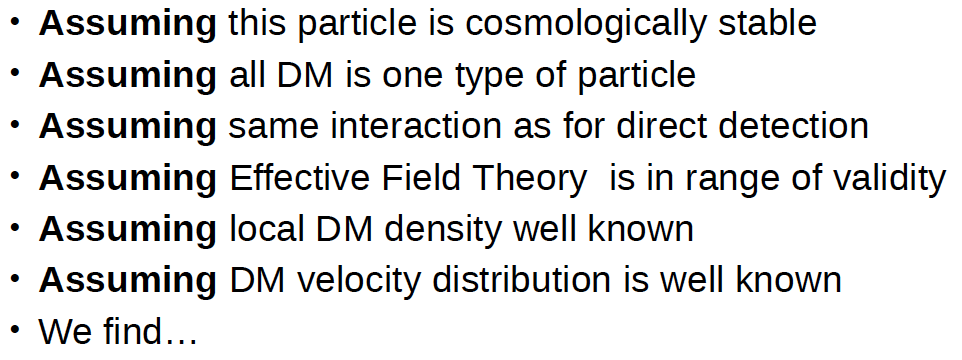}\end{center}
  \caption{The main assumptions allowing DM interpretation of the collider WIMP search.
}\label{fig:Assuming}
\end{figure}

	Finally, it is reasonable to end this section with the slide (given in 
Fig.~\ref{fig:Assuming}) from one of a honest discussions of WIMP-DM search at the LHC, 
	which collects the set of assumptions concerning the main subject. 
	It is rather difficult to refrain from a "criminal" thought
	that here the points are assumed that in fact one should prove experimentally.
	
	New papers are expected from the LHC on the subject, but it is clear now
	that they can add very little to solution of the DM problem and  
	one should wait for success of direct DM search experiments.
%%%%%%

%%%%%%%%%%%%%%%%%%%%%%%%%%%%%%
\section{Exotic DM searches} \label{sec:ExoticsDMsearches}
%\section{Exotic DM searches} \label{sec:ExoticsDMsearches}
	The lack of clear evidence for Physics beyond SM (BSM),
	the absence of any SUSY signals and any hints of WIMPs in the first phase of the LHC,
	insufficient sensitivity of traditional direct and indirect techniques (excluding DAMA/LIBRA) 
	to observe galactic halo DM in a laboratory 
	have all cast some doubts on linking a true DM mass with the electroweak scale
\cite{deSimone:2014pda}.	

	This point, together with some semi-confirmed hints on the BSM physics 
	from rare decays, low-energy measurements, astrophysical observations and cosmological analyses, 
	has motivated one to look for new 
	strategies for DM searches at colliders, 
	and for other DM candidates and DM-related models, like, for example, dark photon,
	light dark matter (LDM) particles with masses $0.1\div 10$ GeV$/c^2$
\cite{Fernandez:2014eja} and many others. 
	{\em Only some of them} are briefly discussed in this section.
	
	From the other side,  people traditionally believe that such a situation 
	suggests a need for much higher energies.
	One expects the future 100--TeV collider  
	would be exciting for BSM-searches, including the search for a DM candidate 
\cite{Livio:2014gda}.

\subsection{Dark photon, dark gauge boson searches}	
%%%%%%%%%%%%%%%%%%%%%%%%%%% 
	One believes that the dark sector is complicated and 
	could be charged under a new Abelian group U$^\prime$(1). 
	A relevant light new mediator, the so-called dark photon $A'$ (with mass $m^{}_{A'}$), 
	connects the dark sector to the SM one. 
	The "dark photon" and "dark boson" are typical names 
	of some objects which mediate interaction between the dark sector and the visible one.
	In fact, any experimental evidence in favor of these dark forces  
	has today nothing to do with detection of galactic true DM particles. 
	Only further model-dependent assumptions can connect these objects with the DM problem.

\smallskip 
	Results of the special search for the dark photon $A'$ in 
	$\pi^0$ decays by the NA48/2 experiment at CERN was reported in                 
\cite{Goudzovski:2014rwa}.
 	One looked for the dark photon via the "signal" decay chain 
	$K^\pm \to\pi^\pm \pi^0$, $\pi^0 \to \gamma A'$, and $A'\to e^+e^-$
	on the basis of $5\times 10^6$ fully reconstructed 
	$K^\pm \to\pi^\pm \pi^0$, $\pi^0\to \gamma e^+e^-$ decays 
	in the kinematic range $m_{ee}>10$ MeV/$c^2$ with a negligible background contamination. 
	No signal was observed, and exclusion limits were set on the dark photon mass 
	$m_{A'}$ and the mixing parameter $\epsilon^2$, which connects dark and visible photons. 
	See also \cite{CERNNA48/2:2015lha}.

\smallskip
	With the BABAR detector, using 514~fb$^{-1}$ of data collected,
	an experimental search for the dark photon $A'$ in the reaction 
	$e^+e^-\to \gamma A'$, $A'\to e^+e^-, \mu^+ \mu^-$
	was carried out
\cite{Lees:2014xha}. 
	 No significant deviations from SM were	observed and 
	 upper limits were established
	 on the dark-to-real photon mixing   
	 at the level of $\epsilon \simeq 10^{-4}$--$10^{-3}$ for $0.02< m^{}_{A'} < 10.2$~GeV$/c^2$.
	The range of the parameter space favored by interpretations of 
 	the discrepancy between the calculated and measured anomalous magnetic moments 
	of the positive muons ($g$--2 muon anomaly
\cite{Bennett:2006fi}) was severely constrained.
	Together with the above-mentioned results from NA48, the BABAR results
exclude the entire region favored by the dark-photon scenario for the $g$--2 measurements
\cite{Eigen:2015rea}.

%\smallskip %\hrule\bigskip %\noindent %---------------------------------------------------
	The proposal for searching for a dark photon $A'$ with mass 10--80 MeV$/c^2$ 
	via the decay $\mu^+ \to e^+ \nu_e \bar\nu_\mu +A'$ followed by $A' \to e^+ e^-$
	was formulated for the upcoming Mu3e experiment at the 
	Paul Scherrer Institute 
\cite{Echenard:2014lma}. %{Projections for Dark Photon Searches at Mu3e, Echenard:2014lma}
	The primary goal of the experiment is "traditional" search for the 
	very rare lepton number violating and SM-forbidden decay $\mu^+ \to e^+ e^+ e^-$. 
	With expected $10^{15}$ ($5.5\times 10^{16}$) muon decays in 2015--2016 (2018 and beyond), 
	the Mu3e collaboration has a very good opportunity to reduce substantially the 
	currently unexplored dark photon parameter space, 
	probing the mixing parameter up to $\epsilon^2 \sim 10^{-7} (10^{-8})$. 

%\smallskip %\hrule\bigskip %\noindent %---------------------------------------------------	 
	In searching for a signal of the dark photons in ATLAS at the LHC
	tightly collimated groups of highly-boosted leptons  --- lepton-jets --- were used
\cite{Diehl:2014dda}.
	There is essentially no SM background for them. 
	These jets were predicted in 
\cite{ArkaniHamed:2008qn} and were motivated by a DM model which is consistent with 
	a possible positron excess in cosmic rays. 
	In the model DM is the SUSY LSP, but it can decay into lighter "hidden valley" particles 
	resulting in the production of dark photons which can finally decay into highly-boosted leptons.

	Searches were made for muon jets containing 4 or more muons; 
	pairs of muons jets of 2 or more muons; pairs of electron jets of 2 or more electrons.  
	No significant excess of lepton jets was observed over the expected SM background. 
	For broad applicability, limits expressed in terms of the signal cross section times branching ratio 
	were derived for each pair of the dark gauge coupling parameter, $\alpha_D$, and $m^{}_{A'}$
\cite{Cheung:2009su}.	
%%%%%%%% 150627 %%%%%
	Dissipative hidden sector generic DM model was considered in
\cite{Foot:2014uba,Foot:2014osa}.	 
%%%%%%%%%%%%%%%%%

%\smallskip %\hrule\bigskip %\noindent %---------------------------------------------------
	Another idea to use the lepton-jets for search for light {\em dark force} carriers, 
	$Z'$-bosons, at the LHC was proposed in 
\cite{Kong:2014iwa}.  %{Charged Higgs Probes of Dark Bosons at the LHC,  Kong:2014iwa}
	The GeV-scale dark gauge boson $Z'$ was discussed 
\cite{Kong:2014jwa,Davoudiasl:2014mqa}	
	in the context of some astrophysical anomalies and the 3.6$\sigma$ deviation from SM in 
	the muon $g$--2 measurement
\cite{Bennett:2006fi}.
	A scenario, which could be easily probed at the LHC RUN-II, 
	was studied where a top quark was assumed to undergo exotic decay 
	to a $b$ quark and a charged Higgs followed by $H^\pm\to W^\pm+Z'$.
	The decay products of the dark $Z'$ further form a highly collimated lepton-jet. 
	It is believed that the feature could help find the new boson with the $t\bar{t}$ samples.

	A possibility for dark gauge $Z'$ boson search at LHC was discussed  in 
\cite{Gupta:2015lfa} via registration of events with the $Z'$ resonance decay into dilepton pairs 
	in association with large \met.
	This search channel is considered as a generic probe of TeV-scale dark models with
	involvement of other dark sector particles.  
	Another example of the dark $Z'$ boson search with a collider was considered in 
\cite{Bai:2015nfa}.	
	The GeV-scale $Z'$ boson, produced mainly from DM final state radiation, 
	decays eventually to hadrons or leptons
\cite{Autran:2015mfa}, which form a unique mono-$Z'$ jet (or dilepton) signature. 
	This final state contains significant discovery potential, 
	which has not yet been examined in detail by the LHC experiments
\cite{Autran:2015mfa}.

\smallskip %\hrule\bigskip %\noindent %---------------------------------------------------
	The famous muon $g$--2 anomaly  
\cite{Bennett:2006fi}
	still works as the main motivation for the theoretical and experimental papers about 
	light dark bosons of MeV-GeV scale which could explain the anomaly. 
	As clearly pointed out, for example in 
\cite{Lee:2014tba}, %{Muon g-2 Anomaly and Dark Leptonic Gauge Boson, Lee:2014tba}
	the discussions on the subject {\em are not necessarily linked} to the DM physics.
	Sometimes there is no such a link at all.
	Nevertheless, the word "dark" is widely used, like
	{\em dark} leptonic gauge $Z'$ boson or {\em dark} photon $A'$, 
	for the very suppressed coupling in contrast to the "bright" photon coupling.
	
	Because of active searches in fixed target experiments and rare meson decays,
	the popular dark photon model is practically excluded 
	as a possible solution of the $g$--2 problem, 
	unless an invisibly-decaying dark photon mode is considered (see review in
\cite{Lee:2014tba}). 
	But a severe drawback of the mode 
	is the requirement of very light and low-motivated 
	new DM particles, lighter than the MeV--GeV scale of the dark photon itself. 

	An alternative model with a dark leptonic gauge $Z'$ boson 
	based on the gauged lepton number or U(1)$_{\rm L}$ symmetry was proposed in 
\cite{Lee:2014tba}. 
	Unlike the dark photon, which couples only to a charged particle, 
	the $Z'$ boson can couple to SM neutrinos and charged leptons with the same strength.
	Furthermore, the $Z'$ boson mainly decays right into the SM neutrinos 
	and avoids the severe quarkonium decay constraints 
\cite{Agakishiev:2013fwl,Babusci:2014sta,Lees:2014xha}  
	as it does not couple to quarks.

	There are further aspects to verify the U(1)$_{\rm L}$ 
	dark leptonic gauge boson model, which  
	include the phenomenology of the exotic leptons required for the gauged U(1)$_{\rm L}$, especially
	for the LHC experiments and implications for the neutrino physics
\cite{Lee:2014tba}. 

	In fact, the idea of a dark $Z'$-boson is very popular (see, for example, recent paper 
\cite{Xu:2015wja}).

	Re-analysis of the data from the electron beam-dump experiment E137, conducted at SLAC in 1980--1982,
	allowed one to obtained new constraints on sub-GeV DM and dark photons 
\cite{Batell:2014mga}. %{Strong Constraints on Sub-GeV Dark Sectors from SLAC Beam Dump E137} 	
	It was assumed that the DM candidates can interact with electrons (via a dark photon). 
	Hence, DM can be produced in the electron-nucleus collisions 
	and can be scattered off electrons in the E137 detector.
	The expected result  could be striking: 
	zero-background signature of a high-energy electromagnetic 
	shower that points back to the beam dump. 
	From non-observation of the signal the E137-result has constrained 
	the possibility that invisibly decaying dark photons 
	can explain the 3.6$\sigma$ discrepancy between the measured and 
	the SM value of the muon anomalous magnetic moment
\cite{Bennett:2006fi}. 
	The E137 data also have convincingly demonstrated 
	that (cosmic) backgrounds can be controlled and 
	could serve as a powerful proof-of-principle 
	for future beam-dump searches for sub-GeV dark matter scattering off electrons in the detector
\cite{Batell:2014mga}. 

	A search for a new electrophobic sub-GeV dark boson with a missing-beam-energy method
	was proposed in 
\cite{Gninenko:2014pea}. %{The muon g-2 and searches for a new electrophobic sub-GeV dark boson in a missing-energy experiment at CERN}
	The idea of the search relies on two main assumptions. 
	First, the $Z'$ boson 
	can be produced by $\mu$-beam scattering on nuclei $A$ via the reaction $\mu+A\to \mu+A+Z'$.
	Second, one can "observe" $Z'$ by looking for some excess of events with the 
	large missing $\mu$-beam energy in a detector.
	The missing of the $\mu$-beam energy is expected to be 
	due to the decay $Z'\to \nu\nu$.
	The authors of  
\cite{Gninenko:2014pea}
 	believe that the very specific signature and high quality of 
	the muon beams at CERN SPS allows  
	one to reach a sensitivity in coupling constant $\alpha_d$,   
	which is three orders of magnitude higher than the value required to explain the $g$--2 anomaly. 
	This looks very promising indeed  {\em if}\/ a good enough, better sub-MeV-scale, 
	missing beam energy resolution will be achieved experimentally,
	and one is able to prove independently that the $Z'$ boson was indeed produced before 
	it decayed invisibly.   

	Preliminary estimates of the expected light hidden photon signal rate were presented in
\cite{Gorbunov:2014wqa}
	with respect of the recently proposed fixed target SHiP experiment
\cite{Alekhin:2015oba} 
	exploiting the CERN SPS beam of 400 GeV protons.
	
	An extended program of different ways for the dark photon search (including 
	invisible Higgs boson decays, etc.) with future hadron colliders was considered in 
\cite{Curtin:2014cca}.

\begin{figure}[h!]
\includegraphics[width=0.6\textwidth]{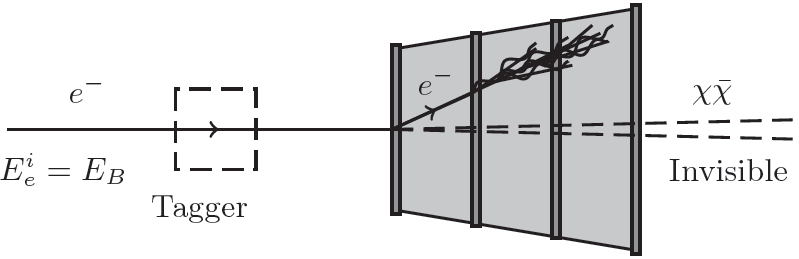}
\caption{Scheme of the fixed-target dark photon search experiment.
	 A single beam electron first passes through an up-stream tagger to fix its energy. 
	 Then it enters the target/calorimeter and emits an $A'$, 
	 which decays somewhere invisibly into a dark pair $\chi\bar{\chi}$ and
	 carries away most of the beam energy.  
	In the final state one has only the electron with very small energy $E^f_e \ll E_B$.
From \cite{Izaguirre:2014bca}.}
\label{fig:2014bca-ScenarioA}  
\end{figure}

	Similar missing energy-momentum approach was described in  
\cite{Izaguirre:2014bca}, %{Testing GeV-Scale Dark Matter with Fixed-Target Missing Momentum Experiments}
	as authors claimed, {\em for detection}\/ of DM and other invisible particles with mass 
	below 1~GeV$/c^2$ in fixed-target accelerator experiments.
	The main idea is to exploit missing energy-momentum measurements 
	and other kinematic features of fixed-target particle production.	
	Several new beam-dump experiments are already 
	aimed to produce light DM candidates and "observe" their scattering in downstream detectors 
\cite{Izaguirre:2013uxa,deNiverville:2012ij,Diamond:2013oda,Batell:2014mga,Battaglieri:2014qoa}.
	A typical setup is given in
Fig.~\ref{fig:2014bca-ScenarioA}.
	One believes that this technique, with $E_{\rm missing} = E^i_e - E^f_e \simeq E_B$,   
	 allows one to discover this kind of events, which under severe assumptions 
(see sections \ref{sec:DM-ATLAS-searches} and \ref{sec:Discussion}) 
	can be interpreted as observation of light DM candidates.
	Furthermore, the method relies on a very small re-scattering probability and it seems 
	challenging to reach relevant sensitivity.
	The sensitivity requires identification of "the DM production events" solely by their kinematics, 
	which in fixed-target electron-nuclear collisions is believed to be quite distinctive 
\cite{Bjorken:2009mm}. 

	Under the assumption that the above-mentioned events 
	are due to an invisibly decaying MeV--GeV-scale dark photon $A'$, 
	this approach can improve present constraints by 2--6 orders of magnitude over the entire $m_{A'}$ range, 
	and sensitivity as low as $\epsilon^2 \sim 10^{-14}$ can be reached
\cite{Izaguirre:2014bca}. 

\smallskip
	The belief in a powerful potential of electron-beam fixed-target experiments
	for discovery of DM and other new WIMPs  
	in the MeV-GeV mass range
\cite{Izaguirre:2013uxa,deNiverville:2012ij,Diamond:2013oda,Batell:2014mga,Battaglieri:2014qoa}
	is realized in a new proposal for a Pilot Dark Matter Search at Jefferson Laboratory
\cite{Izaguirre:2014dua}.
	The physics potential of such an experiment was discussed 
	and highlights of its unique sensitivity to 
	inelastic "exciting" DM and leptophilic DM scenarios were stressed.
	 The first of these seems kinematically inaccessible in traditional direct detection experiments.  
	 
\begin{figure}[h!]
\includegraphics[width=0.6\textwidth]{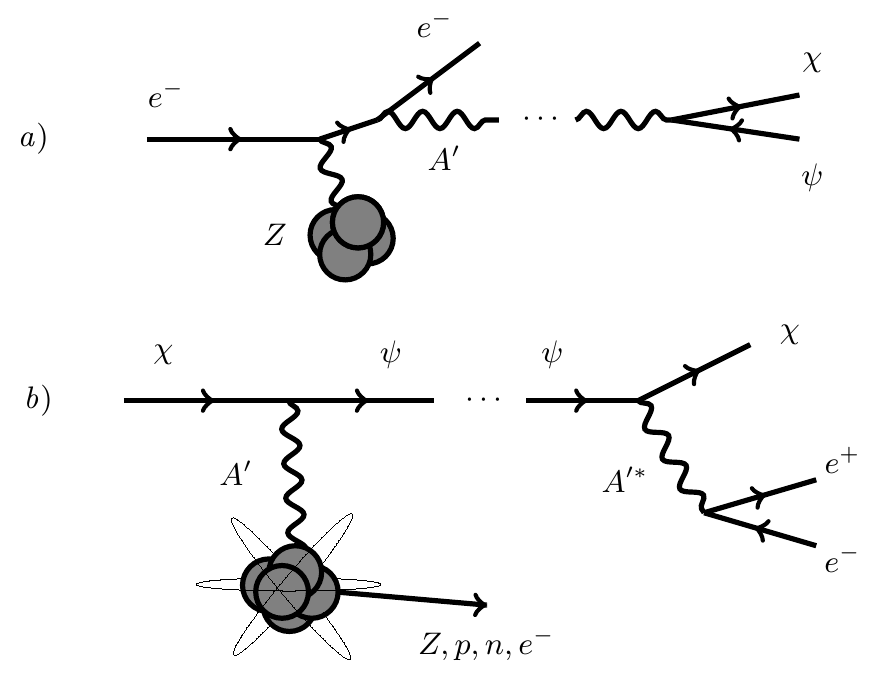}
\caption{a) Fermionic DM pair production via issuing $A'$ boson during electron-nucleus collisions.
In the generic scenario with Dirac and Majorana masses for dark sector fermions,
the $A'$ mediator couples off-diagonally to the mass eigenstates $\chi$ and $\psi$. 
 b) Detector scattering via $A^\prime$ exchange inside the detector. 
 If the mass splitting between dark sector states is negligible, 
 both the incoming and outgoing DM states in the scattering process are invisible and can be 
  treated as the same particle.  
  For order one (or larger) mass splittings, $\chi$ can upscatter into 
  the excited state $\psi$, which promptly decays inside the detector via
 $\psi\to\chi\, e^+e^-$. 
 This  process yields a target (nucleus, nucleon, or electron) recoil $E_R$ 
 and two charged tracks, which is a distinctive, 
 low background signature, so nuclear recoil cuts need not be limiting. 
 Processes analogous to both a) and b) can also exist if DM is a scalar.
 From \cite{Izaguirre:2014dua}.}
\label{fig:2014dua}  
\end{figure}

	The main principles of production and detection of these particles are depicted in 
Fig.~\ref{fig:2014dua}.
	The first stage of this program can be realized at Jefferson Laboratory 
	using the existing plastic-scintillator detector downstream of the Hall D electron beam dump. 
 
\smallskip %\hrule\bigskip %\noindent %---------------------------------------------------
		Letter of Intent for  Beam-Dump eXperiment (BDX) at Jefferson Laboratory  
\cite{Battaglieri:2014qoa} is a concrete realization of the DM program.
	 The BDX will look for the MeV-GeV WIMP DM candidates
	 with a 1 m$^3$ segmented plastic scintillator detector 
	 placed downstream of the beam-dump at one of the high intensity JLab experimental Halls.
	 Up to $10^{22}$ electrons-on-target were expected in a one-year period.
	The BDX will be sensitive to WIMP-nucleon elastic scattering at the level of a 
	1000 counts/year, with very low threshold recoil energies ($\sim$ 1 MeV). 
	Expected sensitivity to WIMP-electron elastic scattering and/or inelastic WIMP would 
	be below 10 counts/year after severely reduced backgrounds.
 
	An existing 0.036 m$^3$ detector prototype based on the same technology 
	will be used to validate simulations with background rate estimates, 
	driving the necessary R\&D towards an optimized detector. 
	A fully realized experiment would be sensitive to large regions of DM parameter space, 
	exceeding the discovery potential of existing and planned experiments by two orders of magnitude 
	in the MeV-GeV DM mass range
\cite{Battaglieri:2014qoa}.

\smallskip %\hrule\bigskip %\noindent %---------------------------------------------------
	Some comments about this typical proposal are in order.
	First, its goal is unusual events with large undetected energy (momentum)
	comparable to the initial electron beam energy. 
	To see the events, one needs very good sensitivity and very strong background reduction.
	The most probable result of the search will be non-observation of any of the events.
	The next point is interpretation of the results.
	 It is no way to prove that the $A'$ boson was indeed produced before it decayed invisibly.   
 	Here one must make a strong assumption on the subject. 
	In fact, the assumption concerns right the point one should prove experimentally.
	Of course, some limit can be set from almost any measurement for 
	almost any measurable value, if one clearly writes down all assumptions. 
	The main question is their adequateness and correctness. 
	
	Finally, such an experiment is much farther from the solution of the DM problem 
	if compared with the LHC DM search program. 
	With the LHC one tries to see a real DM candidate, but here 
	one can see a particle that {\em only}\/ can be or cannot be a DM-SM messenger, nor ever a DM candidate, 
	to say nothing about the true DM galactic feature, like the annual signal modulation.  
	Much more model dependence is needed to connect this search with the DM problem.
	
\smallskip %\hrule\bigskip %\noindent %---------------------------------------------------
	Another step of the DM program at Jefferson Laboratory is DarkLight experiment 
\cite{Balewski:2014pxa} aimed at a precision search for New Physics at low energies.
	The famous dark photon with $m_{A'}=10 \div 100$~MeV/$c^2$ was the main motivation. 
	The DarkLight will precisely study the $ep\to ep+e^+e^-$ reaction via 
	detection of the final state scattered electron, recoiled proton, and $e^+e^-$ pair. 
	The signal would be the reaction $ep\to ep+A'$ followed by $A'\to e^+e^-$.
	To this end a windowless gas target of molecular hydrogen will be irradiated by  
	the 100 MeV electron beam of intensity 5 mA. 
	Phase-I of the experiment was funded and one expects to take data soon.  
	Complete phase-II is under final design and 
	could run within two years after phase-I is completed. 
	The DarkLight experiment to be decisive requires 
	development of a new technology for beam, target, and detector.

	Contrary to the experiment aimed at the invisible dark photon decay, 
	search of the dark photon decay into a well measurable $e^+e^-$-pair
	looks much more promising and better motivated.
		   
\smallskip %\hrule\bigskip %\noindent %---------------------------------------------------
	A proposal to obtain valuable constraints on the dark photon $A'$ as a true DM candidate
	from results of traditional direct DM search experiments was given in  	
\cite{An:2014twa}. %{Direct Detection Constraints on Dark Photon Dark Matter, An:2014twa}
	The absence of an ionization signal in Xe direct detection experiments 
	was used to place a very strong constraint on the dark photon mixing angle, down to $O(10^{-15})$, 
	under the assumptions that 
	the dark photons are long-lived vector states with 0.01--100 keV$/c^2$ masses
	and they comprise the dominant fraction of DM in the Universe.
	
	Here $A'$ photon plays the role of the super-weakly-interacting DM  
	and has certain advantages over axion-like-particle DM with respect to direct detection
\cite{An:2014twa}.

%%%%%%%%%%%%%%%%%%%%%%%%%%%%%%%%%%%%%
\subsection{Exotic DM search with colliders}	 
	The observation of the SUSY or any new BSM model and determination of the DM properties  
	 will require establishing the predicted particle spectrum
\cite{Dutta:2014mya}.
	One can mention several ways of investigating the DM problem at the LHC, additional to the traditional \met-method.
	For example, one can study cascade decays of colored (SUSY) particles, 
	vector-boson-fusion productions of non-colored (SUSY) particles, 
	mono-jet and direct stop production searches
\cite{Dutta:2014mya}. 

\smallskip %\hrule\bigskip %\noindent %---------------------------------------------------
	One believes that some fraction of possible DM candidates
	can couple only to leptons (leptophilic DM
\cite{Fox:2008kb,Kopp:2009et}) and therefore were 
	very weakly constrained from the LHC and direct DM detection searches.
	In such models the interaction is described by effective four-lepton contact operators, 
	which can be probed in $e^+e^-$-collisions. 
	
	The precise data from LEP was used in 
\cite{Freitas:2014jla} %{Leptophilic Dark Matter in Lepton Interactions at LEP and ILC,  Freitas:2014jla}
	to derive limits on this leptophilic DM in a model-independent EFT framework. 
	The bounds turn out to be very competitive with  
	exclusion curves obtained from LHC data from mono-photon events with large \met. 
	Furthermore, the future ILC (International Linear Collider) data  
	allow one to set the strongest limits on 
	TeV-scale leptophilic DM candidates  
\cite{Freitas:2014jla}. 

	The potential of the ILC for the solution of the DM problem was studied in 
\cite{Dreiner:2012xm}.
	Within the EFT approach the reach of the ILC was compared with that of other searches. 
	Low mass WIMPs are as usual a key feature for a collider search including the ILC.
	If it  happens that the WIMP can only couple to 
	leptons or only spin-dependently, the ILC could be especially useful to 
	study and constrain such a set of models. 
	
\smallskip %\hrule\bigskip %\noindent %---------------------------------------------------
	 The next opportunity for a future $e^+e^-$ collider to help with 
	solution of the DM problem was considered in 
\cite{Richard:2014vfa} %{Search for Dark Matter at Colliders, Richard:2014vfa}
	and was motivated by 
	an excess of energetic photons observed in the center of our Galaxy with the Fermi-LAT satellite
\cite{Daylan:2014rsa,Calore:2014xka}. 
	It was shown that if the DM candidates  
	are assumed to be the (Majorana or Dirac) fermions $\chi$ 
	and couple to a $Z'$ boson, one could observe a remarkable DM-related 
	signal at a TeV $e^+ e^-$ collider via the process  $e^+e^-\to \chi\chi+$photon. 
	This result relied on the ability to use 
	highly polarized beams to eliminate the background
$ee \to \nu\nu\gamma$ process due to $W$ exchange.  
It also requires an optimized setup to fully eliminate the contamination 
from $ee\to ee\gamma$.
	Finally, prospects for the DM search at the $e^+e^-$ colliders were presented.

	The overall conclusion of 
\cite{Richard:2014vfa} was that a BSM mediator, for instance a $Z'$ or extra Higgs bosons, 
	is needed to interpret the Fermi-LAT DM evidence. 
	If true, these interpretations predict scenarios which could already be tested at the LHC 
	while a TeV $e^+e^-$-collider should provide an essential tool for a 
	precise measurement of the parameters of the BSM resonances.

\smallskip 
	An $e^+e^-$ collider was also proposed to look for 
	production of a Higgs boson recoiling from a massless invisible system
\cite{Biswas:2015sha}. 
	One believes this can be a quite distinctive signature  of 
	the Higgs boson creation in association with a massless dark photon
	$e^+e^-\to H+A'$. 
	Dark photons can acquire effective couplings to the Higgs boson via loop diagrams.
	The signal and corresponding backgrounds for $H\to \bar{b}b$ were analysed, and the ILC
\cite{Behnke:2013xla} and  FCC-ee 
\cite{Gomez-Ceballos:2013zzn} sensitivities were estimated in a model-independent way
\cite{Biswas:2015sha}.

\smallskip %\hrule\bigskip %\noindent %---------------------------------------------------
	"Untraditional" usage of the positron beam of the DA$\Phi$NE linac at the Laboratori Nazionali di Frascati was proposed
\cite{Raggi:2015gza}	
	for dark photon search in Positron Annihilation into Dark Matter Experiment. 
	The PADME experiment will search for the dark photon $A'$ in the $e^+e^-\to \gamma A'$ process 
	in a positron-on-target experiment configuration.
	After one year a sensitivity in the relative interaction strength $\epsilon^2$ down to $10^{-6}$ 
	is achievable, in the mass region from 2.5~$ <m_{A'}< $~22.5 MeV$/c^2$. 

	The DA$\Phi$NE collider with the detector KLOE was also used to search for the so-called Higgs-strahlung  process 
	$e^+ e^- \to A'+h'$, where $A'$ is the dark photon, and $h'$ is the dark Higgs boson, 
	which decays invisibly.
	No evidence for the signal was observed and, in particular,  upper limits on the kinetic 
	mixing parameter $\epsilon$ in the range $10^{-4}\div 10^{-3}$ were established
\cite{Babusci:2015zda,Curciarello:2015iea}.

\smallskip
	Assuming prompt decays of the dark photon $A^\prime$ and the dark Higgs boson $h^\prime$, 
	the Belle collaboration also performed 
\cite{TheBelle:2015mwa} a search for their 
	production in the Higgs-strahlung channel, 
	$e^+e^- \rightarrow A^\prime h'$, with $h^\prime \rightarrow A^\prime A^\prime$. 
	Analysis of the full set of 977 fb$^{-1}$ Belle data gave no significant signal evidence.
	The 90$\%$ CL upper limits were obtained 
	on the branching fraction times the Born cross section, ${\rm BR} \times \sigma_{\rm Born}$, 
	on the Born cross section, $\sigma_{\rm Born}$, 
	and on the dark photon coupling to the dark Higgs boson times the kinetic mixing between the 
	SM photon and the dark photon, $\alpha_D \times \epsilon^2$.
	These limits improve upon and cover wider mass ranges than previous experiments. 
	For $\alpha_D=1/137$, $m_{h'}<$ 8 GeV/$c^2$, and $m_{A^\prime}<$ 1 GeV/$c^2$ Belle excluded 
	values of the mixing parameter $\epsilon$ above $8 \times 10^{-4}$    
\cite{TheBelle:2015mwa}.

	Further discussions of exotic DM search program with 
	$e^+e^-$ colliders can be found in, for example,
\cite{Essig:2013vha,Chen:2015tia}.

\smallskip %\hrule\bigskip %\noindent %---------------------------------------------------
	For a collider experiment a well-known challenge is to determine the mass of an undetected particle  
	due to the under-constrained kinematics with two missing particles in an event. 
	It is especially difficult at hadron colliders because of the further 
	unknown kinematics of initial partons.  
	There are many attempts to determine the missing particle mass at the LHC, 
	such as endpoint methods (see for example
\cite{Hinchliffe:1996iu,Allanach:2000kt,Gjelsten:2005aw}), 
polynomial methods (see for example \cite{Cheng:2008mg,Cheng:2007xv,Kawagoe:2004rz}), 
$M_{T2}$ methods (see for example \cite{Lester:1999tx,Nojiri:2008ir,Cho:2007qv,Nojiri:2008hy}), 
and the matrix element method (see for example \cite{Alwall:2009sv,Gainer:2013iya}).

\begin{figure}[!ht] \centering
\includegraphics[width=0.3\textwidth]{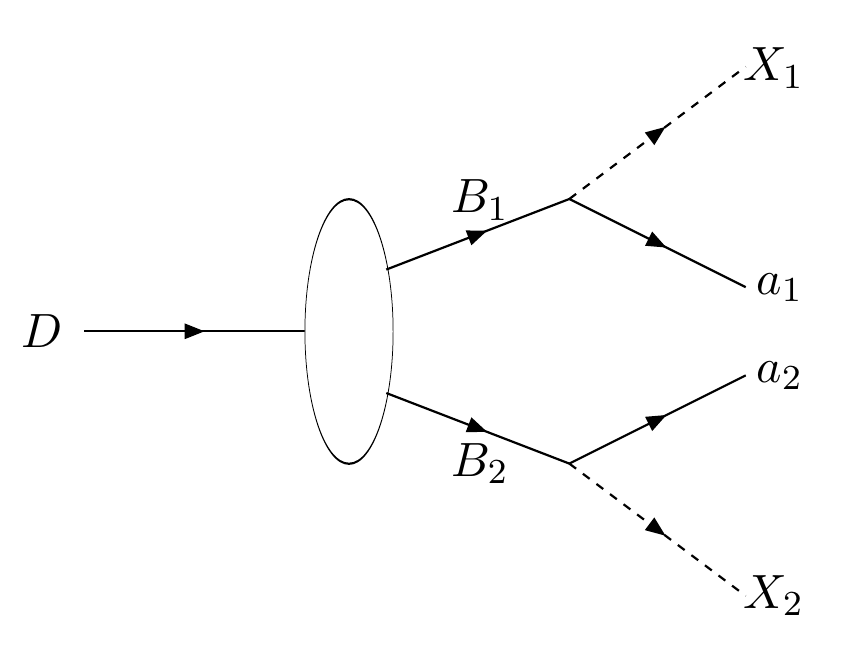}
\caption{The antler decay diagram of a heavy particle $D$ into two visible particles
$a_1$ and $a_2$ and two invisible particles $X_1$ and $X_2$
through on-shell intermediate particles $B_1$ and $B_2$.
From \cite{Christensen:2014yya}.}
\label{fig:2014yya-antler}
\end{figure}

	In view of the scientific program of the future high energy lepton collider, in particular future DM search,   
	the problem of how to determine the mass of undetected DM candidates 
	through the so-called {\em antler topology process}\/ given 
in Fig.~\ref{fig:2014yya-antler} was specially studied in 
\cite{Christensen:2014yya}.
	The {\em antler decay}\/ diagram 
\cite{Han:2009ss} was used for investigation of a resonant decay of a heavy particle $D$ into 
	two intermediate particles ($B_1$ and $B_2$),  
	followed by each $B_i$'s decay into a missing particle $X_i$ and a visible particle $a_i$.
	It was found that the resonant decay through this diagram develops cusps in some kinematic distributions 
	and the cusp positions along with the endpoint positions can determine 
	both the missing particle mass $m_X$ and the intermediate particle mass $m_B$.

	A lepton collider has an advantage due to the  
	well-defined initial state with fixed center-mass energy and center-mass frame,
	which allows one to study many new physics processes by means of the above-mentioned 
	antler topology 
	$e^+e^-\to B_1 B_2\to X_1 a_1 + X_2 a_2$.   
	It was found in
\cite{Christensen:2014yya} that the cusp method  
	appeared to be more stable than the commonly considered energy endpoints 
	against realistic factors (initial state radiation, acceptance cuts, detector resolution)
	and is very efficient for measuring the missing particle mass. 

	Considering as an example the pair production of scalar muons (smuons) 
	as the MSSM process that satisfies the antler topology,
	it was shown that at the 500 GeV ILC mass determination with precision of 0.5 GeV$/c^2$ can be achieved 
	for smuons with a leptonic final state.

	This method can be useful for determination of mass spectrum of any BSM model.
	As an expected by-product of the procedure one can obtain very useful indication 
	of masses of the invisible DM candidates. 
	Nevertheless the proof of true DM nature of the candidates still remains inevitable. 

\smallskip %\hrule\bigskip %\noindent %---------------------------------------------------
	There is a proposal to study mono-jet and \met\/ signature
	from $\gamma p$ collisions at the LHC in context of the DM problem
\cite{Sun:2014ppa}. %{Dark matter searches in jet plus missing energy events in $\gamma p$ collisions at the CERN LHC, Sun:2014ppa}
	In fact, photoproduction of jet+\met\/ final states  
	was simulated in a model independent EFT framework
	assuming a typical LHC forward detector.
	Rather good prospects for constraining the couplings of the quarks to the WIMP DM candidates
	were obtained  
	on the basis of the main reaction $pp\to p\gamma p\to p\chi\chi+$jet
	for an integrated luminosity of 200 fb$^{-1}$
	as a function of 
	the forward detector acceptance.  

\smallskip 
	A very light dark photon from the hidden sector could couple to 
	SM particles and could give a signal via $\nu e$-scattering experiments 
	if the dark photon is the gauge field of a U(1) group.
	This will allow a direct coupling with neutrinos. 
	The new interactions due to existence of the $A'$ boson whose couplings do not contain 
	derivatives lead to the differential cross section being proportional to $1/T^2$, 
	which makes low energy $\nu$-experiments 
	sensitive to the $A'$ boson search in the low mass region. 
	Hence low energy neutrino experiments aimed to measure 
	neutrino-nucleus coherent scattering as well as neutrino magnetic moment
	has an advantage in searching for the $A'$ boson,  
	with mass located much below the electroweak scale. 
	For the higher mass region of $A'$ boson, 
	neutrino experiments with higher incident energy have better sensitivity	
\cite{Bilmis:2015lja}.
	
\smallskip
	It is perhaps not meaningless to recall a "crazy" idea about usage of a very intensive beam of accelerated 
	particles or ions with large enough energy for cold DM particle search
\cite{BednyakovKovalenko:1999,Feng:2006ni}. 
	The detection strategy was in some sense reversed to the traditional direct DM method. 
	The non-relativistic and practically motionless galactic cold DM particles  
	were considered as a target permanently distributed in space around a collider.
	During all working time of the collider 
	this target is irradiated by an intensive beam of relativistic particles.
	One should only wait for a moment when a beam particle suddenly knocks
	one of the DM particles, producing a very unusual event 
	with matter and energy release from "an empty place". 
	
	Unfortunately, numerical estimations showed that 
	one should wait almost as long as man's life (due to very low local relic density).
	Nevertheless, future very intensive beams from CERN's 
	100-TeV Hadron Collider 
\cite{Assadi:2014nea}, 
	International Facility for Antiproton and Ion Research (FAIR)
\cite{Chattopadhyay:2014mha}, or Nuclotron based Ion Collider fAcility (NICA) 
\cite{Kekelidze:2013cua} could perhaps bring some sense to the idea.    

%%%%%%%%%%%%%%%%%%%%%%%%%
\subsection{Exotic DM search with accelerators}
	Search for accelerator-produced light WIMP DM candidates  
	was carried out with MiniBooNE
\cite{Thornton:2014ufa} using $1.86\times10^{20}$  
	protons with 8.9 GeV/$c$ momentum, directed to a steel beam-dump 50 m down-stream. 

	As already mentioned, due to nuclear recoil measurements, current direct DM detection 
	experiments have rather low sensitivity to WIMP masses below about 1 GeV$/c^2$.
	Despite the mandatory improvement of the direct detection sensitivity, it seems reasonable to use an 
	accelerator for production of a "beam" of relativistic (boosted) low-mass WIMP  
	DM candidates, which further can be detected with a 
	proper neutrino detector, due to the similarity of the WIMP- and 
	$\nu$-interaction signatures in the detector (weak neutral current events). 
	Running MiniBooNE in the beam-dump mode reduced the neutrino background 
	by having the beam hit a steel beam-dump instead of hitting the Be target.

	The MiniBooNE experiment, located at FNAL on the Booster Neutrino Beamline, 
	has already accumulated the largest collection of 
	$\nu$ and $\bar\nu$ samples. 
	Being well understood, the setup is well suited for 
	detection of events which could be generated by interaction of  
	accelerator-produced relativistic low-mass WIMPs.
	Preliminary analysis did not show signal events.
	Final results are expected at the end of 2015.

	Nevertheless, one should keep in mind that in this case  
	the accelerator produced WIMPs should have lifetime 
	$c\tau \simeq 0.49$ km to reach unchanged the detector from the collider.

\smallskip %\hrule\bigskip %\noindent %---------------------------------------------------
	Another idea to look for high energy scattering of "dark Dirac fermions" 
	from nuclei was considered in 
\cite{Soper:2014ska}. %{Scattering of Dark Particles with Light Mediators}
	One assumes that when the DM candidate particle $\chi$ is  
	light enough to escape the traditional direct detection, 
	a promising way to look for the $\chi$ is a fixed target experiment
\cite{Batell:2009di,Essig:2010gu}, where $\chi$'s can be pair (hadro)produced 
	by an $s$-channel exchange of a light vector boson.

	Being sufficiently weakly interacting the $\chi$ paricle can 
	pass through shielding (that screens out strongly interacting products) 
	and can be detected in a neutrino-like detector
	by means of interaction similar to neutrino neutral current scattering. 
	The advantage of a fixed target experiment over a colliding one 
	is the much higher luminosity,  
	which becomes decisive when one searches for extremely rare events. 
	
	This idea was tested with the Fermilab experiment E613 data, and 
	limits on a "secluded" (mediator is lighter than $\chi$) DM scenario 
\cite{Batell:2009di,Batell:2009yf} was obtained. 

	A recent review, concerning new limits on light hidden sectors from fixed-target experiments 
	can be found in 
\cite{Morrissey:2014yma}.

\smallskip %\hrule\bigskip %\noindent %---------------------------------------------------
	The idea to use a fixed-target neutrino experiment technique for the laboratory 
	search for light weakly interacting dark sectors was further discussed in
\cite{Kahn:2014sra}. % {DAEdALUS and Dark Matter, Kahn:2014sra}
	It was shown that the DAEdALUS source setup --- 
	an 800 MeV proton beam impinging on a target of graphite and copper --- 
	can improve the present 
	bound on the dark photon $A'$ (produced here mainly from $\pi^0$ decays) 
	by an order of magnitude over 
	much of the accessible parameter space for light DM $\chi$ 
	(produced via $A'\to \chi\bar{\chi}$) 
	when paired with a suitable neutrino detector such as LENA
\cite{Wurm:2011zn} (signal process for detection is $\chi e^- \to \chi e^-$) .
 	
	It was shown in
\cite{Kahn:2014sra} that DAEdALUS was 
	sensitive to DM particles produced from off-shell dark photons and   
	that fixed-target experiments had sensitivity to a much larger range 
	of heavy dark photon masses than previously thought.  
	The mechanism for the DM  production 
	and detection through a dark photon mediator was reviewed
	together with the discussion of the beam-off and beam-on backgrounds, 
	and present the sensitivity to dark photon kinetic mixing for 
	the DAEdALUS/LENA setup in both the on- and off-shell regimes. 

	It appears that intensity frontier experiments like DAEdALUS 
	in conjunction with a large underground neutrino detector such as LENA
	will have unprecedented sensitivity to light (sub-50 MeV) DM $\chi$, 
	light (sub-400 MeV) dark photons $A'$, and other light weakly interacting particles.

	One agrees with the statement from  
\cite{Kahn:2014sra}  
	that both neutrino and accelerator fixed-target DM search experiments 
	share essentially the same signals and backgrounds
	(though often well-separated kinematically), and this fact 
	suggests exciting opportunities for symbiosis between  
	BSM and neutrino physics in the coming years.

\smallskip %\hrule\bigskip %\noindent %---------------------------------------------------
	Another possibility of using a large volume neutrino detector for
	detection of relativistic DM particle candidates was discussed in  
\cite{Berger:2014sqa}, where the Sun was assumed to be a source of these boosted DM particles
\cite{Kong:2014mia}. 
	Their arrival direction from the Sun can be used as a important signature.
	To arrange the energetic DM fraction, one proposed  
	a scenario where thermal DM can be efficiently captured in the Sun 
	and can annihilate into another sort of DM, right the boosted one. 
	At least in models with a multi-component (or non-minimal) structure of DM sector
(for example \cite{Dienes:2015qqa}), 
	annihilations of viable thermal relic DM with masses 1--100 GeV$/c^2$
	can produce other stable DM particles with moderate Lorentz boosts. 

	The detection of this relativistic DM-like particle is expected to be due to 
	its interaction with a target proton (or nucleus) 
	in some (very) large volume terrestrial detector, 
	resulting in an energetic proton recoil track pointing towards the Sun.
	Cherenkov-radiation-based detectors Super-Kamiokande 
\cite{Fukuda:2002uc} and Hyper-Kamiokande
\cite{Abe:2011ts}  were considered as examples for sensitivity study.  

	In particular, one found that by means of  spin-dependent interaction 
	the considered boosted DM candidates could produce detectable signals
	with sensitivity comparable to DM direct detection experiments.  
	Future large-volume liquid Argon neutrino detector
\cite{Badertscher:2010sy} based on ionization signals 
	or neutrino telescopes
\cite{Aartsen:2014oha,Collaboration:2011nsa,Bueno:2007um,Avrorin:2014swy}  
	may significantly extend the sensitivity. 

	Concluding this point, one agrees with
\cite{Berger:2014sqa} that the possibility of detecting some energetic fraction of 
	unknown weakly interacting particles, which have a habit to arrive on the Earth from the Sun 
	can be very crucial for understanding the DM sector structure.
	The very idea to use large-volume neutrino (or proton-decay) detectors 
	for investigation of the DM problem looks  particularly intriguing.

%%%%%%%%%%%%%%%%%%%%
\subsection{Other exotic DM searches}
\begin{figure}[!hb]
\centering
\includegraphics[width=0.95\textwidth]{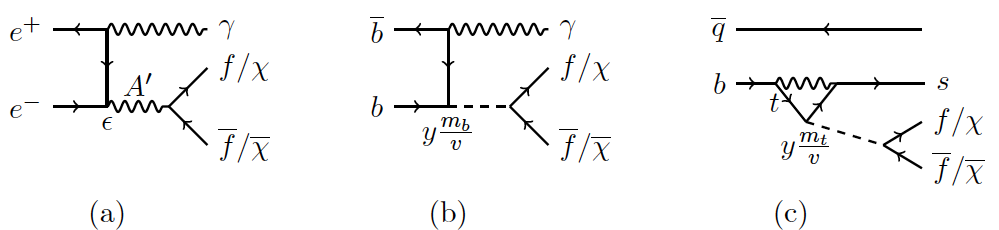} 
  \caption{Production of (a) dark photon in $e^+e^-$-collisions, 
  (b) dark Higgs $\phi$ in Y($b\bar{b}$)-decay, and 
  (c) dark Higgs in penguin B decay. 
The dark photon $A'$ or dark Higgs (dashed line) 
is shown decaying into a pair of SM fermions
$f\bar{f}$ or invisible dark-sector fermions $\chi\bar{\chi}$.
From~\cite{Soffer:2014ona}. }
\label{fig:2014ona-graphs}
\end{figure}

%%%%%%%%%%%%%%%%%%%%%
	Constraints on dark forces from the electron-positron colliders (B factories), 
	fixed-target experiments and flavor-physics measurements at hadron colliders were discussed in  
\cite{Soffer:2014ona}.
	The basic models where DM sector interacts with SM particles 
	via mediation of new dark vector and scalar bosons 
	with masses in the MeV-to-GeV range are reviewed.
	The typical processes are given in 
Fig.~\ref{fig:2014ona-graphs}. 
 	Recently dedicated searches for these low-mass bosons have been conducted,
	and rather tight limits on the parameter spaces of the relevant new-physics models
	have been established.
	Constraints from current measurements rule out
	significant regions of model-parameter space. 
	Higher sensitivities will be achieved by the next generation of B-factory 
	and fixed-target experiments, as well as by RUN II of the LHC.

	One believes that these vector and scalar dark bosons 
	couple to stable DM particles, and all together 
	constitute the so-called dark sector. 

\smallskip %\hrule\bigskip %\noindent %---------------------------------------------------
		Invisible decays of heavy quarkonium states can be considered as another %valuable 
		source of complementary information for constraining properties of possible light DM candidates.
		Contrary to the invisible decays of the not-yet-observed dark photon $A'$ or dark gauge boson $Z'$, 
		invisible decays of such well-observed objects like $\Upsilon$(nS) 
		looks much more reasonable and reliable.
		Furthermore, one believes that in contrast to DM-mono-jet searches at high energy colliders, 
		B- and charm factories are more suitable for light DM candidate search with a lower mass mediator
\cite{Fernandez:2014eja}. 

		Using data from high intensity electron-positron colliders 
		and assuming that the light DM candidate couples universally to all quarks,
		new constraints on the properties of the light DM candidates 
		were obtained from the analysis of invisible quarkonium decays in 
\cite{Fernandez:2014eja}. 
	The analysis was based on the results of the searches for $\Upsilon$(1S) invisible decays 
	performed by Belle
\cite{Tajima:2006nc} and BaBar 
\cite{Aubert:2009ae} operating at the energy of $\Upsilon$(3S) resonance.
	 The transition $\Upsilon$(3S)$\rightarrow\pi^{+}\pi^{-}\Upsilon$(1S) was used 
	 to detect invisible $\Upsilon$(1S) decays and to reconstruct 
	 the presence of the $\Upsilon$(1S) from the $\Upsilon$(1S) peak in the recoil mass distribution, 
	 $M_{\rm rec}$, by tagging $\pi^{+}\pi^{-}$ pairs with kinematics
$M_{\rm rec}^{2} \equiv s + M_{\pi \pi }^{2}-2\sqrt{s}E^{*}_{\pi \pi }$, 
	where $M_{\pi \pi}$ is the invariant mass of the pion system, 
	$E^{*}_{\pi \pi}$ is the energy of the pion system in the center-of-mass frame of the $\Upsilon$(3S), 
	and $\sqrt{s} = 10.3552$ GeV is the $\Upsilon$(3S) resonance energy.
	This tagging allowed one to make sure of the existence of the $\Upsilon$(1S) particle
	(as an initial state for the invisible decay $\Upsilon$(1S)$\to \chi\bar\chi$). 
	Similar searches for invisible decays of $J/\Psi$ were based on the transition 
$\Psi$(2S)$\rightarrow\pi^{+}\pi^{-}J/\Psi$.

	The experimental limits (90\% CL) on the branching ratios for the invisible decays are 
${\rm BR} ( \Upsilon$(1S)$ \rightarrow {\rm invisible}) < 3.0 \times 10^{-4}$ from the BaBar collaboration  	
\cite{Aubert:2009ae}
 and  ${\rm BR} ( J / \Psi \rightarrow {\rm invisible}) < 7.2 \times 10^{-4}$ from the BES collaboration
\cite{Ablikim:2007ek}.
	The SM contribution to these invisible decay modes 
	were found to be negligible~\cite{Chang:1997tq}:
${\rm BR} ( \Upsilon$(1S)$ \rightarrow \nu \bar \nu ) = 9.85 \times 10^{-6}$ and
${\rm BR} ( J / \Psi \rightarrow \nu \bar \nu )= 2.70 \times 10^{-8}$.

\begin{figure}[!h]
\includegraphics[width=0.97\textwidth]{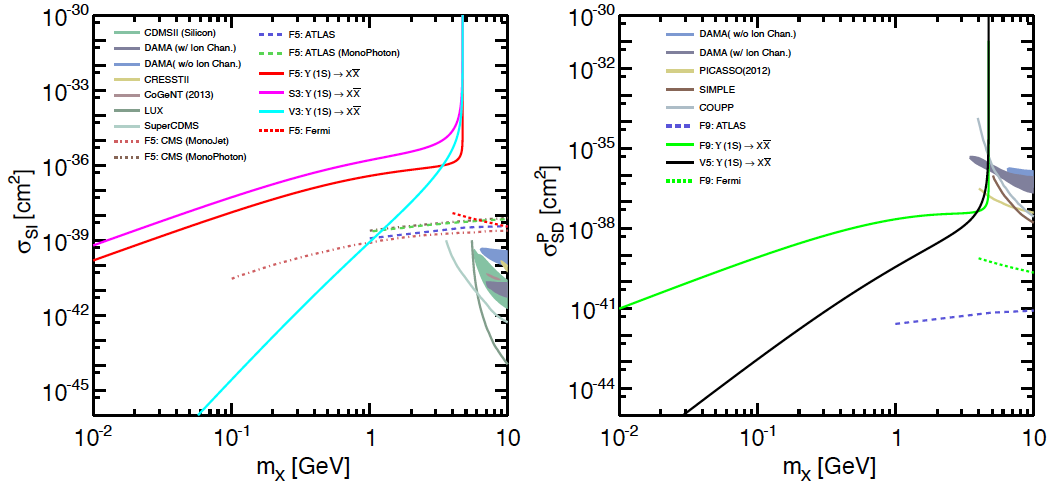} 
\caption{Bounds on the $\chi$-proton spin-independent (left panel) and spin-dependent (right panel) 
	scattering cross section as a function of mass $m_\chi$. 
	The $\chi$ couples universally to quarks through the indicated effective contact operator.
The labeled exclusion curves indicate 90\% CL bounds from limits on invisible decays of $\Upsilon$(1S),
95\% CL bounds from Fermi-LAT constraints on DM annihilation in dwarf spheroidal galaxies,
and 90\% CL bounds from mono-jet searches 
(CMS~\cite{Chatrchyan:2012me,Chatrchyan:2012tea} and  ATLAS~\cite{ATLAS:2012ky}).
The DAMA/LIBRA~\cite{Savage:2008er},  
CRESST II (95\% CL)~\cite{Angloher:2011uu}, 
CoGeNT~\cite{Aalseth:2012if},
CDMS II (Silicon)~\cite{Agnese:2013rvf}, 
SuperCDMS~\cite{Agnese:2014aze},  
LUX~\cite{Akerib:2013tjd},  
SIMPLE~\cite{Felizardo:2011uw}, 
PICASSO~\cite{Archambault:2012pm}, and COUPP~\cite{Behnke:2012ys}
 90\% CL signal regions
are also shown. 
From \cite{Fernandez:2014eja}.}
\label{fig:2014eja-SI-SD-Y1S}
\end{figure}

	In the framework of a low-energy EFT for each of the contact operators 
(Table~\ref{tab:2010ku}) relevant to the invisible  
	  $\Upsilon$(1S)$\rightarrow \chi\bar \chi $ decay
	exclusion limits for the spin-independent and spin-dependent $\chi$-proton cross section 
	were calculated
 \cite{Fernandez:2014eja} and presented in 
Fig.~\ref{fig:2014eja-SI-SD-Y1S}.

	One can see that the invisible-$\Upsilon$(1S) bounds 
	are sensitive to a low $\chi$-mass range 
	significantly below the thresholds of current direct DM detection experiments 
	and complement bounds obtained from $\gamma$-ray searches 
	of dwarf spheroidal galaxies and from mono-X searches at hadron colliders. 
	Unfortunately, a question of applicability of a very low mass region for true 
	DM candidates is still open.  	

	It is worth noting that since the $\Upsilon$(nS) states are non-relativistic,
	their invisible decays can be sensitive to the DM candidate interaction with non-relativistic quarks. 
	These searches have the same footing as the direct DM searches, contrary to 
	the LHC searches with highly relativistic quarks and gluons.

\smallskip  %\hrule\bigskip %\noindent %---------------------------------------------------
	Ignoring of DAMA/LIBRA results forces one to look for any theoretical explanation of signal absence in 
	other direct DM experiments, provided pure experimental explanations of the absence are also ignored.
	On the way a lot of different DM models appear.
	For example, such a model was considered in 
\cite{Fairbairn:2014aqa}, 
	where a DM candidate with a mass in 0.1 -- 1 TeV$/c^2$ interval was coupled to the SM particles 
	via only one vector boson. 
	The key assumption of this exotic model is that the DM candidate mass 
	is taken to be around 1/2 of the mediator mass. 
	Therefore one has a resonant enhancement of the DM-SM interaction and 
	should strongly reduce the couplings to have the correct relic abundance. 
	The main by-product of this procedure is a very low cross section and a zero result of direct detection.
	Prospects for verification of the model  with the LHC, CTA 
\cite{Wood:2013taa,Ibarra:2015tya} and AMS
\cite{Kounine:2012ega} observations of the Galactic center
	were also discussed
\cite{Fairbairn:2014aqa}. 

\smallskip %\hrule\bigskip %\noindent %---------------------------------------------------
	Another exotic simplified model of fermionic DM was considered in  
\cite{Gomez:2014lva}. 
	The DM couples exclusively to the right-handed {\em top quark}\/ via a 
	renormalizable interaction with a color-charged scalar.
	The relic abundance of this DM was computed and constraints 
	were placed on the model parameter space followed by the discussion of 
	prospect for direct detection.
	Furthermore, detailed analysis for the production of the DM candidates 
	at the LHC was performed. 
 	Several kinematic variables were proposed that allow extraction of a clean signal  
	and reduction of  the parameter space of this model during the LHC RUN II. 
	The possibility of detecting this type of DM via its annihilations into $\gamma$-rays
	was also studied.
	Another idea to use the top quark for DM search with a collider was also considered in
\cite{Haisch:2015ioa}.

\smallskip %\hrule\bigskip %\noindent %---------------------------------------------------
        One must say a word about an axion, which was proposed to explain an 
	anomaly in quantum chromodynamics
\cite{Dobrich:2015xca}.
	The electromagnetic signatures of axions have long been sought 
	in the laboratory experiments without any success. 
	String theory suggests an ultralight axion that would be so-called "warm" DM. 
	The mixture of cold and warm DM components (including neutrinos) 
	might resolve some tensions of pure cold DM scenarios. 
	For example, it could explain why there are fewer dwarf galaxies than cold DM predicts
\cite{Livio:2014gda}.

	There are exotic proposals to search for different DM candidates through 
	oscillations in the fine-structure constant using atomic spectroscopy 
\cite{VanTilburg:2015oza}, atoms clocks 
\cite{Arvanitaki:2014faa}, and laser and maser interferometry
\cite{Stadnik:2014tta}.
%%% 150627 %%%%%%%%%%%%%%%%
	Search for composite Dark Atoms is discussed in
\cite{Khlopov:2014wma,Khlopov:2014aia,Belotsky:2014nba,Belotsky:2014haa}. 
%%%%%%%%%%%%%%%%%%%%%%%%

	Finally, there is a common belief
\cite{Livio:2014gda} that a much broader categories of DM particles (substances) should be sought.
	 For example, one can give up DM neutrality and allow DM  
	 to carry a small electric charge or possess some internal states like electron levels of an atom. 
	Precision helioseismology could detect small changes in the solar surface oscillations 
	due to clouds of "millicharged" DM particles scattering off electrons in the solar plasma.
	More spherical DM haloes of distant galaxies could be measured by means of gravitational lensing,
	provided the DM particles can interact electromagnetically, and so on.

\smallskip
{\em Concluding} this section about exotic DM search programs, one should first stress that nowadays
	it is a very hard job to make an attempt to produce a complete review on the subject.
	Almost every day a paper with a new DM model can be found in {\tt arXiv}.
	An obviously incomplete list of recent eprint papers includes 
\cite{Primulando:2015lfa,Ko:2015vaa,Rajaraman:2015xka,Xiang:2015lfa,Martin-Lozano:2015vva,Rossi-Torres:2015eua,Fortes:2015qka,Suzuki:2015sza,Delgado:2015aha,Ghorbani:2015baa,Bishara:2015cha,Kainulainen:2015raa,Dutra:2015vca}.
	
	Furthermore, strictly speaking,  very few of the models and/or searches
	discussed (and not yet) in the section have something real to do 
	with practical detection of true DM particles, constituting the galactic DM halo of our Galaxy.  	
%%%%%%%

%%%%%%%%%%%%%%%%%%%%%%%%%%%%%%%%%%%%%%%
\section{Discussion}\label{sec:Discussion}
 %\section{Discussion}
\subsection{Effective Field Theory and a bit beyond.}
%%%%%%%%%%%%%%%%% 
	The scale of validity of the EFT, in particular the kinematic region where the EFT 
	approach for collider WIMP pair production breaks down, 
has been discussed from the very beginning %(see, for example,
\cite{Bai:2010hh,Cao:2009uw,Goodman:2010ku,Goodman:2011jq,Shoemaker:2011vi,Rajaraman:2011wf,Fox:2011pm,Cheung:2012gi,Cotta:2012nj,Busoni:2013lha,Profumo:2013hqa,Alves:2013tqa,Krauss:2013wfa,Buchmueller:2013dya,Busoni:2014uca,Fedderke:2014wda,deSimone:2014pda,Busoni:2014sya,Busoni:2014haa,D'Eramo:2014aba,Drozd:2015kva,Dudas:2015vka}.   %).
	One of the recent discussion of the problem can be found in the Appendix of  
\cite{Aad:2015zva} 
	where the region of validity of the EFT approach was studied under 
	various assumptions about the underlying new physics.
	
	Very sophisticated attempts to remedy the core of the "model-independent" conception 
	of the EFT with help of a set of "simplified models"  
	can be found, for example, in
\cite{Goodman:2011jq,Frandsen:2011cg,Agrawal:2011ze,An:2012va,Bell:2012rg,Frandsen:2012rk,An:2012ue,Chang:2013oia,An:2013xka,DiFranzo:2013vra,Morgante:2014kra,Malik:2014ggr,Buchmueller:2014yoa,deSimone:2014pda,Dolan:2014ska}. 
	It is difficult to add something quantitatively new into both of these considerations.

Nevertheless some general comments concerning this situation are in order.

---  Indeed, due to complexity of the DM problem one has to incorporate as much as possible 
	useful information, which could help one to solve the problem. 
	On the way the complementarity of all experimental searches for the DM (discussed in section
\ref{sec:DM-detections}) looks inevitable. 
	To arrange quantitatively this complementarity one needs an approach which could describe 
	all relevant observables with a common set of parameters. 
	Only in this case constraints from one experiment can be 
	connected with and/or applied to results of another one.

--- There is a common belief (or prejudice) that such an approach must be as much model-independent as possible. 
	To fulfill this requirement the EFT was proposed and well developed 
\cite{Goodman:2010ku}, where a set of Lorentz-invariant 4-point 
	effective operators was collected for description of all possible (and, perhaps, not-possible) DM-vs-SM couplings.
	For example, in SUSY models the tensor operators look very unusual, or irrelevant at all.  

	The complementarity goal was reached using the same operators in $s$-, $t$-, $u$-channels 
(as in Fig.~\ref{fig:2013ihz-DM-interplay}, section~\ref{sec:DM-detections}). 
	It was believed that the EFT had the undeniable advantage of being independent of the 
	{\em plethora of models of DM}\/
\cite{Busoni:2013lha}.
	Therefore one got a possibility of putting the WIMP constraints from the LHC on the same ground (in the same plot) 
	with the exclusion curves from the direct and indirect DM search experiments.
	Nevertheless, it is still unclear how this model-independent 
	comparison and/or competition could help one to solve the DM problem.  

	Furthermore, in the case of success (an observation of a DM candidate with some/all techniques) 
	one must incorporate observed properties of the DM candidate into a
	well-developed modern (or very new) theory beyond SM. 
	The above-mentioned model-independent results, if being appropriate, 
	should be inevitably incorporated into the new model and therefore get clear {\em model dependence}.

---  If one forgets about inner problems of the EFT, some non-comfortability concerning 
	model-independence of the approach still survives. 
	Indeed, one has  24 different operators (Table~\ref{tab:2010ku}) with 24 different $M_*$ scale parameters which 
	appeared to depend rather differently on the mass of the DM candidate $m_\chi$. 
	Furthermore, these operators are not connected with each other and are used 
	"one-in-time" --- 
	one does not know their interconnection coefficients. Do they interfere constructively or destructively? 

	This situation does not look better than, for example, in the MSSM, where one has 	
	few free parameters for complete numerical descriptions of all above-mentioned observables.
	The parameter manyfold (arbitrariness) of the MSSM is unconsciously 
	substituted by the functional manyfold (arbitrariness) of the EFT in selection of a  
	relevant number of the effective operators.  

	With 24 above-mentioned $M_*$--$m_\chi$ dependences situation looks worse.
	A rhetorical question could arise --- {\em why did not the LHC people include all 24 operators in their analysis?} 
	Maybe the people suspected intuitively something wrong with them?

--- Too many assumptions of the EFT approach were discussed (in previous sections).
	In particular, the key assumption concerning the only one SM-DM operator  
	looks far from being reasonable.  
	This is not necessarily the case for the electroweak interaction which has a $V$--$A$ structure
\cite{Lowette:2014yta}.
	Furthermore, it is very crucial that there are invisible decays of the next-to-lightest 
	(or next-to-next-to-LSP) SUSY particle (sneutrino $\to \chi+\nu$) in many promising SUSY models.

--- It was already many times stressed that validity of the EFT results depends 
	on the momentum transfer (through quarks), $Q$, 
	which should be below the energy scale of the underlying interactions $Q<m_V$ \cite{Busoni:2013lha}. 
With ultraviolet completion $M_*=m_V/\sqrt{g_q g_\chi}$ and perturbative regime of the couplings 
one can obtain, for example, the validity requirements  
$Q_{\rm D1}<4\pi \left( M_*^3/m_q \right)^{1/2}$, 
$Q_{\rm D9}<4\pi M_*$, 
$Q_{\rm C1}<4 \pi M_*^2 / m_q$ \cite{Aad:2014vea}. 
	These $Q$-limits are very different and strongly depend on the type of operator (D1, D9, or C1) and 
	on the details of the relevant parton energy and its distribution. 

--- Furthermore, one can recall 
\cite{deSimone:2014pda} that the rather small expected WIMP-signal rate implies that the scales $M_*$  
	are often smaller than the typical energy of the parton collisions ($\sqrt{s}$).  
	As a result, the interpretation of LHC data in terms of effective operators can lead to erroneous conclusions.
	It can overestimate the WIMP signal because of enhancements proportional to 
$\sqrt{s}/M_*$. 
	Or it can underestimate the signal 
	when the mediator can be produced directly and can give a much better collider signal 
	than the "model-independent" WIMP production.
	The EFT leads to LHC bounds  
	that seem very competitive, but are often only illusory
\cite{deSimone:2014pda}.  

--- Authors of \cite{Bai:2010hh} wrote {\em "A direct detection discovery that is in apparent conflict 
	with mono-jet limits will thus point to a new light state coupling the standard model to the dark sector."}  
	Next, almost the same is found in  
 \cite{Goodman:2010ku}:
	{\em If a direct DM search experiment were to observe a positive signal, the collider constraints would immediately imply a break down of the effective field theory at collider energies, revealing the existence of a light mediator particle}.
	From these statements one can conclude that the collider constraints are useless or ever
	irrelevant for the direct DM search experiments. 
	They help nothing. 
	They are unable to give advice where one should better look for a DM signal.  
	
--- The simplified models are an extra proof of the inconsistence of the EFT. 
	A typical example of development in this direction is a special White Paper 
\cite{Malik:2014ggr}, where one discussed a proposal for the consistent interpretation 
	of DM searches at colliders and in direct detection experiments
	based on the Minimal Simplified Dark Matter (MSDM) model  
\cite{Buchmueller:2014yoa}. 
	Nevertheless, the MSDM also has a lot of variants and has the same level of eclecticism as the EFT. 
	It represents only a potential starting point for going beyond the EFT, 
	further additions to the MSDM model, as well as the consideration of alternative approaches, 
	will be required to develop a 
{\em general strategy}\/ for comparing collider and direct DM experiments in the future.

	Next, even these simplified models can be overly simple from the point of view of the true DM physics. 
	Various  WIMP-SM  interactions are very commonly dictated by interactions with non-hadronic particles. 
	For example, a SUSY neutralino with the correct relic density 
	generically annihilates preferentially into weak vector bosons, 
	but scatters in direct detection primarily through the Higgs boson. 
	All of the approaches above, even if extended to consider interactions with either gauge or Higgs bosons, 
	implicitly assume that for all relevant processes the WIMP interacts dominantly with the same SM field. 
	Ultimately, when a complete model is under consideration, fully focused theoretical vision is the best
\cite{Askew:2014kqa}.

\smallskip
{\em Concluding} this subsection one can point out that  
	using the model-independent EFT approach for interpretation of the LHC DM search results 
	permanently requires bothering whether it is still valid or already not. 
	To some extent the point is still true for the simplified models.

	The \met\/ is the only measured quantity directly connected with invisible particles. 
	Without further {\em model-dependent assumptions}\/ it is impossible
	to make any statement about the nature of the missing particles
\cite{Kong:2013xma}. 

	Perhaps, it would be clever to go back to the well-defined (but still too complicated) SUSY framework.
	In this case one can severely constrain parameter space not of some set of {\em ad hoc}\/ effective operators 
	but of one of very promising SUSY models (MSSM, mSUGRA, etc).
	Experimental rejection, for example, of the MSSM could be a very important result. 

This is {\em the first}\/ hint to go back to the SUSY.

%%%%%%%%%%%%%%%%%%%%%%%%%%%%%%%%%%%%%%%%%%%%%%%%%%
\subsection{It is not possible to find the DM particle with the LHC or any other accelerator.}
	An observation of BSM-excess with the collider \met-signature is very welcome and very important.
	Nevertheless, one is unable to "directly" discover {\em true}\/ DM particles with LHC 
	simply because DM candidates fly away from a detector without any kind of interaction.
	One can judge them only "inderectly" by the missing mass, 
	missing energy and/or missing momentum in an event.
	If one would be able to simultaneously measure missing energy and missing 3-momentum, one 
	would be able to reconstruct the mass of the DM candidate. 
	But even in this case there is no guarantee that 
	this "indirectly detected" particle is the true DM particle ---
	the stable relic particle with clear galactic properties (section~\ref{sec:GalacticDM}). 

	One can stress again that an observation of a DM-like signal 
	(via a collider \met-signature) only proves the production of particle(s) 
	with lifetime $\ge l/c$, where $l$ is the scale of a typical detector. 
	The particles can be the true DM particles only if one can justify     
	an extrapolation from $\simeq$ 100 ns up to the lifetime of the Universe (24 orders of magnitude)
\cite{Bauer:2013ihz,Cushman:2013zza,Gelmini:2015zpa}, but this (undoable) 
	justification is not yet enough to prove the DM nature of the particle. 

	From the collider events with large \met\/, in the case of non-observation of any BSM-signal,   
	one can only set limits on the masses and couplings of 
	a number of different WIMP-like particles that either depart from a detector tracelessly 
	or fail to decay into detectable particles in the detector.	 
	The true DM particles constitute only a subset of this amount of undetected particles, 
	and it is impossible to find model-independent answer to the question 
	of how this subset is large.
	
	In many theories beyond the SM (like SUSY) 
	there are two (or more) massive weakly interacting neutral particles, 
	which are also potential candidates for {\em collider DM particles}, 
	because these next-to-lightest particles also can undergo 
	{\em invisible decays} and can produce \met-signature.
	Limits on such signatures tend to be stronger than bounds from direct collider searches 
	for the DM itself
\cite{Ismail:2014fca}.
 	There is also a multi-component DM scenario, when a model may 
	contain two (or more) different genuine DM particles, 
	whose production in various combinations 
	will inevitably lead at times to asymmetric event topologies
\cite{Kong:2013xma}.

	There is a "hidden problem" with model-independence in the DM search. 
	The results of any model-independent search must afterwards be  
	implemented into one of the modern BSM models of particle physics, for example, supersymmetry. 
	If one day the DM particles is found "in a model-independent way", 
	one inevitably meets a problem to find a proper place for these particles in a new 
	(or still SUSY) model. 
	Today the SUSY looks the best candidate for a model beyond the SM, it   
	unifies all scales, interactions and energies (from extremely high to extremely low), it 
	provides excellent DM particle candidate and so on.
	Instead of SUSY one can consider any adequate model of new physics.

	The only way to "detect DM particle with the LHC" can be realized
	with simultaneous experimental approval of a complete new physics model, 
	such as SUSY, with all particle masses and couplings 
	coherently described by a set of common parameters.
	In this case the "indirectly detected" 
	lightest SUSY particle (LSP) will be the very wanted DM candidate (with proper galactic features,
	relic density, etc).
	 
	There is no contradiction to look for the DM particles at the LHC 
	within some realization of the SUSY model right from the very beginning.
	Often, a more specific model, like MSSM, makes sharper predictions at the cost of generality, 
	but this cost allows verification, or ever complete rejection of the model.
	Any true DM particle discovery with the LHC is possible only together with the %LHC 
	discovery of the SUSY.
This is {\em the second}\/ hint to go back to the SUSY.

\subsection{Still SUSY with RUN-II} %%%%%%%%%%%%%%%%%%%%%%%%%%%%%%
	There is a common belief that the particle physics accelerator 
	experiments can create and study DM particles in the laboratory.
	SUSY models with R-parity conservation provide the stable WIMP in the form of the LSP (Lightest SUSY particle). 
	WIMPs are usually the final product of cascade decays of heavier unstable SUSY particles accompanied 
	by SM particles, in particular, with high transverse momentum $p^{}_{\rm T}$. 
	Therefore one can observe \met\/ generated by escaped WIMP pairs  
	only if the WIMP pair is tagged by a detectable SM particle yield 
	(jet or photon from initial- or final state radiation) with right the same \met.
	Furthermore, other LHC measurements can be used to constrain the 
	underlying SUSY model and hence to extract information about the nature 
	of DM, provided the LSP is a very good DM candidate.  
	
	The only way to prove the existence of the DM particles at a collider
	is to prove it together with the SUSY. 
	Unfortunately, this is a very complicated task. 
	First, one should observe BSM-excesses in as many collider observables as possible; 
	second, one should describe these excesses 
	coherently with one set of the SUSY parameters.
 	If the set of parameters gives correct DM density,
	the LSP can be considered as a true DM particle.
	
	A good example how the LHC helps to learn (within MSSM) 
	something about DM, in particular the relic density, 
	is given in  
 \cite{Barberio:2009xca}.
	The relation of the LHC measurements to DM could have several steps.
	The 1st step is to look for deviations from the SM, for example, in the multi-jet+ \met\/ signatures. 
	If the deviations are observed, the 2nd step is to clarify the SUSY nature of them and
	to establish the SUSY mass scale using relevant inclusive variables. 
	The 3rd step includes determination of model parameters, 
	selection of particular decay chains and use kinematics to determine mass combinations.
	Due to the escaping LSPs  
	only the kinematic endpoints in the invariant mass distributions 
	of visible decay products can be used to estimate undetected particle masses and 
	to derive the SUSY mass spectrum. 
	As a result, one can obtain the MSSM parameters and the unification scale.
	
	At the end of this step one can perform (model-dependent) estimation of the
	relic density, which should have a correct value 
	to allow the LSP to claim to be the cold DM candidate. 
	Furthermore, to prove the DM nature of the LSP one inevitably needs 
	extra information (from direct and indirect experiments) 
	concerning the LSP lifetime and the LSP fraction in the observed astrophysical DM
\cite{Barberio:2009xca}. 
	The LHC observed WIMP DM candidate could be only a part of the astrophysical DM
\cite{Gelmini:2015zpa}.

%%%%%%%%%%%%%%%%%%%%%%%%%%%%%%%%%%%%%%%%%%%%%%
\smallskip	
	The R-parity violation SUSY models can also contribute to the DM problem, 
	but in the models the WIMP DM candidate can be produced not only by pairs 
	but also in single mode, and should still have a lifetime (against decay into SM particles) 
	compared with that of the Universe.
	The prospects for the R-parity violating DM search with accelerators 
	are not yet well discussed, perhaps due to their complication and unclearness.
%%%%%%%%%%%%%%%%%%%%%%%%%%%%%%%%%%%%%%%%%%%%%%%

\smallskip 	
	The situation concerning possible discovery of the SUSY (together with the DM) 
	looks much more promising for the LHC RUN-II (13--14 TeV),  
	due to "a bit large"  mass of the Higgs boson
\cite{Nath:2015dza}.
	In the SM the Higgs boson with mass 125--126 GeV makes the vacuum almost unstable
	and provides very significant indication of New Physics. 
	The most promising candidate for New Physics is the SUSY
\cite{Nath:2015dza}. 
	This value of the Higgs mass  
	requires a large loop correction and a high TeV-size SUSY scale.
	The scale can explain suppression of any SUSY contribution to FCNC (flavor-changing neutral current) processes 
	and non-observation of (heavy strong interacting) sparticles in 7--8 TeV data.
	It also allows for light-mass uncolored sleptons and gauginos (gluino, charginos, neutralinos). 
			
	Hence,  
	contrary to RUN-I (8 TeV) RUN-II has a much better chance of observing these low-mass sparticles,
	in particular the LSP, the best SUSY DM candidate. 
	Furthermore, it was shown that future direct DM experiments such as XENON1T 
	will also be able to explore a large part of the SUSY 
	parameter space consistent with the measured Higgs boson mass
\cite{Nath:2015dza}.
	
	Therefore, it looks today promising, 
	following Pran Nath's paper 
\cite{Nath:2015dza}, 
	to perform at the 13-TeV LHC a comprehensive search for the SUSY, 
	having in mind simultaneous search for the SUSY-DM candidate(s).
	According to John Ellis \cite{Ellis:2015daa}, 
	one should also strongly think about SUSY discovery, which "may not be far away".  

	The discovery of a light Higgs boson at the LHC opens a broad program of studies and measurements to understand the role of this particle in connection with New Physics and Cosmology. SUSY is the best motivated and most thoroughly formulated and investigated model of New Physics which predicts a light Higgs boson and can solve the DM problem 
\cite{Arbey:2015aca}.
	
	This is {\em the third}\/ hint to go back into the SUSY framework.

%%%%%%%%%%%%%%%%%%%%%%%%%%%%%%%
\subsection{The decisive role of direct DM detection technique} 
%%%%%%%%%%%%%%%%%%%%%%%%%%%%%%%%
	With the unique features of the LHC 
	a new collider-DM-search community has appeared and has very quickly matured.
	The community has produced a number of experimental LHC-based DM search papers, 
	accompanied with a huge amount of theoretical papers on the subject 
(see, for example, some of recent papers
\cite{Fedderke:2014wda,Buchmueller:2014yoa,Yu:2014mfa,Kile:2014jea,Ma:2014wea,Bramante:2014tba,Calibbi:2014lga,Queiroz:2014pra,Busoni:2014gta,Buckley:2014fba,Harris:2014hga,Liu:2014rqa,Low:2014cba,Abdallah:2014hon,Han:2014nba,Bhattacharya:2014yha,Blumenthal:2014cwa,Li:2014vza,Agashe:2014yua,Bell:2015sza}).	
	Almost every day a paper with a new collider-related DM model is issued
(see, for example, 
\cite{Primulando:2015lfa,Ibarra:2015fqa,Xiang:2015lfa,Garny:2015wea,Crivellin:2015wva}).
	It looks like the old-fashioned direct-DM-detection community,  
	together with its traditions, achievements and main results, 
	is shifted aside from the main stream on the way to solution of the DM problem. 
	Some precaution here is in order. 

	First of all one should decide  whether an experimental research is aimed at {\em real detection}\/ of a DM particle 
	or its goal is to constrain the parameter space of some SUSY, BSM, EFT, simplified or any other model. 
	These very different goals need very different strategies and will have very different results. 
	This point should be clearly understood right from the very beginning. 

	The above-mentioned discussion shows that with a collider one is unable to detect a true DM particle.
	One can only constraint the relevant parameter space either with the \met-signature, or
	with the endpoint or any other technique.
	Therefore, a collider DM search program inevitably can produce 
	today only exclusion curves as a final result. 
	In the case of an observation of a BSM signal, 
	the curve can only be sharper and more decisive. 

	Due to a common belief in non-yet-observation of the 
	DM signal (DAMA/LIBRA is ignored), 
	all different DM search techniques (section~\ref{sec:DM-detections}) 
	compete with each other only in the space of the exclusion curves.
	The goal of this competition has little sense, 
	because it only shows who is today better in looking for nothing 
	(who is a better excluder).
	
       	A main physical reason to {\em improve}\/ an exclusion curve is usually an attempt
	to constrain a SUSY-like or BSM model. 
	From non-observation of the BSM signal one can push forward a business on 
	reduction of the parameter space of the models.
	But it is not very promising and effective, 
	the manifold and flexibility of these models are too large.
	Every day a new model appears.
        
        The famous model-independent EFT approach together with its modifications 
        do not helps here, 
	they make sense only for comparison of exclusion curves from different 
	search techniques.  
	New forms for a more sophisticated exclusion curve also appear 
\cite{Anderson:2015xaa}, but they still not help a lot with true DM detection, the goal is still 
	competition and consistence between different DM searches.  	

        At the present and foreseeable level of experimental accuracy, 
	simple fighting for the best exclusion curve is almost useless 
	either for real DM detection or for substantial restrictions for SUSY.	
       	One should inevitably go beyond the exclusion curve paradigm and 
	aim at registration of the DM particles. 
	As was discussed above, this can only be done together with a 
	discovery of a SUSY-like model. 
	
	Furthermore, {\em the key message here concerns the point} that  
	only direct DM observation can prove the existence of DM particles.
	Only in a direct DM experiment one can have a chance to see 
	the galactic nature of the true DM particle population, via
	measuring the annual modulation of the recoil signal
(section~\ref{sec:DirectDetectionDM}).
	This signature is inevitable to prove DM existence. 
	There is no way from the key role of direct DM detection.
	The other DM search approaches --- indirect, collider, astrophysical --- 
	can only help the direct DM detection, for example, 
	with mass region search advice, local relic density estimates, etc. 
	 
       New generations of DM experiments right from their beginning 
       {\em should aim at detection}\/ of the DM particles.
       This will require development of new setups, which will be able to register 
       {\em positive signatures} of the DM particle interactions with nuclear targets.

       One should try to obtain a reliable {\em recoil energy spectrum}.
       First, very accurate off-line investigation of the measured spectrum 
       allows one to single out different non-WIMP background
       sources and to perform controllable background subtractions.  
       Second, the spectrum allows one to look for the annual modulation effect,  
       the only currently  available positive signature 
 	of DM particle interactions with terrestrial nuclei.
	
       This effect is not simply a possibility (among many others) 
       of rejecting background, but it is a unique signature 
        which reflects the inner physical properties of the DM interaction with matter. 
	It is a very decisive and eagerly welcomed feature, 
	which is inevitable for the laboratory proof of the existence of the DM population nearby the Earth  
\cite{Bednyakov:2008gv,Bednyakov:2012cu}.
	Furthermore, natural (solar, supernovae, geo-) 
	neutrino background sooner or latter will be irreducible for 
	direct DM search technique
\cite{Gutlein:2014gma}, 
	and only a positive signature (like annual modulation, or directional correlations) 
	will be able to eliminate it.    

%%%%%%

%%%%%%%%%%%
\section{Conclusion}
	One is unable to prove, especially model-independently, a discovery of a dark matter particle with an accelerator.
	What one can do with an accelerator is only to discover evidence for existence of 
	a weakly interacting (massive) particle (WIMP), which could be, or could not be a dark matter (DM) candidate.  
	The WIMP is not yet a true DM particle.
	A true DM particle possesses a galactic signature, which one should clearly demonstrate.
	This signature --- annual modulation of a signal --- is  nowadays accessible 
	only for direct DM detection experiments.
	Therefore, to prove the DM nature of a collider-discovered candidate one must 
	observe it together with this galactic signature in a direct DM search experiment.
	Colliders can only play here a role of a goal-pointer.
	Furthermore, being observed sooner or later, the DM particle must be implemented 
	into a modern theoretical framework, like, for example, the Standard Model today, or
	the supersymmetry (SUSY) in the future.    

	With this point in mind (together with high complexity of the DM problem),  
	a better strategy may be to look for the DM particles openly relying on a SUSY model.
	In this {\em model-dependent}\/ case determination of the DM properties will be 
	a simple by-product of the SUSY model experimental observation.
			
	The current LHC DM search program uses only the missing transverse energy signature 
	for the WIMP search in the data. 
	Large-\met\/ search over any possible final state with ATLAS 
	has its own importance for the SUSY or any other beyond-SM theories.

	Non-observation of any excess above SM expectations forces the LHC 
	experiments to enter into the same fighting for the best exclusion curve, in which almost all direct 
	DM experiment took place. 
	On this way the {\em model-independent}\/ effective field theory approach is commonly used. 
	But this fighting has nothing to do with the real goal to discover a DM particle, 
	especially with an accelerator.

	Obviously, if one fails to find a DM-SUSY candidate (together with a SUSY) at the LHC, 
	one must think about another very new BSM framework (or a new collider).
     It is on the other hand absolutely clear that the SUSY, 
     although in contrast to others being preferred,  
     is not the only candidate for the origin of DM, 
     and other scenarios have also to be investigated. 
	 
\smallskip
	Perhaps, some words as a {\em non-physical}\/ conclusion are in order.
	Every day at least one new paper with a new model concerning the DM subject can be found in the {\tt arXiv}.  
	Obviously, this fact reflects the highest level of interest in the DM problem, 
	and one should not stop (or ever strongly reduce) issuing such kind of analysis and papers. 
	A lot of very useful achievements are connected with these works  --- detector, hardware, middleware, 
	software development and study; PhD and PostDoc production and defenses;
	outreach, spin-off and by-products; society and people education,  
	common scientific community, good international relations and future society organization, etc. 
	Nevertheless, as a precaution, 
	one should not forget with the above-mentioned papers and achievements about the main 
	fundamental scientific goal --- discovery of galactic dark matter particles. 

\normalsize
%%%%%%%%%%%%%%%%%
\bibliographystyle{JHEP} 
\renewcommand{\baselinestretch}{1.0}
%\bibliography{bibfiles/DM-ATLAS-papers,bibfiles/DM-LHC-papers,bibfiles/DM-Accelerator-papers,bibfiles/Other-papers,bibfiles/DM-papers}

\begin{thebibliography}{100}

\bibitem{Zwicky:1933gu}
F.~Zwicky, {\it {The redshift of extragalactic nebulae}},  {\em Helv.Phys.Acta}
  {\bf 6} (1933) 110--127.

\bibitem{Livio:2014gda}
M.~Livio and J.~Silk, {\it {Broaden the search for dark matter}},  {\em Nature}
  {\bf 507} (2014) 29, [\href{http://xxx.lanl.gov/abs/1404.2591}{{\tt
  arXiv:1404.2591}}].

\bibitem{Drees:2012ji}
M.~Drees and G.~Gerbier, {\it {Mini-Review of Dark Matter: 2012}},
  \href{http://xxx.lanl.gov/abs/1204.2373}{{\tt arXiv:1204.2373}}.

\bibitem{Saab:2012th}
T.~Saab, {\it {An Introduction to Dark Matter Direct Detection Searches and
  Techniques}},  \href{http://xxx.lanl.gov/abs/1203.2566}{{\tt
  arXiv:1203.2566}}.

\bibitem{Bertone:2004pz}
G.~Bertone, D.~Hooper, and J.~Silk, {\it {Particle dark matter: Evidence,
  candidates and constraints}},  {\em Phys. Rept.} {\bf 405} (2005) 279--390,
  [\href{http://xxx.lanl.gov/abs/hep-ph/0404175}{{\tt hep-ph/0404175}}].

\bibitem{Famaey:2015bba}
B.~Famaey, {\it {Dark Matter in the Milky Way}},
  \href{http://xxx.lanl.gov/abs/1501.0178}{{\tt arXiv:1501.0178}}.

\bibitem{Iocco:2015xga}
F.~Iocco, M.~Pato, and G.~Bertone, {\it {Evidence for dark matter in the inner
  Milky Way}},  \href{http://xxx.lanl.gov/abs/1502.0382}{{\tt
  arXiv:1502.0382}}.

\bibitem{Durazo:2015zzzz}
R.~Durazo, X.~Hernandez, and S.~Mendoza, {\it {Evidence for dark matter in the
  inner Milky Way...Really?}},  \href{http://xxx.lanl.gov/abs/1503.0750}{{\tt
  arXiv:1503.0750}}.

\bibitem{McGaugh:2015tha}
S.~McGaugh, F.~Lelli, M.~Pawlowski, G.~Angus, O.~BienaymŽ, et~al., {\it
  {Comment on "Evidence for dark matter in the inner Milky Way''}},
  \href{http://xxx.lanl.gov/abs/1503.0781}{{\tt arXiv:1503.0781}}.

\bibitem{Iocco:2015bja}
F.~Iocco, M.~Pato, and G.~Bertone, {\it {Reply to Comment on "Evidence for dark
  matter in the inner Milky Way"}},
  \href{http://xxx.lanl.gov/abs/1503.0878}{{\tt arXiv:1503.0878}}.

\bibitem{Sofue:2015xpa}
Y.~Sofue, {\it {Dark Halos of M31 and the Milky Way}},
  \href{http://xxx.lanl.gov/abs/1504.0536}{{\tt arXiv:1504.0536}}.

\bibitem{Kuhlen:2012ft}
M.~Kuhlen, M.~Vogelsberger, and R.~Angulo, {\it {Numerical Simulations of the
  Dark Universe: State of the Art and the Next Decade}},  {\em Phys.Dark Univ.}
  {\bf 1} (2012) 50--93, [\href{http://xxx.lanl.gov/abs/1209.5745}{{\tt
  arXiv:1209.5745}}].

\bibitem{Gelmini:2015zpa}
G.~B. Gelmini, {\it {TASI 2014 Lectures: The Hunt for Dark Matter}},
  \href{http://xxx.lanl.gov/abs/1502.0132}{{\tt arXiv:1502.0132}}.

\bibitem{Hoeneisen:2015rva}
B.~Hoeneisen, {\it {Trying to understand dark matter}},
  \href{http://xxx.lanl.gov/abs/1502.0737}{{\tt arXiv:1502.0737}}.

\bibitem{Frere:2015xba}
J.-M. Frre, {\it {Dark matter variations}},
  \href{http://xxx.lanl.gov/abs/1504.0822}{{\tt arXiv:1504.0822}}.

\bibitem{Kamionkowski:1997xg}
M.~Kamionkowski and A.~Kinkhabwala, {\it {Galactic halo models and particle
  dark matter detection}},  {\em Phys. Rev.} {\bf D57} (1998) 3256--3263,
  [\href{http://xxx.lanl.gov/abs/hep-ph/9710337}{{\tt hep-ph/9710337}}].

\bibitem{Pato:2015dua}
M.~Pato, F.~Iocco, and G.~Bertone, {\it {Dynamical constraints on the dark
  matter distribution in the Milky Way}},
  \href{http://xxx.lanl.gov/abs/1504.0632}{{\tt arXiv:1504.0632}}.

\bibitem{Kolb:1990vq}
E.~W. Kolb and M.~S. Turner, {\it {The Early Universe}},  {\em Front.Phys.}
  {\bf 69} (1990) 1--547.

\bibitem{Feng:2010gw}
J.~L. Feng, {\it {Dark Matter Candidates from Particle Physics and Methods of
  Detection}},  {\em Ann.Rev.Astron.Astrophys.} {\bf 48} (2010) 495--545,
  [\href{http://xxx.lanl.gov/abs/1003.0904}{{\tt arXiv:1003.0904}}].

\bibitem{Christensen:2014yya}
N.~D. Christensen, T.~Han, J.~Song, and Stefanus, {\it {Determining the Dark
  Matter Particle Mass through Antler Topology Processes at Lepton Colliders}},
   {\em Phys.Rev.} {\bf D90} (2014) 114029,
  [\href{http://xxx.lanl.gov/abs/1404.6258}{{\tt arXiv:1404.6258}}].

\bibitem{Cerulli:2012dw}
R.~Cerulli, R.~Bernabei, P.~Belli, F.~Cappella, C.~Dai, et~al., {\it {Technical
  aspects in dark matter investigations}},
  \href{http://xxx.lanl.gov/abs/1201.4582}{{\tt arXiv:1201.4582}}.

\bibitem{Bauer:2013ihz}
D.~Bauer, J.~Buckley, M.~Cahill-Rowley, R.~Cotta, A.~Drlica-Wagner, et~al.,
  {\it {Dark Matter in the Coming Decade: Complementary Paths to Discovery and
  Beyond}},  \href{http://xxx.lanl.gov/abs/1305.1605}{{\tt arXiv:1305.1605}}.

\bibitem{Askew:2014kqa}
A.~Askew, S.~Chauhan, B.~Penning, W.~Shepherd, and M.~Tripathi, {\it {Searching
  for Dark Matter at Hadron Colliders}},  {\em Int.J.Mod.Phys.} {\bf A29}
  (2014) 1430041, [\href{http://xxx.lanl.gov/abs/1406.5662}{{\tt
  arXiv:1406.5662}}].

\bibitem{Zitzer:2015uta}
{\bf VERITAS} Collaboration, B.~Zitzer, {\it {The VERITAS Dark Matter
  Program}},  \href{http://xxx.lanl.gov/abs/1503.0074}{{\tt arXiv:1503.0074}}.

\bibitem{Ibarra:2015tya}
A.~Ibarra, A.~S. Lamperstorfer, S.~L. Gehler, M.~Pato, and G.~Bertone, {\it {On
  the sensitivity of CTA to gamma-ray boxes from multi-TeV dark matter}},
  \href{http://xxx.lanl.gov/abs/1503.0679}{{\tt arXiv:1503.0679}}.

\bibitem{Ackermann:2015zua}
{\bf Fermi-LAT} Collaboration, M.~Ackermann et~al., {\it {Searching for Dark
  Matter Annihilation from Milky Way Dwarf Spheroidal Galaxies with Six Years
  of Fermi-LAT Data}},  \href{http://xxx.lanl.gov/abs/1503.0264}{{\tt
  arXiv:1503.0264}}.

\bibitem{Conrad:2015bsa}
J.~Conrad, J.~Cohen-Tanugi, and L.~E. Strigari, {\it {WIMP searches with gamma
  rays in the Fermi era: challenges, methods and results}},
  \href{http://xxx.lanl.gov/abs/1503.0634}{{\tt arXiv:1503.0634}}.

\bibitem{Choi:2015ara}
{\bf Super-Kamiokande} Collaboration, K.~Choi et~al., {\it {Search for
  neutrinos from annihilation of captured low-mass dark matter particles in the
  Sun by Super-Kamiokande}},  \href{http://xxx.lanl.gov/abs/1503.0485}{{\tt
  arXiv:1503.0485}}.

\bibitem{Kasuya:2015uka}
S.~Kasuya, M.~Kawasaki, and T.~T. Yanagida, {\it {IceCube potential for
  detecting the Q-ball dark matter in gauge mediation}},
  \href{http://xxx.lanl.gov/abs/1502.0071}{{\tt arXiv:1502.0071}}.

\bibitem{Zornoza:2014dma}
{\bf ANTARES} Collaboration, J.~Zornoza and G.~Lambard, {\it {Results and
  prospects of dark matter searches with ANTARES}},  {\em Nucl.Instrum.Meth.}
  {\bf A742} (2014) 173--176, [\href{http://xxx.lanl.gov/abs/1404.0148}{{\tt
  arXiv:1404.0148}}].

\bibitem{Avrorin:2014swy}
{\bf Baikal Collaboration} Collaboration, A.~Avrorin et~al., {\it {Search for
  neutrino emission from relic dark matter in the Sun with the Baikal NT200
  detector}},  \href{http://xxx.lanl.gov/abs/1405.3551}{{\tt arXiv:1405.3551}}.

\bibitem{Hooper:2012gq}
D.~Hooper and W.~Xue, {\it {Possibility of Testing the Light Dark Matter
  Hypothesis with the Alpha Magnetic Spectrometer}},  {\em Phys.Rev.Lett.} {\bf
  110} (2013), no.~4 041302, [\href{http://xxx.lanl.gov/abs/1210.1220}{{\tt
  arXiv:1210.1220}}].

\bibitem{Adriani:2008zr}
{\bf PAMELA} Collaboration, O.~Adriani et~al., {\it {An anomalous positron
  abundance in cosmic rays with energies 1.5-100 GeV}},  {\em Nature} {\bf 458}
  (2009) 607--609, [\href{http://xxx.lanl.gov/abs/0810.4995}{{\tt
  arXiv:0810.4995}}].

\bibitem{Giesen:2015ufa}
G.~Giesen, M.~Boudaud, Y.~Genolini, V.~Poulin, M.~Cirelli, et~al., {\it {AMS-02
  antiprotons, at last! Secondary astrophysical component and immediate
  implications for Dark Matter}},
  \href{http://xxx.lanl.gov/abs/1504.0427}{{\tt arXiv:1504.0427}}.

\bibitem{Evoli:2015vaa}
C.~Evoli, D.~Gaggero, and D.~Grasso, {\it {Secondary antiprotons as a Galactic
  Dark Matter probe}},  \href{http://xxx.lanl.gov/abs/1504.0517}{{\tt
  arXiv:1504.0517}}.

\bibitem{Lin:2015taa}
S.-J. Lin, X.-J. Bi, P.-F. Yin, and Z.-H. Yu, {\it {Implications for dark
  matter annihilation from the AMS-02 $\bar{p}/p$ ratio}},
  \href{http://xxx.lanl.gov/abs/1504.0723}{{\tt arXiv:1504.0723}}.

\bibitem{Hamaguchi:2015wga}
K.~Hamaguchi, T.~Moroi, and K.~Nakayama, {\it {AMS-02 Antiprotons from
  Annihilating or Decaying Dark Matter}},
  \href{http://xxx.lanl.gov/abs/1504.0593}{{\tt arXiv:1504.0593}}.

\bibitem{Chen:2015cqa}
C.-H. Chen, C.-W. Chiang, and T.~Nomura, {\it {Dark matter for excess of AMS-02
  positrons and antiprotons}},  \href{http://xxx.lanl.gov/abs/1504.0784}{{\tt
  arXiv:1504.0784}}.

\bibitem{Abdo:2010nc}
A.~Abdo, M.~Ackermann, M.~Ajello, W.~Atwood, L.~Baldini, et~al., {\it {Fermi
  LAT Search for Photon Lines from 30 to 200 GeV and Dark Matter
  Implications}},  {\em Phys.Rev.Lett.} {\bf 104} (2010) 091302,
  [\href{http://xxx.lanl.gov/abs/1001.4836}{{\tt arXiv:1001.4836}}].

\bibitem{Ackermann:2012qk}
{\bf Fermi-LAT} Collaboration, M.~Ackermann et~al., {\it {Fermi LAT Search for
  Dark Matter in Gamma-ray Lines and the Inclusive Photon Spectrum}},  {\em
  Phys.Rev.} {\bf D86} (2012) 022002,
  [\href{http://xxx.lanl.gov/abs/1205.2739}{{\tt arXiv:1205.2739}}].

\bibitem{Calore:2014nla}
F.~Calore, I.~Cholis, C.~McCabe, and C.~Weniger, {\it {A Tale of Tails: Dark
  Matter Interpretations of the Fermi GeV Excess in Light of Background Model
  Systematics}},  {\em Phys.Rev.} {\bf D91} (2015), no.~6 063003,
  [\href{http://xxx.lanl.gov/abs/1411.4647}{{\tt arXiv:1411.4647}}].

\bibitem{Abramowski:2013ax}
{\bf HESS Collaboration} Collaboration, A.~Abramowski et~al., {\it {Search for
  Photon-Linelike Signatures from Dark Matter Annihilations with H.E.S.S.}},
  {\em Phys.Rev.Lett.} {\bf 110} (2013) 041301,
  [\href{http://xxx.lanl.gov/abs/1301.1173}{{\tt arXiv:1301.1173}}].

\bibitem{Boddy:2015efa}
K.~K. Boddy and J.~Kumar, {\it {Indirect Detection of Dark Matter Using
  MeV-Range Gamma-Ray Telescopes}},
  \href{http://xxx.lanl.gov/abs/1504.0402}{{\tt arXiv:1504.0402}}.

\bibitem{Ghorbani:2014gka}
K.~Ghorbani and H.~Ghorbani, {\it {Scalar Split WIMPs and Galactic Gamma-Ray
  Excess}},  \href{http://xxx.lanl.gov/abs/1501.0020}{{\tt arXiv:1501.0020}}.

\bibitem{Ghorbani:2014qpa}
K.~Ghorbani, {\it {Fermionic dark matter with pseudo-scalar Yukawa
  interaction}},  {\em JCAP} {\bf 1501} (2015) 015,
  [\href{http://xxx.lanl.gov/abs/1408.4929}{{\tt arXiv:1408.4929}}].

\bibitem{Cumani:2015ava}
P.~Cumani, A.~Galper, V.~Bonvicini, N.~Topchiev, O.~Adriani, et~al., {\it {The
  GAMMA-400 Space Mission}},  \href{http://xxx.lanl.gov/abs/1502.0297}{{\tt
  arXiv:1502.0297}}.

\bibitem{Bhattacherjee:2014dya}
B.~Bhattacherjee, M.~Ibe, K.~Ichikawa, S.~Matsumoto, and K.~Nishiyama, {\it
  {Wino Dark Matter and Future dSph Observations}},  {\em JHEP} {\bf 1407}
  (2014) 080, [\href{http://xxx.lanl.gov/abs/1405.4914}{{\tt
  arXiv:1405.4914}}].

\bibitem{Campbell:2013rua}
S.~S. Campbell and J.~F. Beacom, {\it {Combined Flux and Anisotropy Searches
  Improve Sensitivity to Gamma Rays from Dark Matter}},
  \href{http://xxx.lanl.gov/abs/1312.3945}{{\tt arXiv:1312.3945}}.

\bibitem{Bergstrom:2012bd}
L.~Bergstrom, {\it {The 130 GeV Fingerprint of Right-Handed Neutrino Dark
  Matter}},  {\em Phys.Rev.} {\bf D86} (2012) 103514,
  [\href{http://xxx.lanl.gov/abs/1208.6082}{{\tt arXiv:1208.6082}}].

\bibitem{Bringmann:2012ez}
T.~Bringmann and C.~Weniger, {\it {Gamma Ray Signals from Dark Matter:
  Concepts, Status and Prospects}},  {\em Phys.Dark Univ.} {\bf 1} (2012)
  194--217, [\href{http://xxx.lanl.gov/abs/1208.5481}{{\tt arXiv:1208.5481}}].

\bibitem{Regis:2015zka}
M.~Regis, J.-Q. Xia, A.~Cuoco, E.~Branchini, N.~Fornengo, et~al., {\it
  {Particle dark matter searches outside the Local neighborhood}},
  \href{http://xxx.lanl.gov/abs/1503.0592}{{\tt arXiv:1503.0592}}.

\bibitem{Khlopov:2014nva}
M.~Y. Khlopov, {\it {Introduction to the special issue of Modern Physics
  Letters A "Indirect dark matter searches"}},  {\em Mod.Phys.Lett.} {\bf A29}
  (2014) 1402001, [\href{http://xxx.lanl.gov/abs/1411.2150}{{\tt
  arXiv:1411.2150}}].

\bibitem{Barstow:2014dda}
M.~A. Barstow, S.~Casewell, S.~Catalan, C.~Copperwheat, B.~Gaensicke, et~al.,
  {\it {White paper: Gaia and the end states of stellar evolution}},
  \href{http://xxx.lanl.gov/abs/1407.6163}{{\tt arXiv:1407.6163}}.

\bibitem{Feldmann:2013hqa}
R.~Feldmann and D.~Spolyar, {\it {Detecting Dark Matter Substructures around
  the Milky Way with Gaia}},  {\em Mon.Not.Roy.Astron.Soc.} {\bf 446} (2015)
  1000--1012, [\href{http://xxx.lanl.gov/abs/1310.2243}{{\tt
  arXiv:1310.2243}}].

\bibitem{Perez-Garcia:2014dra}
M.~Angeles Perez-Garcia and J.~Silk, {\it {Constraining decaying dark matter
  with neutron stars}},  {\em Phys.Lett.} {\bf B744} (2015) 13--17,
  [\href{http://xxx.lanl.gov/abs/1403.6111}{{\tt arXiv:1403.6111}}].

\bibitem{Fuller:2014rza}
J.~Fuller and C.~Ott, {\it {Dark Matter-induced Collapse of Neutron Stars: A
  Possible Link Between Fast Radio Bursts and the Missing Pulsar Problem}},
  {\em Mon.Not.Roy.Astron.Soc.} {\bf L71} (2015) L75,
  [\href{http://xxx.lanl.gov/abs/1412.6119}{{\tt arXiv:1412.6119}}].

\bibitem{Graham:2015yga}
P.~W. Graham, S.~Rajendran, K.~Van~Tilburg, and T.~D. Wiser, {\it {Towards a
  Bullet-proof test for indirect signals of dark matter}},
  \href{http://xxx.lanl.gov/abs/1502.0382}{{\tt arXiv:1502.0382}}.

\bibitem{Carlson:2015daa}
E.~Carlson and S.~Profumo, {\it {When Dark Matter interacts with Cosmic Rays or
  Interstellar Matter: A Morphological Study}},
  \href{http://xxx.lanl.gov/abs/1504.0478}{{\tt arXiv:1504.0478}}.

\bibitem{Dienes:2015qqa}
K.~R. Dienes, J.~Kumar, B.~Thomas, and D.~Yaylali, {\it {Dark-Matter Decay as a
  Complementary Probe of Multicomponent Dark Sectors}},  {\em Phys.Rev.Lett.}
  {\bf 114} (2015), no.~5 051301,
  [\href{http://xxx.lanl.gov/abs/1406.4868}{{\tt arXiv:1406.4868}}].

\bibitem{Madhavacheril:2014slf}
{\bf ACT} Collaboration, M.~Madhavacheril et~al., {\it {Evidence of Lensing of
  the Cosmic Microwave Background by Dark Matter Halos}},  {\em Phys.Rev.Lett.}
  {\bf 114} (2015), no.~15 151302,
  [\href{http://xxx.lanl.gov/abs/1411.7999}{{\tt arXiv:1411.7999}}].

\bibitem{Cushman:2013zza}
P.~Cushman, C.~Galbiati, D.~McKinsey, H.~Robertson, T.~Tait, et~al., {\it
  {Working Group Report: WIMP Dark Matter Direct Detection}},
  \href{http://xxx.lanl.gov/abs/1310.8327}{{\tt arXiv:1310.8327}}.

\bibitem{Goodman:1985dc}
M.~W. Goodman and E.~Witten, {\it Detectability of certain dark-matter
  candidates},  {\em Phys. Rev.} {\bf D31} (1985) 3059.

\bibitem{Jungman:1996df}
G.~Jungman, M.~Kamionkowski, and K.~Griest, {\it Supersymmetric dark matter},
  {\em Phys. Rept.} {\bf 267} (1996) 195--373,
  [\href{http://xxx.lanl.gov/abs/hep-ph/9506380}{{\tt hep-ph/9506380}}].

\bibitem{Lewin:1996rx}
J.~D. Lewin and P.~F. Smith, {\it Review of mathematics, numerical factors, and
  corrections for dark matter experiments based on elastic nuclear recoil},
  {\em Astropart. Phys.} {\bf 6} (1996) 87--112.

\bibitem{Bednyakov:2002mb}
V.~A. Bednyakov, {\it On possible lower bounds for the direct detection rate of
  susy dark matter},  {\em Phys. Atom. Nucl.} {\bf 66} (2003) 490--493,
  [\href{http://xxx.lanl.gov/abs/hep-ph/0201046}{{\tt hep-ph/0201046}}].

\bibitem{Bednyakov:1998is}
V.~A. Bednyakov and H.~V. Klapdor-Kleingrothaus, {\it About direct dark matter
  detection in next-to-minimal supersymmetric standard model},  {\em Phys.
  Rev.} {\bf D59} (1999) 023514,
  [\href{http://xxx.lanl.gov/abs/hep-ph/9802344}{{\tt hep-ph/9802344}}].

\bibitem{Bednyakov:1997jr}
V.~A. Bednyakov, S.~G. Kovalenko, H.~V. Klapdor-Kleingrothaus, and
  Y.~Ramachers, {\it Is susy accessible by direct dark matter detection?},
  {\em Z. Phys.} {\bf A357} (1997) 339--347,
  [\href{http://xxx.lanl.gov/abs/hep-ph/9606261}{{\tt hep-ph/9606261}}].

\bibitem{Schumann:2014uva}
M.~Schumann, {\it {Dual-Phase Liquid Xenon Detectors for Dark Matter
  Searches}},  {\em JINST} {\bf 9} (2014) C08004,
  [\href{http://xxx.lanl.gov/abs/1405.7600}{{\tt arXiv:1405.7600}}].

\bibitem{Schumann:2015wfa}
M.~Schumann, {\it {Dark Matter 2014}},  {\em Unpublished,~} (2015)
  [\href{http://xxx.lanl.gov/abs/1501.0120}{{\tt arXiv:1501.0120}}].

\bibitem{Aprile:2012nq}
{\bf XENON100 Collaboration} Collaboration, E.~Aprile et~al., {\it {Dark Matter
  Results from 225 Live Days of XENON100 Data}},  {\em Phys.Rev.Lett.} {\bf
  109} (2012) 181301, [\href{http://xxx.lanl.gov/abs/1207.5988}{{\tt
  arXiv:1207.5988}}].

\bibitem{Aprile:2013doa}
{\bf XENON100 Collaboration} Collaboration, E.~Aprile et~al., {\it {Limits on
  spin-dependent WIMP-nucleon cross sections from 225 live days of XENON100
  data}},  {\em Phys.Rev.Lett.} {\bf 111} (2013), no.~2 021301,
  [\href{http://xxx.lanl.gov/abs/1301.6620}{{\tt arXiv:1301.6620}}].

\bibitem{Orrigo:2015cha}
{\bf XENON} Collaboration, S.~Orrigo, {\it {Direct Dark Matter Search with
  XENON100}},  \href{http://xxx.lanl.gov/abs/1501.0349}{{\tt arXiv:1501.0349}}.

\bibitem{Akerib:2013tjd}
{\bf LUX Collaboration} Collaboration, D.~Akerib et~al., {\it {First results
  from the LUX dark matter experiment at the Sanford Underground Research
  Facility}},  {\em Phys.Rev.Lett.} {\bf 112} (2014) 091303,
  [\href{http://xxx.lanl.gov/abs/1310.8214}{{\tt arXiv:1310.8214}}].

\bibitem{Savage:2015xta}
C.~Savage, A.~Scaffidi, M.~White, and A.~G. Williams, {\it {LUX likelihood and
  limits on spin-independent and spin-dependent WIMP couplings with LUXCalc}},
  \href{http://xxx.lanl.gov/abs/1502.0266}{{\tt arXiv:1502.0266}}.

\bibitem{Cao:2014jsa}
{\bf PandaX} Collaboration, X.~Cao et~al., {\it {PandaX: A Liquid Xenon Dark
  Matter Experiment at CJPL}},  {\em Sci.China Phys.Mech.Astron.} {\bf 57}
  (2014) 1476--1494, [\href{http://xxx.lanl.gov/abs/1405.2882}{{\tt
  arXiv:1405.2882}}].

\bibitem{Xiao:2014xyn}
{\bf PandaX} Collaboration, M.~Xiao et~al., {\it {First dark matter search
  results from the PandaX-I experiment}},  {\em Sci.China Phys.Mech.Astron.}
  {\bf 57} (2014) 2024--2030, [\href{http://xxx.lanl.gov/abs/1408.5114}{{\tt
  arXiv:1408.5114}}].

\bibitem{Aprile:2012zx}
{\bf XENON1T} Collaboration, E.~Aprile, {\it {The XENON1T Dark Matter Search
  Experiment}},  {\em Springer Proc.Phys.} {\bf C12-02-22} (2013) 93--96,
  [\href{http://xxx.lanl.gov/abs/1206.6288}{{\tt arXiv:1206.6288}}].

\bibitem{Aprile:2015lha}
{\bf XENON} Collaboration, E.~Aprile et~al., {\it {Lowering the radioactivity
  of the photomultiplier tubes for the XENON1T dark matter experiment}},
  \href{http://xxx.lanl.gov/abs/1503.0769}{{\tt arXiv:1503.0769}}.

\bibitem{Angloher:2011uu}
G.~Angloher, M.~Bauer, I.~Bavykina, A.~Bento, C.~Bucci, et~al., {\it {Results
  from 730 kg days of the CRESST-II Dark Matter Search}},  {\em Eur.Phys.J.}
  {\bf C72} (2012) 1971, [\href{http://xxx.lanl.gov/abs/1109.0702}{{\tt
  arXiv:1109.0702}}].

\bibitem{Angloher:2015eza}
{\bf CRESST} Collaboration, G.~Angloher et~al., {\it {Probing low WIMP masses
  with the next generation of CRESST detector}},
  \href{http://xxx.lanl.gov/abs/1503.0806}{{\tt arXiv:1503.0806}}.

\bibitem{Agnese:2013rvf}
{\bf CDMS Collaboration} Collaboration, R.~Agnese et~al., {\it {Silicon
  Detector Dark Matter Results from the Final Exposure of CDMS II}},  {\em
  Phys.Rev.Lett.} {\bf 111} (2013), no.~25 251301,
  [\href{http://xxx.lanl.gov/abs/1304.4279}{{\tt arXiv:1304.4279}}].

\bibitem{Agnese:2014aze}
{\bf SuperCDMS Collaboration} Collaboration, R.~Agnese et~al., {\it {Search for
  Low-Mass Weakly Interacting Massive Particles with SuperCDMS}},  {\em
  Phys.Rev.Lett.} {\bf 112} (2014), no.~24 241302,
  [\href{http://xxx.lanl.gov/abs/1402.7137}{{\tt arXiv:1402.7137}}].

\bibitem{Agnese:2015sga}
R.~Agnese, A.~Anderson, M.~Asai, D.~Balakishiyeva, D.~Barker, et~al., {\it
  {Improved WIMP-search reach of the CDMS II germanium data}},
  \href{http://xxx.lanl.gov/abs/1504.0587}{{\tt arXiv:1504.0587}}.

\bibitem{Aalseth:2014jpa}
C.~Aalseth, P.~Barbeau, J.~Colaresi, J.~D. Leon, J.~Fast, et~al., {\it {Maximum
  Likelihood Signal Extraction Method Applied to 3.4 years of CoGeNT Data}},
  \href{http://xxx.lanl.gov/abs/1401.6234}{{\tt arXiv:1401.6234}}.

\bibitem{Bernabei:2008yi}
{\bf DAMA Collaboration} Collaboration, R.~Bernabei et~al., {\it {First results
  from DAMA/LIBRA and the combined results with DAMA/NaI}},  {\em Eur.Phys.J.}
  {\bf C56} (2008) 333--355, [\href{http://xxx.lanl.gov/abs/0804.2741}{{\tt
  arXiv:0804.2741}}].

\bibitem{D'Angelo:2015lia}
{\bf DarkSide} Collaboration, D.~D'Angelo, {\it {DarkSide50 results from first
  argon run}},  \href{http://xxx.lanl.gov/abs/1501.0354}{{\tt
  arXiv:1501.0354}}.

\bibitem{Archambault:2012pm}
{\bf PICASSO Collaboration} Collaboration, S.~Archambault et~al., {\it
  {Constraints on Low-Mass WIMP Interactions on $^{19}F$ from PICASSO}},  {\em
  Phys.Lett.} {\bf B711} (2012) 153--161,
  [\href{http://xxx.lanl.gov/abs/1202.1240}{{\tt arXiv:1202.1240}}].

\bibitem{Desai:2004pq}
{\bf Super-Kamiokande Collaboration} Collaboration, S.~Desai et~al., {\it
  {Search for dark matter WIMPs using upward through-going muons in
  Super-Kamiokande}},  {\em Phys.Rev.} {\bf D70} (2004) 083523,
  [\href{http://xxx.lanl.gov/abs/hep-ex/0404025}{{\tt hep-ex/0404025}}].

\bibitem{Abbasi:2009uz}
{\bf ICECUBE Collaboration} Collaboration, R.~Abbasi et~al., {\it {Limits on a
  muon flux from neutralino annihilations in the Sun with the IceCube 22-string
  detector}},  {\em Phys.Rev.Lett.} {\bf 102} (2009) 201302,
  [\href{http://xxx.lanl.gov/abs/0902.2460}{{\tt arXiv:0902.2460}}].

\bibitem{Behnke:2010xt}
E.~Behnke, J.~Behnke, S.~Brice, D.~Broemmelsiek, J.~Collar, et~al., {\it
  {Improved Limits on Spin-Dependent WIMP-Proton Interactions from a Two Liter
  CF$_3$I Bubble Chamber}},  {\em Phys.Rev.Lett.} {\bf 106} (2011) 021303,
  [\href{http://xxx.lanl.gov/abs/1008.3518}{{\tt arXiv:1008.3518}}].

\bibitem{Felizardo:2011uw}
M.~Felizardo, T.~Girard, T.~Morlat, A.~Fernandes, A.~Ramos, et~al., {\it {Final
  Analysis and Results of the Phase II SIMPLE Dark Matter Search}},  {\em
  Phys.Rev.Lett.} {\bf 108} (2012) 201302,
  [\href{http://xxx.lanl.gov/abs/1106.3014}{{\tt arXiv:1106.3014}}].

\bibitem{Bi:2014hpa}
X.-J. Bi, P.-F. Yin, and Q.~Yuan, {\it {Status of Dark Matter Detection}},
  {\em Front.Phys.China} {\bf 8} (2013) 794--827,
  [\href{http://xxx.lanl.gov/abs/1409.4590}{{\tt arXiv:1409.4590}}].

\bibitem{Engel:1992bf}
J.~Engel, S.~Pittel, and P.~Vogel, {\it {Nuclear physics of dark matter
  detection}},  {\em Int.J.Mod.Phys.} {\bf E1} (1992) 1--37.

\bibitem{Ressell:1993qm}
M.~T. Ressell, M.~B. Aufderheide, S.~D. Bloom, K.~Griest, G.~J. Mathews,
  et~al., {\it {Nuclear shell model calculations of neutralino - nucleus
  cross-sections for Si-29 and Ge-73}},  {\em Phys.Rev.} {\bf D48} (1993)
  5519--5535.

\bibitem{Bernabei:2003za}
R.~Bernabei et~al., {\it Dark matter search},  {\em Riv. Nuovo Cim.} {\bf 26}
  (2003) 1--73, [\href{http://xxx.lanl.gov/abs/astro-ph/0307403}{{\tt
  astro-ph/0307403}}].

\bibitem{Ressell:1997kx}
M.~Ressell and D.~Dean, {\it {Spin dependent neutralino - nucleus scattering
  for A approximately 127 nuclei}},  {\em Phys.Rev.} {\bf C56} (1997) 535--546,
  [\href{http://xxx.lanl.gov/abs/hep-ph/9702290}{{\tt hep-ph/9702290}}].

\bibitem{Bednyakov:2004xq}
V.~Bednyakov and F.~Simkovic, {\it {Nuclear spin structure in dark matter
  search: The Zero momentum transfer limit}},  {\em Phys.Part.Nucl.} {\bf 36}
  (2005) 131--152, [\href{http://xxx.lanl.gov/abs/hep-ph/0406218}{{\tt
  hep-ph/0406218}}].

\bibitem{Bednyakov:2006ux}
V.~Bednyakov and F.~Simkovic, {\it {Nuclear spin structure in dark matter
  search: The Finite momentum transfer limit}},  {\em Phys.Part.Nucl.} {\bf 37}
  (2006) S106--S128, [\href{http://xxx.lanl.gov/abs/hep-ph/0608097}{{\tt
  hep-ph/0608097}}].

\bibitem{Bednyakov:2004be}
V.~A. Bednyakov and H.~V. Klapdor-Kleingrothaus, {\it On dark matter search
  after dama with ge-73},  {\em Phys. Rev.} {\bf D70} (2004) 096006,
  [\href{http://xxx.lanl.gov/abs/hep-ph/0404102}{{\tt hep-ph/0404102}}].

\bibitem{Freese:1987wu}
K.~Freese, J.~A. Frieman, and A.~Gould, {\it Signal modulation in cold dark
  matter detection},  {\em Phys. Rev.} {\bf D37} (1988) 3388.

\bibitem{Schneck:2015eqa}
{\bf SuperCDMS} Collaboration, K.~Schneck et~al., {\it {Dark matter effective
  field theory scattering in direct detection experiments}},
  \href{http://xxx.lanl.gov/abs/1503.0337}{{\tt arXiv:1503.0337}}.

\bibitem{Spooner:2007zh}
N.~Spooner, {\it {Direct Dark Matter Searches}},  {\em J.Phys.Soc.Jap.} {\bf
  76} (2007) 111016, [\href{http://xxx.lanl.gov/abs/0705.3345}{{\tt
  arXiv:0705.3345}}].

\bibitem{Catena:2015uua}
R.~Catena and P.~Gondolo, {\it {Global limits and interference patterns in dark
  matter direct detection}},  \href{http://xxx.lanl.gov/abs/1504.0655}{{\tt
  arXiv:1504.0655}}.

\bibitem{Bednyakov:2008gv}
V.~Bednyakov and H.~Klapdor-Kleingrothaus, {\it {Direct Search for Dark Matter
  --- Striking the Balance --- and the Future}},  {\em Phys.Part.Nucl.} {\bf
  40} (2009) 583--611, [\href{http://xxx.lanl.gov/abs/0806.3917}{{\tt
  arXiv:0806.3917}}].

\bibitem{Freese:2012xd}
K.~Freese, M.~Lisanti, and C.~Savage, {\it {Colloquium: Annual modulation of
  dark matter}},  {\em Rev.Mod.Phys.} {\bf 85} (2013) 1561--1581,
  [\href{http://xxx.lanl.gov/abs/1209.3339}{{\tt arXiv:1209.3339}}].

\bibitem{Bednyakov:1999yr}
V.~Bednyakov and H.~Klapdor-Kleingrothaus, {\it {Possibilities of directly
  detecting dark-matter particles in the next-to-minimal supersymmetric
  standard model}},  {\em Phys.Atom.Nucl.} {\bf 62} (1999) 966--974.

\bibitem{DelNobile:2015tza}
E.~Del~Nobile, G.~B. Gelmini, and S.~J. Witte, {\it {Target dependence of the
  annual modulation in direct dark matter searches}},
  \href{http://xxx.lanl.gov/abs/1504.0677}{{\tt arXiv:1504.0677}}.

\bibitem{Belli:2011kw}
P.~Belli, R.~Bernabei, A.~Bottino, F.~Cappella, R.~Cerulli, et~al., {\it
  {Observations of annual modulation in direct detection of relic particles and
  light neutralinos}},  {\em Phys.Rev.} {\bf D84} (2011) 055014,
  [\href{http://xxx.lanl.gov/abs/1106.4667}{{\tt arXiv:1106.4667}}].

\bibitem{Vergados:2003pk}
J.~D. Vergados, {\it Theoretical directional and modulated rates for direct
  susy dark matter detection},  {\em Phys. Rev.} {\bf D67} (2003) 103003,
  [\href{http://xxx.lanl.gov/abs/hep-ph/0303231}{{\tt hep-ph/0303231}}].

\bibitem{Morgan:2004ys}
B.~Morgan, A.~M. Green, and N.~J.~C. Spooner, {\it Directional statistics for
  wimp direct detection},  {\em Phys. Rev.} {\bf D71} (2005) 103507,
  [\href{http://xxx.lanl.gov/abs/astro-ph/0408047}{{\tt astro-ph/0408047}}].

\bibitem{Mohlabeng:2015efa}
G.~Mohlabeng, K.~Kong, J.~Li, A.~Para, and J.~Yoo, {\it {Dark Matter
  Directionality Revisited with a High Pressure Xenon Gas Detector}},
  \href{http://xxx.lanl.gov/abs/1503.0393}{{\tt arXiv:1503.0393}}.

\bibitem{Bednyakov:2008zz}
V.~A. Bednyakov, H.~V. Klapdor-Kleingrothaus, and I.~V. Krivosheina, {\it {New
  constraints on spin-dependent WIMP-neutron interactions from HDMS with
  natural Ge and Ge-73}},  {\em Phys. Atom. Nucl.} {\bf 71} (2008) 111--116.

\bibitem{Vergados:2004hw}
J.~D. Vergados, {\it Direct susy dark matter detection: Theoretical rates due
  to the spin},  {\em J. Phys.} {\bf G30} (2004) 1127--1144,
  [\href{http://xxx.lanl.gov/abs/hep-ph/0406134}{{\tt hep-ph/0406134}}].

\bibitem{Bednyakov:2012cu}
V.~Bednyakov, {\it {One needs positive signatures for detection of Dark
  Matter}},  {\em Phys.Part.Nucl.} {\bf 44} (2013) 220--228,
  [\href{http://xxx.lanl.gov/abs/1207.2899}{{\tt arXiv:1207.2899}}].

\bibitem{Gutlein:2014gma}
A.~GŸtlein, G.~Angloher, A.~Bento, C.~Bucci, L.~Canonica, et~al., {\it {Impact
  of Coherent Neutrino Nucleus Scattering on Direct Dark Matter Searches based
  on CaWO$_4$ Crystals}},  \href{http://xxx.lanl.gov/abs/1408.2357}{{\tt
  arXiv:1408.2357}}.

\bibitem{Bernabei:2014maa}
R.~Bernabei, {\it {Dark Matter Particles in the Galactic Halo}},  {\em Physics}
  {\bf 15} (2014) 10, [\href{http://xxx.lanl.gov/abs/1412.6524}{{\tt
  arXiv:1412.6524}}].

\bibitem{Bernabei:2013xsa}
R.~Bernabei, P.~Belli, F.~Cappella, V.~Caracciolo, S.~Castellano, et~al., {\it
  {Final model independent result of DAMA/LIBRA-phase1}},  {\em Eur.Phys.J.}
  {\bf C73} (2013), no.~12 2648, [\href{http://xxx.lanl.gov/abs/1308.5109}{{\tt
  arXiv:1308.5109}}].

\bibitem{Aalseth:2011wp}
C.~Aalseth, P.~Barbeau, J.~Colaresi, J.~Collar, J.~Diaz~Leon, et~al., {\it
  {Search for an Annual Modulation in a P-type Point Contact Germanium Dark
  Matter Detector}},  {\em Phys.Rev.Lett.} {\bf 107} (2011) 141301,
  [\href{http://xxx.lanl.gov/abs/1106.0650}{{\tt arXiv:1106.0650}}].

\bibitem{Aalseth:2012if}
{\bf CoGeNT Collaboration} Collaboration, C.~Aalseth et~al., {\it {CoGeNT: A
  Search for Low-Mass Dark Matter using p-type Point Contact Germanium
  Detectors}},  {\em Phys.Rev.} {\bf D88} (2013), no.~1 012002,
  [\href{http://xxx.lanl.gov/abs/1208.5737}{{\tt arXiv:1208.5737}}].

\bibitem{Berlin:2015aba}
A.~Berlin, T.~Lin, M.~Low, and L.-T. Wang, {\it {Neutralinos in Vector Boson
  Fusion at High Energy Colliders}},
  \href{http://xxx.lanl.gov/abs/1502.0504}{{\tt arXiv:1502.0504}}.

\bibitem{Bednyakov:1999vh}
V.~Bednyakov and H.~Klapdor-Kleingrothaus, {\it {SUSY spectrum constraints on
  direct dark matter detection}},  {\em Phys.Rev.} {\bf D62} (2000) 043524,
  [\href{http://xxx.lanl.gov/abs/hep-ph/9908427}{{\tt hep-ph/9908427}}].

\bibitem{Bednyakov:2000uw}
V.~Bednyakov, H.~Klapdor-Kleingrothaus, and H.~Tu, {\it {Higgs bosons and the
  indirect search for WIMPs}},  {\em Phys.Rev.} {\bf D64} (2001) 075004,
  [\href{http://xxx.lanl.gov/abs/hep-ph/0101223}{{\tt hep-ph/0101223}}].

\bibitem{Bednyakov:2009zz}
V.~Bednyakov, Y.~Budagov, A.~Gladyshev, D.~Kazakov, E.~Khramov, et~al., {\it
  {On the LHC observation of gluinos from the Egret-preferred region}},  {\em
  Phys.Atom.Nucl.} {\bf 72} (2009) 619--637.

\bibitem{Eifert:2008zza}
{\bf ATLAS Collaboration} Collaboration, T.~Eifert, {\it {Searches for
  supersymmetry at the LHC and its dark matter candidate}}, .

\bibitem{Polesello:2004qy}
G.~Polesello and D.~Tovey, {\it {Constraining SUSY dark matter with the ATLAS
  detector at the LHC}},  {\em JHEP} {\bf 0405} (2004) 071,
  [\href{http://xxx.lanl.gov/abs/hep-ph/0403047}{{\tt hep-ph/0403047}}].

\bibitem{Barberio:2009xca}
{\bf ATLAS} Collaboration, E.~Barberio, {\it {SUSY searches at LHC and Dark
  Matter}}, .

\bibitem{Ismail:2014fca}
A.~Ismail, {\it {Dark matter complementarity in the phenomenological MSSM}},
  {\em AIP Conf.Proc.} {\bf 1604} (2014) 53--65.

\bibitem{Roszkowski:2014iqa}
L.~Roszkowski, E.~M. Sessolo, and A.~J. Williams, {\it {Prospects for dark
  matter searches in the pMSSM}},  {\em JHEP} {\bf 1502} (2015) 014,
  [\href{http://xxx.lanl.gov/abs/1411.5214}{{\tt arXiv:1411.5214}}].

\bibitem{Catalan:2015cna}
M.~E.~C. Catalan, S.~Ando, C.~Weniger, and F.~Zandanel, {\it {Indirect and
  direct detection prospect for TeV dark matter in the MSSM-9}},
  \href{http://xxx.lanl.gov/abs/1503.0059}{{\tt arXiv:1503.0059}}.

\bibitem{Aparicio:2015sda}
L.~Aparicio, M.~Cicoli, B.~Dutta, S.~Krippendorf, A.~Maharana, et~al., {\it
  {Non-thermal CMSSM with a 125 GeV Higgs}},
  \href{http://xxx.lanl.gov/abs/1502.0567}{{\tt arXiv:1502.0567}}.

\bibitem{Kowalska:2015zja}
K.~Kowalska, L.~Roszkowski, E.~M. Sessolo, and A.~J. Williams, {\it
  {GUT-inspired SUSY and the muon g-2 anomaly: prospects for LHC 14 TeV}},
  \href{http://xxx.lanl.gov/abs/1503.0821}{{\tt arXiv:1503.0821}}.

\bibitem{Caron:2015wda}
A.~Achterberg, S.~Caron, L.~Hendriks, R.~Ruiz~de Austri, and C.~Weniger, {\it
  {A description of the Galactic Center excess in the Minimal Supersymmetric
  Standard Model and the Dark Matter signatures for the LHC and direct and
  indirect detection experiments}},
  \href{http://xxx.lanl.gov/abs/1502.0570}{{\tt arXiv:1502.0570}}.

\bibitem{Han:2015zba}
C.~Han, D.~Kim, S.~Munir, and M.~Park, {\it {$\mathcal{O}$(1) GeV dark matter
  in SUSY and a very light pseudoscalar at the LHC}},
  \href{http://xxx.lanl.gov/abs/1504.0508}{{\tt arXiv:1504.0508}}.

\bibitem{Gherghetta:2015ysa}
T.~Gherghetta, B.~von Harling, A.~D. Medina, M.~A. Schmidt, and T.~Trott, {\it
  {SUSY Implications from WIMP Annihilation into Scalars at the Galactic
  Centre}},  \href{http://xxx.lanl.gov/abs/1502.0717}{{\tt arXiv:1502.0717}}.

\bibitem{Arina:2015uea}
C.~Arina, M.~E.~C. Catalan, S.~Kraml, S.~Kulkarni, and U.~Laa, {\it
  {Constraints on sneutrino dark matter from LHC Run 1}},
  \href{http://xxx.lanl.gov/abs/1503.0296}{{\tt arXiv:1503.0296}}.

\bibitem{An:2011uq}
H.~An, P.~B. Dev, Y.~Cai, and R.~Mohapatra, {\it {Sneutrino Dark Matter in
  Gauged Inverse Seesaw Models for Neutrinos}},  {\em Phys.Rev.Lett.} {\bf 108}
  (2012) 081806, [\href{http://xxx.lanl.gov/abs/1110.1366}{{\tt
  arXiv:1110.1366}}].

\bibitem{BhupalDev:2012ru}
P.~Bhupal~Dev, S.~Mondal, B.~Mukhopadhyaya, and S.~Roy, {\it {Phenomenology of
  Light Sneutrino Dark Matter in cMSSM/mSUGRA with Inverse Seesaw}},  {\em
  JHEP} {\bf 1209} (2012) 110, [\href{http://xxx.lanl.gov/abs/1207.6542}{{\tt
  arXiv:1207.6542}}].

\bibitem{Banerjee:2013fga}
S.~Banerjee, P.~S.~B. Dev, S.~Mondal, B.~Mukhopadhyaya, and S.~Roy, {\it
  {Invisible Higgs Decay in a Supersymmetric Inverse Seesaw Model with Light
  Sneutrino Dark Matter}},  {\em JHEP} {\bf 1310} (2013) 221,
  [\href{http://xxx.lanl.gov/abs/1306.2143}{{\tt arXiv:1306.2143}}].

\bibitem{Kakizaki:2015nua}
M.~Kakizaki, E.-K. Park, J.-h. Park, and A.~Santa, {\it {Phenomenological
  constraints on light mixed sneutrino dark matter scenarios}},
  \href{http://xxx.lanl.gov/abs/1503.0678}{{\tt arXiv:1503.0678}}.

\bibitem{Barducci:2015ffa}
D.~Barducci, A.~Belyaev, A.~K.~M. Bharucha, W.~Porod, and V.~Sanz, {\it
  {Uncovering Natural Supersymmetry via the interplay between the LHC and
  Direct Dark Matter Detection}},
  \href{http://xxx.lanl.gov/abs/1504.0247}{{\tt arXiv:1504.0247}}.

\bibitem{Akula:2011dd}
S.~Akula, D.~Feldman, Z.~Liu, P.~Nath, and G.~Peim, {\it {New Constraints on
  Dark Matter from CMS and ATLAS Data}},  {\em Mod.Phys.Lett.} {\bf A26} (2011)
  1521--1535, [\href{http://xxx.lanl.gov/abs/1103.5061}{{\tt
  arXiv:1103.5061}}].

\bibitem{Nath:2015dza}
P.~Nath, {\it {Supersymmetry after the Higgs}},
  \href{http://xxx.lanl.gov/abs/1501.0167}{{\tt arXiv:1501.0167}}.

\bibitem{Goodman:2010yf}
J.~Goodman, M.~Ibe, A.~Rajaraman, W.~Shepherd, T.~M. Tait, et~al., {\it
  {Constraints on Light Majorana dark Matter from Colliders}},  {\em
  Phys.Lett.} {\bf B695} (2011) 185--188,
  [\href{http://xxx.lanl.gov/abs/1005.1286}{{\tt arXiv:1005.1286}}].

\bibitem{Bai:2010hh}
Y.~Bai, P.~J. Fox, and R.~Harnik, {\it {The Tevatron at the Frontier of Dark
  Matter Direct Detection}},  {\em JHEP} {\bf 1012} (2010) 048,
  [\href{http://xxx.lanl.gov/abs/1005.3797}{{\tt arXiv:1005.3797}}].

\bibitem{Goodman:2010ku}
J.~Goodman, M.~Ibe, A.~Rajaraman, W.~Shepherd, T.~M. Tait, et~al., {\it
  {Constraints on Dark Matter from Colliders}},  {\em Phys.Rev.} {\bf D82}
  (2010) 116010, [\href{http://xxx.lanl.gov/abs/1008.1783}{{\tt
  arXiv:1008.1783}}].

\bibitem{Fox:2011pm}
P.~J. Fox, R.~Harnik, J.~Kopp, and Y.~Tsai, {\it {Missing Energy Signatures of
  Dark Matter at the LHC}},  {\em Phys.Rev.} {\bf D85} (2012) 056011,
  [\href{http://xxx.lanl.gov/abs/1109.4398}{{\tt arXiv:1109.4398}}].

\bibitem{Busoni:2014uca}
G.~Busoni, {\it {Limitation of EFT for DM interactions at the LHC}},  {\em PoS}
  {\bf DIS2014} (2014) 134, [\href{http://xxx.lanl.gov/abs/1411.3600}{{\tt
  arXiv:1411.3600}}].

\bibitem{Busoni:2013lha}
G.~Busoni, A.~De~Simone, E.~Morgante, and A.~Riotto, {\it {On the Validity of
  the Effective Field Theory for Dark Matter Searches at the LHC}},  {\em
  Phys.Lett.} {\bf B728} (2014) 412--421,
  [\href{http://xxx.lanl.gov/abs/1307.2253}{{\tt arXiv:1307.2253}}].

\bibitem{Lowette:2014yta}
{\bf CMS Collaboration} Collaboration, S.~Lowette, {\it {Search for Dark Matter
  at CMS}},  \href{http://xxx.lanl.gov/abs/1410.3762}{{\tt arXiv:1410.3762}}.

\bibitem{Mitsou:2014wta}
V.~A. Mitsou, {\it {Overview of searches for dark matter at the LHC}},
  \href{http://xxx.lanl.gov/abs/1402.3673}{{\tt arXiv:1402.3673}}.

\bibitem{ATLAS:2012ky}
{\bf ATLAS Collaboration} Collaboration, G.~Aad et~al., {\it {Search for dark
  matter candidates and large extra dimensions in events with a jet and missing
  transverse momentum with the ATLAS detector}},  {\em JHEP} {\bf 1304} (2013)
  075, [\href{http://xxx.lanl.gov/abs/1210.4491}{{\tt arXiv:1210.4491}}].

\bibitem{Aaltonen:2008hh}
{\bf CDF} Collaboration, T.~Aaltonen et~al., {\it {Search for large extra
  dimensions in final states containing one photon or jet and large missing
  transverse energy produced in $p \bar{p}$ collisions at $\sqrt{s}$ =
  1.96-TeV}},  {\em Phys.Rev.Lett.} {\bf 101} (2008) 181602,
  [\href{http://xxx.lanl.gov/abs/0807.3132}{{\tt arXiv:0807.3132}}].

\bibitem{Aad:2015zva}
{\bf ATLAS Collaboration} Collaboration, G.~Aad et~al., {\it {Search for new
  phenomena in final states with an energetic jet and large missing transverse
  momentum in pp collisions at $\sqrt{s}=8$ TeV with the ATLAS detector}},
  \href{http://xxx.lanl.gov/abs/1502.0151}{{\tt arXiv:1502.0151}}.

\bibitem{Angloher:2002in}
G.~Angloher, S.~Cooper, R.~Keeling, H.~Kraus, J.~Marchese, et~al., {\it {Limits
  on WIMP dark matter using sapphire cryogenic detectors}},  {\em
  Astropart.Phys.} {\bf 18} (2002) 43--55.

\bibitem{Ahmed:2009zw}
{\bf CDMS-II Collaboration} Collaboration, Z.~Ahmed et~al., {\it {Dark Matter
  Search Results from the CDMS II Experiment}},  {\em Science} {\bf 327} (2010)
  1619--1621, [\href{http://xxx.lanl.gov/abs/0912.3592}{{\tt
  arXiv:0912.3592}}].

\bibitem{Angle:2007uj}
{\bf XENON Collaboration} Collaboration, J.~Angle et~al., {\it {First Results
  from the XENON10 Dark Matter Experiment at the Gran Sasso National
  Laboratory}},  {\em Phys.Rev.Lett.} {\bf 100} (2008) 021303,
  [\href{http://xxx.lanl.gov/abs/0706.0039}{{\tt arXiv:0706.0039}}].

\bibitem{Aalseth:2010vx}
{\bf CoGeNT collaboration} Collaboration, C.~Aalseth et~al., {\it {Results from
  a Search for Light-Mass Dark Matter with a P-type Point Contact Germanium
  Detector}},  {\em Phys.Rev.Lett.} {\bf 106} (2011) 131301,
  [\href{http://xxx.lanl.gov/abs/1002.4703}{{\tt arXiv:1002.4703}}].

\bibitem{Aprile:2010um}
{\bf XENON100 Collaboration} Collaboration, E.~Aprile et~al., {\it {First Dark
  Matter Results from the XENON100 Experiment}},  {\em Phys.Rev.Lett.} {\bf
  105} (2010) 131302, [\href{http://xxx.lanl.gov/abs/1005.0380}{{\tt
  arXiv:1005.0380}}].

\bibitem{Akerib:2006rr}
D.~Akerib, M.~Attisha, C.~Bailey, L.~Baudis, D.~A. Bauer, et~al., {\it {The
  SuperCDMS proposal for dark matter detection}},  {\em Nucl.Instrum.Meth.}
  {\bf A559} (2006) 411--413.

\bibitem{Aprile:2009yh}
{\bf XENON100 Collaboration} Collaboration, E.~Aprile and L.~Baudis, {\it
  {Status and Sensitivity Projections for the XENON100 Dark Matter
  Experiment}},  {\em PoS} {\bf IDM2008} (2008) 018,
  [\href{http://xxx.lanl.gov/abs/0902.4253}{{\tt arXiv:0902.4253}}].

\bibitem{Archambault:2009sm}
S.~Archambault, F.~Aubin, M.~Auger, E.~Behnke, B.~Beltran, et~al., {\it {Dark
  Matter Spin-Dependent Limits for WIMP Interactions on F-19 by PICASSO}},
  {\em Phys.Lett.} {\bf B682} (2009) 185--192,
  [\href{http://xxx.lanl.gov/abs/0907.0307}{{\tt arXiv:0907.0307}}].

\bibitem{Lee.:2007qn}
{\bf KIMS Collaboration} Collaboration, H.~Lee et~al., {\it {Limits on
  WIMP-nucleon cross section with CsI(Tl) crystal detectors}},  {\em
  Phys.Rev.Lett.} {\bf 99} (2007) 091301,
  [\href{http://xxx.lanl.gov/abs/0704.0423}{{\tt arXiv:0704.0423}}].

\bibitem{Sciolla:2009fb}
G.~Sciolla, J.~Battat, T.~Caldwell, D.~Dujmic, P.~Fisher, et~al., {\it {The
  DMTPC project}},  {\em J.Phys.Conf.Ser.} {\bf 179} (2009) 012009,
  [\href{http://xxx.lanl.gov/abs/0903.3895}{{\tt arXiv:0903.3895}}].

\bibitem{Gramling:2013wra}
J.~Gramling, {\it {Probing dark matter with monojets in ATLAS at the LHC}},
  {\em PoS} {\bf Corfu2012} (2013) 058.

\bibitem{Diehl:2014dda}
{\bf ATLAS} Collaboration, E.~Diehl, {\it {The search for dark matter using
  monojets and monophotons with the ATLAS detector}},  {\em AIP Conf.Proc.}
  {\bf 1604} (2014) 324--330.

\bibitem{Aad:2012fw}
{\bf ATLAS Collaboration} Collaboration, G.~Aad et~al., {\it {Search for dark
  matter candidates and large extra dimensions in events with a photon and
  missing transverse momentum in $pp$ collision data at $\sqrt{s}=7$ TeV with
  the ATLAS detector}},  {\em Phys.Rev.Lett.} {\bf 110} (2013) 011802,
  [\href{http://xxx.lanl.gov/abs/1209.4625}{{\tt arXiv:1209.4625}}].

\bibitem{Chatrchyan:2012tea}
{\bf CMS Collaboration} Collaboration, S.~Chatrchyan et~al., {\it {Search for
  Dark Matter and Large Extra Dimensions in pp Collisions Yielding a Photon and
  Missing Transverse Energy}},  {\em Phys.Rev.Lett.} {\bf 108} (2012) 261803,
  [\href{http://xxx.lanl.gov/abs/1204.0821}{{\tt arXiv:1204.0821}}].

\bibitem{Chatrchyan:2011nd}
{\bf CMS} Collaboration, S.~Chatrchyan et~al., {\it {Search for New Physics
  with a Mono-Jet and Missing Transverse Energy in $pp$ Collisions at $\sqrt{s}
  = 7$ TeV}},  {\em Phys.Rev.Lett.} {\bf 107} (2011) 201804,
  [\href{http://xxx.lanl.gov/abs/1106.4775}{{\tt arXiv:1106.4775}}].

\bibitem{Aad:2013oja}
{\bf ATLAS Collaboration} Collaboration, G.~Aad et~al., {\it {Search for dark
  matter in events with a hadronically decaying W or Z boson and missing
  transverse momentum in $pp$ collisions at $\sqrt{s} =$ 8 TeV with the ATLAS
  detector}},  {\em Phys.Rev.Lett.} {\bf 112} (2014), no.~4 041802,
  [\href{http://xxx.lanl.gov/abs/1309.4017}{{\tt arXiv:1309.4017}}].

\bibitem{Bai:2012xg}
Y.~Bai and T.~M. Tait, {\it {Searches with Mono-Leptons}},  {\em Phys.Lett.}
  {\bf B723} (2013) 384--387, [\href{http://xxx.lanl.gov/abs/1208.4361}{{\tt
  arXiv:1208.4361}}].

\bibitem{Ahmed:2010wy}
{\bf CDMS-II Collaboration} Collaboration, Z.~Ahmed et~al., {\it {Results from
  a Low-Energy Analysis of the CDMS II Germanium Data}},  {\em Phys.Rev.Lett.}
  {\bf 106} (2011) 131302, [\href{http://xxx.lanl.gov/abs/1011.2482}{{\tt
  arXiv:1011.2482}}].

\bibitem{Aartsen:2012kia}
{\bf IceCube collaboration} Collaboration, M.~Aartsen et~al., {\it {Search for
  dark matter annihilations in the Sun with the 79-string IceCube detector}},
  {\em Phys.Rev.Lett.} {\bf 110} (2013), no.~13 131302,
  [\href{http://xxx.lanl.gov/abs/1212.4097}{{\tt arXiv:1212.4097}}].

\bibitem{Behnke:2012ys}
{\bf COUPP Collaboration} Collaboration, E.~Behnke et~al., {\it {First Dark
  Matter Search Results from a 4-kg CF$_3$I Bubble Chamber Operated in a Deep
  Underground Site}},  {\em Phys.Rev.} {\bf D86} (2012), no.~5 052001,
  [\href{http://xxx.lanl.gov/abs/1204.3094}{{\tt arXiv:1204.3094}}].

\bibitem{Aad:2014iia}
{\bf ATLAS Collaboration} Collaboration, G.~Aad et~al., {\it {Search for
  Invisible Decays of a Higgs Boson Produced in Association with a Z Boson in
  ATLAS}},  {\em Phys.Rev.Lett.} {\bf 112} (2014) 201802,
  [\href{http://xxx.lanl.gov/abs/1402.3244}{{\tt arXiv:1402.3244}}].

\bibitem{Heinemeyer:2013tqa}
{\bf LHC Higgs Cross Section Working Group} Collaboration, S.~Heinemeyer
  et~al., {\it {Handbook of LHC Higgs Cross Sections: 3. Higgs Properties}},
  \href{http://xxx.lanl.gov/abs/1307.1347}{{\tt arXiv:1307.1347}}.

\bibitem{Kanemura:2010sh}
S.~Kanemura, S.~Matsumoto, T.~Nabeshima, and N.~Okada, {\it {Can WIMP Dark
  Matter overcome the Nightmare Scenario?}},  {\em Phys.Rev.} {\bf D82} (2010)
  055026, [\href{http://xxx.lanl.gov/abs/1005.5651}{{\tt arXiv:1005.5651}}].

\bibitem{Djouadi:2011aa}
A.~Djouadi, O.~Lebedev, Y.~Mambrini, and J.~Quevillon, {\it {Implications of
  LHC searches for Higgs--portal dark matter}},  {\em Phys.Lett.} {\bf B709}
  (2012) 65--69, [\href{http://xxx.lanl.gov/abs/1112.3299}{{\tt
  arXiv:1112.3299}}].

\bibitem{Angle:2011th}
{\bf XENON10 Collaboration} Collaboration, J.~Angle et~al., {\it {A search for
  light dark matter in XENON10 data}},  {\em Phys.Rev.Lett.} {\bf 107} (2011)
  051301, [\href{http://xxx.lanl.gov/abs/1104.3088}{{\tt arXiv:1104.3088}}].

\bibitem{Fox:2011px}
P.~J. Fox, J.~Kopp, M.~Lisanti, and N.~Weiner, {\it {A CoGeNT Modulation
  Analysis}},  {\em Phys.Rev.} {\bf D85} (2012) 036008,
  [\href{http://xxx.lanl.gov/abs/1107.0717}{{\tt arXiv:1107.0717}}].

\bibitem{Agnese:2013jaa}
{\bf SuperCDMS Collaboration} Collaboration, R.~Agnese et~al., {\it {Search for
  Low-Mass Weakly Interacting Massive Particles Using Voltage-Assisted
  Calorimetric Ionization Detection in the SuperCDMS Experiment}},  {\em
  Phys.Rev.Lett.} {\bf 112} (2014), no.~4 041302,
  [\href{http://xxx.lanl.gov/abs/1309.3259}{{\tt arXiv:1309.3259}}].

\bibitem{Aad:2014vka}
{\bf ATLAS Collaboration} Collaboration, G.~Aad et~al., {\it {Search for dark
  matter in events with a Z boson and missing transverse momentum in pp
  collisions at $\sqrt{s}$=8 TeV with the ATLAS detector}},  {\em Phys.Rev.}
  {\bf D90} (2014) 012004, [\href{http://xxx.lanl.gov/abs/1404.0051}{{\tt
  arXiv:1404.0051}}].

\bibitem{Bell:2012rg}
N.~F. Bell, J.~B. Dent, A.~J. Galea, T.~D. Jacques, L.~M. Krauss, et~al., {\it
  {Searching for Dark Matter at the LHC with a Mono-Z}},  {\em Phys.Rev.} {\bf
  D86} (2012) 096011, [\href{http://xxx.lanl.gov/abs/1209.0231}{{\tt
  arXiv:1209.0231}}].

\bibitem{Aad:2014vea}
{\bf ATLAS Collaboration} Collaboration, G.~Aad et~al., {\it {Search for dark
  matter in events with heavy quarks and missing transverse momentum in $pp$
  collisions with the ATLAS detector}},
  \href{http://xxx.lanl.gov/abs/1410.4031}{{\tt arXiv:1410.4031}}.

\bibitem{Rogan:2010kb}
C.~Rogan, {\it {Kinematical variables towards new dynamics at the LHC}},
  \href{http://xxx.lanl.gov/abs/1006.2727}{{\tt arXiv:1006.2727}}.

\bibitem{Agrawal:2014una}
P.~Agrawal, B.~Batell, D.~Hooper, and T.~Lin, {\it {Flavored Dark Matter and
  the Galactic Center Gamma-Ray Excess}},  {\em Phys.Rev.} {\bf D90} (2014),
  no.~6 063512, [\href{http://xxx.lanl.gov/abs/1404.1373}{{\tt
  arXiv:1404.1373}}].

\bibitem{Aad:2014tda}
{\bf ATLAS Collaboration} Collaboration, G.~Aad et~al., {\it {Search for new
  phenomena in events with a photon and missing transverse momentum in $pp$
  collisions at $\sqrt{s}=8$ TeV with the ATLAS detector}},  {\em Phys.Rev.}
  {\bf D91} (2015), no.~1 012008,
  [\href{http://xxx.lanl.gov/abs/1411.1559}{{\tt arXiv:1411.1559}}].

\bibitem{Nelson:2013pqa}
A.~Nelson, L.~M. Carpenter, R.~Cotta, A.~Johnstone, and D.~Whiteson, {\it
  {Confronting the Fermi Line with LHC data: an Effective Theory of Dark Matter
  Interaction with Photons}},  {\em Phys.Rev.} {\bf D89} (2014), no.~5 056011,
  [\href{http://xxx.lanl.gov/abs/1307.5064}{{\tt arXiv:1307.5064}}].

\bibitem{Weniger:2012tx}
C.~Weniger, {\it {A Tentative Gamma-Ray Line from Dark Matter Annihilation at
  the Fermi Large Area Telescope}},  {\em JCAP} {\bf 1208} (2012) 007,
  [\href{http://xxx.lanl.gov/abs/1204.2797}{{\tt arXiv:1204.2797}}].

\bibitem{Komatsu:2010fb}
{\bf WMAP Collaboration} Collaboration, E.~Komatsu et~al., {\it {Seven-Year
  Wilkinson Microwave Anisotropy Probe (WMAP) Observations: Cosmological
  Interpretation}},  {\em Astrophys.J.Suppl.} {\bf 192} (2011) 18,
  [\href{http://xxx.lanl.gov/abs/1001.4538}{{\tt arXiv:1001.4538}}].

\bibitem{ATLAS:2014wra}
{\bf ATLAS Collaboration} Collaboration, G.~Aad et~al., {\it {Search for new
  particles in events with one lepton and missing transverse momentum in $pp$
  collisions at $\sqrt{s}$ = 8 TeV with the ATLAS detector}},  {\em JHEP} {\bf
  1409} (2014) 037, [\href{http://xxx.lanl.gov/abs/1407.7494}{{\tt
  arXiv:1407.7494}}].

\bibitem{Chizhov:2009fc}
M.~Chizhov and G.~Dvali, {\it {Origin and Phenomenology of Weak-Doublet Spin-1
  Bosons}},  {\em Phys.Lett.} {\bf B703} (2011) 593--598,
  [\href{http://xxx.lanl.gov/abs/0908.0924}{{\tt arXiv:0908.0924}}].

\bibitem{Chizhov:2008tp}
M.~Chizhov, V.~Bednyakov, and J.~Budagov, {\it {Proposal for chiral bosons
  search at LHC via their unique new signature}},  {\em Phys.Atom.Nucl.} {\bf
  71} (2008) 2096--2100, [\href{http://xxx.lanl.gov/abs/0801.4235}{{\tt
  arXiv:0801.4235}}].

\bibitem{Abazov:2008kp}
{\bf D0 Collaboration} Collaboration, V.~Abazov et~al., {\it {Search for large
  extra dimensions via single photon plus missing energy final states at
  $\sqrt{s}$ = 1.96-TeV}},  {\em Phys.Rev.Lett.} {\bf 101} (2008) 011601,
  [\href{http://xxx.lanl.gov/abs/0803.2137}{{\tt arXiv:0803.2137}}].

\bibitem{Aaltonen:2012jb}
{\bf CDF Collaboration} Collaboration, T.~Aaltonen et~al., {\it {A Search for
  dark matter in events with one jet and missing transverse energy in
  $p\bar{p}$ collisions at $\sqrt{s} = 1.96$ TeV}},  {\em Phys.Rev.Lett.} {\bf
  108} (2012) 211804, [\href{http://xxx.lanl.gov/abs/1203.0742}{{\tt
  arXiv:1203.0742}}].

\bibitem{Chatrchyan:2012me}
{\bf CMS} Collaboration, S.~Chatrchyan et~al., {\it {Search for dark matter and
  large extra dimensions in monojet events in $pp$ collisions at $\sqrt{s}=7$
  TeV}},  {\em JHEP} {\bf 1209} (2012) 094,
  [\href{http://xxx.lanl.gov/abs/1206.5663}{{\tt arXiv:1206.5663}}].

\bibitem{Aad:2011xw}
{\bf ATLAS} Collaboration, G.~Aad et~al., {\it {Search for new phenomena with
  the monojet and missing transverse momentum signature using the ATLAS
  detector in $\sqrt{s}=7$ TeV proton-proton collisions}},  {\em Phys.Lett.}
  {\bf B705} (2011) 294--312, [\href{http://xxx.lanl.gov/abs/1106.5327}{{\tt
  arXiv:1106.5327}}].

\bibitem{Khachatryan:2014rra}
{\bf CMS} Collaboration, V.~Khachatryan et~al., {\it {Search for dark matter,
  extra dimensions, and unparticles in monojet events in proton-proton
  collisions at $\sqrt{s}$ = 8 TeV}},
  \href{http://xxx.lanl.gov/abs/1408.3583}{{\tt arXiv:1408.3583}}.

\bibitem{Khachatryan:2014tva}
{\bf CMS} Collaboration, V.~Khachatryan et~al., {\it {Search for physics beyond
  the standard model in final states with a lepton and missing transverse
  energy in proton-proton collisions at $\sqrt{s}$ = 8 TeV}},
  \href{http://xxx.lanl.gov/abs/1408.2745}{{\tt arXiv:1408.2745}}.

\bibitem{Read:2002hq}
A.~L. Read, {\it {Presentation of search results: The CL(s) technique}},  {\em
  J.Phys.} {\bf G28} (2002) 2693--2704.

\bibitem{Steigman:1984ac}
G.~Steigman and M.~S. Turner, {\it {Cosmological Constraints on the Properties
  of Weakly Interacting Massive Particles}},  {\em Nucl.Phys.} {\bf B253}
  (1985) 375.

\bibitem{Hinshaw:2012aka}
{\bf WMAP} Collaboration, G.~Hinshaw et~al., {\it {Nine-Year Wilkinson
  Microwave Anisotropy Probe (WMAP) Observations: Cosmological Parameter
  Results}},  {\em Astrophys.J.Suppl.} {\bf 208} (2013) 19,
  [\href{http://xxx.lanl.gov/abs/1212.5226}{{\tt arXiv:1212.5226}}].

\bibitem{Ackermann:2013yva}
{\bf Fermi-LAT Collaboration} Collaboration, M.~Ackermann et~al., {\it {Dark
  matter constraints from observations of 25 Milky Way satellite galaxies with
  the Fermi Large Area Telescope}},  {\em Phys.Rev.} {\bf D89} (2014) 042001,
  [\href{http://xxx.lanl.gov/abs/1310.0828}{{\tt arXiv:1310.0828}}].

\bibitem{Abramowski:2011hc}
{\bf HESS} Collaboration, A.~Abramowski et~al., {\it {Search for a Dark Matter
  annihilation signal from the Galactic Center halo with H.E.S.S}},  {\em
  Phys.Rev.Lett.} {\bf 106} (2011) 161301,
  [\href{http://xxx.lanl.gov/abs/1103.3266}{{\tt arXiv:1103.3266}}].

\bibitem{Ade:2013zuv}
{\bf Planck} Collaboration, P.~Ade et~al., {\it {Planck 2013 results. XVI.
  Cosmological parameters}},  {\em Astron.Astrophys.} {\bf 571} (2014) A16,
  [\href{http://xxx.lanl.gov/abs/1303.5076}{{\tt arXiv:1303.5076}}].

\bibitem{Bishara:2015cha}
F.~Bishara, J.~Brod, P.~Uttayarat, and J.~Zupan, {\it {Nonstandard Yukawa
  Couplings and Higgs Portal Dark Matter}},
  \href{http://xxx.lanl.gov/abs/1504.0402}{{\tt arXiv:1504.0402}}.

\bibitem{Dutra:2015vca}
M.~Dutra, C.~A. d.~S. Pires, and P.~S.~R. da~Silva, {\it {Majorana Dark Matter
  in Minimal Higgs Portal Models after LUX}},
  \href{http://xxx.lanl.gov/abs/1504.0722}{{\tt arXiv:1504.0722}}.

\bibitem{Fedderke:2014wda}
M.~A. Fedderke, J.-Y. Chen, E.~W. Kolb, and L.-T. Wang, {\it {The Fermionic
  Dark Matter Higgs Portal: an effective field theory approach}},  {\em JHEP}
  {\bf 1408} (2014) 122, [\href{http://xxx.lanl.gov/abs/1404.2283}{{\tt
  arXiv:1404.2283}}].

\bibitem{Chatrchyan:2014tja}
{\bf CMS} Collaboration, S.~Chatrchyan et~al., {\it {Search for invisible
  decays of Higgs bosons in the vector boson fusion and associated ZH
  production modes}},  {\em Eur.Phys.J.} {\bf C74} (2014) 2980,
  [\href{http://xxx.lanl.gov/abs/1404.1344}{{\tt arXiv:1404.1344}}].

\bibitem{Aad:2015uga}
{\bf ATLAS} Collaboration, G.~Aad et~al., {\it {Search for invisible decays of
  the Higgs boson produced in association with a hadronically decaying vector
  boson in $pp$ collisions at $\sqrt{s}$ = 8 TeV with the ATLAS detector}},
  \href{http://xxx.lanl.gov/abs/1504.0432}{{\tt arXiv:1504.0432}}.

\bibitem{Khachatryan:2015nua}
{\bf CMS} Collaboration, V.~Khachatryan et~al., {\it {Search for the production
  of dark matter in association with top-quark pairs in the single-lepton final
  state in proton-proton collisions at sqrt(s) = 8 TeV}},
  \href{http://xxx.lanl.gov/abs/1504.0319}{{\tt arXiv:1504.0319}}.

\bibitem{deSimone:2014pda}
A.~De~Simone, G.~F. Giudice, and A.~Strumia, {\it {Benchmarks for Dark Matter
  Searches at the LHC}},  {\em JHEP} {\bf 1406} (2014) 081,
  [\href{http://xxx.lanl.gov/abs/1402.6287}{{\tt arXiv:1402.6287}}].

\bibitem{Fernandez:2014eja}
N.~Fernandez, J.~Kumar, I.~Seong, and P.~Stengel, {\it {Complementary
  Constraints on Light Dark Matter from Heavy Quarkonium Decays}},  {\em
  Phys.Rev.} {\bf D90} (2014) 015029,
  [\href{http://xxx.lanl.gov/abs/1404.6599}{{\tt arXiv:1404.6599}}].

\bibitem{Goudzovski:2014rwa}
{\bf NA48/2 Collaboration} Collaboration, E.~Goudzovski, {\it {Search for the
  dark photon in $\pi^0$ decays by the NA48/2 experiment at CERN}},
  \href{http://xxx.lanl.gov/abs/1412.8053}{{\tt arXiv:1412.8053}}.

\bibitem{CERNNA48/2:2015lha}
{\bf CERN NA48/2} Collaboration, {\it {Search for the dark photon in $\pi^0$
  decays}},  \href{http://xxx.lanl.gov/abs/1504.0060}{{\tt arXiv:1504.0060}}.

\bibitem{Lees:2014xha}
{\bf BaBar Collaboration} Collaboration, J.~Lees et~al., {\it {Search for a
  Dark Photon in $e^+e^-$ Collisions a BaBar}},  {\em Phys.Rev.Lett.} {\bf 113}
  (2014), no.~20 201801, [\href{http://xxx.lanl.gov/abs/1406.2980}{{\tt
  arXiv:1406.2980}}].

\bibitem{Bennett:2006fi}
{\bf Muon g-2} Collaboration, G.~Bennett et~al., {\it {Final Report of the Muon
  E821 Anomalous Magnetic Moment Measurement at BNL}},  {\em Phys.Rev.} {\bf
  D73} (2006) 072003, [\href{http://xxx.lanl.gov/abs/hep-ex/0602035}{{\tt
  hep-ex/0602035}}].

\bibitem{Eigen:2015rea}
{\bf BaBar} Collaboration, G.~Eigen, {\it {Direct Searches for New Physics
  Particles at BABAR}},  \href{http://xxx.lanl.gov/abs/1503.0286}{{\tt
  arXiv:1503.0286}}.

\bibitem{Echenard:2014lma}
B.~Echenard, R.~Essig, and Y.-M. Zhong, {\it {Projections for Dark Photon
  Searches at Mu3e}},  \href{http://xxx.lanl.gov/abs/1411.1770}{{\tt
  arXiv:1411.1770}}.

\bibitem{ArkaniHamed:2008qn}
N.~Arkani-Hamed, D.~P. Finkbeiner, T.~R. Slatyer, and N.~Weiner, {\it {A Theory
  of Dark Matter}},  {\em Phys.Rev.} {\bf D79} (2009) 015014,
  [\href{http://xxx.lanl.gov/abs/0810.0713}{{\tt arXiv:0810.0713}}].

\bibitem{Cheung:2009su}
C.~Cheung, J.~T. Ruderman, L.-T. Wang, and I.~Yavin, {\it {Lepton Jets in
  (Supersymmetric) Electroweak Processes}},  {\em JHEP} {\bf 1004} (2010) 116,
  [\href{http://xxx.lanl.gov/abs/0909.0290}{{\tt arXiv:0909.0290}}].

\bibitem{Foot:2014uba}
R.~Foot and S.~Vagnozzi, {\it {Dissipative hidden sector dark matter}},  {\em
  Phys.Rev.} {\bf D91} (2015) 023512,
  [\href{http://xxx.lanl.gov/abs/1409.7174}{{\tt arXiv:1409.7174}}].

\bibitem{Foot:2014osa}
R.~Foot and S.~Vagnozzi, {\it {Diurnal modulation signal from dissipative
  hidden sector dark matter}},  \href{http://xxx.lanl.gov/abs/1412.0762}{{\tt
  arXiv:1412.0762}}.

\bibitem{Kong:2014iwa}
K.~Kong, H.-S. Lee, and M.~Park, {\it {Charged Higgs Probes of Dark Bosons at
  the LHC}},  \href{http://xxx.lanl.gov/abs/1408.4021}{{\tt arXiv:1408.4021}}.

\bibitem{Kong:2014jwa}
K.~Kong, H.-S. Lee, and M.~Park, {\it {Dark decay of the top quark}},  {\em
  Phys.Rev.} {\bf D89} (2014), no.~7 074007,
  [\href{http://xxx.lanl.gov/abs/1401.5020}{{\tt arXiv:1401.5020}}].

\bibitem{Davoudiasl:2014mqa}
H.~Davoudiasl, W.~J. Marciano, R.~Ramos, and M.~Sher, {\it {Charged Higgs
  Discovery in the W plus "Dark" Vector Boson Decay Mode}},  {\em Phys.Rev.}
  {\bf D89} (2014), no.~11 115008,
  [\href{http://xxx.lanl.gov/abs/1401.2164}{{\tt arXiv:1401.2164}}].

\bibitem{Gupta:2015lfa}
A.~Gupta, R.~Primulando, and P.~Saraswat, {\it {A New Probe of Dark Sector
  Dynamics at the LHC}},  \href{http://xxx.lanl.gov/abs/1504.0138}{{\tt
  arXiv:1504.0138}}.

\bibitem{Bai:2015nfa}
Y.~Bai, J.~Bourbeau, and T.~Lin, {\it {Dark Matter Searches with a Mono-Z'
  jet}},  \href{http://xxx.lanl.gov/abs/1504.0139}{{\tt arXiv:1504.0139}}.

\bibitem{Autran:2015mfa}
M.~Autran, K.~Bauer, T.~Lin, and D.~Whiteson, {\it {Mono-Z': searches for dark
  matter in events with a resonance and missing transverse energy}},
  \href{http://xxx.lanl.gov/abs/1504.0138}{{\tt arXiv:1504.0138}}.

\bibitem{Lee:2014tba}
H.-S. Lee, {\it {Muon g-2 Anomaly and Dark Leptonic Gauge Boson}},  {\em
  Phys.Rev.} {\bf D90} (2014) 091702,
  [\href{http://xxx.lanl.gov/abs/1408.4256}{{\tt arXiv:1408.4256}}].

\bibitem{Agakishiev:2013fwl}
{\bf HADES} Collaboration, G.~Agakishiev et~al., {\it {Searching a Dark Photon
  with HADES}},  {\em Phys.Lett.} {\bf B731} (2014) 265--271,
  [\href{http://xxx.lanl.gov/abs/1311.0216}{{\tt arXiv:1311.0216}}].

\bibitem{Babusci:2014sta}
{\bf KLOE-2} Collaboration, D.~Babusci et~al., {\it {Search for light vector
  boson production in $e^+e^- \rightarrow \mu^+ \mu^- \gamma$ interactions with
  the KLOE experiment}},  {\em Phys.Lett.} {\bf B736} (2014) 459--464,
  [\href{http://xxx.lanl.gov/abs/1404.7772}{{\tt arXiv:1404.7772}}].

\bibitem{Xu:2015wja}
F.~Xu, {\it {Dark $Z$ Implication for Flavor Physics}},
  \href{http://xxx.lanl.gov/abs/1504.0741}{{\tt arXiv:1504.0741}}.

\bibitem{Batell:2014mga}
B.~Batell, R.~Essig, and Z.~Surujon, {\it {Strong Constraints on Sub-GeV Dark
  Sectors from SLAC Beam Dump E137}},  {\em Phys.Rev.Lett.} {\bf 113} (2014),
  no.~17 171802, [\href{http://xxx.lanl.gov/abs/1406.2698}{{\tt
  arXiv:1406.2698}}].

\bibitem{Gninenko:2014pea}
S.~Gninenko, N.~Krasnikov, and V.~Matveev, {\it {The muon g-2 and searches for
  a new electrophobic sub-GeV dark boson in a missing-energy experiment at
  CERN}},  \href{http://xxx.lanl.gov/abs/1412.1400}{{\tt arXiv:1412.1400}}.

\bibitem{Gorbunov:2014wqa}
D.~Gorbunov, A.~Makarov, and I.~Timiryasov, {\it {Decaying light particles in
  the SHiP experiment: Signal rate estimates for hidden photons}},  {\em
  Phys.Rev.} {\bf D91} (2015), no.~3 035027,
  [\href{http://xxx.lanl.gov/abs/1411.4007}{{\tt arXiv:1411.4007}}].

\bibitem{Alekhin:2015oba}
S.~Alekhin, W.~Altmannshofer, T.~Asaka, B.~Batell, F.~Bezrukov, et~al., {\it {A
  facility to Search for Hidden Particles at the CERN SPS: the SHiP physics
  case}},  \href{http://xxx.lanl.gov/abs/1504.0485}{{\tt arXiv:1504.0485}}.

\bibitem{Curtin:2014cca}
D.~Curtin, R.~Essig, S.~Gori, and J.~Shelton, {\it {Illuminating Dark Photons
  with High-Energy Colliders}},  {\em JHEP} {\bf 1502} (2015) 157,
  [\href{http://xxx.lanl.gov/abs/1412.0018}{{\tt arXiv:1412.0018}}].

\bibitem{Izaguirre:2014bca}
E.~Izaguirre, G.~Krnjaic, P.~Schuster, and N.~Toro, {\it {Testing GeV-Scale
  Dark Matter with Fixed-Target Missing Momentum Experiments}},
  \href{http://xxx.lanl.gov/abs/1411.1404}{{\tt arXiv:1411.1404}}.

\bibitem{Izaguirre:2013uxa}
E.~Izaguirre, G.~Krnjaic, P.~Schuster, and N.~Toro, {\it {New Electron
  Beam-Dump Experiments to Search for MeV to few-GeV Dark Matter}},  {\em
  Phys.Rev.} {\bf D88} (2013) 114015,
  [\href{http://xxx.lanl.gov/abs/1307.6554}{{\tt arXiv:1307.6554}}].

\bibitem{deNiverville:2012ij}
P.~deNiverville, D.~McKeen, and A.~Ritz, {\it {Signatures of sub-GeV dark
  matter beams at neutrino experiments}},  {\em Phys.Rev.} {\bf D86} (2012)
  035022, [\href{http://xxx.lanl.gov/abs/1205.3499}{{\tt arXiv:1205.3499}}].

\bibitem{Diamond:2013oda}
M.~D. Diamond and P.~Schuster, {\it {Searching for Light Dark Matter with the
  SLAC Millicharge Experiment}},  {\em Phys.Rev.Lett.} {\bf 111} (2013), no.~22
  221803, [\href{http://xxx.lanl.gov/abs/1307.6861}{{\tt arXiv:1307.6861}}].

\bibitem{Battaglieri:2014qoa}
{\bf BDX Collaboration} Collaboration, M.~Battaglieri et~al., {\it {Dark matter
  search in a Beam-Dump eXperiment (BDX) at Jefferson Lab}},
  \href{http://xxx.lanl.gov/abs/1406.3028}{{\tt arXiv:1406.3028}}.

\bibitem{Bjorken:2009mm}
J.~D. Bjorken, R.~Essig, P.~Schuster, and N.~Toro, {\it {New Fixed-Target
  Experiments to Search for Dark Gauge Forces}},  {\em Phys.Rev.} {\bf D80}
  (2009) 075018, [\href{http://xxx.lanl.gov/abs/0906.0580}{{\tt
  arXiv:0906.0580}}].

\bibitem{Izaguirre:2014dua}
E.~Izaguirre, G.~Krnjaic, P.~Schuster, and N.~Toro, {\it {Physics Motivation
  for a Pilot Dark Matter Search at Jefferson Laboratory}},  {\em Phys.Rev.}
  {\bf D90} (2014) 014052, [\href{http://xxx.lanl.gov/abs/1403.6826}{{\tt
  arXiv:1403.6826}}].

\bibitem{Balewski:2014pxa}
J.~Balewski, J.~Bernauer, J.~Bessuille, R.~Corliss, R.~Cowan, et~al., {\it {The
  DarkLight Experiment: A Precision Search for New Physics at Low Energies}},
  \href{http://xxx.lanl.gov/abs/1412.4717}{{\tt arXiv:1412.4717}}.

\bibitem{An:2014twa}
H.~An, M.~Pospelov, J.~Pradler, and A.~Ritz, {\it {Direct Detection Constraints
  on Dark Photon Dark Matter}},  \href{http://xxx.lanl.gov/abs/1412.8378}{{\tt
  arXiv:1412.8378}}.

\bibitem{Dutta:2014mya}
B.~Dutta, {\it {Dark Matter Searches at Accelerator Facilities}},
  \href{http://xxx.lanl.gov/abs/1403.6217}{{\tt arXiv:1403.6217}}.

\bibitem{Fox:2008kb}
P.~J. Fox and E.~Poppitz, {\it {Leptophilic Dark Matter}},  {\em Phys.Rev.}
  {\bf D79} (2009) 083528, [\href{http://xxx.lanl.gov/abs/0811.0399}{{\tt
  arXiv:0811.0399}}].

\bibitem{Kopp:2009et}
J.~Kopp, V.~Niro, T.~Schwetz, and J.~Zupan, {\it {DAMA/LIBRA and leptonically
  interacting Dark Matter}},  {\em Phys.Rev.} {\bf D80} (2009) 083502,
  [\href{http://xxx.lanl.gov/abs/0907.3159}{{\tt arXiv:0907.3159}}].

\bibitem{Freitas:2014jla}
A.~Freitas and S.~Westhoff, {\it {Leptophilic Dark Matter in Lepton
  Interactions at LEP and ILC}},  {\em JHEP} {\bf 1410} (2014) 116,
  [\href{http://xxx.lanl.gov/abs/1408.1959}{{\tt arXiv:1408.1959}}].

\bibitem{Dreiner:2012xm}
H.~Dreiner, M.~Huck, M.~KrŠmer, D.~Schmeier, and J.~Tattersall, {\it
  {Illuminating Dark Matter at the ILC}},  {\em Phys.Rev.} {\bf D87} (2013),
  no.~7 075015, [\href{http://xxx.lanl.gov/abs/1211.2254}{{\tt
  arXiv:1211.2254}}].

\bibitem{Richard:2014vfa}
F.~Richard, G.~Arcadi, and Y.~Mambrini, {\it {Search for Dark Matter at
  Colliders}},  \href{http://xxx.lanl.gov/abs/1411.0088}{{\tt
  arXiv:1411.0088}}.

\bibitem{Daylan:2014rsa}
T.~Daylan, D.~P. Finkbeiner, D.~Hooper, T.~Linden, S.~K.~N. Portillo, et~al.,
  {\it {The Characterization of the Gamma-Ray Signal from the Central Milky
  Way: A Compelling Case for Annihilating Dark Matter}},
  \href{http://xxx.lanl.gov/abs/1402.6703}{{\tt arXiv:1402.6703}}.

\bibitem{Calore:2014xka}
F.~Calore, I.~Cholis, and C.~Weniger, {\it {Background model systematics for
  the Fermi GeV excess}},  \href{http://xxx.lanl.gov/abs/1409.0042}{{\tt
  arXiv:1409.0042}}.

\bibitem{Biswas:2015sha}
S.~Biswas, E.~Gabrielli, M.~Heikinheimo, and B.~Mele, {\it {Higgs-boson
  production in association with a Dark Photon in $e^+ e^-$ collisions}},
  \href{http://xxx.lanl.gov/abs/1503.0583}{{\tt arXiv:1503.0583}}.

\bibitem{Behnke:2013xla}
T.~Behnke, J.~E. Brau, B.~Foster, J.~Fuster, M.~Harrison, et~al., {\it {The
  International Linear Collider Technical Design Report - Volume 1: Executive
  Summary}},  \href{http://xxx.lanl.gov/abs/1306.6327}{{\tt arXiv:1306.6327}}.

\bibitem{Gomez-Ceballos:2013zzn}
{\bf TLEP Design Study Working Group} Collaboration, M.~Bicer et~al., {\it
  {First Look at the Physics Case of TLEP}},  {\em JHEP} {\bf 1401} (2014) 164,
  [\href{http://xxx.lanl.gov/abs/1308.6176}{{\tt arXiv:1308.6176}}].

\bibitem{Raggi:2015gza}
M.~Raggi, V.~Kozhuharov, and P.~Valente, {\it {The PADME experiment at LNF}},
  \href{http://xxx.lanl.gov/abs/1501.0186}{{\tt arXiv:1501.0186}}.

\bibitem{Babusci:2015zda}
{\bf KLOE-2} Collaboration, D.~Babusci et~al., {\it {Search for dark
  Higgsstrahlung in e+ e- -> mu+ mu- and missing energy events with the KLOE
  experiment}},  \href{http://xxx.lanl.gov/abs/1501.0679}{{\tt
  arXiv:1501.0679}}.

\bibitem{Curciarello:2015iea}
{\bf KLOE-2} Collaboration, F.~Curciarello, {\it {Dark Forces at DA$\Phi$NE}},
  \href{http://xxx.lanl.gov/abs/1502.0551}{{\tt arXiv:1502.0551}}.

\bibitem{TheBelle:2015mwa}
{\bf Belle} Collaboration, I.~Jaegle, {\it {Search for the dark photon and the
  dark Higgs boson at Belle}},  \href{http://xxx.lanl.gov/abs/1502.0008}{{\tt
  arXiv:1502.0008}}.

\bibitem{Essig:2013vha}
R.~Essig, J.~Mardon, M.~Papucci, T.~Volansky, and Y.-M. Zhong, {\it
  {Constraining Light Dark Matter with Low-Energy $e^+e^-$ Colliders}},  {\em
  JHEP} {\bf 1311} (2013) 167, [\href{http://xxx.lanl.gov/abs/1309.5084}{{\tt
  arXiv:1309.5084}}].

\bibitem{Chen:2015tia}
N.~Chen, J.~Wang, and X.-P. Wang, {\it {The leptophilic dark matter with $Z'$
  interaction: from indirect searches to future $e^+ e^-$ collider searches}},
  \href{http://xxx.lanl.gov/abs/1501.0448}{{\tt arXiv:1501.0448}}.

\bibitem{Hinchliffe:1996iu}
I.~Hinchliffe, F.~Paige, M.~Shapiro, J.~Soderqvist, and W.~Yao, {\it {Precision
  SUSY measurements at CERN LHC}},  {\em Phys.Rev.} {\bf D55} (1997)
  5520--5540, [\href{http://xxx.lanl.gov/abs/hep-ph/9610544}{{\tt
  hep-ph/9610544}}].

\bibitem{Allanach:2000kt}
B.~Allanach, C.~Lester, M.~A. Parker, and B.~Webber, {\it {Measuring sparticle
  masses in nonuniversal string inspired models at the LHC}},  {\em JHEP} {\bf
  0009} (2000) 004, [\href{http://xxx.lanl.gov/abs/hep-ph/0007009}{{\tt
  hep-ph/0007009}}].

\bibitem{Gjelsten:2005aw}
B.~Gjelsten, D.~Miller, and P.~Osland, {\it {Measurement of the gluino mass via
  cascade decays for SPS 1a}},  {\em JHEP} {\bf 0506} (2005) 015,
  [\href{http://xxx.lanl.gov/abs/hep-ph/0501033}{{\tt hep-ph/0501033}}].

\bibitem{Cheng:2008mg}
H.-C. Cheng, D.~Engelhardt, J.~F. Gunion, Z.~Han, and B.~McElrath, {\it
  {Accurate Mass Determinations in Decay Chains with Missing Energy}},  {\em
  Phys.Rev.Lett.} {\bf 100} (2008) 252001,
  [\href{http://xxx.lanl.gov/abs/0802.4290}{{\tt arXiv:0802.4290}}].

\bibitem{Cheng:2007xv}
H.-C. Cheng, J.~F. Gunion, Z.~Han, G.~Marandella, and B.~McElrath, {\it {Mass
  determination in SUSY-like events with missing energy}},  {\em JHEP} {\bf
  0712} (2007) 076, [\href{http://xxx.lanl.gov/abs/0707.0030}{{\tt
  arXiv:0707.0030}}].

\bibitem{Kawagoe:2004rz}
K.~Kawagoe, M.~Nojiri, and G.~Polesello, {\it {A New SUSY mass reconstruction
  method at the CERN LHC}},  {\em Phys.Rev.} {\bf D71} (2005) 035008,
  [\href{http://xxx.lanl.gov/abs/hep-ph/0410160}{{\tt hep-ph/0410160}}].

\bibitem{Lester:1999tx}
C.~Lester and D.~Summers, {\it {Measuring masses of semiinvisibly decaying
  particles pair produced at hadron colliders}},  {\em Phys.Lett.} {\bf B463}
  (1999) 99--103, [\href{http://xxx.lanl.gov/abs/hep-ph/9906349}{{\tt
  hep-ph/9906349}}].

\bibitem{Nojiri:2008ir}
M.~M. Nojiri and M.~Takeuchi, {\it {Study of the top reconstruction in
  top-partner events at the LHC}},  {\em JHEP} {\bf 0810} (2008) 025,
  [\href{http://xxx.lanl.gov/abs/0802.4142}{{\tt arXiv:0802.4142}}].

\bibitem{Cho:2007qv}
W.~S. Cho, K.~Choi, Y.~G. Kim, and C.~B. Park, {\it {Gluino Stransverse Mass}},
   {\em Phys.Rev.Lett.} {\bf 100} (2008) 171801,
  [\href{http://xxx.lanl.gov/abs/0709.0288}{{\tt arXiv:0709.0288}}].

\bibitem{Nojiri:2008hy}
M.~M. Nojiri, Y.~Shimizu, S.~Okada, and K.~Kawagoe, {\it {Inclusive transverse
  mass analysis for squark and gluino mass determination}},  {\em JHEP} {\bf
  0806} (2008) 035, [\href{http://xxx.lanl.gov/abs/0802.2412}{{\tt
  arXiv:0802.2412}}].

\bibitem{Alwall:2009sv}
J.~Alwall, A.~Freitas, and O.~Mattelaer, {\it {Measuring Sparticles with the
  Matrix Element}},  {\em AIP Conf.Proc.} {\bf 1200} (2010) 442--445,
  [\href{http://xxx.lanl.gov/abs/0910.2522}{{\tt arXiv:0910.2522}}].

\bibitem{Gainer:2013iya}
J.~S. Gainer, J.~Lykken, K.~T. Matchev, S.~Mrenna, and M.~Park, {\it {The
  Matrix Element Method: Past, Present, and Future}},
  \href{http://xxx.lanl.gov/abs/1307.3546}{{\tt arXiv:1307.3546}}.

\bibitem{Han:2009ss}
T.~Han, I.-W. Kim, and J.~Song, {\it {Kinematic Cusps: Determining the Missing
  Particle Mass at Colliders}},  {\em Phys.Lett.} {\bf B693} (2010) 575--579,
  [\href{http://xxx.lanl.gov/abs/0906.5009}{{\tt arXiv:0906.5009}}].

\bibitem{Sun:2014ppa}
H.~Sun, {\it {Dark matter searches in jet plus missing energy events in $\gamma
  p$ collisions at the CERN LHC}},  {\em Phys.Rev.} {\bf D90} (2014) 035018,
  [\href{http://xxx.lanl.gov/abs/1407.5356}{{\tt arXiv:1407.5356}}].

\bibitem{Bilmis:2015lja}
S.~Bilmis, I.~Turan, T.~Aliev, M.~Deniz, L.~Singh, et~al., {\it {Constraints on
  Dark Photon from Neutrino-Electron Scattering Experiments}},
  \href{http://xxx.lanl.gov/abs/1502.0776}{{\tt arXiv:1502.0776}}.

\bibitem{BednyakovKovalenko:1999}
V.~Bednyakov and S.~Kovalenko, {\it {On possibility for dark matter search with
  accelerated beam of particles}},  {\em Unpublished} (1999).

\bibitem{Feng:2006ni}
T.-F. Feng, X.-Q. Li, W.-G. Ma, J.-X. Wang, and G.-B. Zhao, {\it {Detecting the
  ambient neutralino dark matter particles at accelerator}},  {\em HEPNP} {\bf
  30} (2006) 12, [\href{http://xxx.lanl.gov/abs/hep-ph/0610396}{{\tt
  hep-ph/0610396}}].

\bibitem{Assadi:2014nea}
S.~Assadi, C.~Collins, P.~McIntyre, J.~Gerity, J.~Kellams, et~al., {\it {Higgs
  Factory and 100 TeV Hadron Collider: Opportunity for a New World Laboratory
  within a Decade}},  \href{http://xxx.lanl.gov/abs/1402.5973}{{\tt
  arXiv:1402.5973}}.

\bibitem{Chattopadhyay:2014mha}
S.~Chattopadhyay, {\it {Physics at FAIR}},  {\em Nucl.Phys.} {\bf A931} (2014)
  267--276.

\bibitem{Kekelidze:2013cua}
V.~Kekelidze, A.~Kovalenko, R.~Lednicky, V.~Matveev, I.~Meshkov, et~al., {\it
  {Project NICA at JINR}},  {\em Nucl.Phys.} {\bf A904-905} (2013) 945c--948c.

\bibitem{Thornton:2014ufa}
{\bf MiniBooNE Collaboration} Collaboration, R.~Thornton, {\it
  {Accelerator-Produced Dark Matter Search using MiniBooNE}},
  \href{http://xxx.lanl.gov/abs/1411.4311}{{\tt arXiv:1411.4311}}.

\bibitem{Soper:2014ska}
D.~E. Soper, M.~Spannowsky, C.~J. Wallace, and T.~M.~P. Tait, {\it {Scattering
  of Dark Particles with Light Mediators}},  {\em Phys.Rev.} {\bf D90} (2014),
  no.~11 115005, [\href{http://xxx.lanl.gov/abs/1407.2623}{{\tt
  arXiv:1407.2623}}].

\bibitem{Batell:2009di}
B.~Batell, M.~Pospelov, and A.~Ritz, {\it {Exploring Portals to a Hidden Sector
  Through Fixed Targets}},  {\em Phys.Rev.} {\bf D80} (2009) 095024,
  [\href{http://xxx.lanl.gov/abs/0906.5614}{{\tt arXiv:0906.5614}}].

\bibitem{Essig:2010gu}
R.~Essig, R.~Harnik, J.~Kaplan, and N.~Toro, {\it {Discovering New Light States
  at Neutrino Experiments}},  {\em Phys.Rev.} {\bf D82} (2010) 113008,
  [\href{http://xxx.lanl.gov/abs/1008.0636}{{\tt arXiv:1008.0636}}].

\bibitem{Batell:2009yf}
B.~Batell, M.~Pospelov, and A.~Ritz, {\it {Probing a Secluded U(1) at
  B-factories}},  {\em Phys.Rev.} {\bf D79} (2009) 115008,
  [\href{http://xxx.lanl.gov/abs/0903.0363}{{\tt arXiv:0903.0363}}].

\bibitem{Morrissey:2014yma}
D.~E. Morrissey and A.~P. Spray, {\it {New Limits on Light Hidden Sectors from
  Fixed-Target Experiments}},  {\em JHEP} {\bf 1406} (2014) 083,
  [\href{http://xxx.lanl.gov/abs/1402.4817}{{\tt arXiv:1402.4817}}].

\bibitem{Kahn:2014sra}
Y.~Kahn, G.~Krnjaic, J.~Thaler, and M.~Toups, {\it {DAEdALUS and Dark Matter}},
   \href{http://xxx.lanl.gov/abs/1411.1055}{{\tt arXiv:1411.1055}}.

\bibitem{Wurm:2011zn}
{\bf LENA} Collaboration, M.~Wurm et~al., {\it {The next-generation
  liquid-scintillator neutrino observatory LENA}},  {\em Astropart.Phys.} {\bf
  35} (2012) 685--732, [\href{http://xxx.lanl.gov/abs/1104.5620}{{\tt
  arXiv:1104.5620}}].

\bibitem{Berger:2014sqa}
J.~Berger, Y.~Cui, and Y.~Zhao, {\it {Detecting Boosted Dark Matter from the
  Sun with Large Volume Neutrino Detectors}},
  \href{http://xxx.lanl.gov/abs/1410.2246}{{\tt arXiv:1410.2246}}.

\bibitem{Kong:2014mia}
K.~Kong, G.~Mohlabeng, and J.-C. Park, {\it {Boosted dark matter signals
  uplifted with self-interaction}},  {\em Phys.Lett.} {\bf B743} (2015)
  256--266, [\href{http://xxx.lanl.gov/abs/1411.6632}{{\tt arXiv:1411.6632}}].

\bibitem{Fukuda:2002uc}
{\bf Super-Kamiokande} Collaboration, Y.~Fukuda et~al., {\it {The
  Super-Kamiokande detector}},  {\em Nucl.Instrum.Meth.} {\bf A501} (2003)
  418--462.

\bibitem{Abe:2011ts}
K.~Abe, T.~Abe, H.~Aihara, Y.~Fukuda, Y.~Hayato, et~al., {\it {Letter of
  Intent: The Hyper-Kamiokande Experiment --- Detector Design and Physics
  Potential ---}},  \href{http://xxx.lanl.gov/abs/1109.3262}{{\tt
  arXiv:1109.3262}}.

\bibitem{Badertscher:2010sy}
A.~Badertscher, A.~Curioni, U.~Degunda, L.~Epprecht, S.~Horikawa, et~al., {\it
  {Giant Liquid Argon Observatory for Proton Decay, Neutrino Astrophysics and
  CP-violation in the Lepton Sector (GLACIER)}},
  \href{http://xxx.lanl.gov/abs/1001.0076}{{\tt arXiv:1001.0076}}.

\bibitem{Aartsen:2014oha}
{\bf IceCube PINGU} Collaboration, M.~Aartsen et~al., {\it {Letter of Intent:
  The Precision IceCube Next Generation Upgrade (PINGU)}},
  \href{http://xxx.lanl.gov/abs/1401.2046}{{\tt arXiv:1401.2046}}.

\bibitem{Collaboration:2011nsa}
{\bf ANTARES} Collaboration, M.~Ageron et~al., {\it {ANTARES: the first
  undersea neutrino telescope}},  {\em Nucl.Instrum.Meth.} {\bf A656} (2011)
  11--38, [\href{http://xxx.lanl.gov/abs/1104.1607}{{\tt arXiv:1104.1607}}].

\bibitem{Bueno:2007um}
A.~Bueno, Z.~Dai, Y.~Ge, M.~Laffranchi, A.~Melgarejo, et~al., {\it {Nucleon
  decay searches with large liquid argon TPC detectors at shallow depths:
  Atmospheric neutrinos and cosmogenic backgrounds}},  {\em JHEP} {\bf 0704}
  (2007) 041, [\href{http://xxx.lanl.gov/abs/hep-ph/0701101}{{\tt
  hep-ph/0701101}}].

\bibitem{Soffer:2014ona}
A.~Soffer, {\it {Constraints on dark forces from the B factories and low-energy
  experiments}},  \href{http://xxx.lanl.gov/abs/1409.5263}{{\tt
  arXiv:1409.5263}}.

\bibitem{Tajima:2006nc}
{\bf Belle} Collaboration, O.~Tajima et~al., {\it {Search for invisible decay
  of the Upsilon(1S)}},  {\em Phys.Rev.Lett.} {\bf 98} (2007) 132001,
  [\href{http://xxx.lanl.gov/abs/hep-ex/0611041}{{\tt hep-ex/0611041}}].

\bibitem{Aubert:2009ae}
{\bf BaBar} Collaboration, B.~Aubert et~al., {\it {A Search for Invisible
  Decays of the Upsilon(1S)}},  {\em Phys.Rev.Lett.} {\bf 103} (2009) 251801,
  [\href{http://xxx.lanl.gov/abs/0908.2840}{{\tt arXiv:0908.2840}}].

\bibitem{Ablikim:2007ek}
{\bf BES} Collaboration, M.~Ablikim et~al., {\it {Search for the invisible
  decay of J / psi in psi(2S) --- pi+ pi- J / psi}},  {\em Phys.Rev.Lett.} {\bf
  100} (2008) 192001, [\href{http://xxx.lanl.gov/abs/0710.0039}{{\tt
  arXiv:0710.0039}}].

\bibitem{Chang:1997tq}
L.~Chang, O.~Lebedev, and J.~Ng, {\it {On the invisible decays of the Upsilon
  and J / Psi resonances}},  {\em Phys.Lett.} {\bf B441} (1998) 419--424,
  [\href{http://xxx.lanl.gov/abs/hep-ph/9806487}{{\tt hep-ph/9806487}}].

\bibitem{Savage:2008er}
C.~Savage, G.~Gelmini, P.~Gondolo, and K.~Freese, {\it {Compatibility of
  DAMA/LIBRA dark matter detection with other searches}},  {\em JCAP} {\bf
  0904} (2009) 010, [\href{http://xxx.lanl.gov/abs/0808.3607}{{\tt
  arXiv:0808.3607}}].

\bibitem{Fairbairn:2014aqa}
M.~Fairbairn and J.~Heal, {\it {On the complementarity of Dark Matter Searches
  at Resonance}},  {\em Phys.Rev.} {\bf D90} (2014) 115019,
  [\href{http://xxx.lanl.gov/abs/1406.3288}{{\tt arXiv:1406.3288}}].

\bibitem{Wood:2013taa}
M.~Wood, J.~Buckley, S.~Digel, S.~Funk, D.~Nieto, et~al., {\it {Prospects for
  Indirect Detection of Dark Matter with CTA}},
  \href{http://xxx.lanl.gov/abs/1305.0302}{{\tt arXiv:1305.0302}}.

\bibitem{Kounine:2012ega}
A.~Kounine, {\it {The Alpha Magnetic Spectrometer on the International Space
  Station}},  {\em Int.J.Mod.Phys.} {\bf E21} (2012), no.~08 1230005.

\bibitem{Gomez:2014lva}
M.~Gomez, C.~Jackson, and G.~Shaughnessy, {\it {Dark Matter on Top}},
  \href{http://xxx.lanl.gov/abs/1404.1918}{{\tt arXiv:1404.1918}}.

\bibitem{Haisch:2015ioa}
U.~Haisch and E.~Re, {\it {Simplified dark matter top-quark interactions at the
  LHC}},  \href{http://xxx.lanl.gov/abs/1503.0069}{{\tt arXiv:1503.0069}}.

\bibitem{Dobrich:2015xca}
B.~Dšbrich, {\it {Looking for dark matter on the light side}},
  \href{http://xxx.lanl.gov/abs/1501.0327}{{\tt arXiv:1501.0327}}.

\bibitem{VanTilburg:2015oza}
K.~Van~Tilburg, N.~Leefer, L.~Bougas, and D.~Budker, {\it {Search for
  ultralight scalar dark matter with atomic spectroscopy}},
  \href{http://xxx.lanl.gov/abs/1503.0688}{{\tt arXiv:1503.0688}}.

\bibitem{Arvanitaki:2014faa}
A.~Arvanitaki, J.~Huang, and K.~Van~Tilburg, {\it {Searching for dilaton dark
  matter with atomic clocks}},  {\em Phys.Rev.} {\bf D91} (2015), no.~1 015015,
  [\href{http://xxx.lanl.gov/abs/1405.2925}{{\tt arXiv:1405.2925}}].

\bibitem{Stadnik:2014tta}
Y.~Stadnik and V.~Flambaum, {\it {Searching for dark matter and variation of
  fundamental constants with laser and maser interferometry}},  {\em
  Phys.Rev.Lett.} {\bf 114} (2015), no.~16 161301,
  [\href{http://xxx.lanl.gov/abs/1412.7801}{{\tt arXiv:1412.7801}}].

\bibitem{Khlopov:2014wma}
M.~Khlopov, {\it {Dark atoms and puzzles of dark matter searches}},  {\em
  Int.J.Mod.Phys.} {\bf A29} (2014) 1443002.

\bibitem{Khlopov:2014aia}
M.~Y. Khlopov and R.~Shibaev, {\it {Probes for 4th generation constituents of
  dark atoms in Higgs boson studies at the LHC}},
  \href{http://xxx.lanl.gov/abs/1402.0180}{{\tt arXiv:1402.0180}}.

\bibitem{Belotsky:2014nba}
K.~Belotsky, M.~Khlopov, and M.~Laletin, {\it {Dark Atoms and their decaying
  constituents}},  \href{http://xxx.lanl.gov/abs/1411.3657}{{\tt
  arXiv:1411.3657}}.

\bibitem{Belotsky:2014haa}
K.~Belotsky, M.~Khlopov, C.~Kouvaris, and M.~Laletin, {\it {Decaying Dark Atom
  constituents and cosmic positron excess}},  {\em Adv.High Energy Phys.} {\bf
  2014} (2014) 214258, [\href{http://xxx.lanl.gov/abs/1403.1212}{{\tt
  arXiv:1403.1212}}].

\bibitem{Primulando:2015lfa}
R.~Primulando, E.~Salvioni, and Y.~Tsai, {\it {The Dark Penguin Shines Light at
  Colliders}},  \href{http://xxx.lanl.gov/abs/1503.0420}{{\tt
  arXiv:1503.0420}}.

\bibitem{Ko:2015vaa}
P.~Ko, {\it {Dark matter, dark radiation and Higgs phenomenology in the hidden
  sector DM models}},  \href{http://xxx.lanl.gov/abs/1503.0541}{{\tt
  arXiv:1503.0541}}.

\bibitem{Rajaraman:2015xka}
A.~Rajaraman, J.~Smolinsky, and P.~Tanedo, {\it {On-Shell Mediators and
  Top-Charm Dark Matter Models for the Fermi-LAT Galactic Center Excess}},
  \href{http://xxx.lanl.gov/abs/1503.0591}{{\tt arXiv:1503.0591}}.

\bibitem{Xiang:2015lfa}
Q.-F. Xiang, X.-J. Bi, P.-F. Yin, and Z.-H. Yu, {\it {Searches for dark matter
  signals in simplified models at future hadron colliders}},
  \href{http://xxx.lanl.gov/abs/1503.0293}{{\tt arXiv:1503.0293}}.

\bibitem{Martin-Lozano:2015vva}
V.~Martin-Lozano, M.~Peiro, and P.~Soler, {\it {Isospin violating dark matter
  in St\"uckelberg portal scenarios}},
  \href{http://xxx.lanl.gov/abs/1503.0178}{{\tt arXiv:1503.0178}}.

\bibitem{Rossi-Torres:2015eua}
F.~Rossi-Torres and C.~Moura, {\it {Scalar Dark Matter in the light of LEP and
  ILC Experiments}},  \href{http://xxx.lanl.gov/abs/1503.0647}{{\tt
  arXiv:1503.0647}}.

\bibitem{Fortes:2015qka}
E.~Fortes, V.~Pleitez, and F.~Stecker, {\it {Secluded WIMPs, QED with massive
  photons, and the galactic center gamma-ray excess}},
  \href{http://xxx.lanl.gov/abs/1503.0822}{{\tt arXiv:1503.0822}}.

\bibitem{Suzuki:2015sza}
J.~Suzuki, T.~Horie, Y.~Inoue, and M.~Minowa, {\it {Experimental Search for
  Hidden Photon CDM in the eV mass range with a Dish Antenna}},
  \href{http://xxx.lanl.gov/abs/1504.0011}{{\tt arXiv:1504.0011}}.

\bibitem{Delgado:2015aha}
A.~Delgado, M.~Garcia-Pepin, B.~Ostdiek, and M.~Quiros, {\it {Dark Matter from
  the Supersymmetric Custodial Triplet Model}},
  \href{http://xxx.lanl.gov/abs/1504.0248}{{\tt arXiv:1504.0248}}.

\bibitem{Ghorbani:2015baa}
K.~Ghorbani and H.~Ghorbani, {\it {Two-portal Dark Matter}},
  \href{http://xxx.lanl.gov/abs/1504.0361}{{\tt arXiv:1504.0361}}.

\bibitem{Kainulainen:2015raa}
K.~Kainulainen, K.~Tuominen, and J.~VirkajŠrvi, {\it {A model for dark matter,
  naturalness and a complete gauge unification}},
  \href{http://xxx.lanl.gov/abs/1504.0719}{{\tt arXiv:1504.0719}}.

\bibitem{Cao:2009uw}
Q.-H. Cao, C.-R. Chen, C.~S. Li, and H.~Zhang, {\it {Effective Dark Matter
  Model: Relic density, CDMS II, Fermi LAT and LHC}},  {\em JHEP} {\bf 1108}
  (2011) 018, [\href{http://xxx.lanl.gov/abs/0912.4511}{{\tt
  arXiv:0912.4511}}].

\bibitem{Goodman:2011jq}
J.~Goodman and W.~Shepherd, {\it {LHC Bounds on UV-Complete Models of Dark
  Matter}},  \href{http://xxx.lanl.gov/abs/1111.2359}{{\tt arXiv:1111.2359}}.

\bibitem{Shoemaker:2011vi}
I.~M. Shoemaker and L.~Vecchi, {\it {Unitarity and Monojet Bounds on Models for
  DAMA, CoGeNT, and CRESST-II}},  {\em Phys.Rev.} {\bf D86} (2012) 015023,
  [\href{http://xxx.lanl.gov/abs/1112.5457}{{\tt arXiv:1112.5457}}].

\bibitem{Rajaraman:2011wf}
A.~Rajaraman, W.~Shepherd, T.~M. Tait, and A.~M. Wijangco, {\it {LHC Bounds on
  Interactions of Dark Matter}},  {\em Phys.Rev.} {\bf D84} (2011) 095013,
  [\href{http://xxx.lanl.gov/abs/1108.1196}{{\tt arXiv:1108.1196}}].

\bibitem{Cheung:2012gi}
K.~Cheung, P.-Y. Tseng, Y.-L.~S. Tsai, and T.-C. Yuan, {\it {Global Constraints
  on Effective Dark Matter Interactions: Relic Density, Direct Detection,
  Indirect Detection, and Collider}},  {\em JCAP} {\bf 1205} (2012) 001,
  [\href{http://xxx.lanl.gov/abs/1201.3402}{{\tt arXiv:1201.3402}}].

\bibitem{Cotta:2012nj}
R.~Cotta, J.~Hewett, M.~Le, and T.~Rizzo, {\it {Bounds on Dark Matter
  Interactions with Electroweak Gauge Bosons}},  {\em Phys.Rev.} {\bf D88}
  (2013) 116009, [\href{http://xxx.lanl.gov/abs/1210.0525}{{\tt
  arXiv:1210.0525}}].

\bibitem{Profumo:2013hqa}
S.~Profumo, W.~Shepherd, and T.~Tait, {\it {Pitfalls of dark matter crossing
  symmetries}},  {\em Phys.Rev.} {\bf D88} (2013), no.~5 056018,
  [\href{http://xxx.lanl.gov/abs/1307.6277}{{\tt arXiv:1307.6277}}].

\bibitem{Alves:2013tqa}
A.~Alves, S.~Profumo, and F.~S. Queiroz, {\it {The dark $Z^{'}$ portal: direct,
  indirect and collider searches}},  {\em JHEP} {\bf 1404} (2014) 063,
  [\href{http://xxx.lanl.gov/abs/1312.5281}{{\tt arXiv:1312.5281}}].

\bibitem{Krauss:2013wfa}
M.~B. Krauss, S.~Morisi, W.~Porod, and W.~Winter, {\it {Higher Dimensional
  Effective Operators for Direct Dark Matter Detection}},  {\em JHEP} {\bf
  1402} (2014) 056, [\href{http://xxx.lanl.gov/abs/1312.0009}{{\tt
  arXiv:1312.0009}}].

\bibitem{Buchmueller:2013dya}
O.~Buchmueller, M.~J. Dolan, and C.~McCabe, {\it {Beyond Effective Field Theory
  for Dark Matter Searches at the LHC}},  {\em JHEP} {\bf 1401} (2014) 025,
  [\href{http://xxx.lanl.gov/abs/1308.6799}{{\tt arXiv:1308.6799}}].

\bibitem{Busoni:2014sya}
G.~Busoni, A.~De~Simone, J.~Gramling, E.~Morgante, and A.~Riotto, {\it {On the
  Validity of the Effective Field Theory for Dark Matter Searches at the LHC,
  Part II: Complete Analysis for the $s$-channel}},  {\em JCAP} {\bf 1406}
  (2014) 060, [\href{http://xxx.lanl.gov/abs/1402.1275}{{\tt
  arXiv:1402.1275}}].

\bibitem{Busoni:2014haa}
G.~Busoni, A.~De~Simone, T.~Jacques, E.~Morgante, and A.~Riotto, {\it {On the
  Validity of the Effective Field Theory for Dark Matter Searches at the LHC
  Part III: Analysis for the $t$-channel}},  {\em JCAP} {\bf 1409} (2014) 022,
  [\href{http://xxx.lanl.gov/abs/1405.3101}{{\tt arXiv:1405.3101}}].

\bibitem{D'Eramo:2014aba}
F.~D'Eramo and M.~Procura, {\it {Connecting Dark Matter UV Complete Models to
  Direct Detection Rates via Effective Field Theory}},  {\em JHEP} {\bf 1504}
  (2015) 054, [\href{http://xxx.lanl.gov/abs/1411.3342}{{\tt
  arXiv:1411.3342}}].

\bibitem{Drozd:2015kva}
A.~Drozd, J.~Ellis, J.~Quevillon, and T.~You, {\it {Comparing EFT and Exact
  One-Loop Analyses of Non-Degenerate Stops}},
  \href{http://xxx.lanl.gov/abs/1504.0240}{{\tt arXiv:1504.0240}}.

\bibitem{Dudas:2015vka}
E.~Dudas and D.~Ghilencea, {\it {Effective operators in SUSY, superfield
  constraints and searches for a UV completion}},
  \href{http://xxx.lanl.gov/abs/1503.0831}{{\tt arXiv:1503.0831}}.

\bibitem{Frandsen:2011cg}
M.~T. Frandsen, F.~Kahlhoefer, S.~Sarkar, and K.~Schmidt-Hoberg, {\it {Direct
  detection of dark matter in models with a light Z'}},  {\em JHEP} {\bf 1109}
  (2011) 128, [\href{http://xxx.lanl.gov/abs/1107.2118}{{\tt
  arXiv:1107.2118}}].

\bibitem{Agrawal:2011ze}
P.~Agrawal, S.~Blanchet, Z.~Chacko, and C.~Kilic, {\it {Flavored Dark Matter,
  and Its Implications for Direct Detection and Colliders}},  {\em Phys.Rev.}
  {\bf D86} (2012) 055002, [\href{http://xxx.lanl.gov/abs/1109.3516}{{\tt
  arXiv:1109.3516}}].

\bibitem{An:2012va}
H.~An, X.~Ji, and L.-T. Wang, {\it {Light Dark Matter and $Z'$ Dark Force at
  Colliders}},  {\em JHEP} {\bf 1207} (2012) 182,
  [\href{http://xxx.lanl.gov/abs/1202.2894}{{\tt arXiv:1202.2894}}].

\bibitem{Frandsen:2012rk}
M.~T. Frandsen, F.~Kahlhoefer, A.~Preston, S.~Sarkar, and K.~Schmidt-Hoberg,
  {\it {LHC and Tevatron Bounds on the Dark Matter Direct Detection
  Cross-Section for Vector Mediators}},  {\em JHEP} {\bf 1207} (2012) 123,
  [\href{http://xxx.lanl.gov/abs/1204.3839}{{\tt arXiv:1204.3839}}].

\bibitem{An:2012ue}
H.~An, R.~Huo, and L.-T. Wang, {\it {Searching for Low Mass Dark Portal at the
  LHC}},  {\em Phys.Dark Univ.} {\bf 2} (2013) 50--57,
  [\href{http://xxx.lanl.gov/abs/1212.2221}{{\tt arXiv:1212.2221}}].

\bibitem{Chang:2013oia}
S.~Chang, R.~Edezhath, J.~Hutchinson, and M.~Luty, {\it {Effective WIMPs}},
  {\em Phys.Rev.} {\bf D89} (2014), no.~1 015011,
  [\href{http://xxx.lanl.gov/abs/1307.8120}{{\tt arXiv:1307.8120}}].

\bibitem{An:2013xka}
H.~An, L.-T. Wang, and H.~Zhang, {\it {Dark matter with $t$-channel mediator: a
  simple step beyond contact interaction}},  {\em Phys.Rev.} {\bf D89} (2014),
  no.~11 115014, [\href{http://xxx.lanl.gov/abs/1308.0592}{{\tt
  arXiv:1308.0592}}].

\bibitem{DiFranzo:2013vra}
A.~DiFranzo, K.~I. Nagao, A.~Rajaraman, and T.~M. Tait, {\it {Simplified Models
  for Dark Matter Interacting with Quarks}},  {\em JHEP} {\bf 1311} (2013) 014,
  [\href{http://xxx.lanl.gov/abs/1308.2679}{{\tt arXiv:1308.2679}}].

\bibitem{Morgante:2014kra}
E.~Morgante, {\it {On the Validity of the EFT for Dark Matter Searches at the
  LHC}},  \href{http://xxx.lanl.gov/abs/1409.6668}{{\tt arXiv:1409.6668}}.

\bibitem{Malik:2014ggr}
S.~Malik, C.~McCabe, H.~Araujo, A.~Belyaev, C.~Boehm, et~al., {\it {Interplay
  and Characterization of Dark Matter Searches at Colliders and in Direct
  Detection Experiments}},  \href{http://xxx.lanl.gov/abs/1409.4075}{{\tt
  arXiv:1409.4075}}.

\bibitem{Buchmueller:2014yoa}
O.~Buchmueller, M.~J. Dolan, S.~A. Malik, and C.~McCabe, {\it {Characterising
  dark matter searches at colliders and direct detection experiments: Vector
  mediators}},  \href{http://xxx.lanl.gov/abs/1407.8257}{{\tt
  arXiv:1407.8257}}.

\bibitem{Dolan:2014ska}
M.~J. Dolan, C.~McCabe, F.~Kahlhoefer, and K.~Schmidt-Hoberg, {\it {A taste of
  dark matter: Flavour constraints on pseudoscalar mediators}},
  \href{http://xxx.lanl.gov/abs/1412.5174}{{\tt arXiv:1412.5174}}.

\bibitem{Kong:2013xma}
K.~Kong, {\it {Measuring Properties of Dark Matter at the LHC}},  {\em AIP
  Conf.Proc.} {\bf 1604} (2014) 381--388,
  [\href{http://xxx.lanl.gov/abs/1309.6936}{{\tt arXiv:1309.6936}}].

\bibitem{Ellis:2015daa}
J.~Ellis, {\it {The Physics Landscape after the Higgs Discovery at the LHC}},
  \href{http://xxx.lanl.gov/abs/1504.0365}{{\tt arXiv:1504.0365}}.

\bibitem{Arbey:2015aca}
A.~Arbey, M.~Battaglia, and F.~Mahmoudi, {\it {The Higgs boson, Supersymmetry
  and Dark Matter: Relations and Perspectives}},
  \href{http://xxx.lanl.gov/abs/1504.0509}{{\tt arXiv:1504.0509}}.

\bibitem{Yu:2014mfa}
Z.-H. Yu, X.-J. Bi, Q.-S. Yan, and P.-F. Yin, {\it {Tau Portal Dark Matter
  models at the LHC}},  \href{http://xxx.lanl.gov/abs/1410.3347}{{\tt
  arXiv:1410.3347}}.

\bibitem{Kile:2014jea}
J.~Kile, A.~Kobach, and A.~Soni, {\it {Lepton-Flavored Dark Matter}},
  \href{http://xxx.lanl.gov/abs/1411.1407}{{\tt arXiv:1411.1407}}.

\bibitem{Ma:2014wea}
E.~Ma and A.~Natale, {\it {Dark Matter with Flavor Symmetry and its Collider
  Signature}},  {\em Phys.Lett.} {\bf B740} (2014) 80--82,
  [\href{http://xxx.lanl.gov/abs/1410.2902}{{\tt arXiv:1410.2902}}].

\bibitem{Bramante:2014tba}
J.~Bramante, P.~J. Fox, A.~Martin, B.~Ostdiek, T.~Plehn, et~al., {\it {The
  Relic Neutralino Surface at a 100 TeV collider}},
  \href{http://xxx.lanl.gov/abs/1412.4789}{{\tt arXiv:1412.4789}}.

\bibitem{Calibbi:2014lga}
L.~Calibbi, J.~M. Lindert, T.~Ota, and Y.~Takanishi, {\it {LHC Tests of Light
  Neutralino Dark Matter without Light Sfermions}},  {\em JHEP} {\bf 1411}
  (2014) 106, [\href{http://xxx.lanl.gov/abs/1410.5730}{{\tt
  arXiv:1410.5730}}].

\bibitem{Queiroz:2014pra}
F.~S. Queiroz, K.~Sinha, and A.~Strumia, {\it {Leptoquarks, Dark Matter, and
  Anomalous LHC Events}},  \href{http://xxx.lanl.gov/abs/1409.6301}{{\tt
  arXiv:1409.6301}}.

\bibitem{Busoni:2014gta}
G.~Busoni, A.~De~Simone, T.~Jacques, E.~Morgante, and A.~Riotto, {\it {Making
  the Most of the Relic Density for Dark Matter Searches at the LHC 14 TeV
  Run}},  \href{http://xxx.lanl.gov/abs/1410.7409}{{\tt arXiv:1410.7409}}.

\bibitem{Buckley:2014fba}
M.~R. Buckley, D.~Feld, and D.~Goncalves, {\it {Scalar Simplified Models for
  Dark Matter}},  {\em Phys.Rev.} {\bf D91} (2015), no.~1 015017,
  [\href{http://xxx.lanl.gov/abs/1410.6497}{{\tt arXiv:1410.6497}}].

\bibitem{Harris:2014hga}
P.~Harris, V.~V. Khoze, M.~Spannowsky, and C.~Williams, {\it {Constraining Dark
  Sectors at Colliders: Beyond the Effective Theory Approach}},
  \href{http://xxx.lanl.gov/abs/1411.0535}{{\tt arXiv:1411.0535}}.

\bibitem{Liu:2014rqa}
Y.-B. Liu and Z.-J. Xiao, {\it {Constraining dark matter in the LRTH model with
  latest LHC, XENON100 and LUX date}},
  \href{http://xxx.lanl.gov/abs/1409.8000}{{\tt arXiv:1409.8000}}.

\bibitem{Low:2014cba}
M.~Low and L.-T. Wang, {\it {Neutralino dark matter at 14 TeV and 100 TeV}},
  {\em JHEP} {\bf 1408} (2014) 161,
  [\href{http://xxx.lanl.gov/abs/1404.0682}{{\tt arXiv:1404.0682}}].

\bibitem{Abdallah:2014hon}
J.~Abdallah, A.~Ashkenazi, A.~Boveia, G.~Busoni, A.~De~Simone, et~al., {\it
  {Simplified Models for Dark Matter and Missing Energy Searches at the LHC}},
  \href{http://xxx.lanl.gov/abs/1409.2893}{{\tt arXiv:1409.2893}}.

\bibitem{Han:2014nba}
T.~Han, Z.~Liu, and S.~Su, {\it {Light Neutralino Dark Matter: Direct/Indirect
  Detection and Collider Searches}},  {\em JHEP} {\bf 1408} (2014) 093,
  [\href{http://xxx.lanl.gov/abs/1406.1181}{{\tt arXiv:1406.1181}}].

\bibitem{Bhattacharya:2014yha}
A.~Bhattacharya, R.~Gandhi, and A.~Gupta, {\it {The Direct Detection of Boosted
  Dark Matter at High Energies and PeV events at IceCube}},  {\em
  Unpublished,~} (2014) [\href{http://xxx.lanl.gov/abs/1407.3280}{{\tt
  arXiv:1407.3280}}].

\bibitem{Blumenthal:2014cwa}
J.~Blumenthal, P.~Gretskov, M.~KrŠmer, and C.~Wiebusch, {\it {Effective field
  theory interpretation of searches for dark matter annihilation in the Sun
  with the IceCube Neutrino Observatory}},
  \href{http://xxx.lanl.gov/abs/1411.5917}{{\tt arXiv:1411.5917}}.

\bibitem{Li:2014vza}
T.~Li, S.~Miao, and Y.-F. Zhou, {\it {Light mediators in dark matter direct
  detections}},  \href{http://xxx.lanl.gov/abs/1412.6220}{{\tt
  arXiv:1412.6220}}.

\bibitem{Agashe:2014yua}
K.~Agashe, Y.~Cui, L.~Necib, and J.~Thaler, {\it {(In)direct Detection of
  Boosted Dark Matter}},  {\em JCAP} {\bf 1410} (2014), no.~10 062,
  [\href{http://xxx.lanl.gov/abs/1405.7370}{{\tt arXiv:1405.7370}}].

\bibitem{Bell:2015sza}
N.~F. Bell, Y.~Cai, J.~B. Dent, R.~K. Leane, and T.~J. Weiler, {\it {Dark
  matter at the LHC: EFTs and gauge invariance}},
  \href{http://xxx.lanl.gov/abs/1503.0787}{{\tt arXiv:1503.0787}}.

\bibitem{Ibarra:2015fqa}
A.~Ibarra and S.~Wild, {\it {Dirac dark matter with a charged mediator: a
  comprehensive one-loop analysis of the direct detection phenomenology}},
  \href{http://xxx.lanl.gov/abs/1503.0338}{{\tt arXiv:1503.0338}}.

\bibitem{Garny:2015wea}
M.~Garny, A.~Ibarra, and S.~Vogl, {\it {Signatures of Majorana dark matter with
  t-channel mediators}},  \href{http://xxx.lanl.gov/abs/1503.0150}{{\tt
  arXiv:1503.0150}}.

\bibitem{Crivellin:2015wva}
A.~Crivellin, U.~Haisch, and A.~Hibbs, {\it {LHC constraints on gauge boson
  couplings to dark matter}},  \href{http://xxx.lanl.gov/abs/1501.0090}{{\tt
  arXiv:1501.0090}}.

\bibitem{Anderson:2015xaa}
A.~J. Anderson, P.~J. Fox, Y.~Kahn, and M.~McCullough, {\it {Halo-Independent
  Direct Detection Analyses Without Mass Assumptions}},
  \href{http://xxx.lanl.gov/abs/1504.0333}{{\tt arXiv:1504.0333}}.

\end{thebibliography}
\providecommand{\href}[2]{#2}\begingroup\raggedright\endgroup
  
\end{document}